\documentclass[
11pt, 
english, 
singlespacing, 
headsepline, 
]{MastersDoctoralThesis} 

\usepackage[utf8]{inputenc} 
\usepackage[T1]{fontenc} 

\usepackage{lmodern} 
\usepackage[backend=bibtex,style=numeric-comp,sorting=none]{biblatex} 

\usepackage{graphicx}
\usepackage{dcolumn}
\usepackage{bm}
\usepackage{subcaption}
\usepackage{hyperref}
\usepackage{mathtools}
\usepackage{eufrak}
\usepackage{amssymb}
\usepackage{amsmath}
\usepackage{verbatim}
\usepackage{enumitem}
\usepackage{cprotect}
\usepackage{fancyvrb}

\usepackage{slashed}
\def\RE{{\rm Re}} 
\def\IM{{\rm Im}}
\def\Tr{\mathop{\rm Tr}\nolimits}
\def\vecsigma{\boldsymbol{\sigma}}

\def\pv{{\bf{p}}}

\def\kv{{\bf{k}}}

\def\xu{\hat{\bf{x}}}
\def\yu{\hat{\bf{y}}}
\def\zu{\hat{\bf{z}}}
\def\ntil{\tilde{\bf{n}}}
\def\nunit{\hat{\bf{n}}}

\def\ie{{\sl i.e.}}

\def\kt{     { \bf{k}_{\rm T}  }    }

\def\kpt{  \kv'_{\rm T} }
\def\ktkt{\kv^2_{\rm T}}
\def\kptkpt{{\kv'}^2_{\rm T}}

\def\pt{     { \bf{p}_{\rm T}  }    }
\def\ptpt{\pv^2_{\rm T}}

\def\kperp{     { \bf{k}_{\perp}  }    }
\def\kperpkperp{     { \bf{k}^2_{\perp}  }    }

\def\be {b_{\rm{\epsilon}}}
\def\T{_{\rm{T}}}

\def\A{_{\rm{A}}}
\def\B{_{\rm{B}}}
\def\bt{b_{\rm{T}}}
\def\bl{b_{\rm{L}}}

\def\Mx{\rm{\textbf{m}}}
\def\Nx{\rm{\textbf{n}}}
\def\Lx{\rm{\textbf{l}}}

\def\q{\mathfrak{q}}

\def\ktil{\tilde{\textbf{k}}\T}
\def\kptil{\tilde{\textbf{k}}'\T}

\def\h{\mathfrak{h}}
\def\A{_{\rm A}}
\def\B{_{\rm B}}

\def\Sq{\textbf{S}_q}
\def\Sqp{\textbf{S}'_q}
\def\STq{\textbf{S}_{\rm{T}q}}

\def\Deltaqp{\Delta_{q'}(\kpt)}
\def\Deltaqpdag{\Delta^{\dagger}_{q'}(\kpt)}
\def\Gammah{\Gamma_h}
\def\Gammahdag{\Gamma^{\dagger}_h}

\def\trace{\rm{tr}}

\usepackage[autostyle=true]{csquotes} 

\thesistitle{Recursive fragmentation  of a \\ polarized quark} 
\supervisor{Prof. Anna \textsc{Martin} \\ Prof. Xavier \textsc{Artru}} 
\coordinator{Prof. Francesco \textsc{Longo}} 
\degree{Doctor of Philosophy} 
\author{Albi \textsc{Kerbizi}} 
\addresses{} 

\subject{Physics} 
\university{{Universit\`{a} degli Studi di Trieste}} 
\department{\href{https://www.units.it}{}} 
\group{\href{https://www.units.it}{$\,$}} 
\faculty{\href{https://www.units.it}{}} 

\AtBeginDocument{
\hypersetup{pdftitle=\ttitle} 
\hypersetup{pdfauthor=\authorname} 
\hypersetup{pdfkeywords=\keywordnames} 
}

\begin{document}

\DefineShortVerb{\#}
\SaveVerb{verbPythia}#PYTHIA#

\frontmatter 

\pagestyle{plain} 

\begin{titlepage}
\begin{center}

\vspace*{.0\textheight}
{\includegraphics[width=4cm]{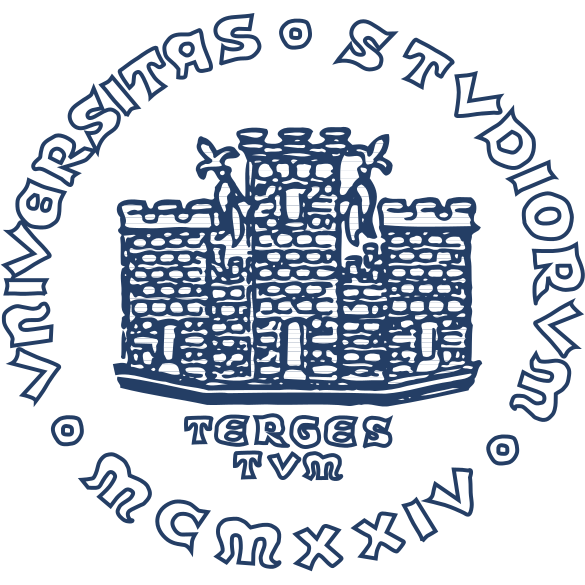}\protect\\}
\vspace{1.5cm}
{\scshape\LARGE \univname\par}\vspace{1.0cm}
\textsc{\Large XXXII Ciclo del dottorato di ricerca in}\\[1.5cm]
\textsc{\Large Fisica}\\[1.0cm]
{\huge \bfseries \ttitle\par}\vspace{1.0cm}
{Settore scientifico-disciplinare: \textbf{FIS/04}}
\vspace{2.5cm}

\begin{minipage}[t]{0.4\textwidth}
\begin{flushleft} \large
\large Dottorando\\
\textbf{\authorname}\\
\vspace{0.2cm}
\large Coordinatore\\
\textbf{\coordname}\\
\vspace{0.2cm}
\large Supervisore di tesi \\
\textbf{\supname} 
\end{flushleft}
\end{minipage}\\[2cm]
 
\vfill
\begin{centering}
{ANNO ACCADEMICO 2018/2019}
\end{centering}
\vfill
\vfill
\end{center}
\end{titlepage}

\begin{titlepage}
\begin{center}

\vspace*{.0\textheight}
{\includegraphics[width=4cm]{logo.png}\protect\\}
\vspace{1.5cm}
{\scshape\LARGE \univname\par}\vspace{1.0cm}
\textsc{\Large XXXII Ciclo del dottorato di ricerca in}\\[1.5cm]
\textsc{\Large Fisica}\\[1.0cm]
{\huge \bfseries \ttitle\par}\vspace{1.0cm}
{Settore scientifico-disciplinare: \textbf{FIS/04}}
\vspace{2.5cm}

\begin{minipage}[t]{0.4\textwidth}
\begin{flushleft} \large
\large PhD Student\\
\textbf{\authorname}\\
\vspace{0.2cm}
\large Coordinator\\
\textbf{\coordname}\\
\vspace{0.2cm}
\large Supervisor \\
\textbf{\supname} 
\end{flushleft}
\end{minipage}\\[2cm]
 
\vfill
\begin{centering}
{ACADEMIC YEAR 2018/2019}
\end{centering}
\vfill
\vfill
\end{center}
\end{titlepage}

\chapter*{Abstract}
\addchaptertocentry{\abstractname} 

This thesis is devoted to the study of the fragmentation process of polarized quarks, which cannot be described in the framework of perturbative Quantum Chromodynamics (pQCD) and is a very interesting topic per se. The work is mainly motivated by the study of the Collins effect which was proposed and is used as a polarimeter to measure the transverse polarization of quarks. The most important application is the extraction of the quark transversity distribution, the third collinear parton distribution function necessary for the description of the nucleon at leading order. Its extraction requires the measurement of the polarization of the quarks in a transversely polarized nucleon. The present phenomenological extractions of the transversity distribution make use of the Collins asymmetry measured in Semi-Inclusive Deep Inelastic Scattering (SIDIS) of leptons on transversely polarized nucleons and of the corresponding asymmetries in $e^+e^-$ annihilation into hadrons, which both give access to the Collins function.

A useful tool for the extraction of the quark transversity distribution and the study of its properties will be a Monte Carlo simulation of the Collins effect. This will be also the case for the extraction of other still unknown chiral odd parton distribution functions like the Boer-Mulders function, coupled in SIDIS to the Collins function.

Furthermore, an event generator including the quark spin degree of freedom in the fragmentation process will be helpful for the interpretation and for the analysis of the experimental data as well as for the design of future experiments. Up to now such an event generator based on a consistent model for this process is not available. The aim of this work is to fill this shortcoming by
including the spin degree of freedom in the hadronization part of Monte Carlo event generators. The main emphasis here lies on the fragmentation of quarks with transverse polarization but longitudinal spin effects like jet-handedness are included as well.

The polarized quark fragmentation model, as formulated by Artru and Belghobsi between 2009 and 2013 and reviewed here, is partly
based on the Lund symmetric string fragmentation model, used in the event
generator \verb|PYTHIA|, which describes the hadronization process as the cascade
decay of a relativistic string, but does not include spin in the quark degrees of
feedom. In the polarized quark fragmentation model, the quark spin is introduced and treated as a full quantum
variable, using Pauli spinors, $2\times2$ density matrices
and transition matrix amplitudes. These amplitudes are inspired by the ${}^3P_0$
model of quark pair production which assumes that the quark-antiquark pairs
produced at the string breakings are in the ${}^3P_0$ state. 
%

The properties and the degrees of freedom of the string+${}^P_0$ model are spelled out in a systematic way. The various relevant functions are written in a form which is suitable for the implementation in Monte Carlo event generators. An important development made here is the inclusion of vector mesons and their decays in the fragmentation chain, propagating the quark spin information according to the rules of Quantum Mechanics by applying existing recipes. The resulting models are translated in this work in recursive Monte Carlo codes of polarized quark fragmentation and the results of the simulations are compared with existing data.

This work is not meant to be a complete study of the spin effects, still it can be regarded as several important initial steps in that direction.

Chapter 1 begins with the introduction to the nucleon structure. A purpose of this work is indeed its application to the study of the partonic transverse spin and transverse momentum structure of the nucleon.

The theoretical formulation of the string+${}^3P_0$ model restricted to pseudoscalar meson emission is reviewed in the first part of Chapter 2. Then the two main variants of the model used in this work are presented in detail. The first variant (M18, published in 2018) includes the so called spin independent correlations between the transverse momenta of successive quarks created at string breakings. The other variant (M19, published in 2019) is free from such correlations and has the advantage of being simpler from the analytical and numerical points of view.%

Both versions are written in a form suitable for numerical simulations and have been implemented in stand alone Monte Carlo programs as described in Chapter 3. 
In the same chapter results of the simulations with the variant M18 are compared with the experimental data.
After tuning the only parameter relevant for the spin effects, both the measured Collins and the dihadron transverse spin asymmetries are well described by this model. In the last part of the chapter the simulation results with the variant M19 are also presented. It turns out to give practically the same results for the transverse spin asymmetries as the previous variant and is more suitable for further extensions and for the implementation in event generators where the hadronization part is based on the Lund Model.
These important results motivated further very recent work, namely the inclusion of vector meson production in the polarized quark fragmentation process (Chapter 5).

In Chapter 4 the interface of M19 with \verb|PYTHIA 8.2|, an essential task to understand the combined effect of the polarized fragmentation process with other physics sub-processes, is presented. The code which has been developed for this purpose and the implementation of the transversity distribution in SIDIS are described in the same chapter. The comparison of the resulting asymmetries from simulations of the polarized SIDIS process with experimental data is also discussed.

Chapter 5 is dedicated to the model M20, in which the vector meson production and their decays are included. The spin density matrix of the vector meson is calculated and utilized for the generation of the angular distribution of the decay hadrons. The quantum mechanical correlations between the relative orientation of the decay products and the next quark in the recursive process, generated at the adjacent string breaking, are taken into account. The effects of vector meson production on observables like the Collins and the dihadron asymmetries, as obtained by stand alone Monte Carlo simulations, are presented here in detail. The string+${}^3P_0$ model in the M20 version is shown to be a promising and powerful model to describe the polarized quark fragmentation.

A summary and an overview of possible future applications and studies which can be performed with this model are given in the last part of the thesis.


%
%

\tableofcontents 

\dedicatory{P\"er gjysh Leksin edhe n\"ena Ndin\"en} 


\mainmatter 

\pagestyle{thesis} 


\chapter{The nucleon structure anihadron its investigation} 

\label{Chapter0}

\section{Introduction}
The partonic structure of protons and neutrons is described at the leading order in the present theory of strong interactions, quantum chromodynamics (QCD), by three parton distribution functions (PDFs): the unpolarized number density $f_1$, the helicity distribution $g_1$ and the transversity distribution $h_1$.
The unpolarized PDF $f_1^q$ is the number density of partons with flavor $q$ that carry a fraction $x$ of the nucleons longitudinal momentum. For an ultra relativistic nucleon, $g_1^q$ and $h_1^q$ give the correlation between the polarization of the partons and that of the parent nucleon. In a longitudinally polarized nucleon $g_1^q$ is the difference between the distributions of partons with spin parallel and antiparallel to the nucleon one. In a transversely polarized nucleon $h_1^q$ is the difference between the distributions of partons with transverse spin parallel and antiparallel to the nucleon one. Transverse direction means orthogonal with respect to the direction of motion of the nucleon.


Among these three leading order quantities, $f_1$ is the best known. It was introduced by Feynman to explain the results of the deep inelastic scattering (DIS) experiments carried out at SLAC in the late 60's. Inelastic scattering at large transverse momentum could be interpreted as elastic scattering off charged and point-like nucleon constituents, named partons by Feynman, and later identified with the quarks and gluons. These measurements led to the important result that quarks account only for half of the nucleon momentum, in contradiction with the non relativistic quark model which described the nucleon structure only in terms of three valence quarks. It brought to the discovery of the gluon contribution to the nucleon momentum, which accounted for the other half, satisfying in this way the momentum sum rule. Precise measurements of these functions both for quarks and for gluons came later with the HERA experiment at DESY and from the neutrino and muon experiments at FNAL and at CERN.

The knowledge of the nucleon spin structure, instead, is poorer than the present knowledge of the momentum structure. Sophisticated techniques were developed at SLAC in the 70's to polarize the electron beam, and pioneering experiments (E80 and E130) were conducted and provided first measurements for the helicity PDF $g_1$ of the proton. A breakthrough occurred however in the late 80's, when the EMC collaboration carried on at CERN DIS experiments scattering high energy polarized muons off a longitudinally polarized proton target. These measurements led to the unexpected finding that the quark contribution to the nucleon spin might be small or even compatible with zero within the large experimental uncertainty \cite{EMC1988}.
This was again in contradiction with the non relativistic quark model which explained the nucleon helicity as the sum of the three valence quark helicities. These results were confirmed later by the SMC and the COMPASS experiments at CERN, by the HERMES experiment at DESY, by the E143 and the E155 experiments at SLAC, and by the CLAS experiment at Jefferson Laboratory\footnote{A recent measurement for the quark contribution to the proton spin gives a value between $0.26$ and $0.36$ \cite{compass-longitudinal}.}. The same measurements were done with comparable accuracy with polarized neutron targets and the result was the same. More details can be found in recent reviews, see e.g. Ref. \cite{Aidala:2012mv}. The EMC finding in 1988 was the beginning of the so called \textit{nucleon spin crisis}. Namely it was impossible to obtain the proton helicity by adding up the contributions of its constituents. It was impossible to fill the gap even adding up the gluon helicity, which also was measured to be small. 

The spin crisis brought to a vivid interest in the community of high energy physicists for the nucleon spin structure. Its QCD description was revisited and extended also to the transverse spin.

Transverse spin effects were reported in the mid 70s when large single spin asymmetries (SSAs) were measured for pions produced in $\pi + p^{\uparrow}\rightarrow \pi + X$ collisions at the Proton Synchrotron at CERN \cite{SSA-PS-CERN} and in $p+p^{\uparrow} \rightarrow \pi+X$ collisions at ZGS in Argonne \cite{SSA-ARGONNE-1,SSA-ARGONNE-2}. In both experiments the target protons were transversely polarized. Transverse spin effects were also observed in the same years in $p+Be\rightarrow \Lambda+X$ collisions at Fermilab, where $\Lambda$s were measured with large transverse polarizations \cite{Lambda-Pol-Fermilab}. A decade later, the E704 experiment at FNAL, carrying out collisions of transversely polarized protons off an unpolarized liquid hydrogen target, observed large SSAs for pions up to $40\%$ in the forward region \cite{SSA-E704}, definitely demonstrating that transverse spin effects survive at high energies at variance with the theoretical prejudice that transverse spin effects had to vanish in hard reactions \cite{kane1978}. Later on, large transverse spin effects in $pp$ scattering were observed at RHIC (BNL) at center of mass energies as large as $500\,\rm{GeV}$.

As explanation for the observed SSAs, Sivers \cite{Sivers1990} suggested that the origin might be in the quark motion inside the nucleon and in particular in a spin-orbit correlation between the intrinsic transverse momentum of quarks and the nucleon transverse polarization. An alternative explanation was based on a non-vanishing transversity PDF.
The distributions of transversely polarized quarks in a transversely polarized nucleon were mentioned by Feynman in 1973 \cite{Feynman-photon-hadron}. The transversity PDF was introduced in 1979 by Ralston and Soper \cite{ralston-soper1979}.
It was rediscovered by Artru and Mekhfi in 1990 \cite{Artru-Mekfhi-transversity}.

The transversity distribution is interestingly different from $g_1$. They would be the same only for nonrelativistc quarks in the nucleon. The difference between $h_1$ and $g_1$ is an indication of the richness of the relativistic nucleon wave function. In the helicity basis, $h_1$ is related to a helicity-flip and therefore represents a probe of the chiral symmetry breaking. It is decoupled from gluons, and this makes its evolution with energy different from that of helicity \cite{Artru-Mekfhi-transversity}.
Still, the transversity PDF is much less known than the helicity distribution. It is not accessible in inclusive DIS due to its chiral odd nature and to build an observable a coupling with another chiral odd object is required.
For this purpose, different processes have been proposed.

The most important is the semi-inclusive deep inelastic scattering (SIDIS) process off a transversely polarized nucleon, namely the process $l+N^{\uparrow}\rightarrow l'+h+X$ where a high energy lepton $l$ scatters off a transversely polarized target nucleon $N$ and the scattered lepton $l'$ is measured in coincidence with at least one hadron $h$. The initial ideas were to look at the production of $\Lambda$s \cite{Baldracchini,Artru-Mekfhi-transversity-measurement}. Their transverse polarizations depend on the transverse polarization of the struck quark and on the chiral-odd function describing how this quark fragments into the observed $\Lambda$. The $\Lambda$ polarization can be measured looking at the decay angular distributions in the weak process $\Lambda^{\uparrow}\rightarrow p+\pi$ with reference axis defined by the polarization of the struck quark. However, this method has a poor efficiency due to the low abundance of $\Lambda$s in quark jets. Also, the polarization transfer is presently unknown.

A different polarimeter for the measurement of the quark transversity distribution was proposed in 1993 by Collins \cite{collins}, who conjectured that a transversely polarized quark produces a jet where the emitted mesons have an asymmetrical distribution in the azimuthal angle, described by the so called \textit{Collins fragmentation function} (Collins FF). It is a chiral odd fragmentaton function, and in the SIDIS process it would be coupled to $h_1$ producing the so called \textit{Collins asymmetry}, namely an azimuthal asymmetry in the distribution of the final state hadrons. The Collins asymmetry could thus give access to the transversity distribution, provided that the Collins FF is different from zero.

In the same period, it was realized that there could be an asymmetry in the azimuthal angle of the relative momentum of a hadron pair produced in the fragmentation of a transversely polarized quark \cite{Collins:1993kq}. The asymmetry in this case is described by the chiral-odd \textit{dihadron fragmentation function} (DiFF), or \textit{interference fragmentation function} (IFF). In the SIDIS process it is coupled to $h_1$ producing a \textit{dihadron asymmetry} which was regarded as an alternative way of accessing $h_1$. This requires the knowledge of the involved IFF, which was completely unknown, like the Collins FF.

For the measurement of the Collins and interference FFs, the $e^+e^-$ annihilation process into hadrons was proposed. Here the intermediate $q\bar{q}$ system is characterized by correlated transverse polarizations and allows to probe both the Collins and the dihadron FFs. By combining measurements from SIDIS and $e^+e^-$ annihilation, it becomes possible to extract $h_1$. 

These FFs, as well as the unpolarized FFs, are fascinating non-perturbative objects hard to be calculated and so far not computed in lattice QCD. A specific model for the fragmentation of polarized quarks and the corresponding Monte Carlo implementation are the subject of this work.


Among the other processes, the Drell-Yan process $p^{\uparrow}\bar{p}^{\uparrow}\rightarrow l^+l^- + X$ is particularly interesting. Since both the proton and the anti-proton are transversely polarized, a direct access to the product of the underlying quark and anti-quark (in $\bar{p}$) transversity PDFs is provided. These measurements were proposed at GSI \cite{Abazov:2004ih, Barone:2005pu} but not realized because of the difficulties in polarizing antiproton beams.

The first data showing the existence of transverse SSAs in SIDIS off a transversely polarized nucleon target were produced in 2004 for a proton by the HERMES Collaboration \cite{HERMES-2004} and for a deuteron by the COMPASS Collaboration \cite{COMPASS-2005,COMPASS-2006}.
In order to disentangle the Collins FF and the transversity function, a step forward occured in 2005 when the BELLE Collaboration measured transverse spin asymmetries in $e^+e^-$ annihilation experiments \cite{BELLE-2005}.

All these achievements on the transverse spin effects marked the beginning of the wide field of the study of transverse spin and transverse momentum structure of hadrons which is presently very active. The study of the nucleon structure is ongoing at other experimental facilities currently in operation, like RHIC for polarized proton proton scattering \cite{rhic}, and Jefferson Laboratory \cite{jlab-ssa} and COMPASS for SIDIS. Drell-Yan experiments with a pion beam off a transversely polarized proton target have been performed in COMPASS \cite{compass-drell-yan}. The BES \cite{besIII} and BABAR \cite{babar} Collaborations carry complementary measurements performing $e^+e^-$ annihilation experiments. In the future other experiments at FNAL, EIC and LHC, dedicated to the study of the nucleon spin structure, are foreseen.

In this Chapter, after a reminder of the present description of the nucleon structure, the observables used to access it and which involve the fragmentation functions, are introduced. They will be used to compare the Monte Carlo simulations with the experimental data.

\section{The three dimensional structure of the nucleon}\label{sec:nucleon structure}
\begin{figure}
\centering
\begin{minipage}{.6\textwidth}
  \centering
  \includegraphics[width=.7\linewidth]{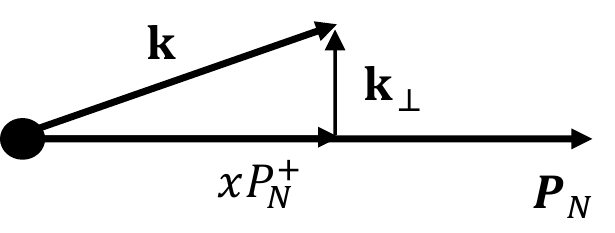}
\end{minipage}%
\caption{Representation of the quark momentum in the parent nucleon.}
\label{fig:partons}
\end{figure}

Quarks (and gluons) are in general not collinear with respect to the parent nucleon direction. In the so called \textit{infinite momentum frame} \cite{Weinberg-infinite-momentum}, namely in the reference frame where the nucleon travels along some direction with infinite momentum, they carry a fraction $x$ of the nucleon momentum and have also transverse motion with respect to the nucleon momentum. The transverse motion is characterized by the \textit{intrinsic} (or \textit{primordial}) transverse momentum that will be indicated with $\kperp$. This is schematically represented in Fig. \ref{fig:partons}.


In this generalized parton model, which includes the parton transverse motion, the three leading order collinear partonic distributions $f_1$, $g_1$ and $h_1$ (which previously were $\kperp$-integrated) are generalized as \textit{transverse momentum dependent} (TMD) parton distributions in $x$ and $\kperp$.
The three-dimensional structure of the nucleon is no more exhausted by $f_1$, $g_1$ and $h_1$. In fact, $\kperp$ represents a natural vector to correlate either with the transverse polarization of the nucleon or with the quark transverse polarization \cite{Barone-et-al-2010}.

The complete description of the (spin $1/2$) nucleon at leading twist, namely at the leading order in the hard scale which characterizes the interaction between the probe and the target, requires eight TMD PDFs \cite{amsterdam-notation,Boer-Mulders-func}. They are listed in Tab. \ref{tab:tmd} according to the quark and to the nucleon polarizations which can be unpolarized ($U$), longitudinally polarized ($L$) or transversely polarized ($T$).

The functions $f_{1T}^{\perp}$, $g_{1T}^{\perp}$, $h_{1L}^{\perp}$, $h_1^{\perp}$ and $h_{1T}^{q\perp}$~\footnote{For these TMD PDFs the "Amsterdam notation" \cite{amsterdam-notation} is used.} are the other five TMD PDFs. The letters $f$, $g$ and $h$ are used to indicate unpolarized, longitudinally polarized and transversely polarized quarks. The subscripts $L$ and $T$ indicate the nucleon longitudinal and transverse polarization respectively. The superscript $\perp$ indicates that the TMD does not survive upon integration over $\kperp$.
The PDFs along the diagonal in Tab. \ref{tab:tmd}, namely $f_1^q$, $g_1^q$ and $h_1^q$, are the only TMDs to survive integration over $\kperp$, reducing to their collinear counterparts.



\begin{table}[!h]
\centering
\begin{tabular}{|l|l|l|l|}
\hline
quark/nucleon &  U & L & T \\
\hline
U & $f_1$ & $\,$ &  $f_{1T}^{\perp}$ \\
\hline
L & $\,$ & $g_{1}$ & $g_{1T}^{\perp}$ \\
\hline
T & $h_1^{\perp}$ & $h_{1L}^{\perp}$ & $h_1,\,h_{1T}^{\perp}$\\
\hline
\end{tabular}
\caption{Leading order TMD parton distribution functions in terms of the quark and of the nucleon polarizations: $U$ (unpolarized), $L$ (longitudinally polarized) and $T$ (transversely polarized).}\label{tab:tmd}
\end{table}

The function $f_{1\rm{T}}^{\perp}(x,\kperpkperp)$ is the time reversal odd (T-odd) \textit{Sivers function} \cite{Sivers1990}. It describes the distribution of unpolarized quarks in a transversely polarized nucleon by correlating the quark transverse momentum with the nucleon transverse polarization. The function $h_{1}^{\perp}(x,\kperpkperp)$ is an other T-odd TMD named \textit{Boer-Mulders function} \cite{Boer-Mulders-func}. It correlates the quark transverse polarization with its transverse momentum and it is chiral odd as the transversity PDF. The two remaining functions $g_{1\rm{T}}^{\perp}(x,\kperpkperp)$ and $h_{1L}^{\perp}(x,\kperpkperp)$ are the T-even \textit{worm-gear} functions. $g_{1\rm{T}}^{\perp}$ describes the distribution of longitudinally polarized quarks in a transversely polarized nucleon. $h_{1L}^{q\perp}$ describes the distribution of transversely polarized quarks in a longitudinally polarized nucleon and it is chiral odd.
The function $h_{1T}^{\perp}$, usually called \textit{pretzelosity}, describes transversely polarized quarks in a transversely polarized nucleon \cite{pretzelosity}. It correlates the nucleons transverse polarization with the quark transverse momentum and transverse polarization, and measures deviations from a cylindrical shape of the polarized quark distribution.

Summarizing, the nucleon description at leading order needs several unknown transverse momentum dependent distribution functions describing the different degrees of freedom of its confined constituents. These functions are non-perturbative objects, cannot be calculated analytically from first principles and are being calculated in lattice QCD. Different tools have been developed to access them experimentally and different processes are currently used for their measurement. 
 
Concerning the quark transversity distribution $h_1^q$, an asymmetry, given by the convolution of $h_1^q$ and of the Collins FF, on the azimuthal distribution of hadrons produced in transversely polarized SIDIS processes is currently used for its extraction. This requires that the Collins FF is known.
Information on this function come from the measurements of the asymmetries in the $e^+e^-$ annihilation process, which involve the convolution between the Collins FF of the quark and of the antiquark.
The extraction of the quark transversity distribution and the interpretation of the experimental data on polarized scattering processes can be greatly helped if one disposes of a solid model for the Collins effect and implement it on well established Monte Carlo event generators like \verb|PYTHIA| \cite{pythia8}. It is such a tool which is developed in this thesis.

 \section{The fragmentation functions and the Collins effect}

\textit{Fragmentation} (or \textit{hadronization}) is the nonperturbative process that brings quarks and gluons to dress into observable hadrons. For an exhaustive review see for instance Ref. \cite{models}. It is usually parametrized by \textit{fragmentation functions} (FFs), which describe the probability that a hadron $h$ is produced in the fragmentation process of a quark $q$, taking away a fraction of the quark momentum. As will be described in the next chapter, the hadrons are produced in the colour field between two ends, e.g. a quark and an anti-quark or a quark and a diquark, which constitute the fragmenting system. FFs can thus be defined in a frame where the two ends are back-to-back and sufficiently well separated. The fragmentation function of an unpolarized quark $q$ into the unpolarized hadron $h$ is usually indicated with $D_{1q}^h(z,\ptpt)$. Indicating with $p$ and $k$ the hadron and the quark momenta, the FFs depend on the longitudinal lightcone momentum fraction $z=p^+/k^+$ of $q$ taken by $h$, and on the transverse momentum $\pt$ of $h$ with respect to the direction of the quark momentum $\hat{\textbf{k}}$ \footnote{The lightcone variables are defined in Appendix \ref{Appendix:lightcone}.}. Fragmentation functions depend also on the so called renormalization and the Collins-Soper scales \cite{Collins:1981uk,Collins:1981va} which characterize the evolution of FFs with the hard scale of the considered scattering process. In the following, scale dependencies are neglected since the present work focuses on the non perturbative.

Fragmentation functions can be \textit{favoured}, if $q$ fragments into a hadron which has $q$ as constituent quark, like
\begin{eqnarray}
u\rightarrow \pi^++X, & s\rightarrow K^- + X, &  d\rightarrow \pi^- + X.
\end{eqnarray}
Alternatively, if $q$ is not a constituent of $h$, the FFs are called \textit{unfavored} as for instance
\begin{eqnarray}
u\rightarrow \pi^- + X, & s\rightarrow K^+ +X, & d\rightarrow \pi^+ + X.
\end{eqnarray}

\begin{figure}
\centering
\begin{minipage}{.5\textwidth}
  \centering
  \includegraphics[width=.8\linewidth]{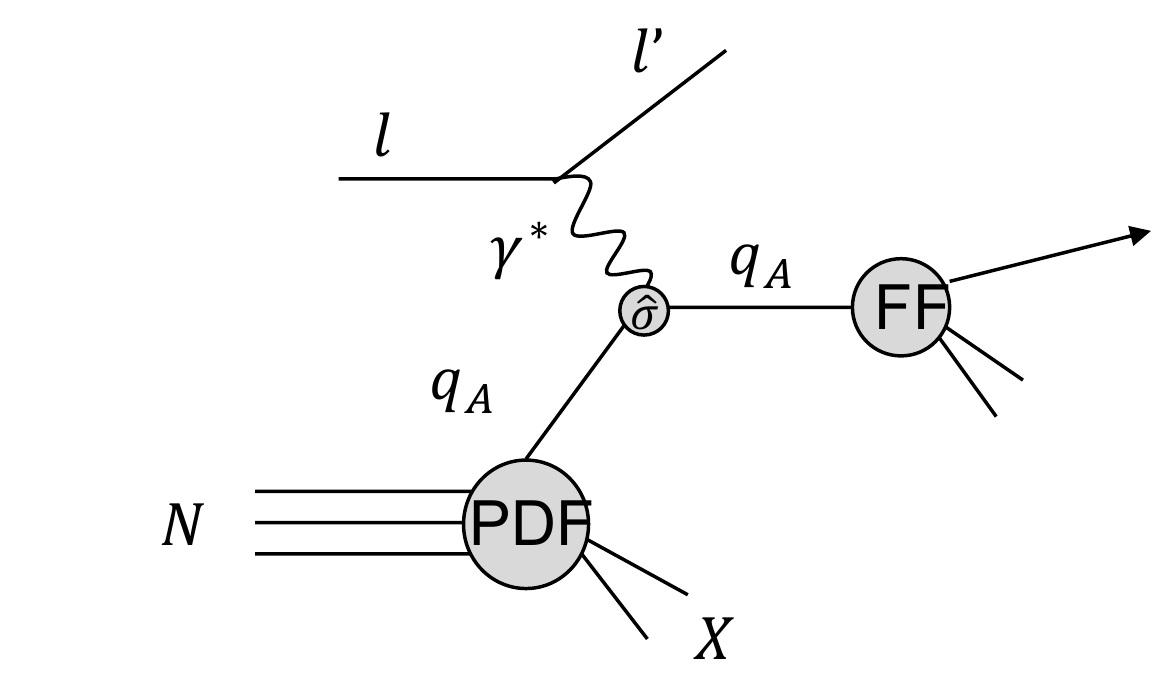}
\end{minipage}%
\begin{minipage}{.5\textwidth}
  \centering
  \includegraphics[width=.8\linewidth]{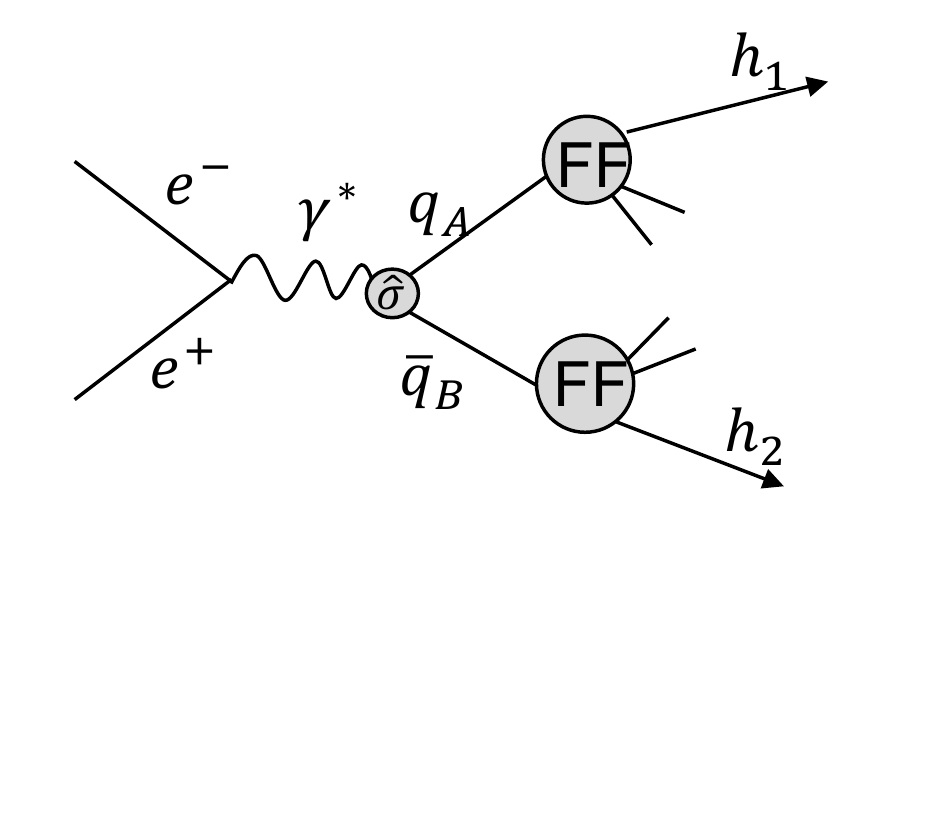}
\end{minipage}
\caption{Graphical representation of the semi inclusive deep inelastic (SIDIS) process (left drawing) and $e^+e^-$ annihilation (right drawing).}
\label{fig:SIDIS}
\end{figure}

Fragmentation functions are believed to be universal, namely the same for all processes and in particular for SIDIS and $e^+e^-$ annihilation process, which are represented diagramatically in Fig. \ref{fig:SIDIS}.

Factorization \cite{Collins-libro} allows to express the cross sections of these processes in terms of factors characterizing sub-processes arising at different energy scales. For instance the SIDIS cross section is factorized as $\sigma^{SIDIS}\sim\rm{PDF}\otimes \hat{\sigma}^{lq_A\rightarrow l'q_A}\otimes\rm{FF}$, namely it is given by the convolution of a PDF, a FF and the hard scattering cross section $\hat{\sigma}^{lq_A\rightarrow l'q_A}$ ($q_A$ indicates the struck quark). The $e^+e^-$ annihilation cross section, instead, is factorized as $\sigma^{e^+e^-}\sim \rm{FF}\otimes \hat{\sigma}^{e^+e^-\rightarrow q_A\bar{q}_B}\otimes \rm{FF}$, namely it is given by the convolution of two FFs with the hard scattering cross section $\hat{\sigma}^{e^+e^-\rightarrow q_A\bar{q}_B}$. Unpolarized $\pt$-integrated FFs have been extracted from $e^+e^-$ annihilation data and are by now rather well known \cite{models}).

Less known are the transverse momentum dependent and polarized FFs, the most important being the Collins FF which characterizes the fragmentation of a transversely polarized quark in unpolarized hadrons and which is usually indicated with $H_{1q}^{\perp h}(z,\ptpt)$.
The Collins FF correlates the transverse polarization of the fragmenting quark $\textbf{S}_{q\rm{T}}$ with the transverse momentum $\pt$ of the produced hadron. The fragmentation of a transversely polarized quark in an unpolarized hadron is described by the function \cite{collins}
\begin{equation}\label{eq:collins FF}
    D_{h/q\uparrow}(z,\pt)=D_{1q}^h(z,\ptpt)+\frac{(\hat{\textbf{k}}\times \pt)\cdot\textbf{S}_{q\rm{T}}}{zm_h}H_{1q}^{\perp h}(z,\ptpt),
\end{equation}
where $m_h$ is the hadron mass.
The peculiarity of this process is that hadrons are emitted asymmetrically with respect to the plane defined by the quark transverse polarization and its direction of motion. From Eq. (\ref{eq:collins FF}), the distribution of the hadron azimuthal angle $\phi_h$ about $\hat{\textbf{k}}$ exhibits a $\sin\phi_C$ modulation, where $\phi_C=\phi_h-\phi_{S_q}$ is the \textit{Collins angle}, $\phi_{S_q}$ being the azimuthal angle of the quark transverse polarization. This is known as the Collins effect. The amplitude of the modulation for a fully polarized quark $q$ is the \textit{analyzing power}
\begin{equation}\label{eq:Collins ap}
    a^{q^{\uparrow}\rightarrow h + X}(z,p_{\rm{T}})=-\frac{p_{\rm{T}}}{zm_h}\frac{H_{1q}^{\perp h}}{D_{1q}^h}.
\end{equation}
Assuming $\hat{\textbf{k}}$ along $\zu$ and $\textbf{S}_{q\rm{T}}$ along $\yu$, if $H_{1q}^{\perp h}<0$ (positive analyzing power), the hadron $h$ is emitted preferentially along $-\xu$. The situation is reversed for $H_{1q}^{\perp h}>0$ (negative analyzing power) or changing the sign of the quark polarization.


Thanks to this mechanism the azimuthal distribution of the produced hadrons can be used as a quark polarimeter. Being chiral odd, in transversely polarized SIDIS the Collins FF is coupled to transversity, producing a particular modulation in the azimuthal distribution of the final hadrons that is known as the \textit{Collins asymmetry}. Knowing the Collins analysing power, the transversity PDF can be extracted from the SIDIS data.

\begin{figure}
\centering
\begin{minipage}{.95\textwidth}
  \centering
  \includegraphics[width=.8\linewidth]{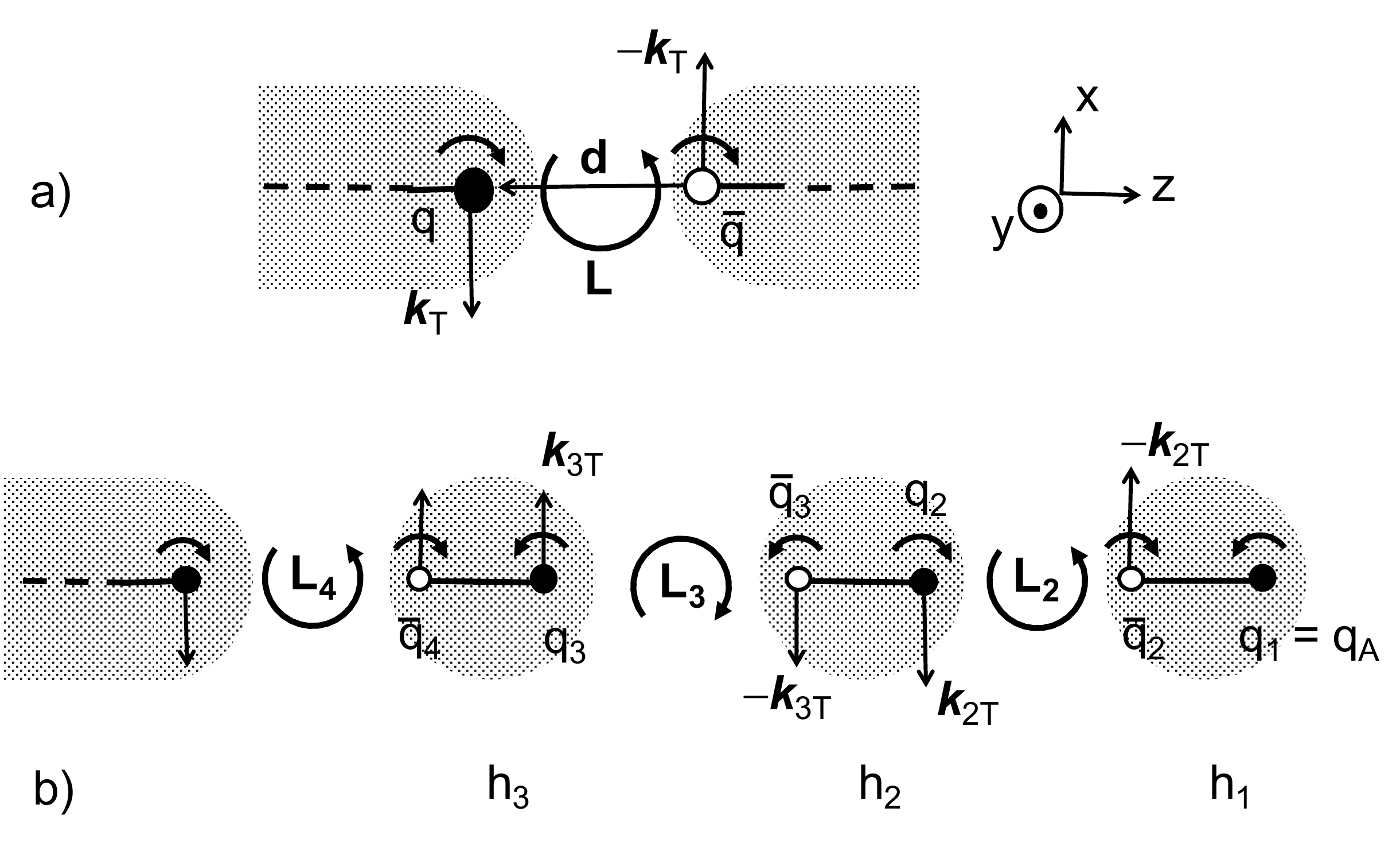}
\end{minipage}%
\caption{Classical $^3P_0$ mechanism at string breaking (a) and its iteration for the production of pseudoscalar mesons resulting in a Collins effect (b).}
\label{fig:classical3P0}
\end{figure}

A model for the Collins effect is based on the ${}^3P_0$ mechanism applied to the string fragmentation model \cite{DS09} and it is the topic studied in this work. In this model the color interaction, for instance between the struck quark and the remnant of the nucleon in a DIS process, is described as a relativistic string stretched between the two color charges. Then the decay of the string via the tunneling of $q\bar{q}$ pairs out of the color field is the hadronization process. Fig. \ref{fig:classical3P0} represents such a string stretched along the $\zu$ axis. In this model the $q\bar{q}$ pairs at each string breaking are produced in the ${}^3P_0$ state, namely they have total orbital angular momentum $L=1$ and total spin $S=1$ opposite to the orbital angular momentum such that $J=L+S=0$. This is shown in Fig. \ref{fig:classical3P0} (a). This model makes an important prediction if it is applied to the fragmentation of a string where the initial quark is transversely polarized as in Fig. \ref{fig:classical3P0} (b) and assuming emission of pseudo-scalar mesons \cite{DS09}. If the initial quark $q\A$ is polarized along $\yu$ axis, in order for the first meson $h_1$ to be pseudo-scalar, the $q_2\bar{q}_2$ pair has to be polarized along $-\yu$. Consequently, the angular momentum $\textbf{L}_2$ is directed along $\yu$, meaning that the quark $q_2$ has transverse momentum, with respect to the string axis, $\rm{\textbf{k}_{2T}}$ along $-\xu$ and the anti-quark $\bar{q}_2$ has opposite transverse momentum $-\rm{\textbf{k}_{2T}}$. The anti-quark is absorbed by the (first rank) meson $h_1$ which has then transverse momentum along the $\xu$ axis. Continuing this mechanism, the second rank meson $h_2$ is emitted along $-\xu$. The third rank is emitted again along $\xu$ and so on. The overall effect is that odd and even rank mesons are emitted to the left and to the right of the plane defined by the initial quark polarization and momentum (directed along $\zu$) vectors. If the initial quark $q\A$ is $u$ then this mechanism produces opposite effects for instance for positive and for negative pions or kaons. Therefore the string+$^3P_0$ model produces a Collins effect. This is however a classical model. Its quantum mechanical formulation will be given in Chapter 2 for the production of pseudoscalar mesons. The extension to vector meson production is described in Chapter 5.

Presently, only two models for the fragmentation of polarized quarks and the Collins effect exist. The second one, which will be shortly presented in Chapter 2, describes single quark jets in the framework of the Nambu–Jona-Lasinio effective field theory \cite{Mate}.


\section{The SIDIS process}
\begin{figure}
\centering
\begin{minipage}{.5\textwidth}
  \centering
  \includegraphics[width=0.7\linewidth]{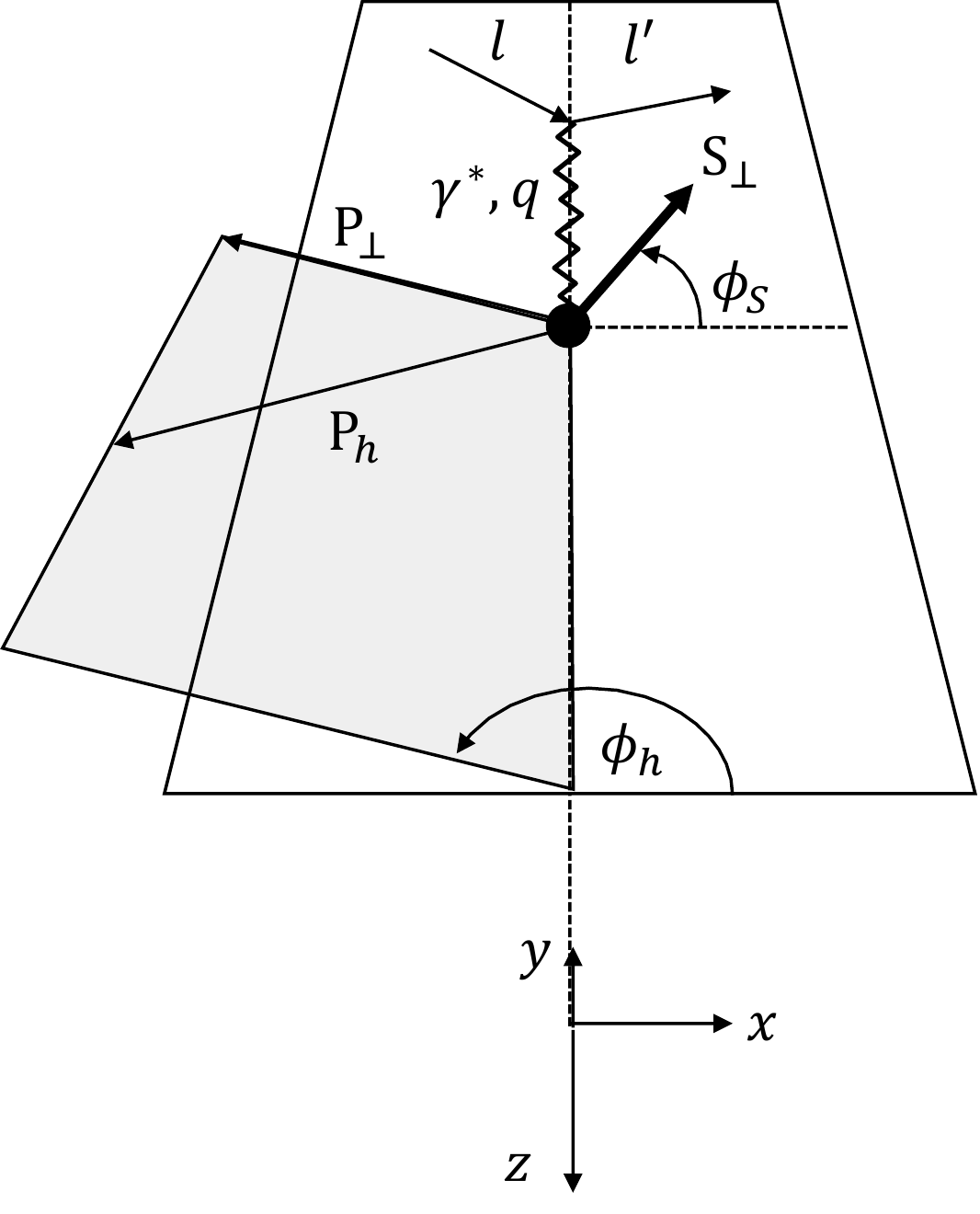}
\end{minipage}%
\begin{minipage}{.5\textwidth}
  \centering
  \includegraphics[width=0.9\linewidth]{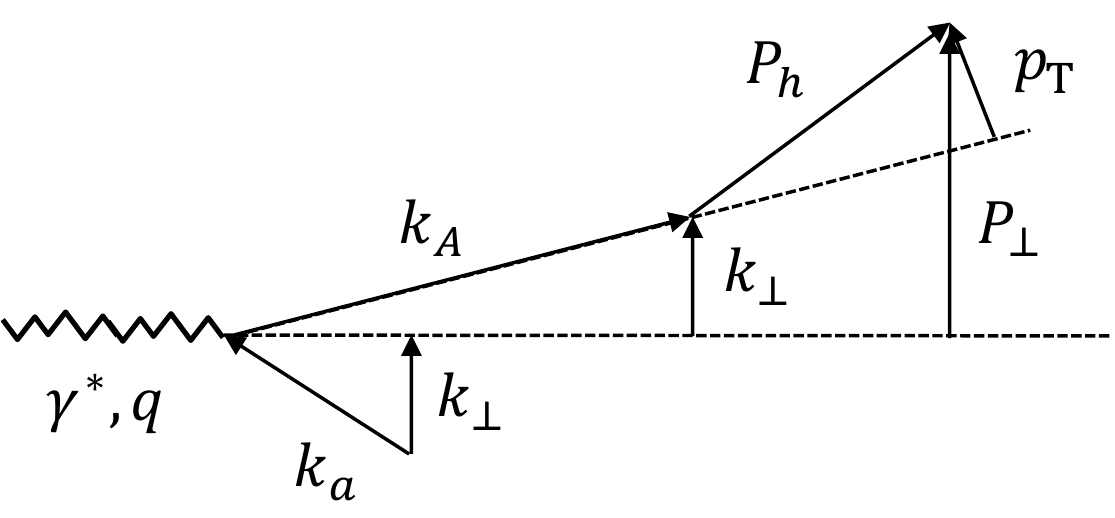}
\end{minipage}%
\caption{Kinematics of the SIDIS process in the $\gamma^*$-$N$ reference system (left panel) and kinematics of the elementary hard process (right panel).}
\label{fig:sidis kin}
\end{figure}

The most powerful tool for the study of the nucleon structure is the SIDIS process, where a high energy lepton with momentum $l$ scatters off a target nucleon $N$ with momentum $P_N$, and in the final state the scattered lepton with momentum $l'$ is measured in coincidence with at least one of the produced hadrons $h$, with momentum $P_h$. The kinematics is represented in Fig. \ref{fig:sidis kin} in the $\gamma^*$-N system (GNS), namely in the frame where the emitted virtual photon momentum $\textbf{q}=\textbf{l}-\textbf{l}'$ is along the $\zu$ axis and the nucleon is at rest. This axis defines the longitudinal direction. The components of vectors transverse and parallel to this axis are indicated with the symbols $\perp$ and $\parallel$. For instance, in this frame the nucleon polarization vector is $\textbf{S}=(\textbf{S}_{\perp},S_{\parallel})$.

The plane containing the lepton momenta $\textbf{l}$, $\textbf{l}'$ (and $\textbf{q}$) defines the \textit{lepton scattering plane}, whereas the plane containing the hadron momentum $\textbf{P}_h$ and $\textbf{q}$ defines the \textit{hadron plane}. The azimuthal angles of $\textbf{S}_{\perp}$ ($\phi_S$) and of the hadron transverse momentum $\textbf{P}_{\perp}$ ($\phi_h$) are defined with respect to the lepton scattering plane, with $\phi_{l'}=0$, as shown in Fig. \ref{fig:sidis kin}. The SIDIS process is described in terms of the invariants $Q^2 = -(l-l')^2$, $x_B = Q^2/2P_N\cdot q$, $y = P_N\cdot q/P_N\cdot l$, of the hadrons fractional energy $z_h = P_N\cdot P_h/P_N\cdot q$, of its transverse momentum $P_{\perp}$ and of the azimuthal angles $\phi_h$ and $\phi_S$.
The leading twist SIDIS cross section in the one photon exchange approximation is \cite{Bacchetta-SIDIS,Diehl-SIDIS}
\begin{eqnarray}\label{eq:pol SIDIS x section}
\nonumber \frac{d\sigma^{l+N\rightarrow l'+h+X}}{dx_Bdydz_hd\phi_h dP_{\perp}^2d\phi_S}&=&\frac{\alpha_{em}^2}{x_B y Q^2}\frac{1+(1-y)^2}{2}\Bigg \{ F_{UU,T} +\varepsilon_2 F_{UU,L} + \varepsilon_1 F_{UU}^{\cos\phi_h}\cos\phi_h \\
\nonumber &+& \varepsilon_2 F_{UU}^{\cos 2\phi_h}\cos 2\phi_h + \lambda_l \varepsilon_3 F_{UU}^{\sin\phi_h}\sin\phi_h \\
\nonumber &+& S_{\parallel} \Big[ \varepsilon_1 F_{UL}^{\sin\phi_h}\sin\phi_h+\varepsilon_2 F_{UL}^{\sin 2\phi_h} \sin 2\phi_h \Big]\\
\nonumber &+& S_{\parallel}\lambda_l \Big[ \varepsilon_4 F_{LL}+\varepsilon_3 F_{LL}^{\cos\phi_h}\cos \phi_h \Big]\\
\nonumber &+& S_{\perp} \Big[ \left(F_{UT,T}^{\sin\phi_{Siv}}+\varepsilon_2 F_{UT,L}^{\sin\phi_{Siv}}\right)\sin\phi_{Siv}\\
\nonumber &+& \varepsilon_2 F_{UT}^{\sin\phi_C}\sin\phi_C + \varepsilon_2 F_{UT}^{\sin\left(3\phi_h-\phi_S\right)}\sin\left(3\phi_h-\phi_S\right)\\
\nonumber &+& \varepsilon_1 F_{UT}^{\sin\phi_S}\sin\phi_S+\varepsilon_1 F_{UT}^{\sin\left(2\phi_h-\phi_S\right)}\sin\left(2\phi_h-\phi_S\right) \Big]\\
\nonumber &+&S_{\perp}\lambda_l\Bigg[ \epsilon_4 F_{LT}^{\cos\left(\phi_h-\phi_S\right)}\cos\left(\phi_h-\phi_S\right)+\varepsilon_3 F_{LT}^{\cos\phi_S}\cos\phi_S \\
&+& \varepsilon_3 F_{LT}^{\cos\left(2\phi_h-\phi_S\right)}\cos\left(2\phi_h-\phi_S\right)\Bigg] \Bigg \},
\end{eqnarray}
where $\lambda_l$ is the helicity of the lepton beam. The azimuthal angle $\phi_C=\phi_h+\phi_S-\pi$ is the Collins angle\footnote{This definition of the Collins angle is obtained from the previous one, $\phi_C=\phi_h-\phi_{S_q}$, when taking into account the reflection of the transverse polarization of the interacting quark about the normal to the lepton scattering plane. Since $\phi_{S_q}=\pi-\phi_{S}$ it is $\phi_C=\phi_h+\phi_S-\pi$.} and $\phi_{Siv}=\phi_h-\phi_S$ is the Sivers angle.

\begin{eqnarray}
\label{eq:epsilon 12}
\varepsilon_1 = \frac{2(2-y)\sqrt{1-y}}{1+(1-y)^2},\,\,\,\,\,\,\,\,\,   \varepsilon_2 = \frac{2(1-y)}{1+(1-y)^2} \\
\label{eq:epsilon 34}
\varepsilon_3 = \frac{2y\sqrt{1-y}}{1+(1-y)^2} ,\,\,\,\,\,\,\,\,\,  \varepsilon_4 = \frac{y(2-y)}{1+(1-y)^2}
\end{eqnarray}
are kinematic factors. In all these formulas $M^2x_B^2/Q^2$ corrections have been neglected.

The cross section is written in terms of $18$ structure functions which depend on $x_B, z_h$ and $P_{\perp}^2$ and are labelled as $F_{XY,Z}^{t(\phi_h,\phi_S)}$. The subscripts $X,Y,Z$, which take the values $U$, $L$ or $T$, label respectively the target, the beam and the virtual photon polarizations. $U$ refers to the unpolarized case, $L$ to the longitudinally polarized and $T$ to the transversely polarized state. The superscript $t(\phi_h,\phi_S)$ refers to the trigonometric function $t$ of the azimuthal angles $\phi_h$ and $\phi_S$.

According to factorization, each structure function is expressed in the parton model in terms of the convolution
\begin{eqnarray}
\nonumber    \mathcal{C}[wfD] &=& \sum_q e_q^2 \int d^2\kperp\, d^2\pt\, \delta^{(2)}(\textbf{P}_{\perp}-z_h\kperp -\pt) w(\kperp,\textbf{P}_{\perp}) \\
    &\times& \, f(x_B,\kperpkperp)\, D(z_h,p^2\T),
\end{eqnarray}
where $w$ represents a generic weight factor and the $Q^2$ dependence of $f$ and $D$ has been omitted for simplicity. For the purposes of this work, the most relevant ones expressed in the GNS are \cite{Anselmino-helicity-formalism}
\begin{eqnarray}
\label{eq: FUU}
F_{UU}&=&\mathcal{C}[f_1\,D_1]\\
\label{eq: F cosphi}
\nonumber F_{UU}^{\cos\phi_h}&=&-\frac{2}{Q}\mathcal{C}\Big[(\hat{\textbf{P}}_{\perp}\cdot\kperp)\,f_1\, D_1 \\
&+& \frac{\kperpkperp(P_{\perp}-z_h\hat{\textbf{P}}_{\perp}\cdot\kperp)}{z_hM_hM}\,h_1^{\perp}\,H_1^{\perp}\Big]\\
\label{eq: F cos2phi}
\nonumber F_{UU}^{\cos 2\phi_h}&=&\mathcal{C}\Big[\frac{(\textbf{P}_{\perp}\cdot\kperp)-2z_h (\hat{\textbf{P}}_{\perp}\cdot\kperp)^2+z_h^2\kperpkperp}{z_hM_hM}\, h_1^{\perp}\,H_1^{\perp}\\
&+&\frac{2(\hat{\textbf{P}}_{\perp}\cdot\kperp)^2-\kperpkperp}{Q^2}\, f_1\, D_1\Big]\\
\label{eq: F collins}
F_{UT}^{\sin\phi_C}&=&\mathcal{C}\Big[\frac{P_{\perp}-z_h (\hat{\textbf{P}}_{\perp}\cdot\kperp)}{z_h M_h}\, h_1\, H_1^{\perp}\Big]\\
\label{eq: F sivers}
F_{UT}^{\sin\phi_{Siv}}&=& \mathcal{C}\Big[ -\frac{\hat{\textbf{P}}_{\perp}\cdot\kperp}{M}\, f_{1T}^{\perp}\, D_1\Big],
\end{eqnarray}
where $F_{UU}=F_{UU,T}+\varepsilon_2 F_{UU,L}$. $\kperp$ is the intrinsic quark momentum and $\pt$ the momentum of the observed hadron with respect to the direction of the fragmenting quark, as shown in the right panel of Fig. \ref{fig:sidis kin}. Neglecting terms of order $\kperpkperp/Q^2$ it is $x\simeq x_B$, $z\simeq z_h$ and $\textbf{P}_{\perp}=z_h\kperp+\pt$.

From the structure functions in Eqs. (\ref{eq: FUU}-\ref{eq: F sivers}), one defines the azimuthal asymmetries
\begin{equation}
    A_{XY}^{t(\phi_h,\phi_S)}=\frac{F_{XY}^{t(\phi_h,\phi_S)}}{F_{UU}},
\end{equation}
which depend on the DIS variables $x_B$ and $Q^2$, and on the hadronic variabls $z_h$ and $P_{\perp}$. In particular, $A_{UU}^{\cos\phi_h}$ and $A_{UU}^{\cos 2\phi_h}$ are often called \textit{unpolarized azimuthal asymmetries}. They contain information on the quark intrinsic transverse momenta and receive contributions from the \textit{Cahn effect} \cite{Cahn-1978,Cahn-1989}, a pure kinematic effect. The Boer-Mulders function appears coupled to the Collins FF in both asymmetries. The Cahn effect contributes mostly to the $\cos\phi_h$ asymmetry whereas the Boer-Mulders function contributes mostly to the $\cos 2\phi_h$ asymmetry. The most recent measurements of these asymmetries come from the COMPASS \cite{COMPASS-unpol-SIDIS} and HERMES \cite{hermes-unpolarized} experiments  and different phenomenological analyses have been performed. However, still there is no definite evidence that the Boer-Mulders function is different from zero \cite{Barone-BoerMulders-2015}.

The amplitude $A_{UT}^{\sin\phi_{Siv}}$ of the $\sin\phi_{Siv}$ modulation is known as the \textit{Sivers asymmetry}. This asymmetry is given by the convolution between the Sivers function $f_{1T}^{\perp}$ and the unpolarized FF $D_{1q}^h$. It has been measured by the COMPASS \cite{COMPASS-Sivers} and the HERMES \cite{HERMES-Sivers} Collaborations and has been found to be different from zero for protons. It allowed to extract the first moment of the Sivers function.

\begin{figure}[bt]
\centering
\begin{minipage}{0.8\textwidth}
  \centering
  \includegraphics[width=0.87\linewidth]{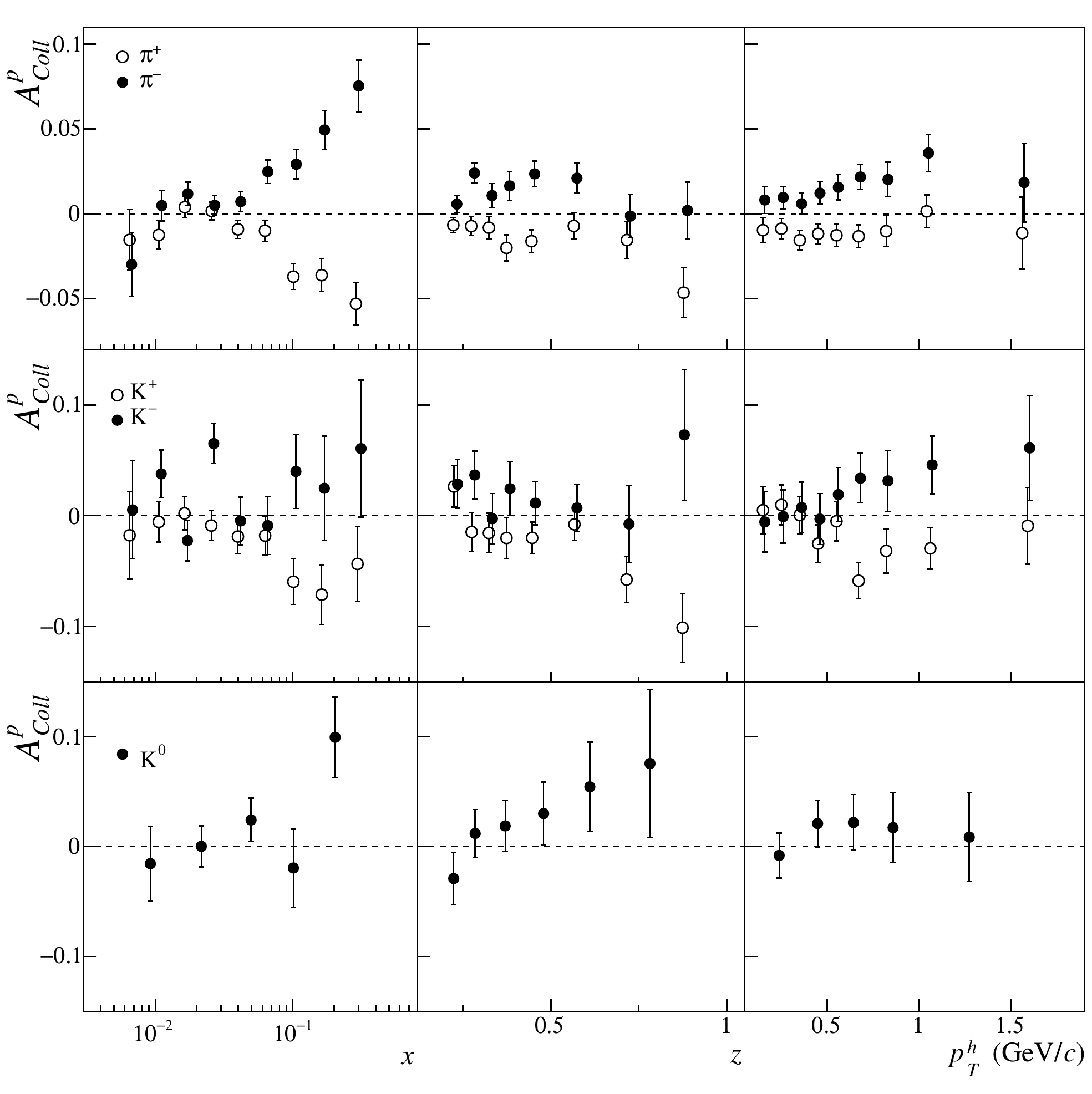}
\end{minipage}%
\caption{Collins asymmetry as measured by the COMPASS Collaboration \cite{COMPASS-collins-sivers} for $\pi^{\pm}$, $K^{\pm}$ and $K^0$ as function of $x_B$, $z_h$ and $P_{\perp}$.}
\label{fig:collins_compass}
\end{figure}
Finally the amplitude $A_{UT}^{\sin\phi_C}$ of the $\sin\phi_C$ modulation is known as the \textit{Collins asymmetry} and in the following it is referred to as $A_{Coll}$. The Collins angle $\phi_C$ is now evaluated in the GNS. This asymmetry is given by the convolution of $h_1^q$ and of $H_{1q}^{\perp h}$ which can be analytically calculated only making assumptions on the transverse momenta dependence of PDFs and FFs. Assuming Gaussian distributions, the Collins asymmetry can be written as \cite{M.B.B}
\begin{eqnarray}
    A_{Coll}(x_B,z_h) = G(z)\frac{\sum_q e_q^2 \, x_Bh_1^q(x_B)\, H_{1q}^{\perp h(1/2)}(z_h)}{\sum_q e_q^2 x_Bf_1^q(x_B)\,D_{1q}^h(z_h)},
\end{eqnarray}
where $G(z)=(1+z_h^2\langle \kperpkperp\rangle/\langle \ptpt\rangle)^{-1/2}$ and the "half-moment" of the Collins FF being defined as
\begin{equation}
    H_{1q}^{\perp h(1/2)}(z_h)=\int d\ptpt \frac{\rm{p_T}}{z_hm_h} \, H_{1q}^{\perp h }(z,\ptpt).
\end{equation}
Neglecting the quark intrinsic transverse momentum, the convolution operation becomes a product and the asymmetry can be expressed as
\begin{eqnarray}
    A_{Coll}(x_B,z_h,P_{\perp}) = \frac{\sum_q e_q^2 \, h_1^q(x_B)\, \frac{P_{\perp}}{z_h m_h}\,H_{1q}^{\perp h}(z_h,P_{\perp})}{\sum_q e_q^2 f_1^q(x_B)\,D_{1q}^h(z_h,P_{\perp})}.
\end{eqnarray}
With a polarized proton target the dominant contribution is that of the $u$ quark, and the Collins asymmetry can be written as
\begin{eqnarray}
A_{Coll}^p(z_h,P_{\perp})\simeq -\frac{h_1^u}{f_1^u}\,a^{u\uparrow\rightarrow h+X}(z_h,P_{\perp}) .
\end{eqnarray}

The Collins asymmetry has been measured in SIDIS by HERMES \cite{hermes-ssa} and COMPASS \cite{COMPASS-collins-sivers} experiments for charged hadrons, identified pions and kaons, on proton and deuteron targets. As an example Fig. \ref{fig:collins_compass} shows $A_{Coll}^p$ as function of $x_B$, $z_h$ and $P_{\perp}$ as measured by the COMPASS Collaboration \cite{COMPASS-collins-sivers}. The asymmetry is clearly different from zero, meaning that both $h_1$ and $H_{1}^{\perp}$ are different from zero. It has opposite sign for positive and negative pions. As already seen, in the $u$-dominance hypothesis, this effect is expected in the string+${}^3P_0$ model. As function of $x_B$ the Collins asymmetry is compatible with zero at $x_B\lesssim 0.03$, indicating that $h_1^q$ is a valence object. Similar features show up also for charged kaons. In particular, the asymmetry for positive kaons is somewhat larger than the corresponding asymmetry for positive pions, but the statistical uncertainties are large.
The same kinematic dependences for pions have been observed by the HERMES Collaboration \cite{HERMES-2004}.



\section{The $e^+e^-$ annihilation process}\label{se:e+e- annihilation}
\begin{figure}
\centering
\begin{minipage}{0.48\textwidth}
  \centering
  \includegraphics[width=1.0\linewidth]{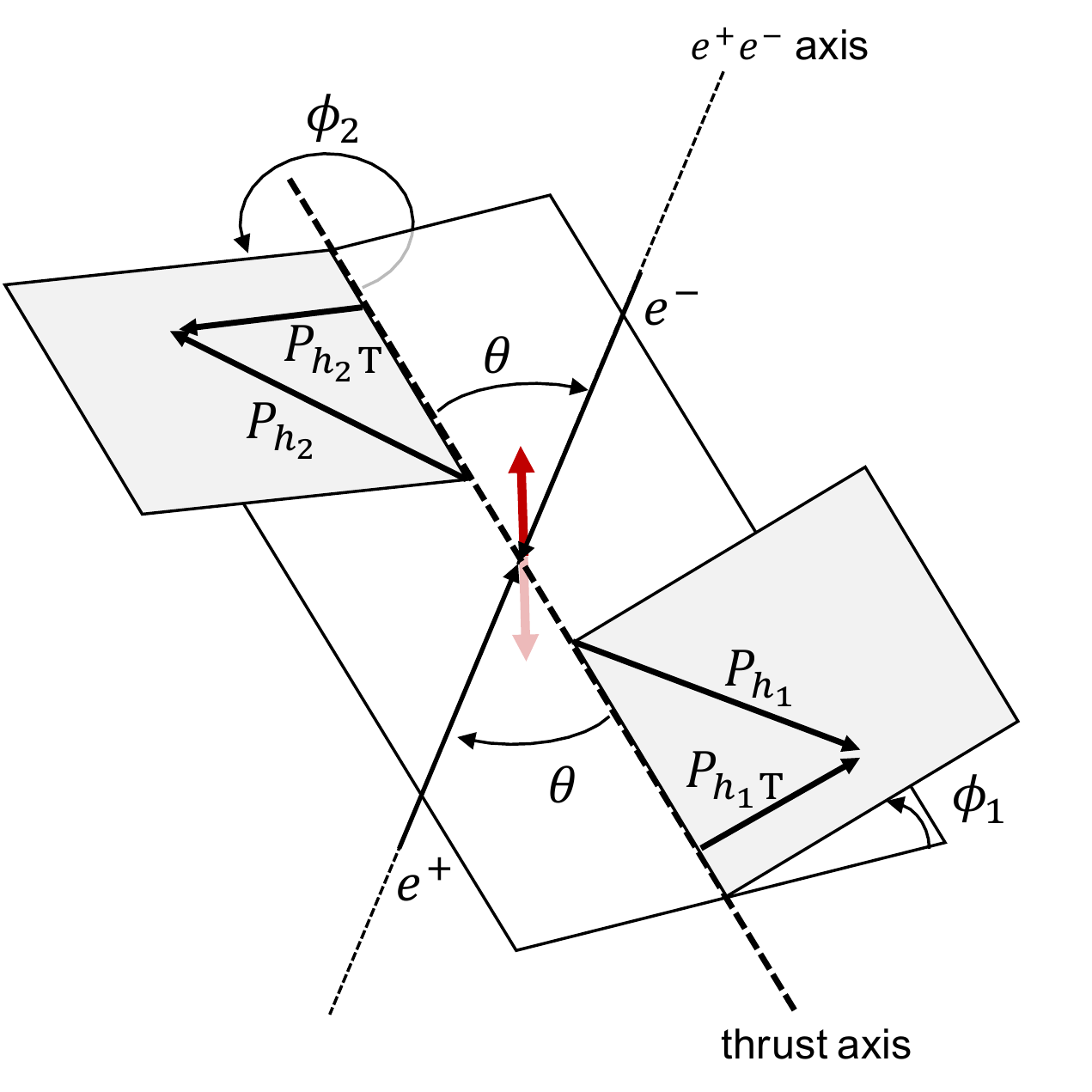}
\end{minipage}
\begin{minipage}{0.48\textwidth}
  \centering
  \includegraphics[width=1.0\linewidth]{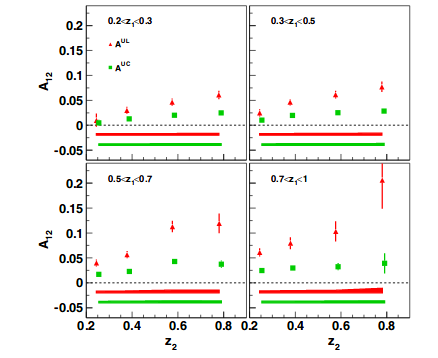}
\end{minipage}
\caption{Left: kinematics of the $e^+e^-$ annihilation process in the center of mass frame. Right: $A_{12}$ asymmetry as function of $z_2$ in four different bins of $z_1$ as measured by BELLE Collaboration \cite{BELLE-2008}.}
\label{fig:e+e-_kin}
\end{figure}

An independent source of information for the Collins FF is the observation of hadron pairs in the $e^+e^-\rightarrow q\bar{q}\rightarrow h_1h_2 + X$ process, shown in the left panel of Fig. \ref{fig:e+e-_kin}. The two hadrons must lie in two different hemispheres, namely one in the quark jet and the other in the anti-quark jet. In this process the intermediate virtual photon decays in a correlated $q\bar{q}$ spin state. This correlation is transferred through the Collins effect to the observed hadron pair producing a $\cos\left(\phi_1+\phi_2\right)$ modulation. The angles $\phi_1$ and $\phi_2$ are the azimuthal angles of $h_1$ and $h_2$ around the $q\bar{q}$ axis respectively. The $\phi=0$ half-plane contains the $e^+$ beam momentum. 
The resulting $\cos\left(\phi_1+\phi_2\right)$ asymmetry is \cite{BELLE-2005,BELLE-2008,Boer:2008-e+e-}
\begin{equation}\label{eq:a12}
    A_{e^+e^-}(z_1,z_2) = \frac{\langle \sin^2\theta\rangle}{\langle1+\cos^2\theta\rangle} \frac{\sum_{q}e_q^2 \, H_{1q}^{\perp h_1 (1/2)}(z_1)H_{1\bar{q}}^{\perp h_2 (1/2)}(z_2) }{\sum_{q}e_q^2 \, D_{1q}^{h_1}(z_1)D_{1q}^{h_2}(z_2)},
\end{equation}
where $\theta$ is the angle between the $e^+e^-$ axis and the $q\bar{q}$ axis (approximated experimentally with the thrust axis) and $z_{1}$ and $z_2$ are the fractional energies of $h_1$ and $h_2$, defined as the ratios between the hadrons energy and half of the center of mass energy $\sqrt{s}$.
This asymmetry is strictly related to the difference of the $A_{12}$ asymmetries for "unlike-sign" and "like-sign" pairs first measured by the BELLE Collaboration \cite{BELLE-2008}. As an example, in the right panel of Fig. \ref{fig:e+e-_kin} the measured asymmetries $A_{12}$ are shown as function of $z_2$ in four different bins of $z_1$. Clearly the asymmetry is different from zero, confirming that the Collins FF is different from zero. The $A_{e^+e^-}$ asymmetry increases with $z_h$, as expected, and in agreement with the $z_h$ dependence of the Collins asymmetry. The Collins FF, extracted using Eq. (\ref{eq:a12}) up to a sign, can then be used together with the Collins asymmetry measured in SIDIS in order to extract the quark transversity distribution (see e.g. \cite{M.B.B,Anselmino-transversity-extraction}).

The $e^+e^-\rightarrow q\bar{q}\rightarrow h_1h_2 + X$ process has not been simulated in this work. The BELLE data have however been used to get the value of the free parameter which gives the size of the Collins analysing power (see sec. \ref{sec: free parameters and kinematical distributions M18 M19}).

\section{The dihadron fragmentation function}
\begin{figure}
\centering
\begin{minipage}{.6\textwidth}
  \centering
  \includegraphics[width=0.8\linewidth]{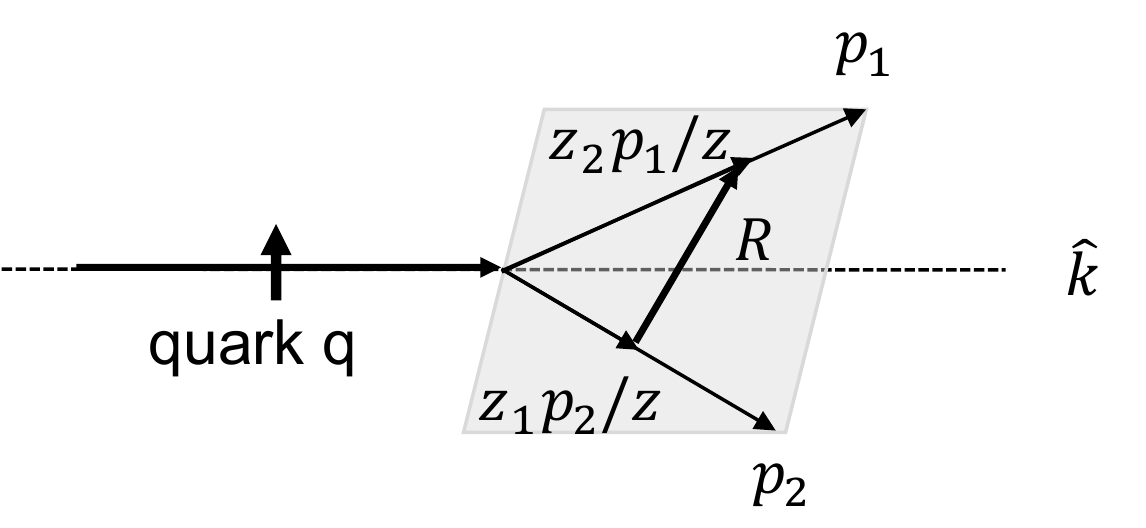}
\end{minipage}%
\caption{Kinematics for the fragmentation of a quark into a hadron pair.}
\label{fig:dihadron_kin}
\end{figure}

The alternative method to access the transversity distribution proposed in Ref. \cite{Collins:1993kq} uses as polarimeter the fragmentation of a transversely polarized quark into a pair of unpolarized mesons ($\pi\pi$, $\pi K$ or $K\bar{K}$) in the same jet, namely the process $q^{\uparrow}\rightarrow h_1 h_2+X$. Such process is described by the leading twist fragmentation function
\begin{equation}\label{eq:dihadron effect}
    D_{h_1h_2/q^{\uparrow}}(z,\textbf{R}\T)=D_{1q}^{h_1h_2}(z,M_{h_1h_2})+H_{1q}^{\sphericalangle h_1h_2}(z,M_{h_1h_2})\frac{(\hat{\textbf{k}}\times \textbf{R}\T)\cdot\textbf{S}_{q\rm{T}}}{M_{h_1h_2}},
\end{equation}
where $z=z_1+z_2$ and $M_{h_1h_2}$ is the invariant mass of the pair. $D_{1q}^{h_1h_2}$ is the unknown spin averaged \textit{dihadron} fragmentation function. The vector $\textbf{R}\T$ is usually defined as $\textbf{R}\T=\left(z_2\textbf{p}_{1\rm{T}}-z_1\textbf{p}_{2\rm{T}}\right)/z$, $\textbf{p}_{1(2)\rm{T}}$ being the transverse momenta of the hadrons with respect to the fragmenting quark direction $\hat{\textbf{k}}$. A schematic representation of this process is shown in Fig. \ref{fig:dihadron_kin}.

The function $H_{1q}^{\sphericalangle h_1h_2}$ is the DiFF, or IFF. It was assumed to come from the interference between competing amplitudes for the production of the hadron pair characterized by different phases.
The phase difference may arise e.g. by the interference between the amplitude for the production of the hadron pair from the decay of a resonance and the amplitude for the direct production of the pair \cite{Collins-Ladinsky}, or from the interference between the amplitudes related to the production of two resonances \cite{Jaffe-Jin-Tang}.
$H_{1q}^{\sphericalangle h_1h_2}$ is also a chiral-odd object.
A non-vanishing $H_{1q}^{\sphericalangle h_1h_2}$ produces a $\sin\phi_{RS_q}$ modulation, where $\phi_{RS_q}=\phi_R-\phi_{S_q}$. It is similar to the Collins one and its amplitude, i.e the analyzing power, is
\begin{equation}\label{eq:dihadron ap}
    a^{q^{\uparrow}\rightarrow h_1h_2 + X}(z,M_{h_1h_2})= -\frac{R\T}{M_{h_1h_2}} \frac{H_{1q}^{\sphericalangle h_1h_2}(z,M_{h_1h_2}^2)}{D_{1q}^{h_1h_2}(z,M_{h_1h_2}^2)}.
\end{equation}
However, this effect is also expected from the classical string+${}^3P_0$ mechanism. Indeed a single hadron Collins effect produces a dihadron effect due to the local compensation of transverse momenta. Thus, in this model, the ${}^3P_0$ mechanism is sufficient to produce both Collins and dihadron effects, without invoking resonances. This will be shown in more detail in the following.

The dihadron FF is used as a quark polarimeter to access $h_1^q$ in SIDIS off transversely polarized nucleons. In the process $l+N^{\uparrow}\rightarrow l' + h_1h_2 +X$, $H_{1q}^{\sphericalangle h_1h_2}$ is coupled to $h_1^q$ and gives the dihadron asymmetry which can be written as \cite{M.B.B}

\begin{equation}\label{eq: dihadron asymmetry}
    A^{h_1h_2}(x_B,z,M_{h_1h_2}^2)= \frac{\sum_q e_q^2\,h_1^q(x)\,\frac{\rm{R_{\perp}}}{M_{h_1h_2}}H_{1q}^{\sphericalangle h_1h_2}(z,M^2_{h_1h_2})}{\sum_q e_q^2 \,f_1^q(x)\,D_{1q}^{h_1h_2}(z,M^2_{h_1h_2})}.
\end{equation}
This asymmetry is proportional to the amplitude of the $\sin\phi_{RS}$, where $\phi_{RS}=\phi_R+\phi_S-\pi$ and $\phi_R$ and $\phi_S$ are the azimuthal angles of $\textbf{R}_{\perp}$ and of the nucleon polarization $\textbf{S}_{\perp}$ in the GNS. At leading order in $k_{\perp}/Q$ it is $\textbf{R}_{\perp}\simeq \textbf{R}_{\rm{T}}$.
The advantage of this asymmetry is that it does not involve $\kperp$ and thus it is the product, and not the convolution, of $h_1^q$ and $H_{1q}^{\sphericalangle h_1h_2}$. In addition the dihadron asymmetry is expected not to be diluted at large $Q^2$ by the gluon radiation which would change randomly the direction of the fragmenting quark momentum but not the relative vector $\textbf{R}_{\perp}$ \cite{Artru-transverse-spin}. For a proton target the main contribution comes again from the $u$ quark and the dihadron asymmetry can be written as
\begin{eqnarray}
A^{h_1h_2}(x_B,z,M_{h_1h_2}^2)\simeq -\frac{h_1^{u}}{f_1^u}a^{u^{\uparrow}\rightarrow h_1h_2 + X}(z,M_{h_1h_2}).
\end{eqnarray}

Besides the relative dihadron effect giving the dihadron asymmetry, there is also a global dihadron effect in the azimuthal angle of the total transverse momentum of the pair $\textbf{p}_{1\rm{T}}+\textbf{p}_{2\rm{T}}$. This effect has been integrated out in Eq. (\ref{eq:dihadron effect}) since it is expected to be blurred due to the intrinsic quark transverse momentum and gluon radiation.

The dihadron asymmetry has been measured in SIDIS off a transversely polarized proton target by HERMES \cite{hermes-dihadron} and off transversely polarized proton and deuteron targets by COMPASS \cite{compass-dihadron} using pairs of oppositely charged hadrons and identifying $h_1$ with the positive hadron. The results are similar to those of the Collins asymmetry with a somehow larger absolute value.

Analogously to the Collins FF, the IFF $H_{1q}^{\sphericalangle h_1h_2}$ can be measured in the process $e^+e^-\rightarrow (h_1h_2) (\bar{h}_1\bar{h}_2) + X$, namely from the $e^+e^-$ annihilation in two hadron pairs $h_1h_2$ and $(\bar{h}_1\bar{h}_2)$ belonging to the quark and to the antiquark jet respectively, through the Artru-Collins asymmetry \cite{Artru-Collins}.
Thus, the dihadron asymmetry can be used in combination with $e^+e^-$ annihilation data to the extract the quark transversity distribution \cite{M.B.B,Radici:2015mwa}.

The analysing power of both the Collins and the dihadron FF have been calculated from simulated events and compared with the existing data, as described in Chapter 3. The Collins and the dihadron asymmetries have also been simulated using \verb|PYTHIA| (see Chapter 4).


\chapter{The polarized quark fragmentation model} 

\label{Chapter1}

\begin{figure}[h]
  \centering
  \includegraphics[width=.35\textwidth]{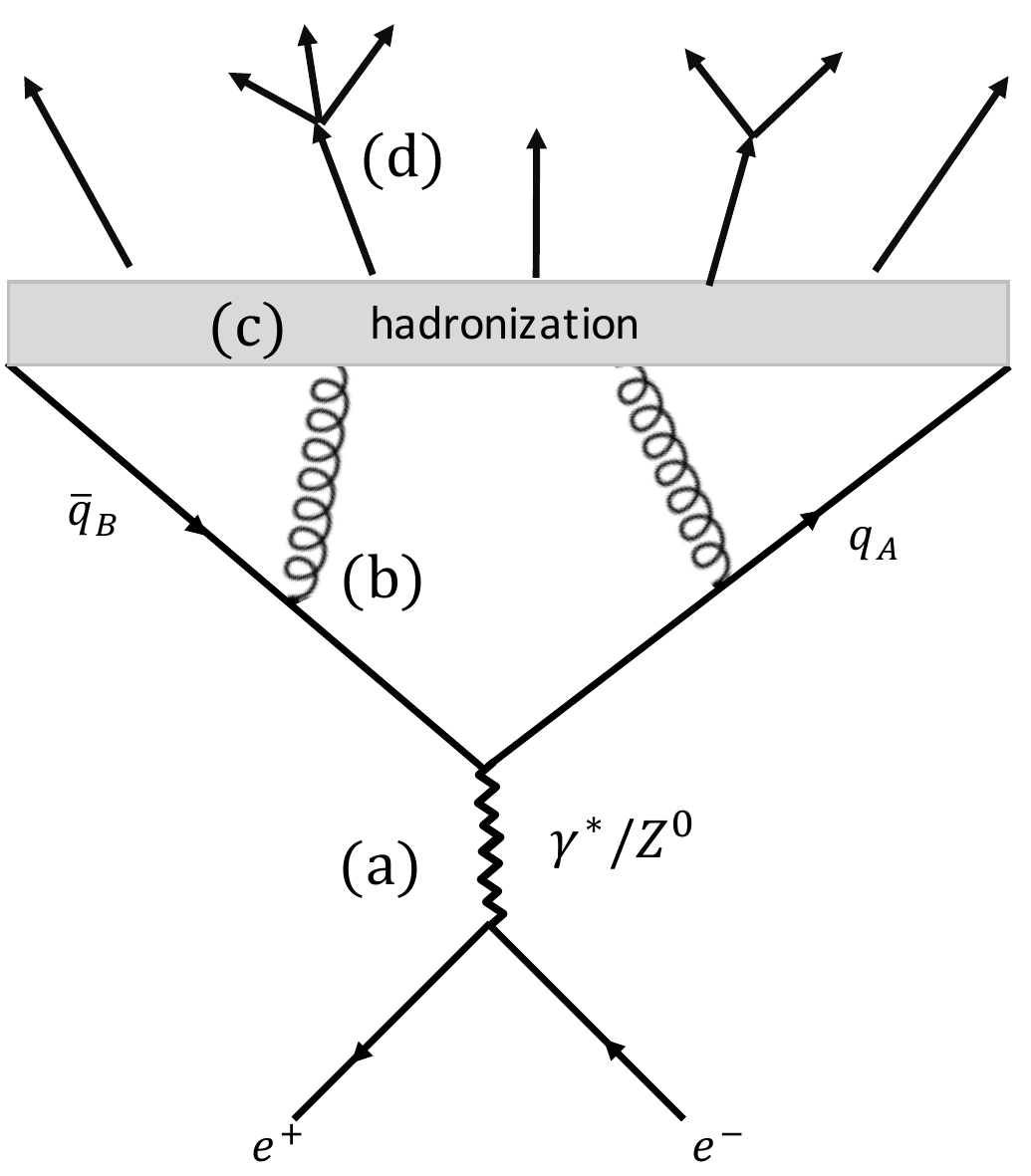}
\caption{The $e^+e^-$ annihilation process into hadrons.}
\label{fig:e+e-}
\end{figure}

In high energy collisions, from two colliding initial particles, typically many others are produced in the final state. Given the large number of particles involved, the description of the process is a too complex problem to be solved theoretically and only the general features can most of time be predicted. On the other hand, from the experimental point of view a detailed description of the process is required. To reach a sophisticated enough description of the collision process Monte Carlo simulations are generally used.
A high energy process involves typically physics at different time scales. For instance, in the $e^+e^-$ annihilation event represented in Fig. \ref{fig:e+e-}, one can identify four stages. The intermediate $\gamma^*/Z^0$ decays into a $q_A\bar{q}_B$ pair on a small time scale, of the order $t_{ann.}\sim 1/\sqrt{s}$ (assuming $c=\hbar=1$), where $\sqrt{s}$ is the center of mass energy of the event (for $\sqrt{s}=10\,\rm{GeV}$ it is $t_{ann.}\sim 10^{-2}\,\rm{fm}$).
At this stage the quark and the anti-quark are essentially free and, while moving in opposite directions in the center of mass frame, they may also emit gluons, and these gluons may branch into more gluons or quark-antiquark pairs, enlarging the number of partons involved in the process which constitute an overall color singlet system. This phase (b) is described perturbatively within pQCD. At larger time scales, phase (c) arises. Here the whole partonic system hadronizes into the final state hadrons. The time scale involved in this phase varies with the rapidity of the produced hadrons. Namely, those with lowest rapidity are produced on a typical time of $1-2\,\rm{fm}$ whereas those with the largest rapidity are produced at a slightly larger time\footnote{In the string fragmentation model, the hadrons with largest rapidity are produced around at $t_{had}=\sqrt{s}/(2\kappa)$, $\kappa\simeq 0.2\,\rm{GeV}^2$ being the tension of the string at rest, where the hadronization also stops. See, e.g., Fig. \ref{fig:yoyo} and Fig. \ref{fig: space-time history}.}. The produced hadrons may be stable particles as pions or kaons, or resonances like vector mesons. At phase (d) the resonances decay into stable particles.
Among the different phases, the hadronization part (phase (c)) is the more difficult to be described, since at present no analytical treatment of confinement exists. For the description of the hadronization process various models with different sophistication levels have been developed and have been implemented in event generators.

\begin{figure}[tb]
\centering
  \includegraphics[width=1.\linewidth]{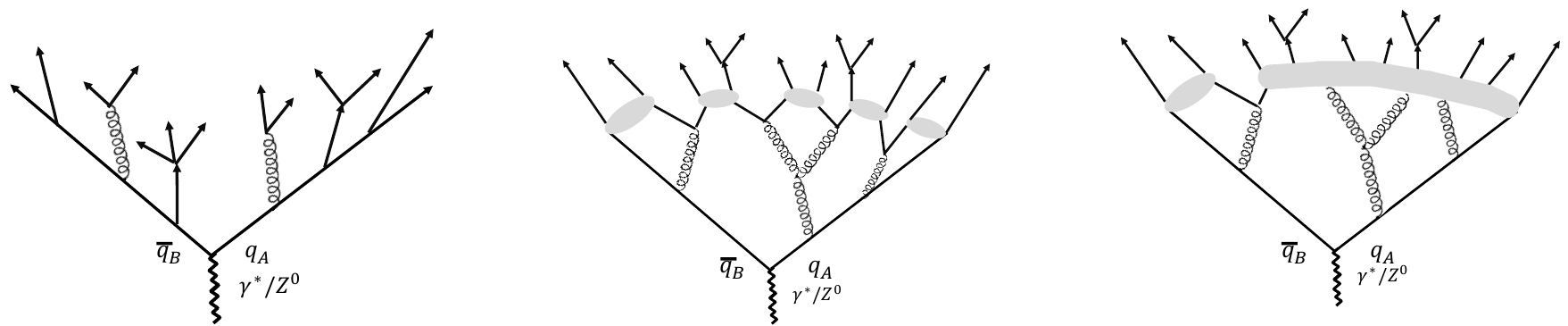}
\caption{Different fragmentation models: independent fragmentation (left), cluster fragmentation (middle) and string fragmentation (right).}
\label{fig:models}
\end{figure}

The \textit{Independent Fragmentation Model} (IFM) or \textit{Field-Feynman model} \cite{Field-Feynman}, treats $q\A$ and $\bar{q}\B$ as two uncorrelated objects which hadronize independently, as can be seen in the left drawing of Fig. \ref{fig:models}. Each jet is generated recursively, for instance $q\A$ is split in a hadron $h_1$ and a leftover quark $q_2$ by generating a $q_2\bar{q}_2$ pair. This splitting occurs according to a probability distribution function for the fraction of the $q\A$ energy taken by $h_1$. $q_2$ is then treated in the same manner and in this way the entire jet is generated. The original model lacks of confinement and does not conserve in a natural way overall momentum and quantum numbers, which need specific prescriptions to be satisfied. The Field-Feynman model has been recently extended for the pseudoscalar meson emission taking into account the quark spin through the spin density matrix formalism within the framework of the Nambu–Jona-Lasinio effective field theory \cite{Mate}.

In the \textit{Cluster Fragmentation Model} (CFM), implemented in \verb|HERWIG| \cite{herwig}, the whole partonic system is correlated in a color neutral configuration, as represented in the middle diagram of Fig. \ref{fig:models}. Gluons are split into quark-antiquark pairs and nearby quark antiquarks all over the system are joined to form color singlet clusters, the internal structure of which is not specified. Each cluster is then allowed to decay into smaller mass clusters representing the final observed hadrons according to some probability distribution based on the phase space.

In the \textit{String Fragmentation Model} (SFM), which confines in a color neutral dipole the initial $q\A\bar{q}\B$ pair, the attractive chromoelectric force field\footnote{The soft interaction between the color charges is described by the potential
\begin{equation}
    V_{QCD}(r)\simeq -\frac{4}{3}\frac{\alpha_s}{r}+\kappa r,
\end{equation}
where $r$ is the distance between $q\A$ and $\bar{q}\B$, $\alpha_s$ the strong coupling constant. At large distances, the Coulomb term is neglected and the interaction is described by the linear attractive term which gives a constant force.} is replaced by a relativistic string with $q\A$ and $\bar{q}\B$ placed at the endpoints, as shown in the right drawing of Fig. \ref{fig:models}.
The strength of the attractive force acting on the color charges is the string tension at rest $\kappa \simeq 1\,\rm{GeV}/\rm{fm}$, or equivalently $\kappa\simeq 0.2\,\rm{GeV}^2$ (taking $\hbar=c=1$) \cite{Andersson:1979-lambda0pol}.
As the charges move apart in opposite directions in the center of mass frame, they lose energy at the constant rate $\kappa$ stretching the string. The energy stored in the string rises linearly at the same rate. In this model gluons are not always split in quark-antiquark pairs but in general produce complicated string configurations. Finally, the fragmentation process is viewed as the breaking (or the decay) of the string into smaller string pieces through the tunneling of new $q\bar{q}$ pairs in the force field. The string breakings happen at space-like separations, allowing the string decay process to be treated recursively. The remaining string pieces represent the observed hadrons or resonances which subsequently decay into stable particles. 
The first SFM was proposed in 1973 by Artru and Mennessier \cite{Artru-Mennessier}. They introduce a universal constant $\mathcal{P}$ as the probability per unit space-time area for a string breaking point to occur. The string decay is treated in analogy with the radioactive exponential decay where the time variable is replaced by the space-time area in the past lightcone of the breaking point. This model already reproduced the main features of the jets but with a continuous mass spectrum for the produced hadrons. A SFM model with the actual particle masses was developed by the Lund group in 1983 \cite{LundModel-article}. This model is \textit{left-right} (LR) symmetric, namely invariant under the exchange of $q\A\leftrightarrow \bar{q}\B$. It is referred to in the following as the \textit{Symmetric Lund Model} (SLM) or simply as the \textit{Lund Model} (LM). It is presently implemented in event generators like \verb|LEPTO| \cite{lepto} and \verb|PYTHIA| \cite{pythia8} and has proven to be very successful in the description of experimental data from $e^+e^-$, DIS and pp experiments.

An interesting comparison from the theoretical point of view among the different class of models is given in Ref. \cite{collins-rogers}. All these models but that in Ref. \cite{Mate} neglect the quark spin degree of freedom. 

The goal of this work is the development of a MC program for the simulation of the polarized quark fragmentation process by using a model which treats the quark spin in a systematical and consistent way, suitable to be included in event generators.
To this aim, the SLM supplemented with the ${}^3P_0$ mechanism and restricted to pseudoscalar meson production has been used. Many papers on this topic already exist \cite{DS09,DS11,DS13}. In this chapter all the material is reviewed and organized in order to produce a consistent analytical formulation, which is the basis of the MC development.

In section \ref{sec:spinless SLM}, the basic concepts of the \textit{yo-yo} mass-less relativistic string and of the SLM in the spinless case are reviewed. In section \ref{sec: introduction of spin} the string$+{}^3P_0$ model is described. In section \ref{sec:possible choices} two specific choices leading to two different analytical formulations of the model (M18 and M19) are discussed. They are recent developments made in the context of this research project and both of them have been implemented in Monte Carlo codes as described in Chapter \ref{chapter 3}. The newest development consisting in the introduction of vector mesons in the fragmentation chain is described in Chapter \ref{chapter5}.





\section{The spinless case in SLM}\label{sec:spinless SLM}
\subsection{The kinematics of the \textit{yo-yo}}
The SLM is based on the dynamics of the $1+1$ dimensional relativistic string with massless ends \cite{Artru-Mennessier,patrascioiu}. To describe such motion in space-time, the time axis is indicated with $t$ and the longitudinal space axis, which coincides with the string axis, is indicated with $z$. Then, the motion in the $(t,\zu)$ plane of the $q\A\bar{q}\B$ system in Fig. \ref{fig:e+e-} is described by the hamiltonian
\begin{equation}\label{eq:string hamiltonian}
    H=|k\A^{mec,z}|+|k_{\rm{\bar{B}}}^{mec,z}|+\kappa |z_A-z_{\bar{B}}|
\end{equation}
where $k\A^{mec,z}$ and $k_{\rm{\bar{B}}}^{mec,z}$ are the longitudinal components of the mechanical energy-momentum $k\A^{mec}=(E\A^{mec},k\A^{z,mec})$ of $q\A$ and $k_{\rm{\bar{B}}}^{mec}=(E_{\rm{\bar{B}}}^{mec},k_{\rm{\bar{B}}}^{z,mec})$ of $\bar{q}\B$ which are assumed to be mass-less, and $z\A$ and $z\B$ are the positions of $q\A$ and $\bar{q}\B$ along the $\zu$ axis at time $t$. The initial energy of each quark is $\sqrt{s}/2$, $\sqrt{s}$ being the center of mass energy of the $q\A\bar{q}\B$ system. In the center of mass of the $q\A\bar{q}\B$ system it is also $k\A^{z,mec}+k_{\rm{\bar{B}}}^{z,mec}=0$.
Concentrating on the $q\A$ side, from Eq. (\ref{eq:string hamiltonian}) one derives the Hamilton's equation $v\A^{z}=\rm{sign}(k\A^{z,mec})$, namely the quark can move along the $+\zu$ or $-\zu$ axis with longitudinal velocity $v\A^{z}$ equal in absolute value to the velocity of light ($c=1$). Assuming $q\A$ has initially $v\A^z=+1$, then from Eq. (\ref{eq:string hamiltonian}) the quark longitudinal momentum obeys the Hamilton's equation
\begin{eqnarray}\label{eq: dk/dt}
    \frac{dk\A^{mec,z}}{dt} = -\kappa,
\end{eqnarray}
i.e. it loses longitudinal momentum at the rate given by the string tension. Equation (\ref{eq: dk/dt}) is invariant under boosts along the $\zu$ axis. The same is also true for $\bar{q}\B$ which has opposite momentum with respect to $q\A$ and thus $v^z_{\rm{\bar{B}}}=-1$. From Eq. (\ref{eq: dk/dt}) one can see also that the mechanical energy of $q\A$ decreases with $z\A$ at the same rate as the longitudinal momentum, namely
\begin{equation}\label{eq: dE/dt}
    \frac{dE\A^{mec}}{dz\A}=\frac{dE\A^{mec}}{dk\A^{mec,z}}\frac{dk\A^{mec,z}}{dt}\frac{dt}{dz\A} = -\kappa.
\end{equation}
From the solutions of Eqs. (\ref{eq: dk/dt}-\ref{eq: dE/dt}), the quark mechanical momentum is
\begin{eqnarray}\label{eq: k(t) and E(z)}
    k\A^{mec}=\kappa\left(z_0 - z\A,t_0 - t \right),
\end{eqnarray}
where $z_0 = t_0 = \sqrt{s}/2\kappa$, $z_0$ being obtained from Eq. (\ref{eq:string hamiltonian}) when $z\A=-z_{\rm{\bar{B}}}=z_0$. For $\bar{q}\B$ a similar equation holds with $k_{\rm{\bar{B}}}^{mec,z}=-k\A^{mec,z}$. This shows also that energy-momentum and space-time coordinates are linearly related through the string tension, meaning that one can associate an energy-momentum vector to each point in space-time. Clearly this is a classical picture.

\begin{figure}
\centering
\begin{minipage}{.95\textwidth}
  \centering
  \includegraphics[width=.8\linewidth]{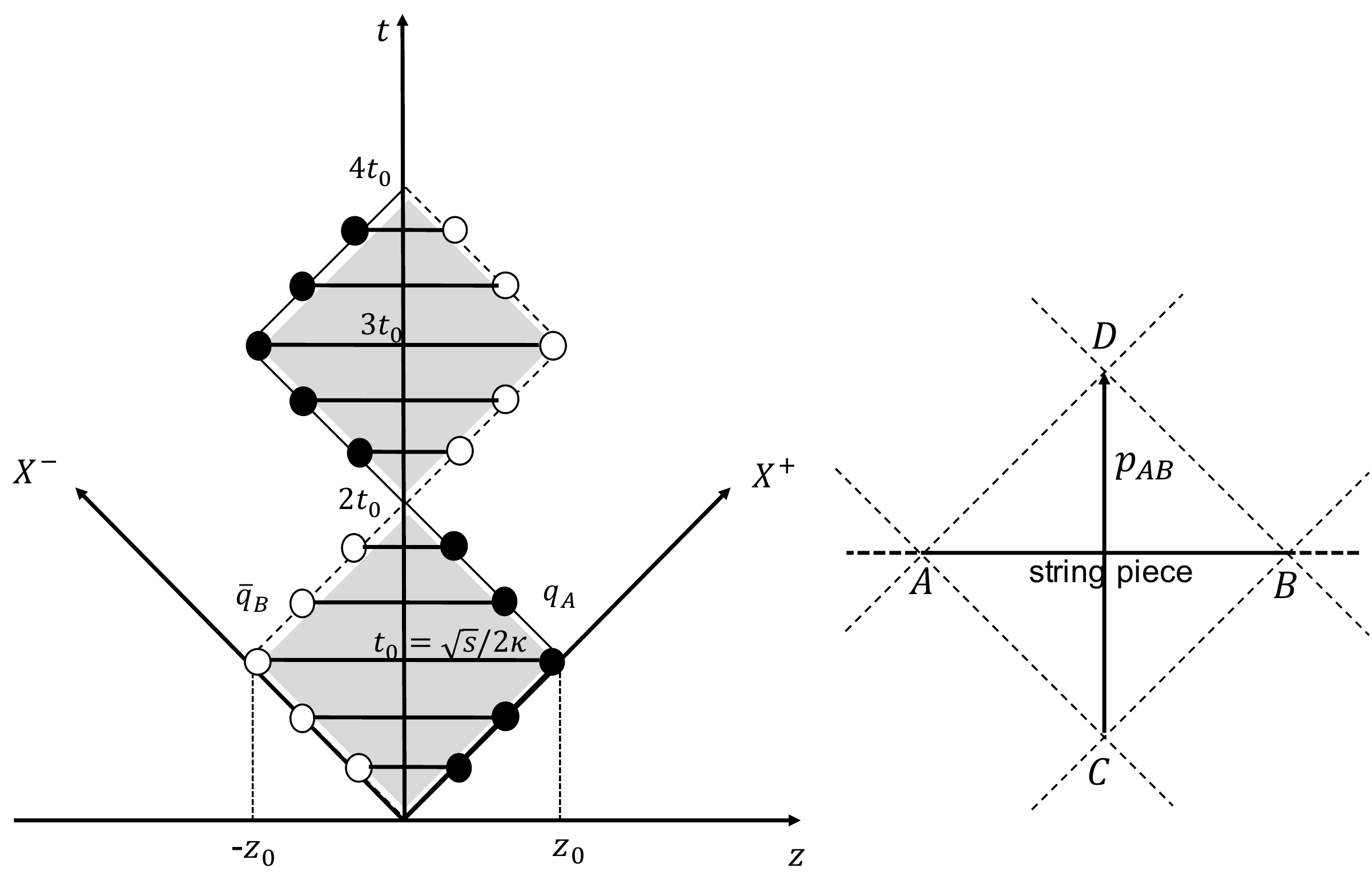}
\end{minipage}%
\caption{Motion of the $1+1$ dimensional yo-yo in its rest frame (left) and momentum of a string piece (right).}
\label{fig:yoyo}
\end{figure}

The motion of the $q\A\bar{q}\B$ string in space-time is shown in Fig. \ref{fig:yoyo},
where the lightcone axes are indicated with $X^{\pm}=t\pm z$.
After being produced at the origin with energy $\sqrt{s}/2$, the quark and the anti-quark move along the lightcone in opposite directions losing energy-momentum due to the string tension. Meanwhile the string between them grows and reaches its maximum elongation at the time $t_0$, $q\A$ and $\bar{q}\B$ being at positions $\pm z_0$. The quarks have lost completely their energies which are now entirely in the string, whose length is $2z_0 = \sqrt{s}/\kappa$. The space-time points with coordinates $(t_0,\pm z_0)$ are \textit{turning-points} where the velocities of the quark and of the anti-quark are inverted. For $t>t_0$ the $q\A$ and $\bar{q}\B$ longitudinal momenta are thus inverted and they are pulled towards the origin taking energy-momentum from the string. At time $2t_0$ the quark and the anti-quark cross the origin with energy $\sqrt{s}/2$ and continue their motion in the $-\zu$ and $+\zu$ directions respectively. At time $3t_0$ an other inversion of the velocities occurs and finally, at the time $4t_0$, $q\A$ and $\bar{q}\B$ are back at their initial configuration. They are confined in the interval $(-z_0,+z_0)$ and this oscillatory motion is referred to as the \textit{yo-yo} motion.

Two general properties can be understood from the study of the string equation of motion and a summary is presented in Ref. \cite{Artru-HowStringsWork}. The energy-momentum (called in the following also simply momentum) of a string piece between the space-time points $A$ and $B$, as shown in the right panel of Fig. (\ref{fig:yoyo}) for a yo-yo, can be obtained as
\begin{equation}
    p_{AB} = \kappa \overrightarrow{CD}.
\end{equation}
The space-time points $C$ and $D$ are obtained from the intersection between the future and past light-paths passing through $A$ and $B$, and the vector $\overrightarrow{CD}$ points from $C$ to $D$.
The momentum flow to the string from right to left along the line which joins $C$ with $D$ is $\kappa \overrightarrow{AB}$ \cite{Artru-HowStringsWork}.

\begin{figure}
\centering
\begin{minipage}{.5\textwidth}
  \centering
  \includegraphics[width=.6\linewidth]{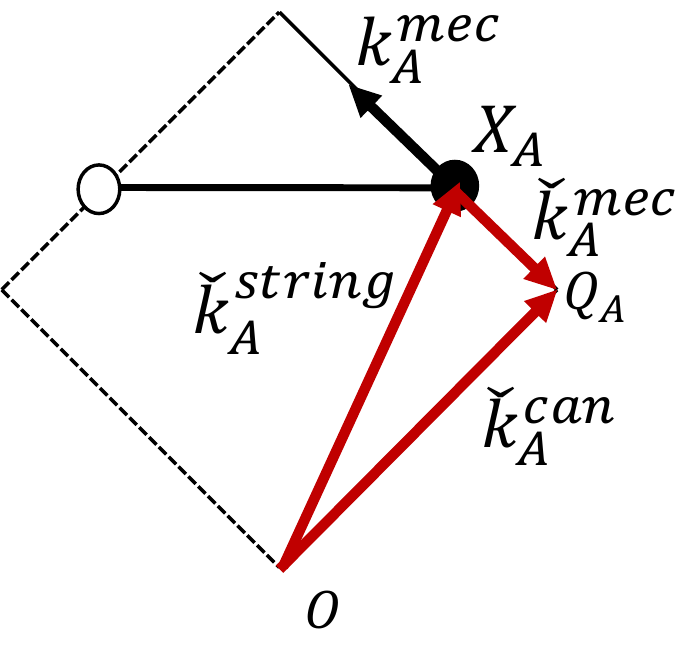}
\end{minipage}%
\caption{Representation of the quark momentum in the string surface.}
\label{fig:momenta}
\end{figure}

It is useful, for the following, to represent the quark momenta in the space-time surface swept by the string. This can be done in a simple way by defining the dual of a vector $k$ as $\check{k}=(k^z,k^0)$, namely as the vector with exchanged time and longitudinal components or equivalently as the reflection about the $X^+$ axis. Then the mechanical momentum of $q\A$ in Eq. (\ref{eq: k(t) and E(z)}) has the geometrical interpretation shown in Fig. \ref{fig:momenta}. The dual vector $\check{k}\A^{mec}$ at the space-time point $X_A$ is
\begin{equation}\label{eq: kA check}
    \check{k}\A^{mec} = \kappa \overrightarrow{X\A Q\A}.
\end{equation}
Namely the dual of the mechanical quark momentum at $X\A$ is given by the length of the vector joining $X\A$ with the turning point $Q\A$.
The mechanical momentum is then the reflection of $\check{k}_A^{mec}$ about $X^+$ axis.

The \textit{canonical} dual momentum of $q_A$ is
\begin{equation}\label{eq:decomposition of momenta}
    \check{k}^{can}\A=\kappa\overrightarrow{O Q\A}=\check{k}^{mec}\A+\check{k}^{string}\A,
\end{equation}
where $\check{k}\A^{string}=\kappa \overrightarrow{O X\A}$ is the dual of the momentum flow to the string from right to left along the line $OX\A$. This decomposition is shown in Fig. \ref{fig:momenta}. It is reminiscent of the relation $p_{mec}=p_{can}+eA$ between the mechanical momentum $p^{mec}$ and the canonical momentum $p^{can}$ of an electric charge $e$ which moves in the potential $A$. Thus the term $\check{k}_A^{string}$ plays the role of a (linear) 2-potential.

\begin{figure}
\centering
\begin{minipage}{.8\textwidth}
  \centering
  \includegraphics[width=.8\linewidth]{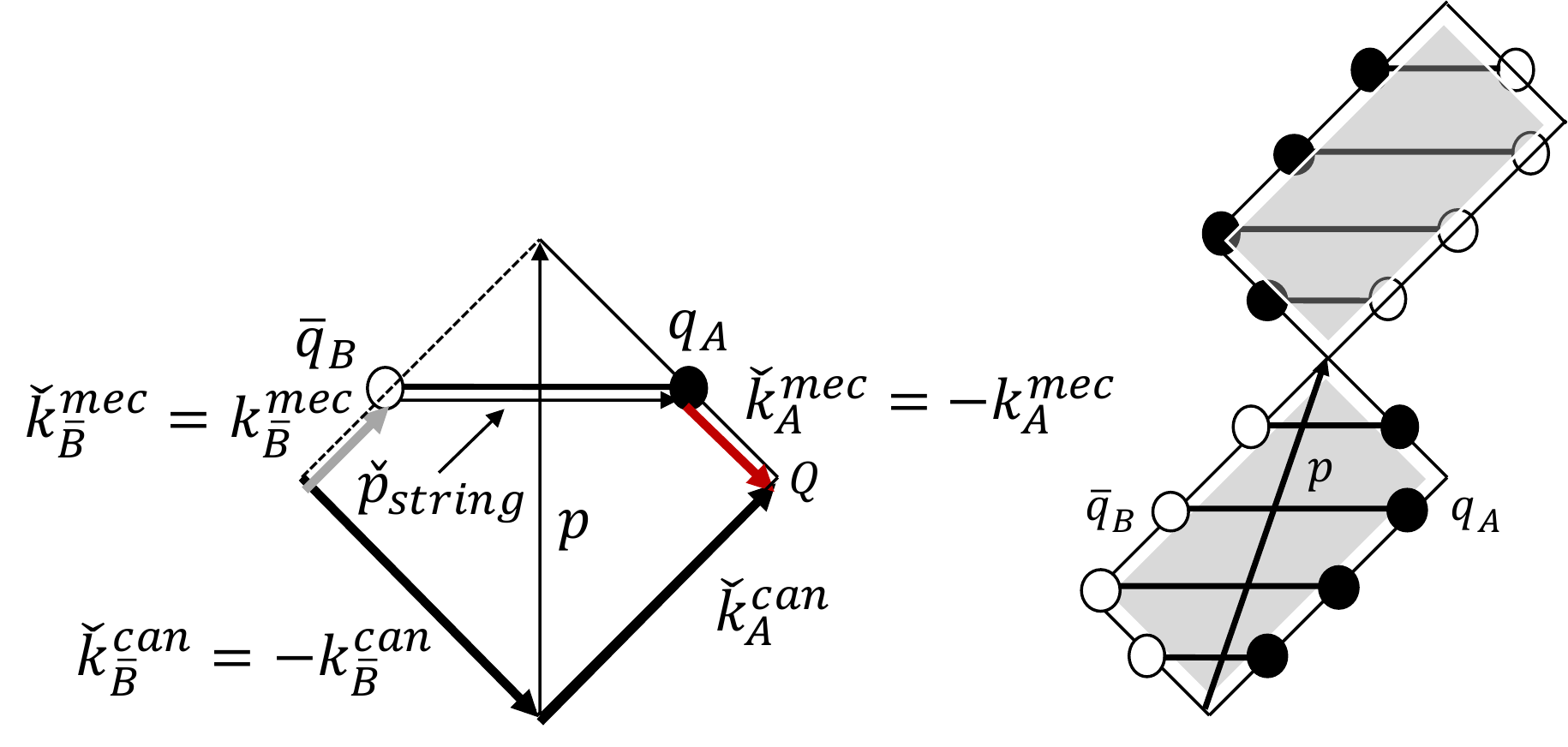}
\end{minipage}
\caption{Decomposition of the hadron momentum (left) and boosted yo-yo along the $\zu$ axis (right).}
\label{fig:yo-yo hadron}
\end{figure}

Finally, in the string fragmentation model, the yo-yo in Fig. \ref{fig:yoyo} represents a hadron $h$ at rest with quark content $q\A\bar{q}\B$ and with momentum
\begin{equation}
    p = k_A^{mec} + k_{\bar{\rm{B}}}^{mec}+p_{string} = k_A^{can}+k^{can}_{\bar{\rm{B}}},
\end{equation}
i.e. the hadron momentum is given by the sum of the quark mechanical momenta and the momentum $p_{string}$ of the string piece between them. It is also given by the sum of the quark canonical momenta, as shown in Fig. \ref{fig:yo-yo hadron}
The mass of the yo-yo hadron is determined by the mass-shell condition
\begin{equation}\label{eq:m_h^2=k^2 2A}
   m_h^2= k^{can\,+}\A k^{can \, -}_{\bar{\rm{B}}} = \kappa^2\, 2A,
\end{equation}
where $k^{\pm}$ indicate the lightcone components of the momentum vector $k$ and $A$ is the area of the space-time region swept by the string during half oscillation.
Hence, the mass corresponds to the area swept by the yo-yo in a complete oscillation and it is represented by the shaded region in Fig. \ref{fig:yoyo}.
A hadron moving along the $\zu$ axis is a boosted yo-yo as represented in the right picture of Fig. \ref{fig:yo-yo hadron}.

These basic properties of the yo-yo allow to study quantitatively its decay process arising when its mass is large enough and this process is introduced in the next section.

\subsection{The Lund string decay}
\begin{figure}
\centering
\begin{minipage}{.95\textwidth}
  \centering
  \includegraphics[width=.8\linewidth]{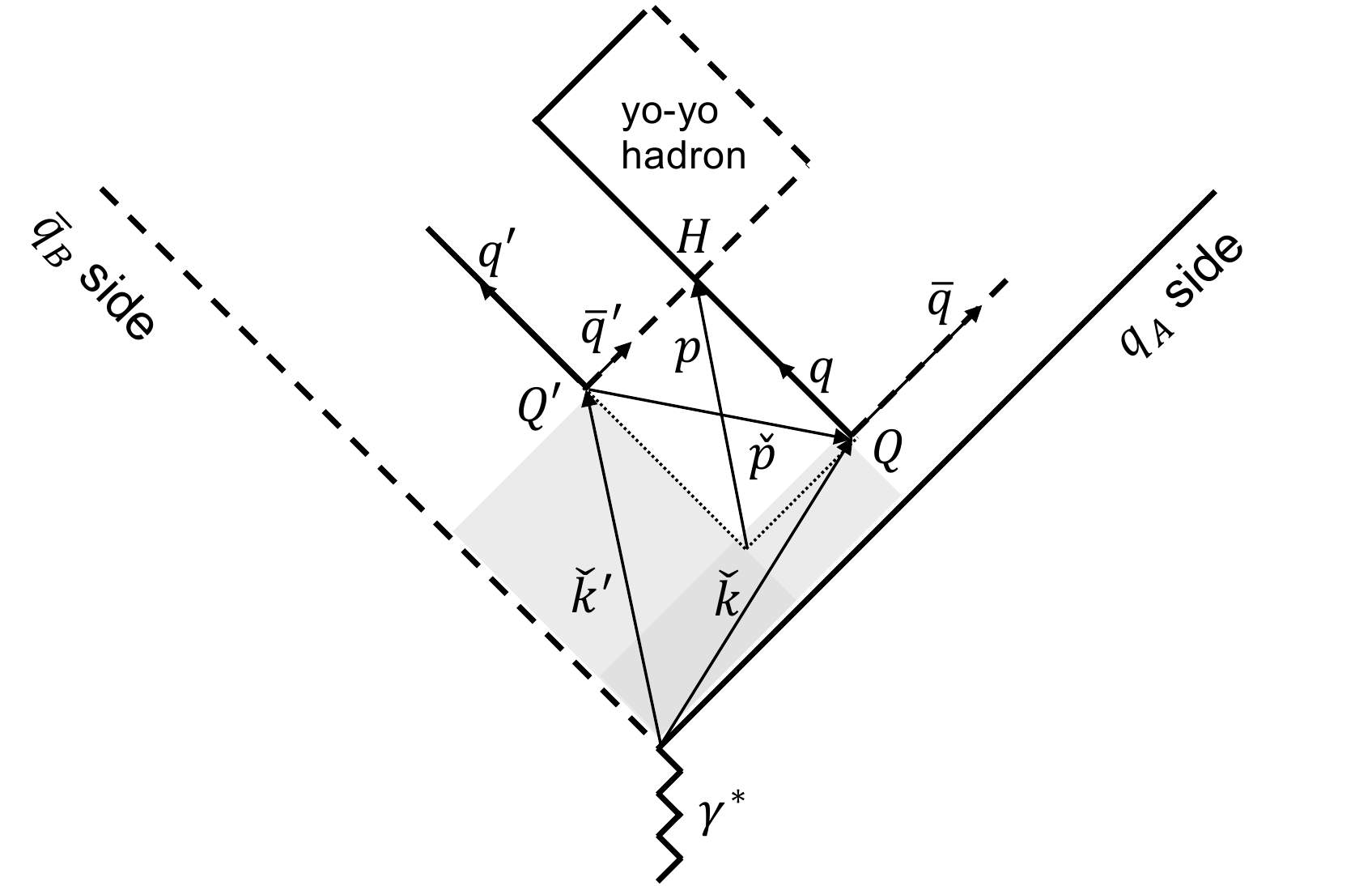}
\end{minipage}%
\caption{String breaking and hadron formation.}
\label{fig: hadron formation}
\end{figure}

If the $q\A\bar{q}\B$ yo-yo string is produced with a sufficiently large $\sqrt{s}$, rather than oscillating it may fragment into smaller strings, which represent stable hadrons or resonances. The string fragmentation process happens through the creation of string breaking points, like $Q$ and $Q'$ in Fig. \ref{fig: hadron formation}. Once a breaking point $Q$ occurs, there is no more string in its future light-cone, so any other breaking point $Q'$ must occur outside this space-time region.
At each breaking point a quark-antiquark pair of the same color as the chromoelectric force field is produced. Classically, at the string breaking points only massless quarks (anti-quarks) without transverse momenta can be generated. They are produced with vanishing energy and then pulled apart in the direction of the initial anti-quark (quark) by the force field. For instance the $\bar{q}$ of the $q\bar{q}$ pair produced in the space-time point $Q$ is pulled towards the $q\A$ side whereas the $q$ is pulled towards the $\bar{q}\B$ side. The quark $q$ takes energy from the $\bar{q}\B q$ string piece and travels along $X^-$. At the space-time point $H$, it meets the anti-quark $\bar{q}'$, produced in $Q'$ together with $q'$, which travels along $X^+$, pulled towards the $q\A$ side by the force field. The remaining $q'$ is pulled in the direction of $\bar{q}\B$.

This process divides the $q\A\bar{q}\B$ string into the string pieces $\bar{q}\B q'$, $\bar{q}q\A$ and $\bar{q}'q$ which is the yo-yo meson $h$. Requiring the momentum of $h$ to be time-like imposes $Q$ and $Q'$ to be at space-like separations and as a consequence the time ordering of $Q$ and $Q'$ is irrelevant. Indeed, Fig. \ref{fig: hadron formation} indicates that the string breaking at $Q$ happens before that at $Q'$. However, since $Q$ and $Q'$ are at space-like separations it is always possible to find a boosted system where this sequence of events is reversed. The process of string fragmentation is in fact invariant under boosts along the string axis, a property called \textit{longitudinal invariance} \cite{Artru-Mennessier}.

\begin{figure}
\centering
\begin{minipage}{.75\textwidth}
  \centering
  \includegraphics[width=.75\linewidth]{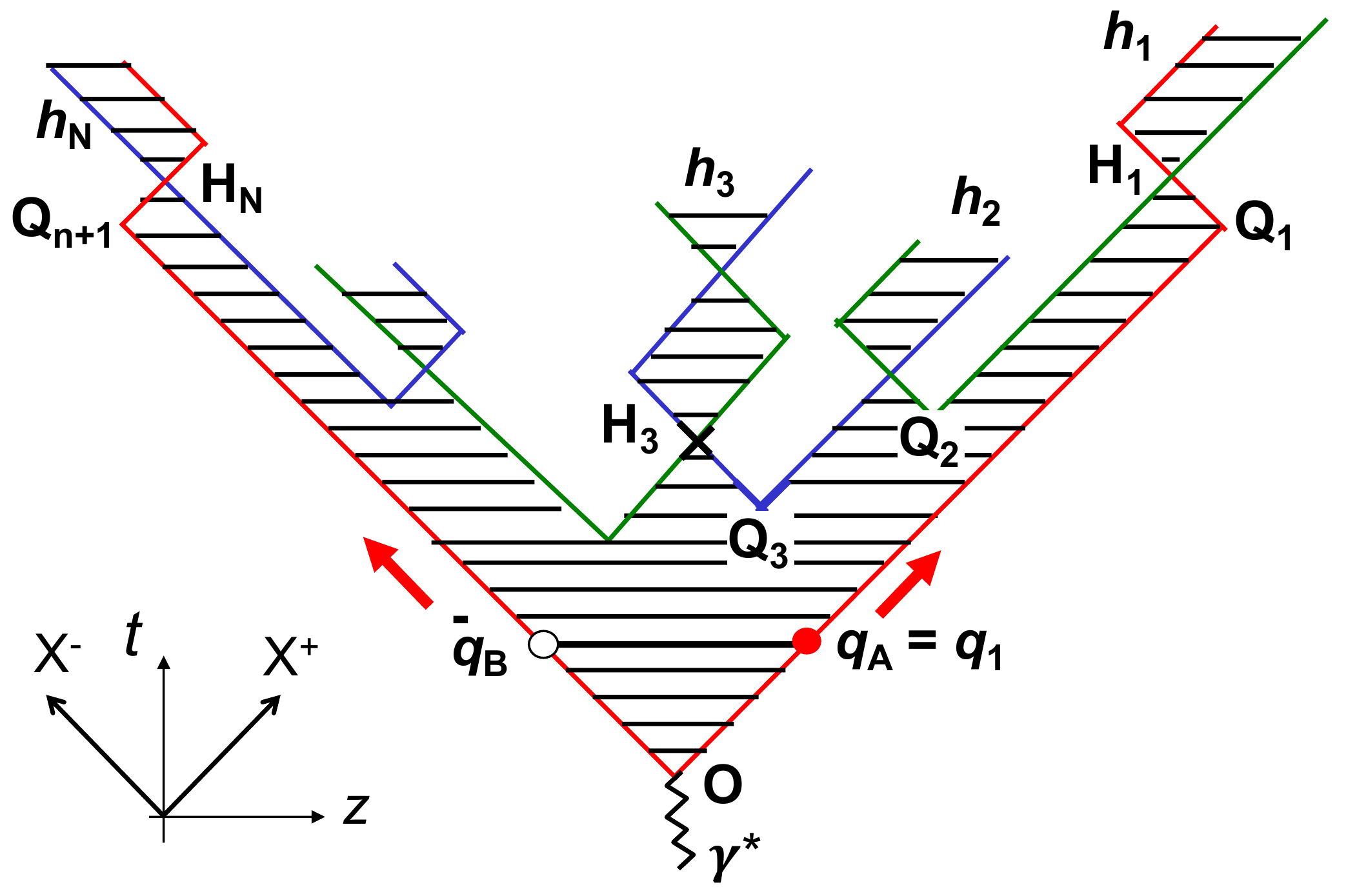}
\end{minipage}%
\caption{Space-time history of the decay process of a relativistic string.}
\label{fig: space-time history}
\end{figure}

The process of string breaking shown in Fig. \ref{fig: hadron formation} can be generalized as in Fig. \ref{fig: space-time history}, which represents a string fragmenting into the hadrons $h_1\dots h_N$. The string breaking points and the subsequent hadron formation occur all over the space-time surface swept by the string. The breaking points $Q_2,\dots , Q_N$ form an a-causal chain together with the turning points $Q_1\equiv Q\A$ and $Q_{N+1}\equiv Q_{\bar{\rm{B}}}$ of $q\A$ and $\bar{q}\B$. Quarks $q\A$ and $\bar{q}\B$ know about the string decay only after the turning points. Hadrons $h_1,\dots , h_N$ are emitted at $H_1,\dots , H_N$ with momenta $p_1,\dots , p_N$. The produced hadrons are ordered according to their ranks, the first rank being $h_1$ that contains the initial quark $q\A$. This ordering corresponds on the average to the ordering in rapidity space.

The momentum $p$ of a hadron $h$ can be read from Fig. \ref{fig: hadron formation}. Denoting with $X_Q$ and $X_{Q'}$ the coordinates of the points $Q$ and $Q'$, one can see that the lightcone components of $p$ are
\begin{eqnarray}\label{eq: p^+ p^-}
p^+ = \kappa (X_Q^+ - X_{Q'}^+ ), &&  p^- = \kappa (X_{Q'}^- - X_{Q}^-),
\end{eqnarray}
which can also be written in terms of the dual of the quark canonical momenta as
\begin{equation}
    \check{p}=\check{k}-\check{k}'.
\end{equation}
Here the superscript "can" has been suppressed and the canonical momenta are indicated with $k$ and $k'$. Only these will be used in the following.
A convenient parametrization of the hadron momentum is obtained using the longitudinal momentum fraction $Z_+=p^+/k^+$, which is invariant with respect to boosts along the string axis. The component $p^-$ is given by the mass-shell condition $p^- =m_h^2/(Zk^+)\,\,$\footnote{Alternatively one can use the negative momentum fraction $Z_- = p^- / \check{k}'^-$ together with $p^+ = m_h^2/p^-$.}.

Due to causality, the string fragmentation in Fig. \ref{fig: space-time history} can also be seen as the set of steps $p_1^+=Z_{1+}\check{k}^+\A,\cdots , p_N^+=Z_{N+}\check{k}^+_N$ in momentum space along the positive light-cone each accompanied by one step along the negative light-cone due to the mass-shell conditions $p_1^- = m_{h_1}^2/p_1^+, \dots , p_N^- = m_{h_N}^2/p_N^+$. The lightcone momenta obey the sum rules
\begin{eqnarray}\label{eq: sum Z_i = 1}
    \sum_i p_{i}^+ = k_A^+, &&  \sum_i p_{i}^- = k^{-}_{\bar{\rm{B}}},
\end{eqnarray}
as required by momentum conservation. All these steps can be seen as the recursive repetition of the elementary splitting
\begin{equation}\label{eq: q->h+q'}
q\rightarrow h(q\bar{q}') + q',
\end{equation}
namely a quark $q$ emits the hadron $h$, which flavour content is $q\bar{q}'$, leaving the quark $q'$ which is the next one to split. The elementary splitting is commonly described by a \textit{splitting function} which defines the energy-momentum sharing between $h$ and $q'$ and can be obtained from a stochastical description of the string fragmentation process, as discussed in the next section.

\subsubsection*{The Lund Symmetric Splitting function for massless quarks}
A quantitative description of the string fragmentation process and in particular of the elementary splitting, is given by the Lund Model (LM) \cite{Lund1983} where it is viewed as a two-step process.
Referring to Fig. \ref{fig: hadron formation}, first occurs a random string breaking in $Q$ with probability $H_q\,d^2X_Q$. The function $H_q$ is the unknown single quark density in space-time and depends only on $X_Q^2$ (the proper time of $Q$) due to the longitudinal invariance. Considering the probabilistic description of the string decay in momentum space, $H_q$ is commonly written in terms of the momentum space variable $\Gamma_Q =\kappa^2 X_Q^2$. Then probability to have a breaking point $Q$ in momentum space is\footnote{A constant factor arising from the transformation from space-time coordinates to momentum space variables has been neglected. It is $d^2X_Q=(1/2)dX_Q^+ dX_Q^- = dX_Q^2dy_Q/2=d\Gamma_Q\,dy_Q/(2\kappa^2)$.}
\begin{equation}\label{eq:H d^2XQ}
    H_q(\Gamma_Q)d\Gamma_Q\, dy_Q,
\end{equation}
where $y_Q=log(X_Q^+/X_Q^-)/2$. This variable is not invariant but additive under longitudinal boosts, hence $H_q$ can not depend on it.

Then, after $Q$, there is a second string breaking in $Q'$ such that the hadron $h$ is born in $H$. This occurs with probability
\begin{equation}\label{eq:f dZ}
    f_{q\rightarrow h + q'}(Z_+)dZ_+.
\end{equation}
The function $f_{q\rightarrow h + q'}$ is the unknown splitting function related to the elementary splitting. Due to longitudinal invariance it depends on $Z_+$ and implicitly on $p^-$ due to the mass-shell condition.

The joint probability of having the breaking point at $Q$ and emitting the hadron $h$ with one further step from $Q$ to $Q'$ (i.e. from right to left) is thus
\begin{equation}\label{eq: HF right -> left}
    H_q(\Gamma_Q)d\Gamma_Qdy_Q\,f_{q\rightarrow h + q'}(Z_+)dZ_+.
\end{equation}

Alternatively the same process can be thought to occur also from left to right, namely first there is a string breaking at $Q'$ and then a second string breaking at $Q$ and the hadron is formed in $H$. In this case the relevant variables are $\Gamma_{Q'}$, $y_{Q'}$ and $Z_-$ and the analogue of the joint probability is
\begin{equation}\label{eq: left -> right}
     H_{q'}(\Gamma_{Q'})d\Gamma_{Q'}dy_{Q'}\,f_{q'\rightarrow h + q}(Z_-)dZ_-.
\end{equation}

The main constraint in the SLM is the requirement that the probability of hadron emission occuring from right to left is the same as the probability for the same process to occur from left to right. This is known as \textit{Left-Right} (LR) symmetry and is the core assumption of the SLM \cite{LundModel-article}. Formally it is written as
\begin{equation}\label{eq: left-right prob}
    H_q(\Gamma_Q)d\Gamma_Q\,f_{q\rightarrow h + q'}(Z_+)dZ_+ =  H_{q'}(\Gamma_{Q'})d\Gamma_{Q'}\,f_{q'\rightarrow h + q}(Z_-)dZ_-,
\end{equation}
with a flat rapidity distribution of the breaking points, thus requiring $dy_Q = dy_{Q'}$. This equation can be simplified and written in terms of $Z_+$ and $Z_-$, using the relations
\begin{eqnarray}\label{eq: Gamma Q Gamma Q'}
    \Gamma_Q = (1-Z_-)\check{k}^+\check{k}'^- && \Gamma_{Q'} = (1-Z_+) \check{k}^+\check{k}'^-,
\end{eqnarray}
which can be deduced from Fig. \ref{fig: hadron formation}.
Then using also the mass-shell condition written as $m_h^2=Z_+Z_-\check{k}^+\check{k}'^-$, Eq. (\ref{eq: Gamma Q Gamma Q'}) become
\begin{eqnarray}\label{eq: Gamma Q Gamma Q' Z+ Z-}
    \Gamma_Q=\frac{1-Z_-}{Z_+Z_-}m_h^2 && \Gamma_{Q'}=\frac{1-Z_+}{Z_+Z_-}m_h^2.
\end{eqnarray}
Using also the relation
\begin{equation}
    d\Gamma_QdZ_+Z_+^{-1} = d\Gamma_{Q'}dZ_-Z_-^{-1},
\end{equation}
that follows from the equation above, the LR symmetry expression becomes
\begin{equation}\label{eq: HfZ}
    H_q(\Gamma_Q)f_{q\rightarrow h + q'}(Z_+)Z_+=H_{q'}(\Gamma_{Q'})f_{q'\rightarrow h + q}(Z_-)Z_-,
\end{equation}
where now $\Gamma_Q$ and $\Gamma_{Q'}$ are given by Eq. (\ref{eq: Gamma Q Gamma Q' Z+ Z-}).
 Taking the logarithm on both sides and defining $\log H_q = h_q$ and $\log Zf_{q\rightarrow h + q'} = g_{q\rightarrow h + q'}$, Eq. (\ref{eq: HfZ}) becomes
\begin{equation}\label{eq: hq + gq}
    h_q(\Gamma_Q)+g_{q\rightarrow h + q'}(Z_+)=h_{q'}(\Gamma_{Q'})+g_{q'\rightarrow h + q}(Z_-).
\end{equation}
Then, differentiating both sides first with respect to $Z_+$, which gives
\begin{equation}\label{eq: dH + dg}
    \frac{\partial h_q(\Gamma_Q)}{\partial Z_+} + \frac{dg_{q\rightarrow h+q'}}{dZ_+} = \frac{\partial h_{q'}(\Gamma_{Q'})}{\partial Z_+}
\end{equation}
and then with respect to $Z_-$, Eq. (\ref{eq: hq + gq}) becomes
\begin{equation}\label{eq: partial ^2 hq}
    \frac{\partial^2 h_q(\Gamma_Q)}{\partial Z_-\partial Z_+}  = \frac{\partial^2 h_{q'}(\Gamma_{Q'})}{\partial Z_-\partial Z_+}.
\end{equation}
Furthermore, from Eq. (\ref{eq: Gamma Q Gamma Q' Z+ Z-}) one can obtain the relations
\begin{eqnarray}\label{eq: derivatives Gamma_Q}
    \frac{\partial ^2 \Gamma_Q}{\partial Z_+\partial Z_-} = \frac{m_h^2}{(Z_+Z_-)^2}, && \frac{\partial ^2 \Gamma_{Q'}}{\partial Z_+\partial Z_-} = \frac{m_h^2}{(Z_+Z_-)^2}
\end{eqnarray}
which allow to re-write Eq. (\ref{eq: partial ^2 hq}) as
\begin{equation}\label{eq: equation for hq}
    \Gamma_Q\frac{d^2h_q}{d^2\Gamma_Q}+\frac{dh_q}{d\Gamma_Q} = \Gamma_{Q'}\frac{d^2h_{q'}}{d^2\Gamma_{Q'}}+\frac{dh_{q'}}{d\Gamma_{Q'}}.
\end{equation}
Since the left hand side of this differential equation depends only on the $\Gamma_Q$ and the right hand side depends only on $\Gamma_{Q'}$, it means that each side must be a constant, which is called $-b$. Then, the solution of Eq. (\ref{eq: equation for hq}) is \cite{LundModel-article}
\begin{equation}\label{eq: H solution}
    H_q(\Gamma_Q) \propto \Gamma_Q^{a_q}e^{-b \Gamma_Q}.
\end{equation}
Namely the function $H_q$ depends on the two unknown integration constants $a_q$ and $b$. The parameter $a_q$ may depend on the quark flavour $q$ whereas $b$ is a universal constant linked to the probability of having a string breaking per unit space-time area \cite{Artru-Mennessier,XA1984}. 

Using Eq. (\ref{eq: H solution}), the differential equation in Eq. \ref{eq: dH + dg} can be solved for $f_{q\rightarrow q'+h}$ and it gives \cite{Lund1983}
\begin{equation}\label{eq: f solution no kT}
    f_{q\rightarrow h + q'}(Z_+) \propto Z_+^{-1} \left(\frac{1-Z_+}{Z_+}\right)^{a_{q'}}\left(\frac{Z_+}{m_h^2}\right)^{a_q}e^{-b m_h^2/Z_+}.
\end{equation}
This function is known as the \textit{Lund Symmetric Splitting Function} (LSSF) and it gives the probability that the hadron $h$ is emitted in the elementary splitting in Eq. (\ref{eq: q->h+q'}) with longitudinal momentum fraction $Z_+$. It depends on the integration constants $a_q$, $a_{q'}$ and $b$.



\subsubsection*{Quark masses and transverse momenta}
The quark masses and transverse momenta at string breaking were neglected in the previous section. They can be taken into account assuming that the $q\bar{q}$ pair is produced by a mechanism similar to the Schwinger mechanism of $e^+e^-$ pair creation in a strong electric field \cite{Schwinger-Mechanism}.

Then the $q$ and the $\bar{q}$ tunnel out of the force field as virtual particles at the same point, with compensating transverse momenta $\kt$ and $-\kt$ with respect to the string axis. Afterwards they are pulled apart by the force field in opposite directions and become real at the distance $d=2\sqrt{m_q^2+\ktkt}/\kappa$. The field energy between them has been converted in their transverse energies. The transverse momenta of the quarks are absorbed by the hadrons they will constitute.

\begin{figure}
\centering
\begin{minipage}{.75\textwidth}
  \centering
  \includegraphics[width=.75\linewidth]{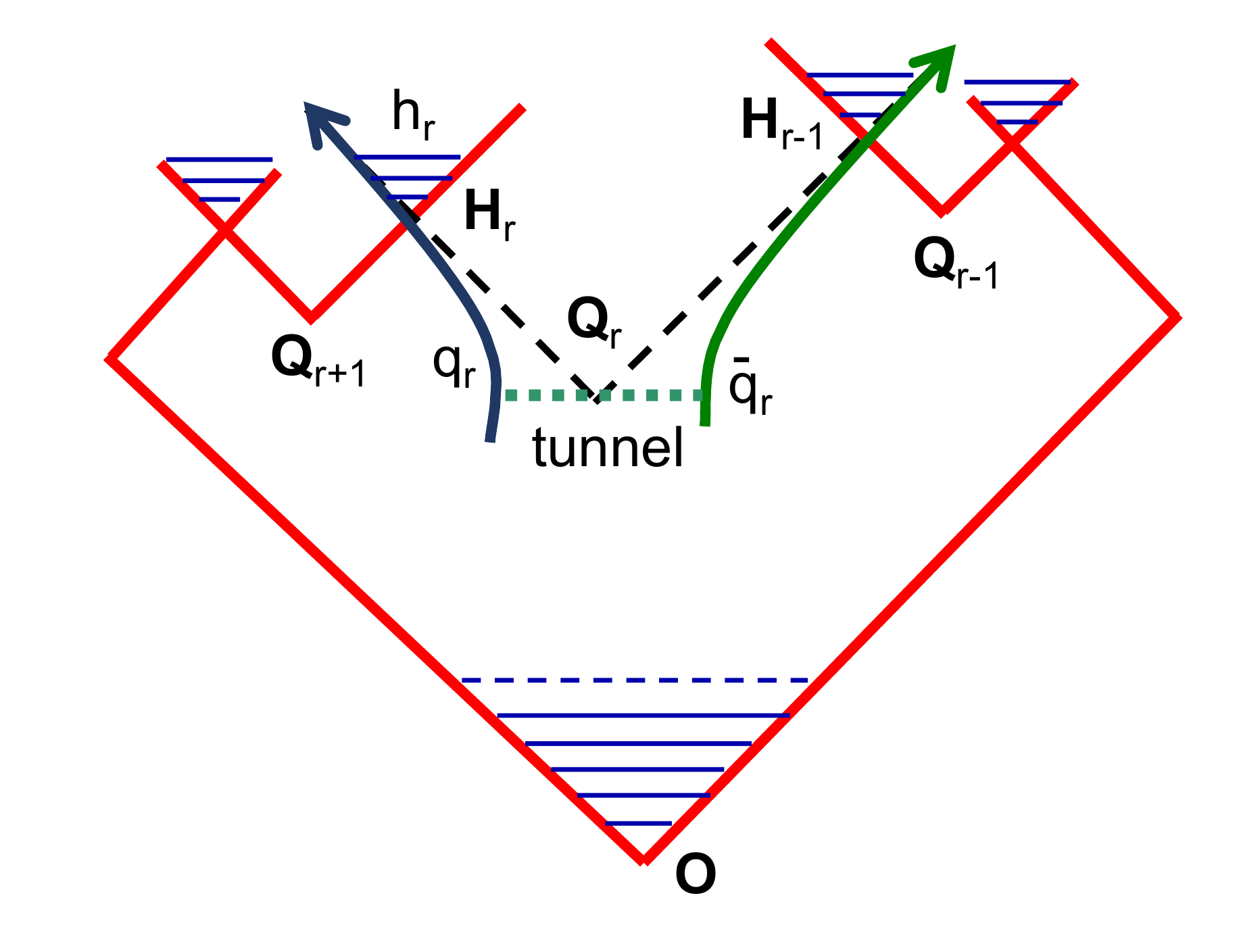}
\end{minipage}%
\caption{Analogue of the Schwinger mechanism for string breaking with quark masses and transverse momenta.}
\label{fig: tunnel}
\end{figure}

This makes the picture of the string breaking in Fig. \ref{fig: hadron formation} approximate. If the quark is massive but with $\kt=0$ it follows an hyperbolic trajectory as shown in Fig. (\ref{fig: tunnel}). For $\kt\neq 0$ the motion is more complicated, but still with the dashed lines as asymptotes. Nevertheless, the space-time point $Q$, defined as the intersection between the asymptotes of the $q$ and of the $\bar{q}$ motions, is what matters.

With these considerations, the tunneling probability of a $q\bar{q}$ is \cite{Lund1983,Casher-Nussinov}
\begin{equation} \label{eq: P Schwinger}
    P(q\bar{q})\,d^2\kt \propto e^{-\pi m_q^2/\kappa}\, e^{-\pi\ktkt/\kappa}\,d^2\kt.
\end{equation}
This formula concerns the flavor and transverse momenta of quarks produced at the string breaking, and it is complementary to $H_q$ in Eq. (\ref{eq: H solution}) which deals with the longitudinal momenta.

The first exponential suppresses large quark masses. Note that the transverse energy $\sqrt{m_q^2+\ktkt}$ appears instead of the mass. In order to take into account the transverse momenta, the hadron mass $m_h$ in Eqs. (\ref{eq:m_h^2=k^2 2A}, \ref{eq: derivatives Gamma_Q}, \ref{eq: f solution no kT}) is also replaced with the transverse energy $\varepsilon_h=\sqrt{m_h^2+\ptpt}$ \cite{Lund1983}. The transverse momentum of $h$ is $\pt=\kt-\kpt$. The LSSF with quark transverse momenta thus becomes
\begin{equation}\label{eq: f solution}
    f_{q\rightarrow h + q'}(Z,\pt;\kt) \propto Z^{-1} \left(\frac{1-Z}{Z}\right)^{a_{q'}}\left(\frac{Z}{\varepsilon_h^2}\right)^{a_q}e^{-b \varepsilon_h^2/Z},
\end{equation}
where $Z_+$ has been simply replaced with $Z$.

Using current quark masses for $m_q$, the probability of producing heavy quarks is vanishingly small meaning that only $u,d,s$ flavours can be produced at string breakings \footnote{For instance $P(c\bar{c})/P(u\bar{u})\simeq 10^{-11}$ and even smaller for $b\bar{b}$. Heavy quarks can appear only as endponits of the initial string, for instance in processes like $e^+e^-\rightarrow c\bar{c}\rightarrow$ hadrons or $e^+e^-\rightarrow b\bar{b}\rightarrow$ hadrons. For these quarks the splitting function is slightly different with respect to the LSSF, as shown in Ref. \cite{Bowlere+e-}.}.
For the light quarks, Eq. (\ref{eq: P Schwinger}) gives the relative probabilities
\begin{equation}\label{eq:uds prob}
    P(u\bar{u}):P(d\bar{d}):P(s\bar{s})=\alpha:\alpha:1-2\alpha.
\end{equation}
Hence strange quarks are suppressed with respect to $u$ and $d$ quarks, and in Monte Carlo simulations the suppression factor $P(s\bar{s})/P(u\bar{u}\, \rm{or}\, d\bar{d})$ is typically about $0.33$.

The second exponential in Eq. (\ref{eq: P Schwinger}) suppresses large quark transverse momenta and, due to the uncertainty relation $\Delta k_x \Delta x>\hbar/2$, implies a lower limit $\Delta x> \sqrt{\pi/(2\kappa)}$ on the transverse localization of the quark. Using for the string tension the value $\kappa=0.2\,\rm{GeV}^2$ \cite{Andersson:1979-lambda0pol}, the typical width of the quark transverse momentum according to Eq. (\ref{eq: P Schwinger}) is $\sqrt{\langle \ktkt \rangle} = \sqrt{\kappa /\pi} \simeq 0.25\, (\rm{GeV}/c)$, where the $c$ factor has been restored. In Monte Carlo simulations this factor is usually replaced by a phenomenological parameter, fitted to the experimental data. 


Note that to obtain Eq. (\ref{eq: P Schwinger}) a constant color field extended all over the space-time has been assumed. A finite extention of the color field changes slightly the quark production probabilities but for Monte Carlo simulations this turns out to be a small effect \cite{Andersson-ThreeDimensionalModel}.

\subsubsection*{The implementation of the LSSF in PYTHIA}
The LSSF is the default option for the generation of the hadron momentum in string fragmentation in the event generator \verb|PYTHIA| \cite{pythia8}. \verb|PYTHIA| is a general purpose and complete event generator, capable of simulating in detail different processes like DIS, $e^+e^-$ annihilation and $pp$ scattering. Each of these processes comes with its own complications and are treated with great care. However they all have in common the hadronization process, which is based on the symmetric Lund model where strings are stretched between color charges produced in different stages of the event generation. There are many string configurations possible. A string can be stretched between a quark and an anti-quark like in $e^+e^-$ event or between a quark and a di-quark for instance in a DIS event. There can also be gluons between the string end points.

However, regardless of the complications of the particular string configuration, each string is fragmented recursively repeating the splitting in Eq. (\ref{eq: q->h+q'}), starting from its endpoints. In the spirit of LR symmetry, each splitting is taken from the quark side of or from the anti-quark side, with equal probability generating in this way the quark and the anti-quark jets. The two jets are then joined on the average in the central rapidity region by a dedicated recipe. The whole procedure ensures conservation of momentum, of charge and the involved quantum numbers.

In particular, concentrating on the production of mesons, each splitting consists in the generation of the flavour of a new $q'\bar{q}'$ pair. Afterwards the type of the emitted hadron $h=q\bar{q}'$ is determined. As for the momentum $p$ of $h$, it is obtained first generating $\kpt$ according to a double exponential distribution
\begin{eqnarray}\label{eq:kT' Pythia}
d^2\kpt \left(p_0 e^{-\kptkpt/\sigma_0^2}+p_1e^{-\kptkpt/\sigma_1^2}\right),
\end{eqnarray}
with $p_0\gg p_1$, which allows to calculate $\pt=\kt-\kpt$, and then generating $Z$ according to the LSSF
\begin{equation}\label{eq: Z Pythia}
    dZZ^{-1} (1-Z)^a\exp\left(-b\varepsilon_h^2/Z\right).
\end{equation}
Since the momentum $k$ of $q$ is known, the energy and longitudinal component of $p$ are obtained from $p^+=Zk^+$ and $p^-=\varepsilon_h^2/p^+$.

The two steps of the generation of $p$ can be gathered in a joint splitting function of $\kpt$ and $Z$. The splitting function of the "\verb|PYTHIA| recipe" is thus
\begin{eqnarray}\label{eq:F Pythia}
\nonumber    F_{q',h,q}(Z,\pt)\frac{dZ}{Z} d^2\kpt &\propto& \frac{dZ}{Z} \left[\left(1-Z\right)/\varepsilon_h^2\right]^a\exp\left(-b\varepsilon_h^2/Z\right)N_a^{-1}(\varepsilon_h^2)
    \\
    &\times& d^2\kpt\left(p_0 e^{-\kptkpt/\sigma_0^2}+p_1e^{-\kptkpt/\sigma_1^2}\right),
\end{eqnarray}
where the normalization function $N_a(\varepsilon_h^2)$ is defined as
\begin{equation}\label{eq: Na}
    N_a(\varepsilon_h^2) = \int_0^1 dZ\,Z^{-1} \left(\frac{1-Z}{\varepsilon_h^2}\right)^a\exp{\left(-b\varepsilon_h^2/Z\right)},
\end{equation}
with a flavour-independent $a_q\equiv a_{q'}= a$.
In \verb|PYTHIA|, first is generated $\kpt$ according to the $Z$-integrated splitting function of Eq. (\ref{eq:F Pythia}) and then is generated $Z$.


\subsection{The symmetric splitting function}
To summarize what has been described in the previous sections, for given a quark $q$ with momentum $k$ which transverse component is $\kt$, the probability
\begin{equation}\label{eq:splitt prob}
dP_{q\rightarrow h+q'} \equiv F_{q',h,q}(Z,\pt;\kt)dZZ^{-1}d^2\pt
\end{equation}
for the hadron $h$ to be emitted with longitudinal momentum fraction $Z=p^+/k^+$ and with transverse momentum $\pt=\kt-\kpt$, where $\kpt$ is the transverse momentum of $q'$, is normalized according to
\begin{eqnarray}\label{eq:normalization}
\sum_{h}\int_0^1\frac{dZ}{Z}\int d^2\pt F_{q',h,q}(Z,\pt;\kt)=1
\end{eqnarray}
and the phase space element $dZZ^{-1}d^2\pt$ is the same as the invariant phase space factor $d^3p/p^0$.
The most general form of the splitting function $F_{q',h,q}$ allowed by LR symmetry is \cite{kerbizi-2018}
\begin{eqnarray}\label{eq: F_q'hq}
\nonumber    F_{q',h,q}(Z,\pt;\kt)&=&\left(\frac{1-Z}{Z}\right)^{a_{q'}(\kptkpt)}\left(\frac{Z}{\varepsilon_h^2}\right)^{a_q(\ktkt)}\,\exp\left(-\bl\varepsilon_h^2/Z\right)\\
    &\times& w_{q',h,q}(\kptkpt,\ptpt,\ktkt)\,u_{q}(\ktkt)^{-1},
\end{eqnarray}
which depends on the longitudinal momentum fraction $Z$, on the quark $q$ ($q'$) transverse momenta $\kt$ ($\kpt$) and on the hadron momentum $\pt$.

The first line has the same analytical form as the longitudinal splitting function given in Eq. (\ref{eq: f solution}) and depends on the hadrons transverse energy squared $\varepsilon_h^2 = m_h^2+\ptpt$ through the substitution $m_h^2\rightarrow \varepsilon_h^2$. 
The inputs of the splitting function are the parameter $a_q(\ktkt)$ and the function $w_{q',h,q}(\kptkpt,\ptpt,\ktkt)$ which depend on the quark flavours $q$ and $q'$, on the hadron type $h$ and on their transverse momenta.
As shown in Ref. \cite{DS11}, in a semi classical approach the factor $a_q(\ktkt)$ is produced by the quantum quark actions which give $a_q(\ktkt)=\alpha_{out}(0)-2\bl(m_q^2+\ktkt)$ , $\alpha_{out}$ being the Regge intercept \cite{XA1984} and $\log\left[\bl(m_q^2+\ktkt)\right]$ the quark action along the hyperbola shown in Fig. \ref{fig: tunnel}. The parameter $\bl$ is the same as $b$ in Eq. (\ref{eq:F Pythia}).

The function $w_{q',h,q}$ gathers the transverse degrees of freedom of $q$ and $q'$ and in order to satisfy the LR symmetry must be symmetrical under the transformation $\lbrace q,\kt,h\rbrace\rightleftharpoons \lbrace q',\kpt ,\bar{h}\rbrace$. In the following it will be taken of the factorized form
\begin{equation}\label{eq: w_q'hq}
    w_{q',h,q}(\kptkpt,\ptpt,\ktkt) = |C_{q',h,q}\,\check{g}(\varepsilon_h^2)\,f\T(\kptkpt)f\T(\ktkt)|^2,
\end{equation}
where the factor $C_{q',h,q}$ takes into account the suppression of $s$ quarks with respect to $u$ and $d$ and is proportional to the hadron $h$ wave function in flavor space $\langle h | q\bar{q}'\rangle$. The function $f\T$ is a fast decreasing function of the quark transverse momentum. It can be a single exponential
as in Eq. (\ref{eq: P Schwinger}) replacing $\pi/\kappa$ with a phenomenological parameter\footnote{As initially proposed in the Field-Feynman independent fragmentation model \cite{Field-Feynman}.}, or it can be a double exponential as in \verb|PYTHIA|. $f\T$ is however a phenomenological function and alternative forms are possible.
The function $\check{g}$ is an input to the model. It depends on $\varepsilon_h^2$ and is therefore symmetric in $\kt$ and $\kpt$. The $\varepsilon_h$ dependence of $\check{g}$ mixes with that in the exponential factor $\exp(-\bl \varepsilon_h^2/Z)$ and depending on the particular choice of the functional form, there may be spin-independent correlations between $\kt$ and $\kpt$ in the fragmentation process \cite{Andersson-kt-correlations,artru-belghobsi-essma}, as will be discussed in detail in the following. In the implementation of the SLM in \verb|PYTHIA| it is $\check{g}^2(\varepsilon_h^2)=1/N_a(\varepsilon_h^2)$, where $N_a$ is the normalization function in Eq. (\ref{eq: Na}).

Finally, the function $u_q(\ktkt)$ is introduced to normalize the splitting function and it depends on the quark $q$ flavour and its transverse momentum $\kt$. Using the normalization condition in Eq. (\ref{eq:normalization}) it is
\begin{eqnarray}\label{eq: uq unpolarized}
  \nonumber  u_q(\ktkt)&=&\sum_h u_{q,h}(\ktkt)=\sum_h\int d^2\kpt\,w_{q',h,q}(\kptkpt,\ptpt,\ktkt) \\
  &\times& \int_0^1 dZZ^{-1} \left(1/Z-1\right)^{a_{q'}}\left(Z/\varepsilon_h^2\right)^{a_q} \exp\left(-\bl\varepsilon_h^2/Z\right).
\end{eqnarray}
For a generic choice of $\check{g}$, the probability of producing the hadron type $h$, obtained integrating Eq. (\ref{eq: F_q'hq}) on $Z$ and $\pt$, may depend on $\kt$ due to the $\kt$ dependence of $u_q$. For the choice $\check{g}^2=1/N_a$, $u_q$ is simply a constant.

\subsection{The string decay in multiperipheral form}\label{sec: string decay in mult form}
For the inclusion of the quark spin in the string fragmentation picture and for the implementation of the resulting polarized fragmentation model in a MC code, it is useful to introduce the analogy between string fragmentation and the multiperipheral model \cite{XA1984}.

\begin{figure}
\centering
\begin{minipage}{.95\textwidth}
  \centering
  \includegraphics[width=.95\linewidth]{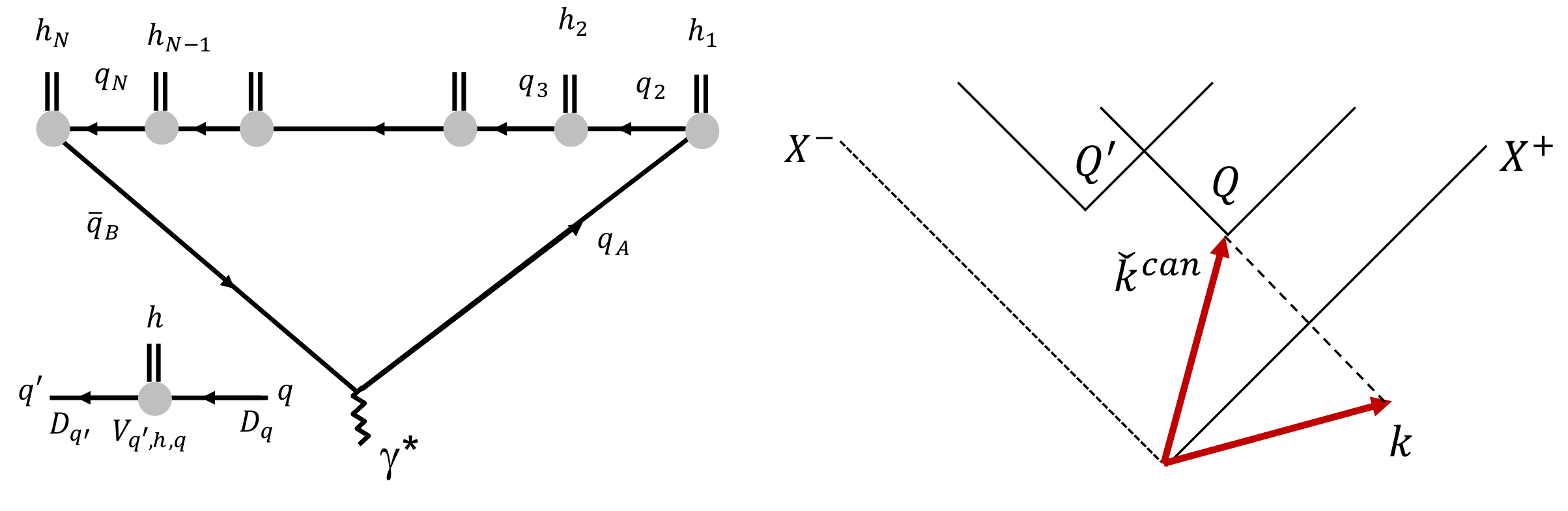}
\end{minipage}%
\caption{Multiperipheral diagram (left) and identification of the canonical quark momenta in the string decay diagram with the momenta of the multiperipheral diagram (right).}
\label{fig:multiperipheral}
\end{figure}

Indeed, the string decay in Fig. \ref{fig: space-time history} can be viewed as a multiperipheral diagram with quark exchanges represented in the left panel of Fig. \ref{fig:multiperipheral}. The equivalence between the two diagrams holds if the canonical quark momenta in the string decay are identified with the quark momenta of the multiperipheral diagram, as shown in right panel of Fig. \ref{fig:multiperipheral}.
Taking $X^-$ as time axis, both processes can be thought as the set of splittings
\begin{eqnarray}\label{eq:set of splittings}
q\A\rightarrow h_1+q_2,\, q_2\rightarrow h_2+q_3,\dots , q_r\rightarrow h_r + q_{r+1},\dots q_N\rightarrow h_N+q\B,
\end{eqnarray}
or as the recursive application of the elementary splitting $q\rightarrow h + q'$
each described by the splitting function as in Eq. (\ref{eq:splitt prob}).

The splitting function can also be defined using the basic elements of the multiperipheral diagram, namely the quark propagator $D_q(k)$ and the vertex function $V_{q',h,q}(k',k)$.
Indeed, introducing the transition amplitude
\begin{equation}\label{eq:T}
    T_{q',h,q}(p,k)=V_{q',h,q}(k',k)\,D_q(k)
\end{equation}
that describes the elementary splitting, the corresponding splitting function can be written as
\begin{equation}\label{eq:F = TT*}
    F_{q',h,q}=T_{q',h,q}\,T^*_{q',h,q}.
\end{equation}
By probability considerations, the splitting function can also be expressed as
\begin{equation}\label{eq: F = Nqq'/Nq}
    F_{q',h,q} = \frac{dN_{q',q}(k',k)}{d^4k'd^4k}\,\left(\frac{dN_q(k)}{d^4k}\right)^{-1},
\end{equation}
where $dN_{q',q}(k',k)/d^4k'd^4k$ and $dN_q(k)/d^4k$ are the double and the single quark densities of the multiperipheral chain in momentum space.
By comparing Eq. (\ref{eq: F = Nqq'/Nq}) with Eq. (\ref{eq:F = TT*}) one obtains that in the multiperipheral model the square of the vertex function is related to the double quark density according to
\begin{equation}\label{eq:V^2}
    \frac{dN_{q',q}(k',k)}{d^4kd^4k'} =2\delta(p^2-m_h^2) V_{q',h,q}(k',k)V^*_{q',h,q}(k',k),
\end{equation}
and that the inverse of the quark propagator squared is related to the single quark density according to
\begin{equation}\label{eq:D^-2}
    \frac{dN_q(k)}{d^4k}=\left(D_q(k)\,D^*_q(k)\right)^{-1}.
\end{equation}
The delta function in Eq. (\ref{eq:V^2}) takes into account the mass shell condition for the emitted hadron $h$. Finally, the propagator and the vertex are related by the normalization condition in Eq. (\ref{eq:normalization}) which gives
\begin{equation}\label{eq: U = int V*V}
    \frac{1}{D_q(k)\,D^*_q(k)}\equiv U(q) =\int \frac{d^3p}{p^0} V_{q',h,q}(k',k)\,V^*_{q',h,q}(k',k).
\end{equation}

\begin{figure}
\centering
\begin{minipage}{.8\textwidth}
  \centering
  \includegraphics[width=.8\linewidth]{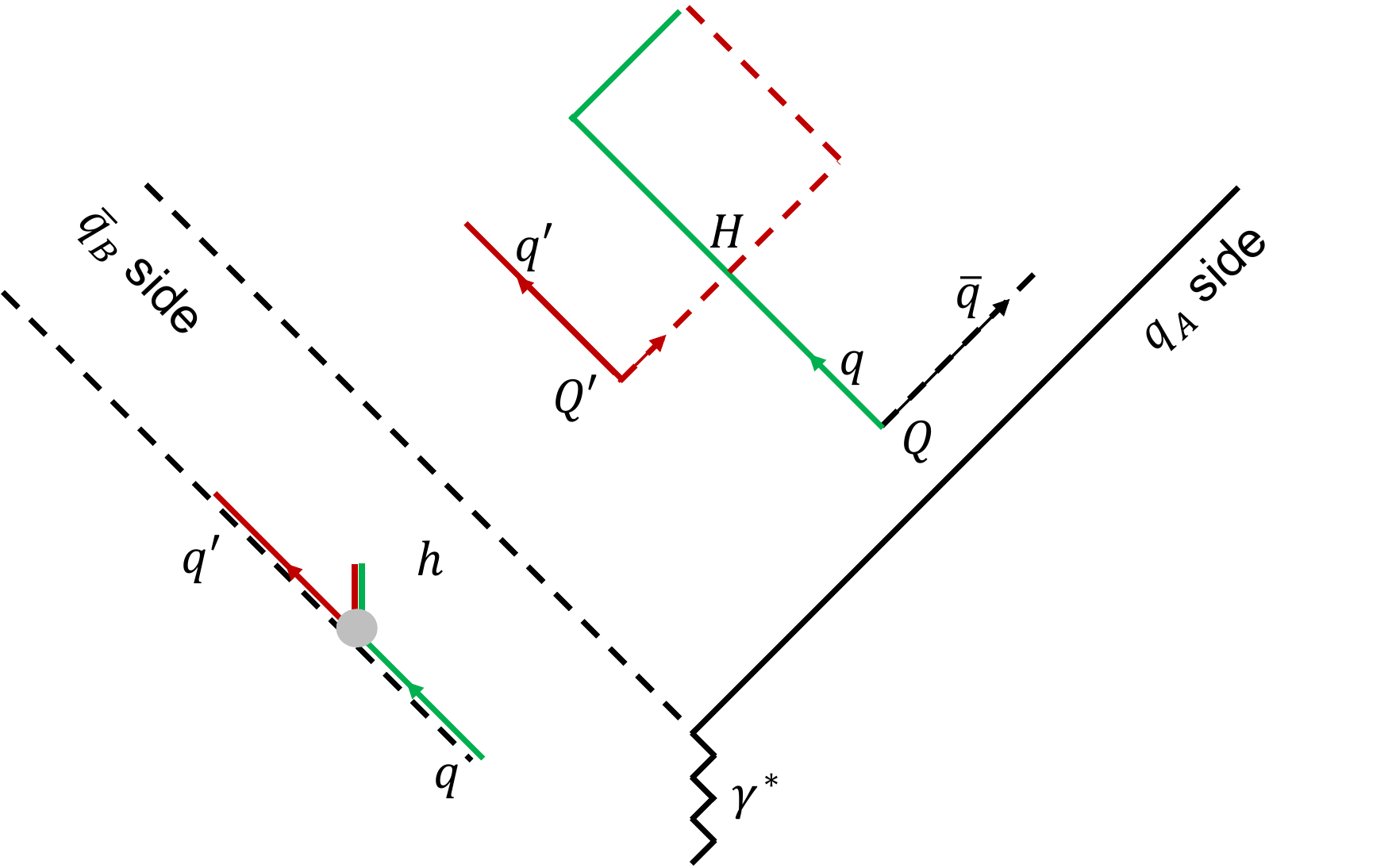}
\end{minipage}%
\caption{String decay diagram viewed as a multiperipheral diagram.}
\label{fig:identification}
\end{figure}

The hadron emission vertex $V_{q',h,q}$ and the quark propagator $D_q$ can be extracted from the string decay diagram by noticing that the quark anti-quark and the hadron production points $Q'$ and $H$ (see Fig. \ref{fig: hadron formation}) occur at the same $X^-$ coordinate. One can view this point and the nearby quark lines as a vertex and as propagators of the corresponding multiperipheral diagram, as shown in Fig. \ref{fig:identification}, by identifying the canonical momenta of the string diagram with the quark momenta of the corresponding multiperipheral diagram. In addition, from Eq. (\ref{eq:V^2}), the square of the vertex in the multiperipheral model corresponds to the double quark density in momentum space, which in the LM is $F_{q',h,q}H_q$. Hence, by rewriting Eq. (\ref{eq:D^-2}) in terms of the quark canonical momenta, it is
\begin{equation}\label{eq: V^2 string}
    |V_{q',h,q}|^2 =\left(\frac{k'^+}{p^+}\right)^{a_{q'}}\bigg\rvert \frac{k^-}{p^-}\bigg\lvert^{a_q}e^{-\bl|k'^-k^+|}\,w_{q',h,q}(\kptkpt,\ptpt,\ktkt).
\end{equation}
The inverse of the quark propagator squared, being identified with the single quark density in momentum space, is simply the function $H_q^{-1}$ given in Eq. (\ref{eq: H solution}). Then writing it in terms of the quark canonical momentum momentum $k$, one obtains the propagator
\begin{equation}\label{eq: D^-2 string}
    |D_{q}(k)|^{-2} = |k^+k^-|^{a_q}e^{-\bl|k^+k^-|}\,u_q(\ktkt)\equiv U_q(k).
\end{equation}

As a final consideration on the properties of the LM, looking at Eqs. (\ref{eq: V^2 string}-\ref{eq: D^-2 string}) and remembering that the string axis is taken as $\zu$ axis, one can note that the vertex and the propagator as obtained from the string decay diagram are invariant with respect to the subgroup of transformations generated by
\begin{itemize}[noitemsep,topsep=0pt,parsep=0pt,partopsep=0pt]
    \item[(S1)] the rotations about the string axis
    \item[(S2)] the Lorentz boosts along the string axis
    \item[(S3)] the reflections about any plane containing the string axis.
\end{itemize}
This is due to the fact that the string axis breaks space isotropy and defines a privileged direction. The invariance under the full Lorentz group is not needed, and in extending the LM to the inclusion of quark spin, only symmetries $S1$-$S3$ and the LR symmetry will be required.

\section{The quark spin in the string fragmentation process}\label{sec: introduction of spin}
The spin of the fragmenting quark in the string decay process has been neglected throughout all the previous section. Its role and propagation along the fragmentation chain is the topic of this work. In this section, this new part is treated in detail, up to the calculation of the spin dependent splitting matrix. The discussion of the different variants of the model is the subject of section \ref{sec: general pol F of the string + 3P0}.

As already mentioned, the transverse momenta of $q$ and $\bar q$ created at each string breaking compensate, so that the total transverse momentum is conserved (assuming that the string does not have transverse vibrations). However due to energy conservation the $q$ and the $\bar q$ are produced on separate points along the string axis, therefore they possess some relative orbital angular momentum. 
This breaks the total angular momentum conservation.
The ${}^3P_0$ mechanism explained in Fig. \ref{fig:classical3P0} can restore this conservation. In the following section this mechanism is presented in more detail and the quantum mechanical formulation of the ${}^3P_0$ model, which is at the basis of this work, is also introduced.

\subsection{The $string+ ^3P_0$ mechanism}
As shown in Fig \ref{fig:classical3P0}, at each string breaking, the $q\bar{q}$ pair tunnels out of the force field becoming real at the distance
\begin{equation}
    \textbf{d}=\textbf{r}_q-\textbf{r}_{\bar{q}}=-\frac{2}{\kappa}\sqrt{m_q^2+\ktkt}\,\, \zu.
\end{equation}
At this point, they lay along the string axis with vanishing longitudinal momenta and compensating transverse momenta $\kt$ and $-\kt$. Their relative distance is oriented from $\bar{q}$ to $q$ and is fixed by energy conservation. The piece of string between the $q$ and the $\bar{q}$ has been converted into their transverse energies.

The quark pair also possesses the relative orbital angular momentum
\begin{equation}\label{eq: distance d}
    \textbf{L} = \textbf{d}\times \kt
\end{equation}
which average value is
\begin{equation}
    \langle L\rangle=\frac{2}{\kappa }\langle\sqrt{m_q^2+\ktkt}\,\,\rm{k}\T\rangle\simeq \frac{2}{\kappa}\langle \rm{k}^2\T\rangle.
\end{equation}
Using $\langle \rm{k}^2\T\rangle=\kappa/\pi$ as from Eq. (\ref{eq: P Schwinger}), the average angular momentum turns out to be $\langle L\rangle \simeq 1$ \cite{TheLundModel-Book}, meaning that the $q\bar{q}$ pair tunnels out in $P$ wave.
Since the force field, does not have angular momentum, unless there are transverse excitations, it means that the quarks must be polarized in order to compensate for the produced orbital angular momentum.
Then, the total angular momentum $\textbf{J}=\textbf{s}+\textbf{L}$ is conserved if the $q\bar{q}$ pair is in a triplet state with the spins $\textbf{s}_q$ and $\textbf{s}_{\bar{q}}$ aligned and opposite to $\textbf{L}$, namely
\begin{eqnarray}\label{eq: corr 3P0}
 \langle \textbf{s}_q\cdot \textbf{s}_{\bar{q}}\rangle > 0, & \langle \textbf{s}_{q}\cdot \textbf{L}\rangle <0, & \langle \textbf{s}_{\bar{q}}\cdot \textbf{L}\rangle <0.
\end{eqnarray}
Then $q\bar{q}$ pair happens thus to be in a $^3P_0$ state, which is characterized by the vacuum quantum numbers $J^{PC}=0^{++}$. This wave function implies that the spins and the transverse momenta of the produced quarks are correlated and in particular it is
\begin{eqnarray}\label{eq:classical string+3P0 correlations}
 \langle \kt \times \textbf{s}_{q}\rangle \cdot \zu > 0, & \langle \kt\times\textbf{s}_{\bar{q}}\rangle \cdot \zu > 0,
\end{eqnarray}
as can be seen using Eq. (\ref{eq: corr 3P0}) and Eq. (\ref{eq: distance d}). If the emitted mesons are pseudoscalars, the internal $^1S_0$ wave function requires the further correlation
\begin{equation}
    \langle \textbf{s}_q\cdot\textbf{s}_{\bar{q}'}\rangle < 0,
\end{equation}
where the content of the meson is $q\bar{q}'$.


The observation that the $q\bar{q}$ pairs at string breaking tunnel out in the $^3P_0$ state, has been proposed initially by the Lund group and was used for the description of the large $\Lambda^0$ hyperon polarizations observed in $pp$ collisions at the ISR experiment \cite{Andersson:1979-lambda0pol}.
The classical string+$^3P_0$ model with initial transversely polarized quark has been used for the description of the single spin asymmetry observed in pion production in the process $pp^{\uparrow}\rightarrow \pi + X$ \cite{XA-JCZ}. However in that case the Collins effect has been considered only for the first rank meson. A complete treatment of the polarized string decay in Fig. \ref{fig:classical3P0} requires the use of amplitudes instead of probabilities and a quantum mechanical formulation of the $^3P_0$ model, capable of describing also the dynamics of spin transfer from one breaking point to the next. Moreover, the use of amplitudes allows to fulfill automatically the conditions necessary for the preservation of positivity.

The inclusion of the quark spin in a covariant formulation of the polarized fragmentation process would necessarily require the use of Dirac spinors. However, Pauli spinors are enough to preserve the LR symmetry and the symmetries $S1$-$S3$. 
Thus the model presented here is obtained starting from amplitudes formulated using Pauli spinors. These can be considered as projections of Dirac spinors on the two-dimensional space of on mass-shell spinors, solutions of $(k^{mec,0}-\alpha_z\,k^{mec,z})\psi=0$, $k^{mec}$ being defined in Eq. (\ref{eq:decomposition of momenta}) and $\alpha_z$ is the Dirac matrix $\alpha_z=\gamma^0\gamma^z$\cite{artru-private}.

The spin part of the $^3P_0$ operator $T_{^3P_0}$ acting on the vacuum and producing the $q\bar{q}$ pair can be written as
\begin{equation}\label{eq: 3P0 amplitude}
   \langle q\bar{q}|T_{^3P_0}|0\rangle \propto \bar{v}(k_{\bar{q}},\textbf{S}_{\bar{q}}) \,u(k_q,\textbf{S}_q),
\end{equation}
where $u(\textbf{k}_q,\textbf{S}_q)$ and $\bar{v}(\textbf{k}_{\bar{q}},\textbf{S}_{\bar{q}})$ are the Dirac on mass-shell spinors of $q$ and of $\bar{q}$ respectively. The vector $\textbf{S}_{i}=(\textbf{S}_{i\,\rm{T}},S_{i\,\rm{L}})$ with $i=q,\bar{q}$ will be referred to as the \textit{polarization vector}. It is the vector used to parametrize the quark spin density matrix in helicity space, and it is not the space part of a covariant spin vector. The reduction of the quark and anti-quark Dirac spinors $u$ and $\bar{v}$ to Pauli space has been given in Ref. \cite{DS09} and it is
\begin{eqnarray}\label{eq:reduction spinors}
 u(\textbf{k},\textbf{S}_q)\rightarrow \chi(\textbf{S}_q), &   \bar{v}(\textbf{k}_{\bar{q}},\textbf{S}_{\bar{q}}) = u(\textbf{k}_{\bar{q}},-\textbf{S}_{q})\gamma_5\rightarrow -\chi^{\dagger}(-\textbf{S}_{\bar{q}})\sigma_z,
\end{eqnarray}
where $\gamma_5$ is reduced to $\sigma_z$.
Using these rules, the amplitude of Eq. (\ref{eq: 3P0 amplitude}) is reduced to
\begin{equation}\label{eq: isotropic 3P0}
    -\chi^{\dagger}(-\textbf{S}_{\bar{q}})\,\sigma_z \boldsymbol{\sigma}\cdot \textbf{k}\,\chi(\textbf{S}_q)
\end{equation}
where $\boldsymbol{\sigma}\cdot\textbf{k}$ is the isotropic $^3P_0$ operator in momentum space. The reduction of the anti-quark spinor brings a $\sigma_z$ matrix and the effective $^3P_0$ operator in Pauli space is
\begin{equation}
 \sigma_z \boldsymbol{\sigma}\cdot \textbf{k} = k_z\,\textbf{1}+\sigma_z\boldsymbol{\sigma}\T\cdot\kt.
\end{equation}
This operator must satisfy symmetries $S1$-$S3$. The second term does respect these symmetries, but it is not the case for the term proportional to the identity matrix. It is not invariant under longitudinal boosts and cannot appear in the parameterization of the $^3P_0$ operator. Hence, the most general form allowed for the $^3P_0$ operator is \cite{DS09}
\begin{equation}\label{eq:3P0 propagator}
    \mu_q(k^+k^-,\ktkt)+\sigma_z\boldsymbol{\sigma}\T\cdot\kt.
\end{equation}
The parameter $\mu_q$ is required to be complex and with $\IM(\mu)>0$ in order to reproduce the spin effects of the classical string+$^3P_0$ mechanism. $-\mu$ can be considered as a \textit{complex mass} and in general can depend on the quark flavours, on $k^+k^-$ and on the quark transverse momentum squared. In the following it is considered to be constant.
Equation (\ref{eq:3P0 propagator}) can be regarded also as the analogue of the Feynman propagator $m_q+\gamma\cdot k$ in the subspace where $\alpha_z = {\rm sign}(k_z^{mec})$, the analogue of the quark mass $m_q$ being the $-\mu_q$ \cite{artru-private}.

\subsection{The quark spin in the multiperipheral formalism}
To describe the reaction represented by the diagram in Fig. \ref{fig:multiperipheral} with polarized quarks, the vertex $\mathcal{V}$ and the propagator $\mathcal{D}$ of Eq. (\ref{eq:T}) are transformed into $2\times 2$ density matrices acting in spin $\otimes$ momentum space. The functions $w_{q',h,q}$ in Eq. (\ref{eq: V^2 string}) and $u_q$ in Eq. (\ref{eq:D^-2}) become also density matrices, hermitian and positive definite.

\begin{figure}
\centering
\begin{minipage}{.95\textwidth}
  \centering
  \includegraphics[width=.95\linewidth]{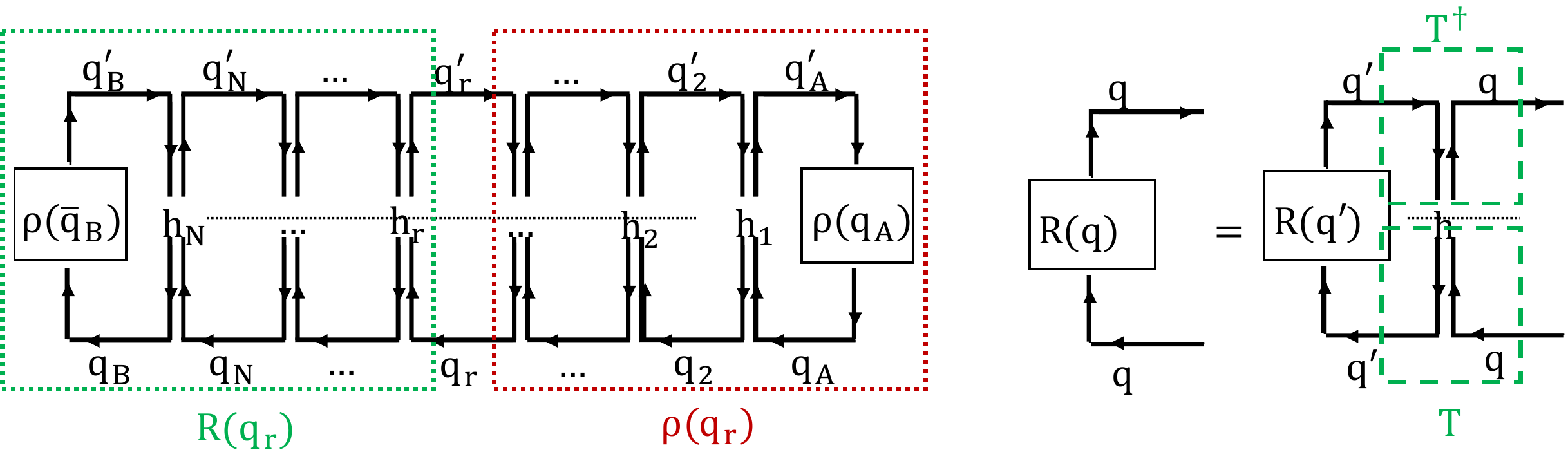}
\end{minipage}%
\caption{Unitary diagram corresponding to the squared amplitude of the multiperipheral diagram (left panel) and reppresenation of the integral equation for the cross section matrix $\mathcal{R}$ (right panel). See Eqs. (\ref{eq:dsigma/dPhir})-(\ref{eq:R(q) integral equation}).}
\label{fig:unitary}
\end{figure}

The recursive model implies the ladder approximation for the reaction described by the multiperipheral diagram in Fig. \ref{fig:multiperipheral}, \ie, it is based on ladder unitarity diagrams like in Fig. \ref{fig:unitary}. 
However the same hadronic final state can be also obtained by other multiperipheral diagrams differing by permutations of the hadrons. The interferences between such diagrams, represented by non-ladder unitarity diagrams, are neglected. 
The justification for this is that the multiperipheral diagrams for which the rank ordering differ too much from the rapidity ordering have a small amplitude. The amplitude for the reaction in Eq. (\ref{eq:set of splittings}) is
\begin{eqnarray}\label{eq:multiperipheral amplitude}
\nonumber \langle \textbf{S}_B| \mathcal{M}_N(\q\A\bar{\q}\B\rightarrow \h_1\,\h_2\dots\,\h_N)|\textbf{S}\A\rangle = \langle \textbf{S}\B|\mathcal{D}(\q\B)\mathcal{V}(\q\B,\h_N,\q_N) \dots\\
 \dots \mathcal{D}(\q_2)\mathcal{V}(\q_2,\h_1,\q\A)\mathcal{D}(\q\A)|\textbf{S}\A\rangle,
\end{eqnarray}
where the gothic letters are adopted to gather different variables, i.e. $\q=\lbrace q,k\rbrace$, where $q$ is the quark flavour, and $\h=\lbrace h,s_h,p\rbrace$ where $h$ is the hadron type and $| s_h\rangle$ its spin state in some choosen basis. With this notation the quark propagator is $\mathcal{D}(\q)=\mathcal{D}_q(k)$ and the vertex is $\mathcal{V}(\q',\h,\q)=V_{q',h,s_h,q}(k',k)$. These are also the input to the amplitude in Eq. (\ref{eq:multiperipheral amplitude}). The Pauli spinor of the initial quarks are $|\textbf{S}\A\rangle$ and $\langle \textbf{S}_B|=-\langle \textbf{S}_B| \sigma_z$ as in Eq. (\ref{eq:reduction spinors}), and $\textbf{S}=(\textbf{S}\T,S_{\rm{L}})$ is the polarization vector with transverse component $\textbf{S}\T$ and longitudinal component $S_{\rm{L}}$.

Taking the square of the amplitude in Eq. (\ref{eq:multiperipheral amplitude}), the inclusive distribution of the first $r$ hadrons is
\begin{eqnarray}\label{eq:dsigma/dPhir}
\nonumber    \frac{d\sigma(\q_A+\bar{\q}_B)}{d\Phi_r} &=& \sum_{N>r}\delta^{(4)}(k\A+\bar{k}\B-\sum_{i=1}^{r} p_i)\rm{tr}\left(\mathcal{M}_{N-r} \rho(q_{r+1})\mathcal{M}_{N-r}^{\dagger}\rho(\bar{q}\B)\right)\\
    &=& \rm{tr}\left[\mathcal{R}(\q_{r+1})\rho(q_{r+1})\right],
\end{eqnarray}
where $d\Phi_r$ is the phase space factor of the first $r$ hadrons. The corresponding unitary diagram is shown in the left panel of Fig. \ref{fig:unitary}.
The matrix $\rho(q_{r+1}) = \mathcal{M}_r\rho(q\A)\mathcal{M}_r^{\dagger}$ is the (not-normalized) spin density matrix of the quark $q_{r+1}$, $\mathcal{M}_r$ being the analog of $\mathcal{M}_N$ in Eq. (\ref{eq:multiperipheral amplitude}) for the emission of the first $r$ hadrons. The matrix $\mathcal{R}(\q_r)$ is the \textit{cross section matrix} \cite{DS11} of the process $q_r+\bar{q}\B\rightarrow h_{r+1}\dots h_N$. It contains information on the "future" emissions of $h_{r+1}\dots h_N$ and is given by
\begin{equation}\label{eq:R(qr) definition}
    \mathcal{R}(\q_r) = \sum_{N>r}\delta^{(4)}(k_{r+1}+\bar{k}\B-\sum_{i=r+1}^N p_i)\rm{tr}\left(\mathcal{M}_{N-r}^{\dagger}\rho(\bar{q}\B)\mathcal{M}_{N-r} \right).
\end{equation}
The hadronization cross section matrix satisfies the ladder recursive integral equation \cite{DS13}
\begin{equation}\label{eq:R(q) integral equation}
    \mathcal{R}(\q) = \int d\h\,\mathcal{D}^{\dagger}(\q) \mathcal{V}^{\dagger}(\q',\h,\q)\mathcal{R}(\q') \mathcal{V}(\q',\h,\q)\mathcal{D}(\q),
\end{equation}
where $\q_r,\h_r,\q_{r+1}$ have been replaced with $\q,\h,q'$ and the non-homogeneous term arising from the possibility for the $q\A\bar{q}\B$ system to resonate as a mesonic state has been neglected. This integral equation can be converted into an integral in the limit of large $m_X^2=(k+k_{\bar{B}})^2\simeq |k_{\bar{B}}^-k^+|^2$ assuming the Regge behaviour \cite{DS11}
\begin{eqnarray}\label{eq:integral eq for R}
\nonumber    \mathcal{R}(\q)&\simeq& |m_X|^{\alpha_R}\mathcal{B}(\q)\\
    &=& |m_X|^{\alpha_R}\,\left[b_{0q}(\ktkt)+b_{1q}(\ktkt)\boldsymbol{\sigma}\cdot\tilde{\textbf{n}}(\kt)\right]
\end{eqnarray}
where $\alpha_R$ is the Regge intercept and $b_{0q},b_{1q}$ are the Regge residue functions, with $b_{0q}>0$, $b_{1q}(0)=0$ and $|b_{1q}/b_{0q}|<1$. The $\mathcal{B}$ matrix respects the symmetries $S1$-$S3$. The integral equation in Eq. (\ref{eq:integral eq for R}) is rewritten for the $\mathcal{B}$ matrix as
\begin{equation}\label{eq: B old}
    \mathcal{B}(\q)=\int d\h \left|\frac{k'^+}{k^+} \right|^{\alpha_R}\,\mathcal{D}^{\dagger}(\q) \mathcal{V}^{\dagger}(\q',\h,\q)\mathcal{B}(\q') \mathcal{V}(\q',\h,\q)\mathcal{D}(\q).
\end{equation}
The resolution of such equation can be avoided observing that the amplitude in Eq. (\ref{eq:multiperipheral amplitude}) is invariant under two \textit{renormalization} procedures which do not change the final multi-hadron distributions \cite{DS13}. The first procedure consists in re-scaling only the propagator and the vertex, leaving invariant the quark spin states and hence spin density matrices. The transformation is
\begin{eqnarray}\label{eq: transformation 1}
 \mathcal{D}(\q)\rightarrow |k^-k^+|^{\lambda}\mathcal{D}(\q), && \mathcal{V}(\q',\h,\q)\rightarrow |k'^+k^-|^{\lambda} \mathcal{V}(\q',\h,\q),
\end{eqnarray}
where $\lambda$ is an arbitrary real parameter. The effect on the overall amplitude in Eq. (\ref{eq:multiperipheral amplitude}) is the multiplication by the factor $(k\A^+k_{\bar{B}}^-)^{\lambda}$, which is fixed and does not change the inclusive distribution of the final hadrons.
Applying such transformation to Eq. (\ref{eq: B old}), the Regge intercept $\alpha_R$ is shifted to the value $\alpha_R\rightarrow \alpha_R+2\lambda$. Hence by choosing $\lambda = -\alpha_R/2$ the factor $|k'^+/k^+|^{\alpha_R}$ in Eq. (\ref{eq: B old}) disappears.
                                                           
The second set of transformations consists in transforming simultaneously Pauli spinors, propagators and vertices, namely
\begin{eqnarray}\label{eq: transformation 2}
 |\textbf{S}\rangle \rightarrow \Lambda^{-1}(\q)|\textbf{S}\rangle, && \rho(q)\rightarrow \Lambda^{-1}(\q)\rho(q){\Lambda^{\dagger}}^{-1}(\q) \\ \mathcal{D}(\q)\rightarrow \Lambda(\q)\mathcal{D}(\q)\Lambda(\q), && \mathcal{V}(\q',\h\,q)\rightarrow \Lambda^{-1}(\q')\mathcal{V}(\q',\h,\q)\Lambda^{-1}(\q).
\end{eqnarray}
$\Lambda(\q)$ is an arbitrary matrix in spin space with $\Lambda(q,\kt=0)=1$. Using Eq. (\ref{eq: transformation 1}) with $\lambda = -\alpha_R/2$ and Eq. (\ref{eq: transformation 2}), the new equation for $\mathcal{B}$ is
\begin{equation}\label{eq: B unit}
    \Lambda^{\dagger}(\q)\mathcal{B}(\q)\Lambda(\q) = \mathcal{D}^{\dagger}(\q)\left(\int d\h\, \mathcal{V}^{\dagger}(\q',\h,\q)\,\Lambda^{\dagger}(\q')\mathcal{B}(\q')\Lambda(\q')\, \mathcal{V}(\q',\h,\q)\right)\mathcal{D}(\q).
\end{equation}
By making the choice $\Lambda(\q)=\eta(\q)\mathcal{B}^{-1/2}$ ($\mathcal{B}$ is positive definite), with $\eta$ a unitary matrix, the new $\mathcal{B}$ matrix becomes the identity matrix. Equation (\ref{eq: B unit}) translates then to the relation
\begin{equation}\label{eq: spin U}
    \frac{1}{\mathcal{D}(\q)\mathcal{D}^{\dagger}(\q)} \equiv \mathcal{U}(\q)= \int \frac{d^3\textbf{p}}{p^0}\, \mathcal{V}^{\dagger}(\q',\h,\q) \mathcal{V}(\q',\h,\q),
\end{equation}
which is the generalization of Eq. (\ref{eq: U = int V*V}) to the spin dependent case. Hence with the renormalization procedure proposed in Ref. \cite{DS13}, the only input of the multiperipheral amplitude is the vertex. The propagator instead is obtained from the solution of Eq. (\ref{eq: spin U}), which is
\begin{equation}\label{eq: D = U^-1/2}
    \mathcal{D}(\q) = \mathcal{U}^{-1/2}(\q).
\end{equation}
The matrix $\mathcal{U}$ is hermitian and positive definite, as can be seen from Eq. (\ref{eq: B unit}). In principle $\mathcal{U}^{-1/2}$ in Eq. (\ref{eq: D = U^-1/2}) can still be multiplied by a unitary matrix which commutes with $\mathcal{U}$ but it can be reabsorbed in $\mathcal{D}$ using again the transformation in Eq. (\ref{eq: transformation 2}). Hence without loss of generality one can take as solution that in Eq. (\ref{eq: D = U^-1/2}).

The invariance under reflections about the $(x,z)$ or $(y,z)$ planes, requires the matrices $\mathcal{U}(\q)$ and $\mathcal{D}(\q)$ to have the form
\begin{equation}\label{eq:U(q)=U0+U1 + tilde definition}
    \mathcal{U}(\q) = U_0(\q)+U_1(\q)\boldsymbol{\sigma}\cdot\tilde{\textbf{n}},
    \mathcal{D}(\q) = D_0(\q)+D_1(\q)\boldsymbol{\sigma}\cdot\tilde{\textbf{n}},
\end{equation}
where the tilde symbol indicates the cross product $\tilde{\textbf{n}}=\zu\times\textbf{n}$, and the unit vector $\textbf{n}$ is defined as $\textbf{n}(\kt)=\kt/|\kt|$. $U_0(\q)$ and $U_1(\q)$ are real functions due to the hermiticity of $\mathcal{U}$. In addition they have the further property $U_0(\q)\geq|U_1(\q)|$ due to the positivity of $\mathcal{U}$.

The generalization of the double quark density in Eq. (\ref{eq:V^2}), is given by the density operator
\begin{equation}\label{eq: density N}
    \langle i',j'|\mathcal{N}(\q',\h,\q)|i,j\rangle = 2\delta(p^2-m_h^2)\langle i',j'|\mathcal{V}_{pt}^{\dagger}|\h\rangle \langle \h|\mathcal{V}_{pt}|i,j\rangle
\end{equation}
acting on the quark spin space. The indices $i$ and $j$ label the spin states of $q$ and of $q'$. It is a density matrix in spin space and a classical density in momentum space. Equation (\ref{eq: density N}) makes use of the partial transpose of the vertex operator, defined as
\begin{equation}
    \langle \h|\mathcal{V}_{pt}|i,j\rangle = \langle j| \mathcal{V}(\q',\h,\q)|i\rangle.
\end{equation}

The \textit{splitting matrix} which generalizes Eq. (\ref{eq:T}) and which describes the transition from $q$ to $q'$ in spin space is
\begin{equation}\label{eq:T=VD spin}
    \mathcal{T}(\q',\h,\q)=\mathcal{V}(\q',\h,\q)\mathcal{D}(\q).
\end{equation}
The \textit{polarized splitting function} generalizing Eq. (\ref{eq:F = TT*}) is
\begin{equation}\label{eq:F=TrhoT pol}
    F_{q',h,q}(Z,\pt;\kt)=\rm{tr}\left[T(\q',\h,\q)\rho(q)T^{\dagger}(\q',\h,\q)\right]
\end{equation}
where $\rho(q)=(\textbf{1}+\textbf{S}_q\cdot\boldsymbol{\sigma})/2$ is the spin density matrix of the quark q and $\textbf{S}_q$ the related polarization vector. The normalization condition of the splitting function given in Eq. (\ref{eq:normalization}) for the polarized case is generalized to
\begin{equation}\label{eq:normalization general}
    \sum_{h,s_h} \int dZZ^{-1}\,d^2\pt\,F_{q',h,q}(Z,\pt;\kt)=1.
\end{equation}
Finally the splitting matrix in Eq. (\ref{eq:T=VD spin}) allows to calculate the spin density matrix of the quark $q'$, namely
\begin{equation}\label{eq:rho'=TrhoT}
    \rho(q')=\frac{T(\q',\h,\q)\rho(q)T^{\dagger}(\q',\h,\q)}{\rm{tr}\left[numerator\right]},
\end{equation}
which is normalized to unit trace. The splitting matrix in Eq. (\ref{eq:F=TrhoT pol}) and the rule for the calculation of the spin density matrix of $q'$ in Eq. (\ref{eq:rho'=TrhoT}) are the basis for the recursive simulation of polarized quark jets, once the explicit form of the input vertex $\mathcal{V}(\q',\h,\q)$ or equivalently the splitting matrix $T(\q',\h,\q)$ is given.

\subsection{The splitting matrix of the $string+\, ^3P_0$ model}
The vertex and the propagator of the multiperipheral diagram can be modelled by considering the string decay formalism of the SLM generalized to the spin dependent case as shown in Ref. \cite{DS13}. The multiperipheral diagram form of the string decay, however, is not a standard Feynman diagram due to the existence of the string itself which acts as an external field \cite{collins-rogers} (see also Eq. (\ref{eq:decomposition of momenta})). Using Eq. (\ref{eq: V^2 string}) and omitting a phase factor $\exp[-i{\cal S}(quarks) - i{\cal S}(string)]$ which arises from the string ($\mathcal{S}(string)$) and quark ($\mathcal{S}(quarks)$) actions after taking the square root, the vertex operator of the polarized string decay can be parameterized as \cite{DS13,kerbizi-2018}
\begin{equation}\label{eq: V spin}
    \mathcal{V}(\q',\h,\q)=\left(\frac{k'^+}{p^+}\right)^{a_{q'}/2}\bigg\rvert \frac{k^-}{p^-}\bigg\lvert^{a_q/2}e^{-\bl|k'^-k^+|/2}\,g(\q',\h,\q)
\end{equation}
where $g(\q',\h,\q)=g_{q',h,s_h,q}(\kpt,\kt)$ is a $2\times 2$ matrix acting on the quark spin space.
This allows to re-write the single quark density $\mathcal{U}(\q)$ in Eq. (\ref{eq: D^-2 string}) as
\begin{eqnarray}\label{eq: U = k+k- part x uq}
\nonumber  \mathcal{U}(\q)&=&\sum_{h,s_h}\int dZZ^{-1}\,d^2\kpt\, \left(\frac{k'^+}{p^+}\right)^{a_{q'}}\bigg\rvert \frac{k^-}{p^-}\bigg\lvert^{a_q}e^{-\bl|k'^-k^+|}\,g^{\dagger}(\q',\h,\q)g(\q',\h,\q)\\
 &=& e^{-\bl|k^+k^-|}|k^+k^-|^{a_q}\,u_q(\kt),
\end{eqnarray}
with
\begin{equation}\label{eq:uq}
    u_q(\kt) = \sum_{h,s_h}\int dZZ^{-1}\,d^2\kpt\, \left(\frac{1-Z}{Z}\right)^{a_{q'}}\left(\frac{Z}{\varepsilon_h^2}\right)^{a_q}e^{-\bl\varepsilon_h^2/Z}\,g^{\dagger}(\q',\h,\q)g(\q',\h,\q),
\end{equation}
where $g^{\dagger}g$ replaces the function $w$ in Eq. (\ref{eq: uq unpolarized}).

The polarized quark propagator obtained generalizing Eq. (\ref{eq:D^-2}) is
\begin{equation}\label{eq: D =k+k- part x dq}
    \mathcal{D}(\q)=|k^+k^-|^{-a_q/2}e^{\bl|k^+k^-|/2}d_q(\kt),
\end{equation}
$d_q(\kt)$ being a $2\times 2$ matrix in quark spin space. Then using Eqs. (\ref{eq: U = k+k- part x uq}-\ref{eq: D =k+k- part x dq}), the  relation between the quark propagator and the vertex operator in the renormalized approach expressed by Eq. (\ref{eq: D = U^-1/2}) now becomes
\begin{equation}\label{eq:dq*dq = uq}
    \left[d_q^{\dagger}(\kt)d_q(\kt)\right]^{-1} = u_q(\kt).
\end{equation}

The $^3P_0$ mechanism is included in the matrix $g(\q',\h,\q)$ which is taken of the form \cite{kerbizi-2018}
\begin{equation}\label{eq:g matrix}
    g(\q',\h,\q)=C_{q',h,q}\,\check{g}(\varepsilon_h^2)\,\Delta_{q'}(\kpt)\Gamma_{h,s_h}(\kpt,\kt)\Delta_q(\kt),
\end{equation}
where the matrix $\Delta_q$ contains the $^3P_0$ operator given in Eq. (\ref{eq:3P0 propagator}), namely it is
\begin{equation}
    \Delta_q(\kt)=f\T(\ktkt)\left(\mu_q+\sigma_z\boldsymbol{\sigma}\cdot\kt\right).
\end{equation}
Notice that the matrix $g(\q',\h,\q)$ in Eq. (\ref{eq:g matrix}) is manifestly LR symmetric. The factor $C_{q',h,q}$, the function $\check{g}$ and the function $f\T$ have the same meaning as introduced in the spin-less SLM in Eq. (\ref{eq: w_q'hq}).
In particular $f\T$ is taken as the exponential
\begin{equation}\label{eq:fT exponential}
    f\T(\ktkt)=e^{-\bt\ktkt/2}\left(\bt/\pi\right)^{1/2},
\end{equation}
where the free parameter $\bt$ controls the width of the quark transverse momentum at string breaking. This function is inspired by the Schwinger mechanism in Eq. (\ref{eq: P Schwinger}). The results of the classical string+$^3P_0$ mechanism are reproduced for $\IM(\mu_q)>0$.

$\Gamma_{h,s_h}$ is a $2\times 2$ matrix representing the hadron emission vertex. It may depend on the hadron type $h$ and on its spin state $s_h$, and is a polynomial in $\sigma_x$, $\sigma_y$, $\sigma_z$, $k_x$,  $k_y$,  $k_z$, $k'_x$, $k'_y$,  $k'_z$. For pseudo-scalar mesons, and to zero order in the quark momenta, it is \cite{DS09}
\begin{equation}\label{eq:coupling ps}
    \Gamma_h=\sigma_z,
\end{equation}
which is the analogue of $\gamma_5$. In Ref. \cite{spin16} another choice, $\Gamma_{h}(\kt,\kpt)=\mu\sigma_z+\boldsymbol{\sigma}\cdot\pt$ with $\Delta=f\T$, was made. Such choice is approximate and does not take into account the full vertex.
The inclusion of vector meson requires a more detailed treatment and will be presented Chapter 5.

Equation (\ref{eq:g matrix}) allows to re-write $u_q(\kt)$ of Eq. (\ref{eq:uq}) in the factorized form
\begin{equation}\label{eq: u = delta u delta}
    u_q(\kt)=\Delta^{\dagger}_q(\kt)\hat{u}_q(\kt)\Delta_q(\kt),
\end{equation}
having defined
\begin{eqnarray}\label{eq: hat uq}
 \nonumber   \hat{u}_q(\kt) &=& \sum_{h,s_h}\hat{u}_{q,h,s_h}(\kt)\\
\nonumber &=&\sum_{h,s_h}\,|C_{q',h,q}|^2\int dZZ^{-1} d^2\kpt\,\left(\frac{1-Z}{\varepsilon_h^2}\right)^a\,e^{-\bl\varepsilon_h^2/Z}\,\check{g}^2(\varepsilon_h^2)\\
    &\times& \Gamma^{\dagger}_{h,s_h}\Delta^{\dagger}_{q'}(\kpt)\Delta_{q'}(\kpt)\Gamma_{h,s_h}\\
    &\equiv& \hat{u}_{0q}(\ktkt)+\hat{u}_{1q}(\ktkt)\boldsymbol{\sigma}\cdot\tilde{\textbf{n}(\kt)}.
\end{eqnarray}
where the parameter $a_q$ is taken constant and the same for all flavors, namely $a_q\equiv a$. This will always be the case in the following. The matrix $\hat{u}_q$ is hermitian and positive definite, hence $\hat{u}_{0q}$ and $\hat{u}_{1q}$ are real functions with $\hat{u}_{0q}\geq |\hat{u}_{1q}|$.

Using Eq. (\ref{eq: u = delta u delta}), the solution of Eq. (\ref{eq:dq*dq = uq}) can be taken as \cite{kerbizi-2018}
\begin{equation}\label{eq: d = delta u}
    d_q(\kt)=\Delta_q^{-1}(\kt)\hat{u}_q^{-1/2}(\kt).
\end{equation}
In general the solution of $d_q$ in Eq. (\ref{eq: d = delta u}) is defined up to a unitary matrix which commutes with $\Delta_q^{-1}$ and $\hat{u}_q^{-1/2}$. The inverse square root of $\hat{u}_q$ exists due to the above properties and it is
\begin{eqnarray}
\hat{u}^{-1/2}_q(\kt)=\hat{d}_{0,q}(\ktkt)+\hat{d}_{1,q}(\ktkt)\vecsigma\cdot\ntil(\kt),
\end{eqnarray}
with
\begin{eqnarray}
    \hat{d}_{0,q}(\ktkt) = \frac{\sqrt{\hat{u}_{+,q}}+\sqrt{\hat{u}_{-,q}}}{2\sqrt{\hat{u}_{+,q}\hat{u}_{-,q}}}, && \hat{d}_{1,q}(\ktkt) = \frac{\sqrt{\hat{u}_{-,q}}-\sqrt{\hat{u}_{+,q}}}{2\sqrt{\hat{u}_{+,q}\hat{u}_{-,q}}}
\end{eqnarray}
and
\begin{eqnarray}
    \hat{u}_{\pm,q}=\hat{u}_{0,q}\pm\hat{u}_{1,q}.
\end{eqnarray}

Finally the splitting matrix of the string$+\,^3P_0$ model is obtained using Eq. (\ref{eq:T=VD spin}), Eqs. (\ref{eq: V spin}-\ref{eq:g matrix}) and Eq. (\ref{eq: d = delta u}) and it is \cite{kerbizi-2018}
\begin{eqnarray}\label{eq: T final}
\nonumber    T_{q',h,q}&=&C_{q',h,q}\,\check{g}(\varepsilon_h^2)\left[(1-Z)/\varepsilon_h^2\right]^{a/2}\,\exp{\left[-\bl\varepsilon_h^2/(2Z)\right]}\\
    &\times& \Delta_{q'}(\kpt)\Gamma_{h,s_h}\hat{u}_q^{-1/2}(\kt).
\end{eqnarray}
This is the basic ingredient for the simulation of the polarized fragmentation process and will be used in the next sections for the calculation of the explicit form of the polarized splitting function and for the propagation of the spin information from $q$ to $q'$.

\section{The splitting function of the $string+\, ^3P_0$ model}\label{sec: general pol F of the string + 3P0}
The final expression for the splitting matrix in Eq. (\ref{eq: T final}) can be used for the calculation of the polarized splitting function together with Eq. (\ref{eq:F=TrhoT pol}). The polarized splitting function of the string+$^3P_0$ model is then \cite{kerbizi-2018}
\begin{eqnarray}\label{eq:general pol F}
  F_{q',h,q}(Z,\pt;\kt,\textbf{S}_q)&=&|C_{q',h,q}|^2 \, \check{g}^2(\varepsilon_h^2)\left(\frac{1-Z}{\varepsilon_h^2}\right)^a \exp\left(-\bl\frac{\varepsilon_h^2}{Z}\right)\\
\nonumber  &\times&  \rm{tr}\big[ \Delta_{q'}(\kpt)\Gamma_{h,s_h}\hat{\rho}_{int}(q)\Gamma^{\dagger}_{h,s_h} \Delta^{\dagger}_{q'}(\kpt)\big],
\end{eqnarray}
having introduced the intermediate spin density matrix
\begin{equation}\label{eq:rhoint}
    \hat{\rho}_{int}(q) =\hat{u}_q^{-1/2}(\kt)\rho(q) \hat{u}_q^{-1/2}(\kt).
\end{equation}

The splitting function is practically used for the generation of the hadron species $h$ and of its momentum, namely of $Z$ and $\pt$. The choice of the input function $\check{g}$ appearing in Eq. (\ref{eq:general pol F}) has an important role in that it affects both the probability of generating the hadron species and the spin-independent correlations between $\kt$ and $\kpt$. 

The probability $P_{q\rightarrow h}$ to generate the hadron type $h$ is obtained integrating the splitting function on $Z$ and $\pt$. It gives
\begin{eqnarray}\label{eq: P q->h}
\nonumber    P_{q\rightarrow h}(\kt,\Sq) &=& \rm{tr}\left[\hat{u}_{q,h}(\kt)\hat{u}^{-1}_q(\kt)\rho(q)\right]\\
    &=&\rm{tr}\left[\hat{u}_{q,h}(\kt)\hat{\rho}_{int}(q)\right].
\end{eqnarray}
The matrix $\hat{u}_{q,h}$ is the same as $\hat{u}_{q,h,s_h}$ defined in Eq. (\ref{eq: hat uq}) with the index $s_h$ suppressed. The normalization condition for $P_{q\rightarrow h}$ follows from the normalization of the spin density matrix of $q$, i.e. $\sum_h P_{q\rightarrow h}=\rm{tr}\rho(q)=1$. Thus, in general this probability does not depend only on the isospin wave function of $h$, i.e. on the coefficient $C_{q',h,q}$.
For a generic choice of the input function $\check{g}$, $P_{q\rightarrow h}$ may depend on the transverse momentum of $q$ and on its polarization vector $\Sq$, due to the matrix $\hat{u}_{q,h}\hat{u}^{-1}_q$, that in principle is not calculable analytically. In order to satisfy symmetries $S1$-$S3$, $P_{q\rightarrow h}$ must be of the form
\begin{equation}
    P_{q\rightarrow h}=p_{q\rightarrow h}(\ktkt)+\delta p_{q\rightarrow h}(\ktkt)\STq \cdot \ktil,
\end{equation}
where $p_{q\rightarrow h}(\ktkt)$ and $\delta p_{q\rightarrow h}(\ktkt)$ are functions that depend on the choice of $\check{g}$. For a generic choice of $\check{g}$ the functions $p_{q\rightarrow h}$ and $\delta p_{q\rightarrow h}$ are non-trivial and in principle not calculable analytically.

In addition, the $\kt$ and $\kpt$ dependence of the splitting function in Eq. (\ref{eq:general pol F}) is not factorized for a generic $\check{g}$, leading to spin-independent correlations between the quark transverse momenta. The correlation coefficient
\begin{equation}
\xi = \langle \kt\cdot\kpt\rangle/\langle \ktkt\rangle
\end{equation}
depends on $Z$ and may be positive or negative. Such correlations add to the $Z$-averaged spin-mediated correlations required by the $^3P_0$ mechanism, which favours the quark transverse momenta to be in opposite directions as can be seen from Fig. \ref{fig:classical3P0}, namely $\xi <0$. The $Z$-independent and spin-independent $\kt$-$\kpt$ correlations are governed by the factor $\check{g}^2(\varepsilon_h^2)N_a(\varepsilon_h^2)$ (the function $N_a$ is introduced in Eq. (\ref{eq: Na})), as can be seen integrating over $Z$ the first line of the splitting function in Eq. (\ref{eq:general pol F}) at fixed $\pt$.

\subsection{Possible choices}\label{sec:possible choices}
The function $\check{g}$ in general is a LR symmetric function of the transverse momenta $\kt$ and $\kpt$. It may also depend on the hadron mass $m_h$. The symmetric string model does not indicate a precise form for this function and different choices are possible.

The most simple choice is \cite{kerbizi-2019}
\begin{equation}\label{eq:choice-c1}
\rm{(C_1)} \,\,\,\,\,\,\,\,\,\,   \check{g}^2(\varepsilon_h^2)  = 1/{N_a(\varepsilon_h^2)}.
\end{equation}
With this choice there are no spin-independent $\kt$-$\kpt$ correlations, indeed $\check{g}^2N_a=1$ and the $Z$-integrated splitting function does not depend on $\kt$.
The matrices $\hat{u}_q$ and $\hat{u}_{q,h}$ in Eq. (\ref{eq: hat uq}) becomes proportional to the unit matrix and $P_{q\rightarrow h}$ does not depend on $\kt$ and $\STq$.
Since $\hat{u}_q$ is unit, the intermediate spin density matrix introduced in Eq. (\ref{eq:rhoint}) is proportional to the true spin density matrix $\rho(q)$. This property makes the description of the spin transfer mechanism simple allowing explicit analytic calculations.
This choice is also implicit in the implementation of the SLM in \verb|PYTHIA|, thus it is very suitable for the implementation of the spin effects in this event generator \cite{kerbizi-lonnblad} as will be shown in Chapter 4.

The possibility
\begin{eqnarray}\label{eq:choice-c2}
\rm{(C_2)} \,\,\,\,\,\,\,\,\,\,\check{g}^2(\varepsilon_h^2)  = (\varepsilon_h^2)^a,
\end{eqnarray}
has also been explored in detail \cite{kerbizi-2018}.
It leads to spin-independent and $Z$-dependent $\kt$-$\kpt$ correlations with $\xi(Z)=\bl/(\bl+Z\bt)$, namely Eq. (\ref{eq:choice-c2}) favours the consecutive quark transverse momenta to be aligned. The matrix $\hat{u}^{-1/2}_q$ is not calculable analytically and has to be tabulated for the implementation of the model in a simulation program. The analysis of the spin transfer mechanism along the fragmentation chain is complicated due to the intermediate spin density matrix $\rho_{int}$ in Eq. (\ref{eq:rhoint}).  The probability $P_{q\rightarrow h}$ depends on $\kt$ and $\STq$. Its precise form cannot be calculated analytically and requires the tabulation of the functions $p_{q\rightarrow h}$ and $\delta p_{q\rightarrow h}$. The dependence of $P_{q\rightarrow h}$ on $\kt$ and $\STq$ has been neglected in the Monte Carlo simulation of Ref. \cite{kerbizi-2018} introducing a breaking of the LR symmetry.
At large $\ptpt$, for the same $f\T$ and $\bl$, the choice $C_2$ gives smaller transverse momenta than $C_1$, indeed
\begin{equation}
    \frac{\check{g}_{C_1}(\varepsilon_h^2)}{\check{g}_{C_2}(\varepsilon_h^2)}\simeq \frac{\bl^{(a+1)/2}e^{\bl\varepsilon_h^2/2}\varepsilon_h^{2a+1}}{\sqrt{\Gamma(a+1)}\varepsilon_h^{2a}}\sim \varepsilon_h e^{\bl\varepsilon_h^2},
\end{equation}
namely $\check{g}$ for $C_1$ grows exponentially at large transverse momenta.

The choices $C_1$ and $C_2$ can be generalized to
\begin{equation}\label{eq:c3}
    \rm{(C_3)} \,\,\,\,\,\,\,\,\,\,\check{g}^2(\varepsilon_h^2)  = \exp{(-\be\varepsilon_h^2)}/{N_a(\varepsilon_h^2)},
\end{equation}
and
\begin{equation}\label{eq:c4}
    \rm{(C_4)} \,\,\,\,\,\,\,\,\,\,\check{g}^2(\varepsilon_h^2)  = (\varepsilon_h^2)^a\,\exp{(c\bl\varepsilon_h^2)}
\end{equation}
respectively. They differ with respect to $C_1$ and $C_2$ by an exponential of the transverse energy.

$C_3$ introduces the spin-independent and $Z$-independent $\kt$-$\kpt$ correlations through the new phenomenological parameter $\be\geq -\bt/2$. It gives $\xi = \be/(\be+\bt)$, hence the correlation coefficient may be positive or negative depending on the sign of $\be$. For vanishing $\be$ we recover $C_1$. With this choice, the matrix $\hat{u}_{q,h}\hat{u}_q^{-1}$ is not proportional to identity and can be calculated analytically. The probability $P_{q\rightarrow h}$ depends on $\kt$ and $\STq$, and can also be calculated analytically.

Choice $C_4$ is a generalization of $C_1$, depends on the new parameter $c\leq 1+ 2\bt/\bl$ and gives $\xi(Z)= \bl(1+Zc)/[\bl(1+Zc)+Z\bt]$. Hence the spin-independent $\kt$-$\kpt$ correlation foreseen by choice $C_2$ is reinforced for $c>0$ and weakened for $c<0$. This choice is characterized by the same complications as $C_2$.

Inspite of the different possibilities for the input function $\check{g}$, it turns out that the bulk predictions of the different models are the same if they reproduce the same $\langle \ptpt\rangle$ observed experimentally \cite{kerbizi-2019}. This condition means for instance that by re-tuning their respective parameters, choices $C_3$ and $C_4$ are expected to give the same results. 
The spin-independent $\kt$-$\kpt$ correlations can be translated to correlations between the transverse momenta of hadrons with different ranks, which can in principle be observed experimentally if the correlation between rank and rapidity orderings are well known.
However this would not be enough to tell if the spin-independent correlations have to be taken into account, since they mix with the spin mediated ones. Among the different possibilities $C_1$ is the most simple and suitable for our studies. This choice has been adopted also for the interface of the $^3P_0$ model and for the inclusion of vector mesons.

\subsection{The simplest string +$\,^3P_0$ model}\label{sec:simple splitt function}
In this Section the choice $C_1$, namely $\check{g}^2=1/N_a(\varepsilon_h^2)$, is explored. The resulting model of the polarized fragmentation process, which has been published in Ref. \cite{kerbizi-2019}, is called M19 throughout the thesis. This choice simplifies notably the matrix $\hat{u}_q$, indeed using Eq. (\ref{eq: hat uq}) it is
\begin{equation}\label{eq:hat uq ps explicit}
    \hat{u}_q=\sum_h \hat{u}_{q,h} = \textbf{1}\,\sum_h |C_{q',h,q}|^2 \left(|\mu|^2+\langle \ktkt\rangle_{f\T}\right),
\end{equation}
where, for a generic function $A(\ktkt)$, the weighting operation
\begin{equation}\label{eq:weight fT}
    \langle A\rangle_{f\T} = \int d^2\kt A(\ktkt)f^2\T(\ktkt)
\end{equation}
has been defined for a normalized $f\T$. Hence the matrix $\hat{u}_{q,h}\hat{u}_q^{-1}$ is proportional to identity, meaning that the probability of emitting the hadron type $P_{q\rightarrow h}$ does not depend either on $\kt$ nor on $\STq$. With this choice Eq. (\ref{eq: P q->h}) becomes
\begin{equation}\label{eq:p_q->h simple 3P0}
    P_{q\rightarrow h} =  \frac{|C_{q',h,q}|^2}{\sum_H |C_{q',H,q}|^2},
\end{equation}
namely the probability of emitting hadron of type $h$ in flavour space depends only on $C_{q',h,q}$.
The intermediate spin density matrix $\hat{\rho}_{int}$ introduced in Eq. (\ref{eq:rhoint}) is also proportional to the true quark spin density matrix.

Using Eq. (\ref{eq:general pol F}) and Eq. (\ref{eq:choice-c1}), the splitting function of the \textit{simplified} $^3P_0$ \textit{model} is
\begin{eqnarray}\label{eq:F_explicit simple 3P0}
F_{q',h,q}(Z,\pt;\kt,\textbf{S}_q)&=& \frac{|C_{q',h,q}|^2}{\sum_H |C_{q',H,q}|^2}\, \left(\frac{1-Z}{\varepsilon_h^2}\right)^a \frac{\exp{(-\bl \varepsilon_h^2/Z)}}{N_a(\varepsilon_h^2)} \\
 &\times& \frac{|\mu|^2+\kptkpt}{|\mu|^2+\langle \ktkt\rangle_{f\T}} f_{\T}^2(\kptkpt)
\nonumber \, \left[ 1-\frac{2\IM(\mu)\,\rm{k'}_{\rm{T}}}{|\mu|^2+\kptkpt}\textbf{S}_{q}\cdot\tilde{\textbf{n}}(\kpt) \right],
\end{eqnarray}
and satisfies the normalization condition in Eq. (\ref{eq:normalization general}).
The factor containing the scalar product $\STq\cdot \ktil$ is the source of the Collins effect at the quark level, which is then translated to the emitted hadron $h$.
From the practical point of view, with this choice of $\check{g}$, it is most convenient to generate first the hadron type using Eq. (\ref{eq:p_q->h simple 3P0}), then $\pt=\kt-\kpt$ according to the $Z$-integrated splitting function and finally $Z$. This is in fact what is done in \verb|PYTHIA| \cite{Pythia6}.

The exponential form of the function $f\T$ in Eq. (\ref{eq:F_explicit simple 3P0}) produces also an exponential distribution in the hadron transverse momentum $\ptpt$.
An alternative form of $f\T$ that produces a tail in the $\ptpt$ distribution (at fixed $\langle \ptpt\rangle)$ is
\begin{equation}\label{eq:fT_alpha}
    f_{\rm{T}}(\ktkt)\propto \frac{\exp(-\bt\ktkt/2)}{\big( |\mu|^2+\ktkt\big)^{\alpha}}.
\end{equation}
It is inspired from the the Feynman propagator of the exchanged quarks in the multiperipheral diagram ($\gamma\cdot k +m_q)/(k^2-m_q^2)$ where the analog of $k^2-m_q^2$ is $-(\ktkt+|\mu|^2)$, and depends on the phenomenological parameter $\alpha$. The analogy with the Feynman propagator would suggest $\alpha=1$ but since here the regime is not that of perturbative QCD, $\alpha$ is of phenomenological nature and any value could be allowed. $\alpha = 0$ brings back to an exponential form for $f\T$. For a given $\langle \ktkt\rangle$, taking $\alpha$ positive and decreasing $\bt$ extends the tail in $\ktkt$. It has been verified that the implementation of Eq. (\ref{eq:fT_alpha}) in Monte Carlo simulations gives only slightly different results, supporting the working choice $\alpha=0$. 

\subsubsection*{Study of the quark spin transfer}
The choice $C_1$ in Eq. (\ref{eq:choice-c1}) allows for a simple analytical calculation of the polarization transfer from $q$ to $q'$ in the elementary splitting $q\rightarrow h + q'$. According to Eq. (\ref{eq:rho'=TrhoT}) and using Eq. (\ref{eq: T final}) with choice $C_1$, the spin density matrix of $q'$ after the emission of a pseudo-scalar meson is
\begin{equation}
    \rho(q') = \frac{\Deltaqp\,\Gammah\,\rho(q)\,\Gammahdag\,\Deltaqpdag}{\trace\left[numerator\right]}.
\end{equation}
The transverse and longitudinal components of the polarization vector $\Sqp$ are
\begin{eqnarray}\label{eq:S_q'T}
\nonumber &&\textbf{S}_{q'\rm{T}}=\frac{1}{N}\big[-(|\mu|^2+\kptkpt)\,\textbf{S}_{q\rm{T}}+2(\textbf{S}_{q\rm{T}}\cdot\,\kpt)\kpt+2\IM(\mu)\,\rm{k'_T}\,\tilde{\textbf{n}}(\kpt)\\
&&\,\,\,\,\,\,\,\,\,\,\,\,\,\,\,\,\,\,\,\,\,\,\,\,-2\RE{\mu}\,S_{q\rm{L}}\,\kpt\big],
\end{eqnarray}
\begin{eqnarray}\label{eq:S_q'L}
S_{q'\rm{L}}=\frac{1}{N}\big[(|\mu|^2-\kptkpt)\,S_{q\rm{L}}-2\RE{\mu}\,\textbf{S}_{q\rm{T}}\cdot\kpt]
\end{eqnarray}
where $N$ is the normalization of $\rho(q')$ given by
\begin{equation}\label{eq:N}
    N=|\mu|^2+\kptkpt-2\IM{\mu}\,\rm{k'_T}\,\textbf{S}_{q\rm{T}}\cdot \tilde{\textbf{n}}(\kpt).
\end{equation}
Equation (\ref{eq:S_q'T}) tells that the transverse polarization of $q'$ has different contributions: a $\kpt$-dependent amount of the transverse polarization of $q$ is transferred to $q'$, also $\kpt$ alone is a source of polarization for $q'$. There is also conversion (worm-gear effect) from transverse to longitudinal polarization and viceversa, as can be seen from Eqs. (\ref{eq:S_q'T}-\ref{eq:S_q'L}). However, the total degree of polarization is conserved: if $\Sq^2=1$ then also $\textbf{S}'^2_q=1$, namely if $q$ is in a pure state, then also $q'$ is in a pure state, owing to the fact that the emitted pseudo-scalar meson does not carry away spin information.

The correlation between the quark transverse momentum and its transverse polarization, expected from the classical string$+\,^3P_0$ model and given in Eq. (\ref{eq:classical string+3P0 correlations}) can be obtained from Eq. (\ref{eq:S_q'T}), taking the vector product between $\kpt$ and $\Sqp$, and then projecting along $\zu$. For $\IM(\mu)>0$, it is 
\begin{equation}
    \langle \kpt\times\Sqp\rangle\cdot \zu = \frac{2\IM(\mu)\langle\ktkt\rangle_{f\T}}{|\mu|^2+\langle \ktkt\rangle_{f\T}}>0,
\end{equation}
integrating separately the numerator and the denominator for fixed $\STq$.

Due to this correlation, integrating over the transverse momentum of $q'$ produces a leakage of spin information. Indeed, the transverse and the longitudinal quark polarizations decay at two different rates quantified by the \textit{depolarization factors} $D^{ps}_{\rm{TT}}$ and $D^{ps}_{\rm{LL}}$ respectively. They can be obtained from Eqs. (\ref{eq:S_q'T}-\ref{eq:S_q'L}) integrating the respective numerators and denominators separately. Namely, they are
\begin{eqnarray}\label{eq:DTT}
\textbf{S}_{q'\rm{T}}&=&-\frac{|\mu|^2}{|\mu|^2+\langle \ktkt\rangle_{f\T}}\textbf{S}_{q\rm{T}}\equiv D^{ps}_{\rm{TT}}\,\,\textbf{S}_{q\rm{T}}
\end{eqnarray}
\begin{eqnarray}\label{eq:DLL}
S_{q'\rm{L}}=\frac{|\mu|^2-\langle \ktkt\rangle_{f\T}}{|\mu|^2+\langle \ktkt\rangle_{f\T}}\,S_{q\rm{L}}\equiv D^{ps}_{\rm{LL}}\,S_{q\rm{L}},
\end{eqnarray}
and for $f\T$ in Eq. (\ref{eq:fT exponential}) one obtains
\begin{eqnarray}\label{eq: DTT DLL}
    D^{ps}_{\rm{TT}}=\frac{-\bt|\mu|^2}{1+\bt|\mu|^2}, && D^{ps}_{\rm{LL}}=\frac{\bt|\mu|^2-1}{\bt|\mu|^2+1}.
\end{eqnarray}
Hence the polarization decay along the fragmentation chain depends on the complex mass and on the parameters of the function $f\T$. The results in Eq. (\ref{eq: DTT DLL}) correspond to those obtained with the elementary model of Ref. \cite{DS09}.
Note also that $D_{\rm{TT}}<0$, meaning that the quark transverse polarization is flipped from one splitting to the next, giving the Collins effect on alternate sides for even and rank mesons as expected from the classical string+$^3P_0$ model.

\subsubsection*{Study of the positivity bounds}
Supposing the polarization of $q'$ to be measured by a \textit{gedanken} quark polarimeter which accepts only the spin state $|+\check{\textbf{S}}_{q'}/2\rangle$, one can define an \textit{all-polarized} splitting function. Encoding the polarization vector $\check{\textbf{S}}_{q'}$ in the \textit{acceptance} spin density matrix $\check{\rho}(q')$, the splitting function in Eq. (\ref{eq:F=TrhoT pol}) is generalized to
\begin{equation}\label{eq: all pol F}
    F_{q',h,q}=\trace\left[T(\q',\h,\q) \rho(q)T^{\dagger}(\q',\h,\q)\check{\rho}(q')\right].
\end{equation}
In this equation the vector $\check{\textbf{S}}_{q'}$ is imposed and does not depend on the involved momenta nor on the polarization of $q$.
The all-polarized splitting function of the simplified string $+\,^3P_0$ model is
\begin{eqnarray}\label{eq:F_doubly_polarized}
 \nonumber  F_{q',h,q}(Z,\pt,\check{\textbf{S}}_{q'};\kt,\textbf{S}_q)&=& \frac{|C_{q',h,q}|^2}{\sum_H |C_{q',H,q}|^2}
\nonumber \left(\frac{1-Z}{\varepsilon_h^2}\right)^a \frac{\exp{(-\bl \varepsilon_h^2/Z)}}{N_a(\varepsilon_h^2)}\\
  &&\times \frac{|\mu|^2+\kptkpt}{|\mu|^2+\langle \ktkt\rangle_{f\T}} f_{\T}^2(\kptkpt) \times \frac{1}{2}\,C(\textbf{S}_q,\check{\textbf{S}}_{q'}).
  \end{eqnarray}
To write down explicitly the correlation function $C(\textbf{S}_q,\check{\textbf{S}}_{q'})$, it is convenient to project the quark polarization vectors on the right-handed basis with axes
\begin{eqnarray}\label{eq: MNL states}
    \left(\rm{\textbf{m}},\rm{\textbf{n}},\rm{\textbf{l}}\right) = \left(\hat{\textbf{k}}'\T,\hat{\textbf{z}}\times \hat{\textbf{k}}'\T,\zu\right).
\end{eqnarray}
$C(\textbf{S}_q,\check{\textbf{S}}_{q'})$ can then be decomposed as
\begin{eqnarray}\label{eq:C(Sq,Sq')}
\nonumber C(\textbf{S}_q,\check{\textbf{S}}_{q'})&=& 1+C_{\rm{n0}}S_{q\rm{n}}+C_{\rm{0n}}\check{S}_{q'\rm{n}}\\
\nonumber  &&+C_{\rm{nn}}S_{q\rm{n}}\check{S}_{q'\rm{n}}+C_{\rm{mm}}S_{q\rm{m}}\check{S}_{q'\rm{m}}\\
 \nonumber  &&+C_{\rm{ml}}S_{q\rm{m}}\check{S}_{q'\rm{l}}+C_{\rm{lm}}S_{q\rm{l}}\check{S}_{q'\rm{m}}\\
  &&+C_{\rm{ll}}S_{q\rm{l}}\check{S}_{q'\rm{l}},
\end{eqnarray}
The coefficients $C_{i,j}$, for $i,j=\rm{0,m,n,l}$ that refer to unpolarized and projections along the $\rm{\textbf{m}},\rm{\textbf{n}},\rm{\textbf{l}}$ axes respectively, describe the dynamics of the different spin transfer possibilities from $q$ to $q'$. They determine the polarization vector $\Sqp$ of Eqs. (\ref{eq:S_q'T}-\ref{eq:S_q'L}) by the equation
\begin{equation}\label{eq: Sq' = grad C}
    \textbf{S}_{q'}=\frac{\nabla_{\check{\textbf{S}}_{q'}} C(\textbf{S}_q,\check{\textbf{S}}_{q'})} {C(\textbf{S}_q,\textbf{0})}.
\end{equation}
The non vanishing ones allowed by parity invariance are
\begin{eqnarray}\label{eq:Cij}
&&C_{\rm{n0}}=-\frac{2\IM{\mu}\,\rm{k'_T}}{|\mu|^2+\kptkpt}=-C_{\rm{0n}}\\
&&C_{\rm{nn}}=-1\\
&&C_{\rm{mm}}=\frac{-|\mu|^2+\kptkpt}{|\mu|^2+\kptkpt}=-C_{\rm{ll}}\\
&&C_{\rm{ml}}=-\frac{2\RE{\mu}\,\rm{k'_T}}{|\mu|^2+\kptkpt}=C_{\rm{lm}}.
\end{eqnarray}
The coefficient $C_{\rm{n0}}$ is responsible for the Collins effect. By analogy with the correlations between a polarized quark inside a polarized parent nucleon (see Eq. (\ref{eq:q(x,kT,S)})), $C_{\rm{n0}}$ corresponds to a Sivers-like correlation, $C_{\rm{0n}}$ to a Boer-Mulders like correlation, $C_{\rm{nn}}$ and $C_{\rm{mm}}$ to transversity- and pretzelosity- like and finally $C_{\rm{ml}}$ and $C_{\rm{lm}}$ to worm-gear-like.

These coefficients saturate the positivity conditions \cite{X.A_et_al_spin_observables}
\begin{eqnarray}\label{eq:positivity_conditions}
 (1\pm C_{\rm{nn}})^2\geq (C_{\rm{0n}}\pm C_{\rm{n0}})^2 + (C_{\rm{ll}}\pm C_{\rm{mm}})^2 + (C_{\rm{lm}}\mp C_{\rm{ml}})^2,
\end{eqnarray}
expected for a quantum mechanical model. The inequality in Eq. (\ref{eq:positivity_conditions}) is derived for the process $N\rightarrow q+X$ where $N$ represents the nucleon, $q$ the quark and $X$ the remnant state. In that case the inequality is not saturated, because there are many possible states $|X\rangle$. The splitting $q\rightarrow h+q'$ has the same spin structure. In this case the inequality must be saturated because $h$ is pseudoscalar and has no spin.

In this respect the model presented here is different from the recursive model of Ref. \cite{Mate}, which also includes the quark spin but where the inequality in Eq. (\ref{eq:positivity_conditions}) is not automatically saturated. A comparison between the two models from a different perspective can be found in Ref. \cite{collins-rogers}.


\subsection{The more general string+${}^3P_0$ model}\label{sec:M18}
The spin-independent $\kt$-$\kpt$ correlations can be included in the splitting function using choice $C_2$ in Eq. (\ref{eq:choice-c2}). This choice of the function $\check{g}$ has been published in Ref. \cite{kerbizi-2018} and the resulting model is called M18 throughout the thesis. With Eq. (\ref{eq:choice-c2}) and with $f\T$ in Eq. (\ref{eq:fT exponential}), the corresponding functions $\hat{u}_{0q}(\kt\kt)$ and $\hat{u}_{1,q}(\ktkt)$ which parameterize the matrix of Eq. (\ref{eq: hat uq}) are
\begin{eqnarray}\label{eq:u0q}
 \hat{u}_{0,q}(\ktkt)&=&\pi\sum_H |C_{q',h,q}|^2\int_0^1 \frac{dZ}{\bl+Z\bt}(1-Z)^a\exp{(-\bl m_h^2/Z)}\\
\nonumber &\times& \exp{\left[\bl\bt\ktkt/(\bl+Z\bt)\right]}\left[|\mu|^2+\frac{Z}{\bl+Z\bt}+\frac{\bl^2\ktkt}{(\bl+Z\bt)^2}\right]
\end{eqnarray}
\begin{eqnarray}\label{eq:u1q}
 \hat{u}_{1,q}(\ktkt)&=&-\pi\sum_H |C_{q',h,q}|^2\int_0^1 \frac{dZ}{\bl+Z\bt}(1-Z)^a\exp{(-\bl m_h^2/Z)}\\
\nonumber &\times& \exp{\left[\bl\bt\ktkt/(\bl+Z\bt)\right]}\frac{2\IM(\mu)Z\rm{k_T}}{\bl+Z\bt}.
\end{eqnarray}
The matrix $\hat{u}^{-1/2}_q$ is therefore non trivial and in the simulations of Ref. \cite{kerbizi-2018} has been tabulated. Also, the probability of generating the hadron type has the general form of Eq. (\ref{eq: P q->h}) where the functions $p_{q\rightarrow h}(\ktkt)$ and $\delta p_{q\rightarrow h}(\ktkt)$ are not calculable analytically and have to be tabulated as well. In order to overcome this complication, in the Monte Carlo simulation of Ref. \cite{kerbizi-2018} the probability of generating the hadron type has been approximated with
\begin{equation}
    P_{q\rightarrow h}\simeq \frac{|C_{q',h,q}|^2}{\sum_H|C_{q',H,q}|^2},
\end{equation}
at the price of a slight violation of the LR symmetry.

Using Eq. (\ref{eq:general pol F}) with choice $C_2$ for $\check{g}$, the splitting function for pseudo-scalar meson emission is
\begin{eqnarray}\label{eq:pol F correlations}
\nonumber    F_{q',h,q}(Z,\kpt;\kt,\Sq)&=& |C_{q',h,q}|^2 (1-Z)^a\exp(-\bl m_h^2/Z)\\
\nonumber &\times& \exp(-\bt\xi(Z)\ktkt) \\
\nonumber &\times& \exp\left[-\frac{\bl}{Z\xi(Z)}(\kpt-\xi(Z)\kt)^2\right]\\
&\times& [|\mu|^2+\kptkpt-2\IM(\mu)\textbf{S}_{q,int}\cdot\ntil(\kpt)].
\end{eqnarray}
The vector $\textbf{S}_{int}$ depends on $\kt$ and is extracted from the intermediate spin density matrix introduced in Eq. (\ref{eq:rhoint}), namely
\begin{equation}
    \textbf{S}_{q,int}=\frac{\trace (\hat{\rho}_{int}(q)\boldsymbol{\sigma})}{\trace \hat{\rho}_{int}(q)}.
\end{equation}
As already introduced in Eq. (\ref{eq:choice-c2}), the spin independent and $Z$-dependent $\kt$-$\kpt$ correlations in this case are described by the correlation coefficient $\xi(Z)=\bl/(\bl+Z\bt)$. This correlation favours $\kpt$ to be aligned with $\kt$ and is opposite to the $\kt$-$\kpt$ pure spin-mediated correlation produced by the $^3P_0$ mechanism.

With the choice $C_2$ made here it is more convenient to generate first $Z$ and then $\pt$. 
The $\kpt$-integrated splitting function in Eq. (\ref{eq:pol F correlations}) is
\begin{eqnarray}\label{eq: Z distribution}
\nonumber && dZ\xi(Z)(1-Z)^a\exp\left[-\bl m_h^2/Z-\bt\xi(Z)\ktkt\right]\\
&\times& \left[|\mu|^2+\bl^{-1}Z\xi(Z)+\xi^2(Z)\ktkt-2\IM(\mu)\rm{k_T}\xi(Z)\textbf{S}_{int}\cdot\ntil(\kt)\right].
\end{eqnarray}
The $Z$ distribution depends exponentially on $\ktkt$. Indeed for large $\ktkt$, larger values of $Z$ are favoured by the exponential term. More precisely, the first rank hadron $h_1$ is emitted by the initial quark which does not possess $\kt$, namely in the splitting $q\A(\textbf{0}_{\rm{T}},k_A^+)\rightarrow h_1(\textbf{p}_{1\rm{T}},Z_1k\A^+)+q_2(\textbf{k}_{2\rm{T}},(1-Z_1)k\A^+)$. For this first splitting only the mass of $h_1$ enters the exponential in Eq. (\ref{eq: Z distribution}). The next splitting is $q_2(\textbf{k}_{2\rm{T}},k_2^+)\rightarrow h_2(\textbf{p}_{2\rm{T}},Z_2k_2^+)+q_3(\textbf{k}_{3\rm{T}},(1-Z_2)k_2^+)$ and $Z_2$ is shifted towards larger values with respect to $Z_1$ because of non vanishing $\textbf{k}^2_{2\rm{T}}$. The other splittings are similar to that of $q_2$ and are characterized by the same $Z$ distributions.

The hadron transverse momentum $\pt$ is generated according to the distribution
\begin{eqnarray}\label{eq: pT distribution}
d^2\pt\exp{\left[-\frac{\bl}{Z\xi(Z)}(\kpt-\xi(Z)\kt)^2\right]}\times [|\mu|^2+\kptkpt-2\IM(\mu)\textbf{S}_{int}\cdot\kptil],
\end{eqnarray}
with $\pt=\kt-\kpt$. The term $\textbf{S}_{int}\cdot\kptil$ for $\IM(\mu)>0$ pushes $\kpt$ in the direction $\zu\times\textbf{S}_{int}$ and is responsible for the Collins effect.

In addition, as already mentioned, with the choice of $\check{g}$ made here the transverse momenta of $q$ and $q'$ are correlated. The exponential factor in Eq. (\ref{eq: pT distribution}) favours the relation $\ptpt\sim \left[1-\xi(Z)\right]^2\ktkt$. Furthermore because $0<\xi(Z)<1$ for every value of $Z$, $\kpt$ is favoured to be aligned with $\kt$. This correlation acts against the one generated by the $^3P_0$ mechanism.

Finally, due the strong dependence of the $\pt$ distribution on $\kt$ through the exponential factor in Eq. (\ref{eq: pT distribution}), the $\ptpt$ distribution of the first rank hadron is characterized by a smaller mean value than the $\ptpt$ distribution of larger ranks, which all have similar distributions.

The spin density matrix of the quark $q'$ can be calculated using Eq. (\ref{eq:rho'=TrhoT}) together with Eq. (\ref{eq: T final}),
\begin{equation}
    \rho(q')=\frac{(\mu+\sigma_z\vecsigma\cdot \kpt)\Gamma_{h,s_h}\hat{\rho}_{int}(q)\Gamma^{\dagger}_{h,s_h}(\mu^*-\sigma_z\vecsigma\cdot\kpt)}{\trace\left[numerator\right]}.
\end{equation}
In this case the presence of the intermediate spin density matrix complicates the analytical study of the positivity conditions as done for choice $C_1$. However they are expected to hold since the model is formulated at the amplitude level. 
\chapter{Simulation of the quark fragmentation and first results}\label{chapter 3}
The present chapter is dedicated to the implementation in a stand alone Monte Carlo program of the string+${}^3P_0$ model described in Chapter 2. This is the first necessary step in view of the interface with complete event generators (Chapter 4) and further improvements of the model (Chapter 5). Quark jets, with given flavor, momentum and spin density matrix are generated. The flavor of the fragmenting quark can be $u$, $d$ or $s$ and only pseudoscalar mesons ($\pi$, $K$, $\eta^0$ and $\eta'$) are produced. The energy of the initial quarks can either be fixed at some value or chosen event-by-event by reading the values from an external file. The initial quark can be unpolarized, transversely or longitudinally polarized. The output consists in a file which contains the relevant information of the produced hadrons, which can then be read and analyzed by a different program.

Both M18 and M19 models, described in section \ref{sec:simple splitt function} and in section \ref{sec:M18} respectively, have been implemented in MC codes, which have the same structure, except from some preliminary tasks for M18, as it will be made clear below. The MC program is described in section \ref{sec:comparison M18 and M19} the comparison between M18 and M19 is presented.

\section{The Monte Carlo program}\label{sec:structure of stand alone MC Chapter 3}
\subsection{The structure of the program}\label{sec:structure of stand alone MC Chapter 3 subsection}
In the initialisation the flavor, and the spin density matrix of the fragmenting quark $q_A$ are chosen. 

In order to compare the simulation results with measurements either from the SIDIS process or from $e^+e^-$ annihilation process, the initial kinematics is defined in two different ways.
For the lepton-nucleon DIS process, the initial kinematics is defined in each event by reading the $x_B$ and $Q^2$ values in a file of DIS events. This is done in the center of mass system of the exchanged virtual photon $\gamma^*$ and of the proton, with the $\zu$ axis along the virtual photon momentum $\textbf{q}_{\gamma^*}$. In the case where $q\A$ has no intrinsic transverse momentum $\textbf{k}_{\perp}$, this axis is also the string axis. The total lightcone momenta are $P^+=P^-=W$. $W$ is the invariant mass of the hadronic system and it is linked to $x_B$ and $Q^2$ by $W^2=P^+P^-=(1/x_B-1)Q^2+M_N^2$. The phase space cuts $Q^2>1\,(\rm{GeV}/c)^2$, $W>5\,\rm{GeV}/c^2$ and $0.2<y<0.9$ have been applied on the DIS events.

In the case of the $e^+e^-$ annihilation process, $W$ coincides with the center of mass energy and is the same for all the events in the simulation.

In the center of mass reference system, the fragmenting quark $q\A$ travels along the forward lightcone with momentum $k_A^+=P^+$. The jet initiated by $\bar{q}\B$ (or the target remnant in the DIS case) which travels along the backward lightcone with momentum $k_{\bar{B}}^-=P^-$ is not considered. The initial $q_A$ and $\bar{q}_B$ are thus taken to be massless. The momenta along the backward lightcone are kept track and used for the exit condition, as explained below.

As seen in the previous chapter, in M18 the two functions $\hat{u}_{0q}$ and $\hat{u}_{1q}$ defined in Eq. (\ref{eq:u0q}) and in Eq. (\ref{eq:u1q}) are needed to calculate the intermediate quark spin density matrix $\hat{\rho}_{int}(q)$. In the initialisation these functions are calculated and tabulated. They allow to construct the matrix $\hat{u}^{-1/2}$ as function of the recurrent quark transverse momentum which is used for the calculation of $\hat{\rho}_{int}(q)$ at each step of the generation process according to Eq. (\ref{eq:rhoint}). For M19 this preliminary task is not needed, because in this variant $\hat{\rho}_{int}(q)$ reduces to the true quark spin density matrix $\rho(q)$.

\subsubsection*{The recursive algorithm}
\begin{figure}
\centering
\begin{minipage}{0.8\textwidth}
  \centering
  \includegraphics[width=0.8\linewidth]{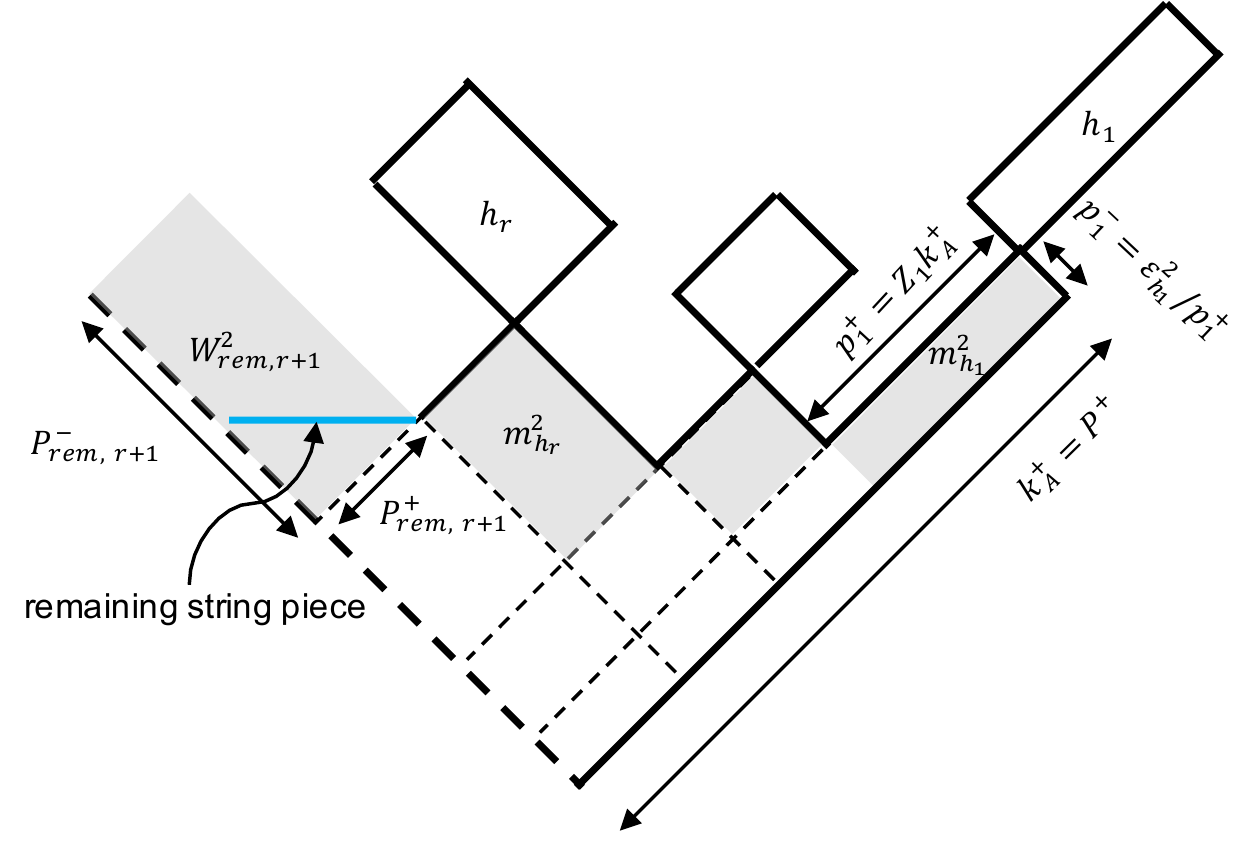}
\end{minipage}
\caption{Graphical representation of the Monte Carlo procedure and of the exit condition.}
\label{fig:exit condition}
\end{figure}

The quark jet is generated by repeating recursively the splitting $q\rightarrow h + q'$ ($q\equiv q_A$ in the first splitting) following the steps:
\begin{itemize}
    \item [(1)] generate a $q'\bar{q}'$ pair
    \item [(2)] form the pseudoscalar meson $h=q\bar{q}'$ and identify its type and mass $m_h$
    \item [(3)] for M18: generate $Z$ according to the $\pt$-integrated splitting function given in Eq. (\ref{eq: Z distribution}) and calculate $p^+=Zk^+$. For M19: generate $\kpt$ according to the $Z$-integrated splitting function given in Eq. (\ref{eq:F_explicit simple 3P0}) and construct $\pt=\kt-\kpt$ (for $q_A$, $\kt=0$).
    \item [(4)] for M18: generate $\pt$ according to the function given in Eq. (\ref{eq: pT distribution}) at the generated value of $Z$, and calculate $\kpt=\kt-\pt$. For M19: generate $Z$ according to the splitting function in Eq. (\ref{eq:F_explicit simple 3P0}) evaluated at the generated value of $\pt$. 
    \item [(5)] calculate $p^-$ imposing the mass shell condition $p^+p^-=m_h^2+\ptpt$ and test the remaining backward lightcone momentum $P^{-}_{rem}$
    \item [(6)] test the exit condition and if it is not satisfied continue with the next step, otherwise the current hadron is removed and the fragmentation chain ends
    \item [(7)] calculate the hadron momentum $p=(E_h,\pt,p_L)$, where $E_h=(p^++p^-)/2$ and $p_L=(p^+-p^-)/2$, and store it in the event record
    \item [(8)] calculate the spin density matrix of $q'$ using Eq. (\ref{eq:rho'=TrhoT}) and return to step 1.
\end{itemize}
The steps $1$-$8$ are iterated until the exit condition is satisfied. The process is represented graphically in Fig. \ref{fig:exit condition}. More information on steps (1), (2), (5) and (6) are given in the following.

\subsubsection*{Generation of the quark flavor and hadron type}
The generation of the quark flavor in step (1) is done taking into account the $s$ quark suppression required by the tunneling mechanism. In particular the values $P(u\bar{u}):P(d\bar{d}):P(s\bar{s})=3/7:3/7:1/7$ are used, which correspond to $\alpha=3/7$ in Eq. (\ref{eq:uds prob}).

The identification of the hadron type corresponding to the state $(q\bar{q}')$ in step (2) uses probabilities obtained from the isospin wave function, i.e. $\pi^+=u\bar{d}$, $\pi^-=d\bar{u}$, $\pi^0=(u\bar{u}+d\bar{d})/\sqrt{2}$, $K^+=u\bar{s}$, $K^-=s\bar{u}$, $K^0=d\bar{s}$, $\bar{K}^0=s\bar{d}$. In addition $\eta^0$ meson production is suppressed with respect to $\pi^0$ to take into account the mass difference. To do this the treatment of $\eta$ and $\eta'$ mesons is simplified (no mixing is taken into account). In practice, each time a $u\bar{u}'$ or $d\bar{d}'$ configuration is encountered, this flavor diagonal configuration is identified with a $\pi^0$ with probability $0.5$ or with an $\eta^0$ with probability $p_{\eta}=1/2-r_{\eta}$ with $r_{\eta}$ set to $1/4$ in order to reproduce the experimental result $N(\eta^0)/N(\pi^0)\simeq 0.57$ \cite{Field-Feynman}. Such flavor diagonal configuration is otherwise rejected with probability $r_{\eta}$ and a new $q\bar{q}'$ pair is generated. If an $s\bar{s}$ state is encountered it is identified with $\eta$ or $\eta'$ with probabilities $2/3$ and $1/3$ respectively.
The recipe for the generation of the quark flavor and the identification of the hadron type gives for the coefficients $|C_{q'hq}|^2$ the expression
\begin{eqnarray}\label{eq: Cq'hq matrix}
|C_{q',h,q}|^2 = \begin{pmatrix}
\alpha(1-p_{\eta}) & \alpha & 1-2\alpha \\
\alpha & \alpha(1-p_{\eta}) & 1-2\alpha \\
1-2\alpha & 1-2\alpha & (1-2\alpha)^2/\alpha
\end{pmatrix}_{qq'} \times |\langle h|q\bar{q}'\rangle|^2.
\end{eqnarray}
The matrix is defined in the SU(3) flavor basis $(u,d,s)$ and $\langle h|q\bar{q}'\rangle$ is the projection of the isospin wave function $|h\rangle$ of $h$ along the state $|q\bar{q}'\rangle$. The invariance under the exchange $q \leftrightarrow q'$ and $h\leftrightarrow \bar{h}$ ensures that the LR symmetry is preserved in the splitting in quark flavor space.

\subsubsection*{Test on the remaining backward lightcone momentum}
Once $r$ splittings have been generated, the momentum of the remaining string piece is $P_{rem,r+1}=k_{r+1}+k_{\bar{B}}$. It has lightcone and transverse components
\begin{eqnarray}
\nonumber P_{rem,r+1}^{+}=P^+_{rem, r}-p_r^+, \\
\nonumber P_{rem,r+1}^{-}=P^-_{rem, r}-\varepsilon_h^2/p_r^+, \\
\textbf{P}_{\rm{T},r+1}=\textbf{P}_{\rm{T},r}-\textbf{p}_{\rm{T},r}.
\end{eqnarray}
If the next hadron is generated with a very low $Z$ then $P_{rem,r+1}^{-}$ may become negative. In this case the hadron is rejected and a new one is tried.

\subsubsection*{Exit condition}
The recursive algorithm is ended if the squared mass of the remaining string piece $W_{r+1}^2=P_{rem,r+1}^{+}P_{rem,r+1}^{-}-\textbf{P}_{\rm{T},r+1}^2$ falls below a given mass $M_R^2$. The quantity $W_{r+1}^2$ is represented in Fig. \ref{fig:exit condition}. In this case the last generated hadron is erased. In our simulations, we take $M_R=1.5\,\rm{GeV}/c^2$ to account for the production of one (not simulated) baryon, as required by baryon number conservation in a DIS process. We have checked that the relevant observables, are not sensitive to this value of $M_R$.

\subsection{Values of the free parameters and kinematical distributions}\label{sec: free parameters and kinematical distributions M18 M19}
From Eq. (\ref{eq:F_explicit simple 3P0}) or Eq. (\ref{eq:pol F correlations}) it can be seen that the model is based on five free parameters, namely $a$, $\bl$, $\bt$ and the complex mass $\mu=\RE(\mu)+i\IM(\mu)$. The parameters $a$, $\bl$ and $\bt$ come from the Lund Model (see Eq. (\ref{eq:F Pythia})) whereas $\mu$ has been introduced via the quantum mechanical ${}^3P_0$ operator in Eq. (\ref{eq:3P0 propagator}). The values of the parameters $a$, $\bl$, $\bt$ and $|\mu|^2$ have been tuned comparing the results from simulations of unpolarized quark jets performed with M18, with the $\ptpt$ distributions of charged hadrons measured in SIDIS off an unpolarized deuteron target \cite{compass-pt2} and with a set of unpolarized $\ptpt$-integrated fragmentation functions obtained from global fits \cite{kniehl2001testing}. The tuning has been performed in order to have a reasonable agreement between simulations and unpolarized data, namely reasonable values for the free parameters. No real fit to the data has been done, since a good quantitative agreement was not expected and practically not needed.

It has been observed that the slope of the $\ptpt$ distributions is sensitive to the values of $\bl$ and $\bt$, whereas its shape for $\ptpt\rightarrow 0$ is sensitive to $|\mu|^2$. 
The fragmentation functions are sensitive to $\bl$, which mostly affects the $\langle z_h\rangle$, and $a$ which affects the large fractional energies. The parameter $a$ has almost no effect on the $\ptpt$ distributions. With the values $a=0.9$, $\bl=0.5\,(\rm{GeV}/c^2)^{-2}$, $\bt=5.17\,(\rm{GeV}/c)^{-2}$ and $|\mu|^2=0.75\,(\rm{Gev}/c^2)^2$ a satisfactory qualitative agreement with the data could be achieved.

The transverse spin asymmetries are sensitive to the value of the parameter $\IM(\mu)$, which is the only parameter of the model needed for the transverse spin dependence of the fragmentation process. It has been fixed comparing the Collins analysing power as obtained from simulations with that extracted from $e^+e^-$ annihilation data. More specifically the mean value of the Collins analysing power for positive pions in transversely polarized $u$ jets from simulations has been compared with the mean value $0.258\pm 0.006$ obtained in Ref. \cite{M.B.B} from BELLE data.
Having fixed $|\mu|^2$ from the unpolarized $\ptpt$ distributions, $\RE(\mu)$ is then also fixed. The value of the complex mass used in all the simulations is $\mu=(0.42+i\,0.76)\,\rm{GeV}/c^2$.

In this section the kinematic spin-independent distributions obtained from the simulations carried on with M18 and the chosen values of the parameters are described.

\begin{figure*}[tb]
 \renewcommand{\thesubfigure}{a}
\begin{subfigure}[t]{.5\textwidth}
  \includegraphics[width=0.8\linewidth]{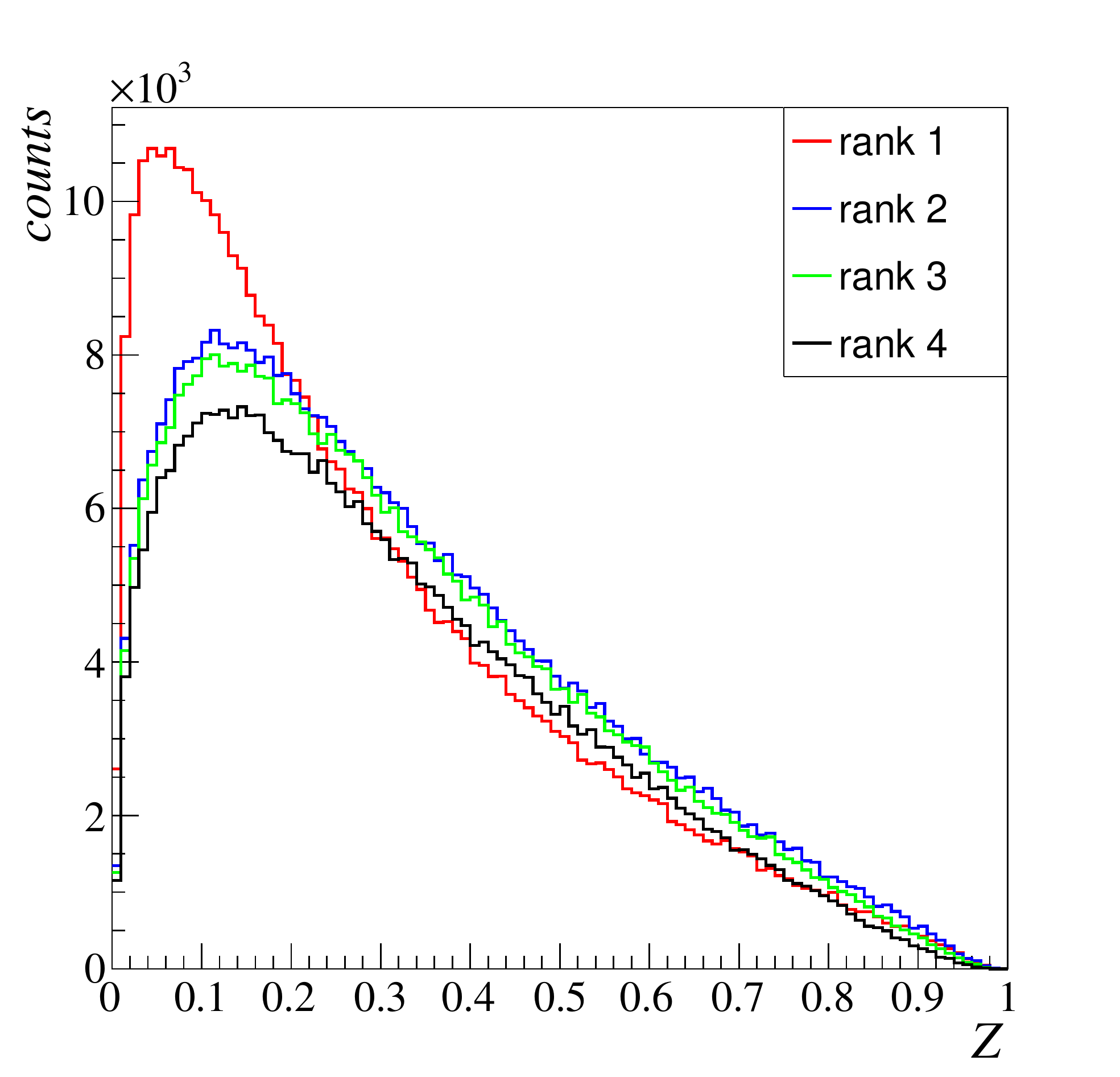}
 \caption{}
  \label{fig:Z distribution}
  \end{subfigure}
\renewcommand{\thesubfigure}{b}
\begin{subfigure}[t]{.5\textwidth}
	\includegraphics[width=0.8\linewidth]{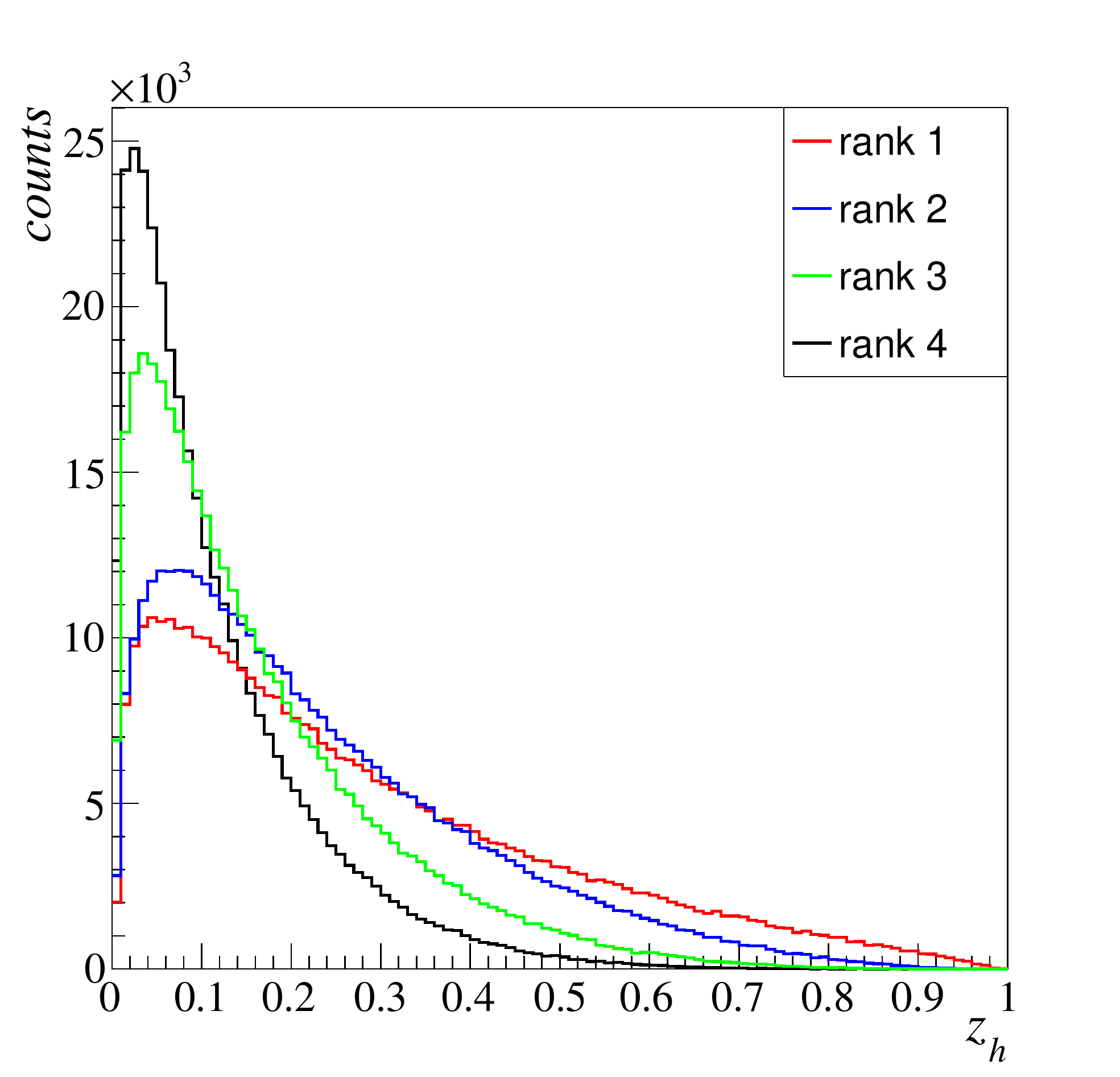}
	\caption{}
	\label{fig:zh distribution}
\end{subfigure}
\caption{Distributions of $Z$ (a) and $z_h$ (b) for the hadrons if rank 1 to 4 produced in $u$ quark jets.}
\end{figure*}

Figure \ref{fig:Z distribution} shows the distributions of the splitting variable $Z$ for the hadrons of rank 1 to 4 generated in the fragmentation chain of a $u$ quark.
As can be clearly seen, the distribution of the first rank hadron is shifted towards smaller values of $Z$ with respect to the distributions of higher rank hadrons. This is due to the $\ktkt$ dependent exponential in the $\pt$-integrated splitting function given in Eq. (\ref{eq: Z distribution}), and for the first rank it is $\ktkt=0$. As expected, the distributions of higher rank hadrons are similar.

The $Z$ and $z_h$ distributions are shown in Fig. \ref{fig:Z distribution} and in Fig. \ref{fig:zh distribution} for the hadrons of rank 1 to 4. By definition $Z$ and $z_h$ coincide for the first rank hadron, hence the corresponding distributions are the same.
For larger ranks the shapes of the $z_h$ distributions change sensibly due to the relation $z_{h_r}\simeq Z_r(1-Z_{r-1})\cdot\cdot\cdot(1-Z_2)(1-Z_1)$, which pushes $z_h$ towards smaller values as the rank increases.

\begin{figure}[tb]
 \renewcommand{\thesubfigure}{a}
\begin{subfigure}[t]{.5\textwidth}
  \includegraphics[width=0.8\linewidth]{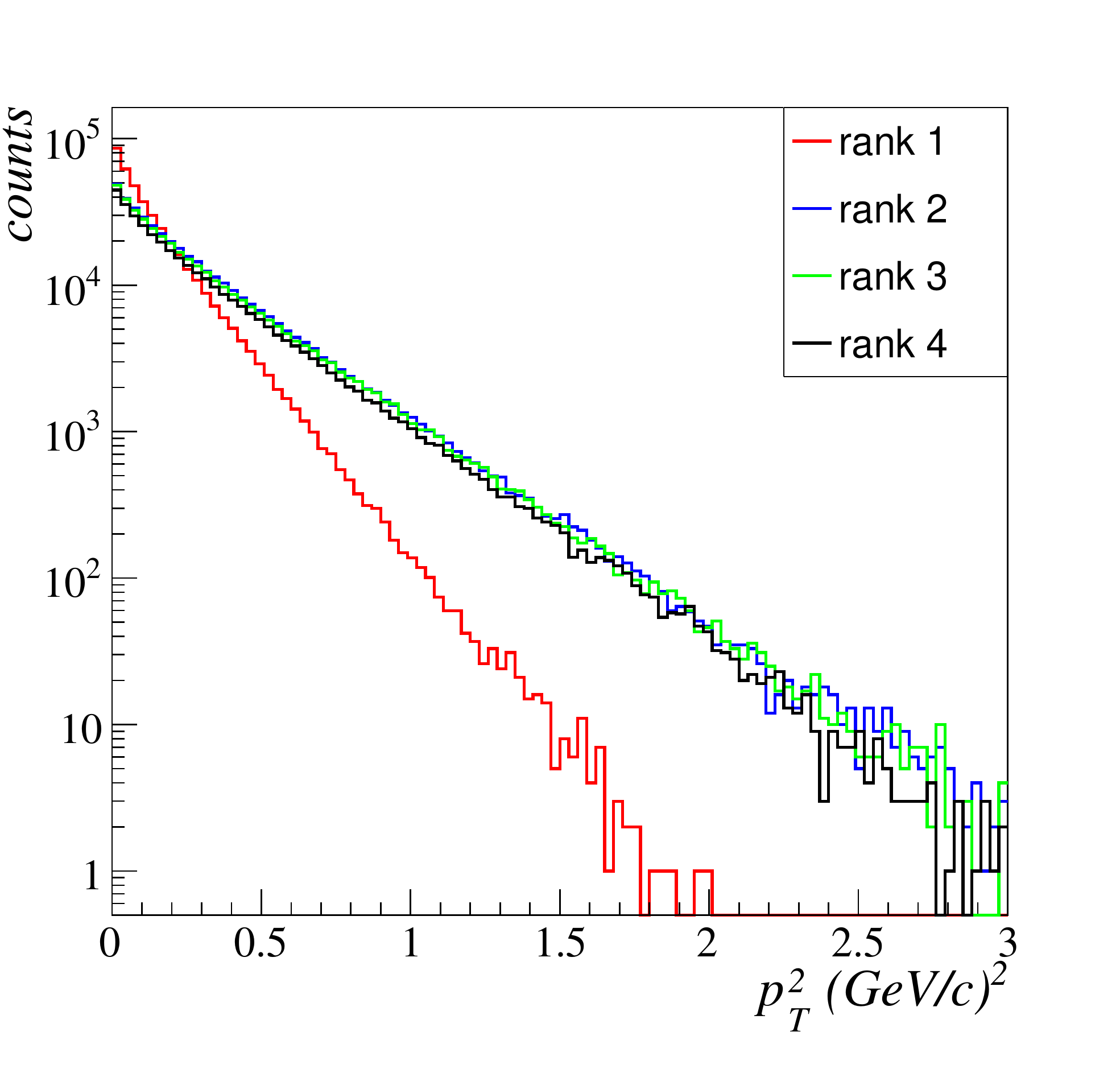}
 \caption{}
  \label{fig:pt distribution}
  \end{subfigure}
\renewcommand{\thesubfigure}{b}
\begin{subfigure}[t]{.5\textwidth}
  \includegraphics[width=0.8\linewidth]{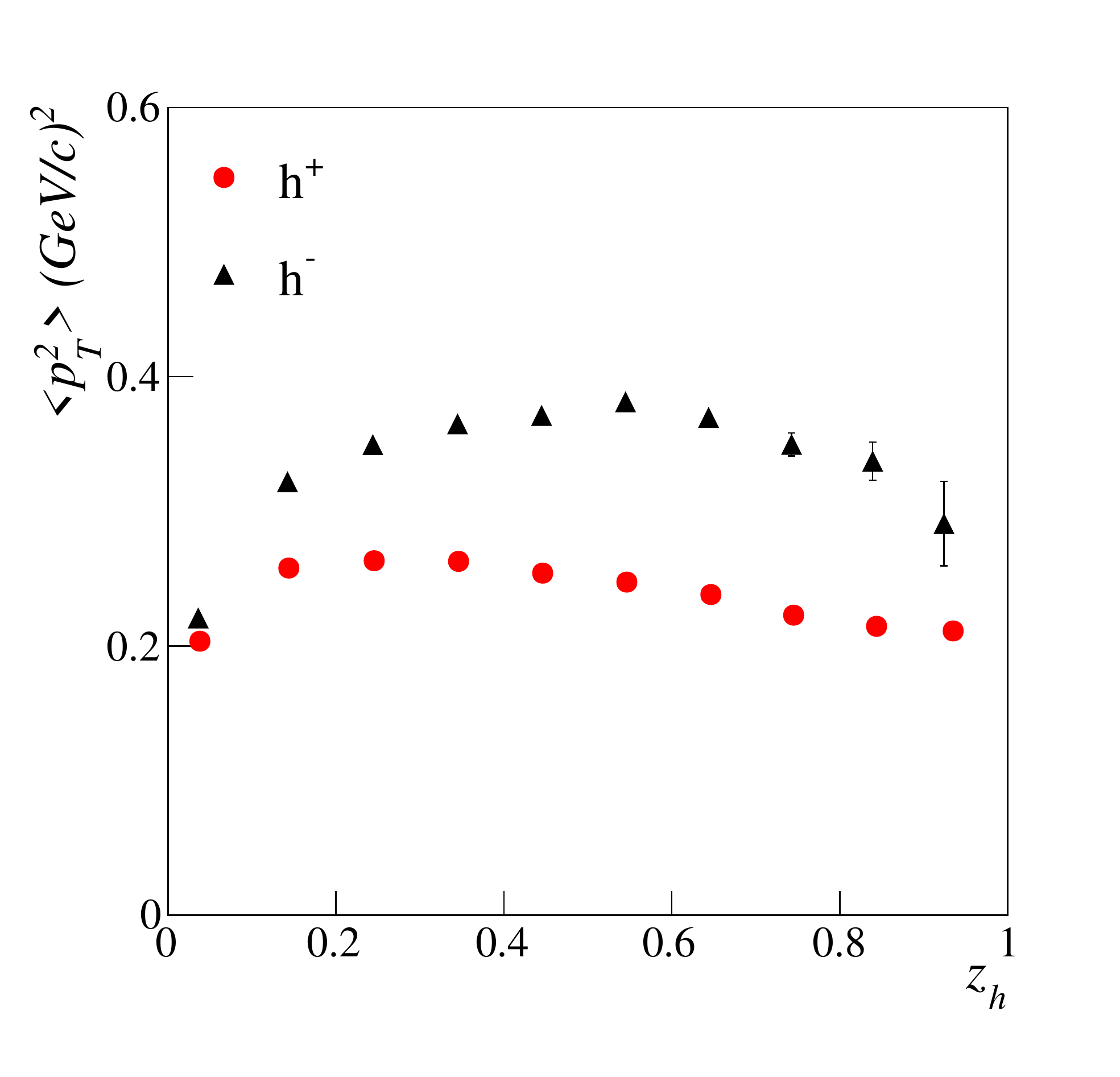}
 \caption{}
  \label{fig:pt2 zh}
  \end{subfigure}
  \renewcommand{\thesubfigure}{c}
\begin{subfigure}[b]{1.0\textwidth}
\hspace{0.8cm}
  \includegraphics[width=0.75\linewidth]{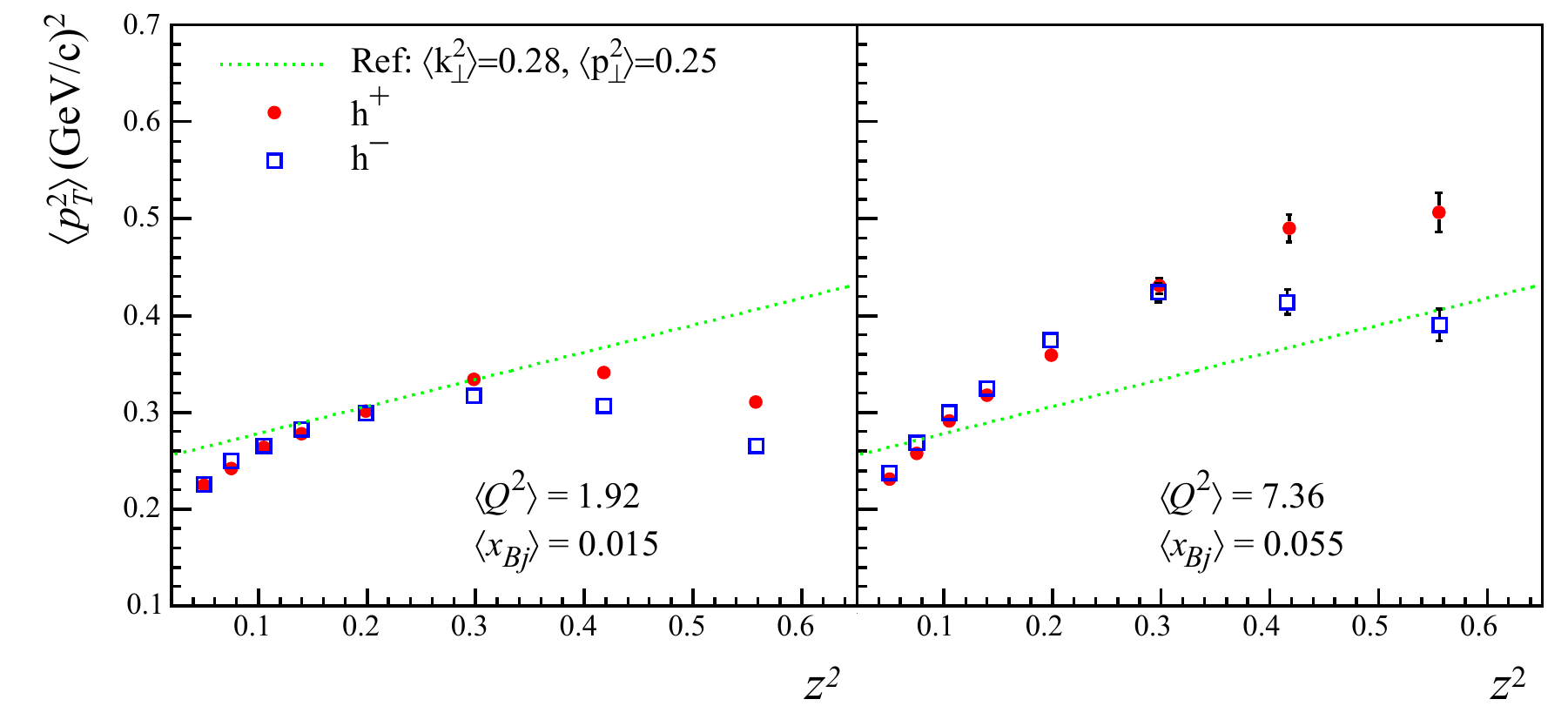}
 \caption{}
  \label{fig:pt2 zh compass}
  \end{subfigure}
  \caption{(a): $\ptpt$ distribution for hadrons of rank 1 to 4 produced in u quark jets. (b): $\langle \ptpt\rangle$ as function of $z_h$ for positive and negative hadrons produced in $u$ quark jets. (c): $\langle \ptpt\rangle$ as function of $z_h^2$ for hadrons produced in SIDIS off a deuteron target \cite{COMPASS-pT2-2013}.}
  \label{fig:fig 3.3}
  \end{figure}

The $\kptkpt$ distributions for the different splittings turn out to be very much the same and, since the initial quark has vanishing $\textbf{k}_{1 \rm T}$, the transverse momentum squared $\textbf{k}^2_{2 \rm T}$ of the left-over quark in the first splitting coincides with the transverse momentum squared $\textbf{p}^2_{1 \rm T}$ of the first hadron.
As a consequence the transverse momentum distribution of the first rank hadron is softer than the distributions of higher rank hadrons, as shown in Fig. \ref{fig:pt distribution}. The slope of the $\ptpt$ distribution of the first rank hadron is larger than the slope of the corresponding distribution of the second rank hadron. Indeed, $\langle \textbf{p}^2_{1 \rm T}\rangle=\langle \textbf{k}^2_{2 \rm T}\rangle$ whereas $\langle \textbf{p}^2_{2 \rm T}\rangle =\langle \textbf{k}^2_{2 \rm T}\rangle+\langle \textbf{k}^2_{3 \rm T}\rangle-2\langle \textbf{k}_{2 \rm T}\cdot\textbf{k}_{3 \rm T}\rangle\geq 2\langle\textbf{k}^2_{1\rm T}\rangle$. As a consequence it is $\langle\ptpt(h^+)\rangle<\langle\ptpt(h^-)\rangle$ in jets initiated by $u$ quarks with vanishing intrinsic transverse momentum, as shown in Fig. \ref{fig:pt2 zh} as function of $z_h$. This difference between the $\langle \ptpt\rangle$ of hadrons of rank $1$ and rank $2$ is a well known feature of the recursive fragmentation models, and decreases when the quark primordial transverse momentum is included. The decrease of $\langle\ptpt\rangle$ at small $z_h$ is due to the factor $\exp(-\bl\ptpt/Z)$ in the splitting function in Eq. (\ref{eq:pol F correlations}).
At variance with the simulation results, the data seem to suggest similar values for $\langle\ptpt(h^+)\rangle$ and $\langle\ptpt(h^-)\rangle$ for $z_h\lesssim 0.55$, as shown in Fig. \ref{fig:pt2 zh compass}. The difference with the simulation results is not due to the use of a deuteron target in Ref. \cite{COMPASS-pT2-2013}.

\section{Results on the transverse spin asymmetries}

In order to study the transverse spin effects measured in SIDIS off transversely polarized protons and in $e^+e^-$ annihilation, fragmentation events have been generated for initial quarks fully polarized along a fixed $\hat{\textbf{y}}$ axis orthogonal to the string axis. Only results for $u$ quark fragmentations are shown in the following. It has been verified that the results for pion production in $d$ quark fragmentation are related to those of the $u$ quark fragmentation as required by isospin symmetry. Namely, the $\pi^+$ and $\pi^-$ analysing powers are exchanged. This is not true for kaons, since both $K^+$ and $K^-$ are unfavored in $d$ quark jets and both of them have positive analysing power. For jets initiated by $s$ quarks $K^+$ is favored and it has positive analysing power, whereas $K^-$ and charged pions are unfavored and they have negative analysing power.

In the SIDIS case the $x_B$ and $Q^2$ values of a sample of real COMPASS events have been used to fix the initial kinematics of the simulation event-by-event.


For the study of the asymmetries in the azimuthal distributions of the hadrons produced in $e^+e^-$ annihilation the events have been generated with a fixed c.m. energy $W=10\,GeV$ corresponding to the BELLE energy.

This section is dedicated to the results for the single hadron and the dihadron transverse spin asymmetries obtained with M18 assuming the intrinsic transverse momentum of the fragmenting quark to be zero. The MC results are compared with the COMPASS data and BELLE data, which are in quite good agreement, but more precise, with the corresponding results from HERMES \cite{hermes-ssa} and Jefferson Laboratory experiments \cite{jlab-ssa} and from BaBar \cite{babar} and BESIII \cite{besIII} experiments respectively.

\subsection{Single hadron transverse spin asymmetries}\label{sec:single hadron transverse spin asymmetry M18}

The azimuthal distribution of hadrons produced in the fragmentation of transversely polarized quarks, already introduced in Chapter 1, is given by
\begin{equation}\label{eq:sin phiC}
\frac{d^4N_h}{dz_h\,d^2\pt\,d\phi_C}\propto 1+a^{q\A\uparrow\rightarrow h+ X}(z_h,p_{\rm T})S_{\rm AT}\sin\phi_C,
\end{equation}
where $S_{\rm AT}$ is the fragmenting quark transverse polarization and $\phi_C=\phi_h-\phi_{\textbf{S}_{\rm A}}$ is the Collins angle. The analysing power $a^{q\A\uparrow \rightarrow h+ X}$ is proportional to the ratio between the Collins FF $H_{1q}^{\perp h}$ and the unpolarized quark FF $D_{1q}^h$ and is given in Eq. (\ref{eq:Collins ap}).

The $\sin\phi_C$ modulation is the only one observed in the simulated events. The fact that no other modulation is present is due to the formulation of the model at the amplitude level, which guarantees also that the positivity constraints are satisfied.
Using simulated events, the analysing power $a^{q\A\uparrow\rightarrow h+ X}$ is calculated as $2\langle\sin\phi_C\rangle$, in general as a function of $z_h$ and $p_{\rm T}$.

\begin{figure*}[tb]\centering
\begin{minipage}{.8\textwidth}
  \includegraphics[width=0.9\textwidth]{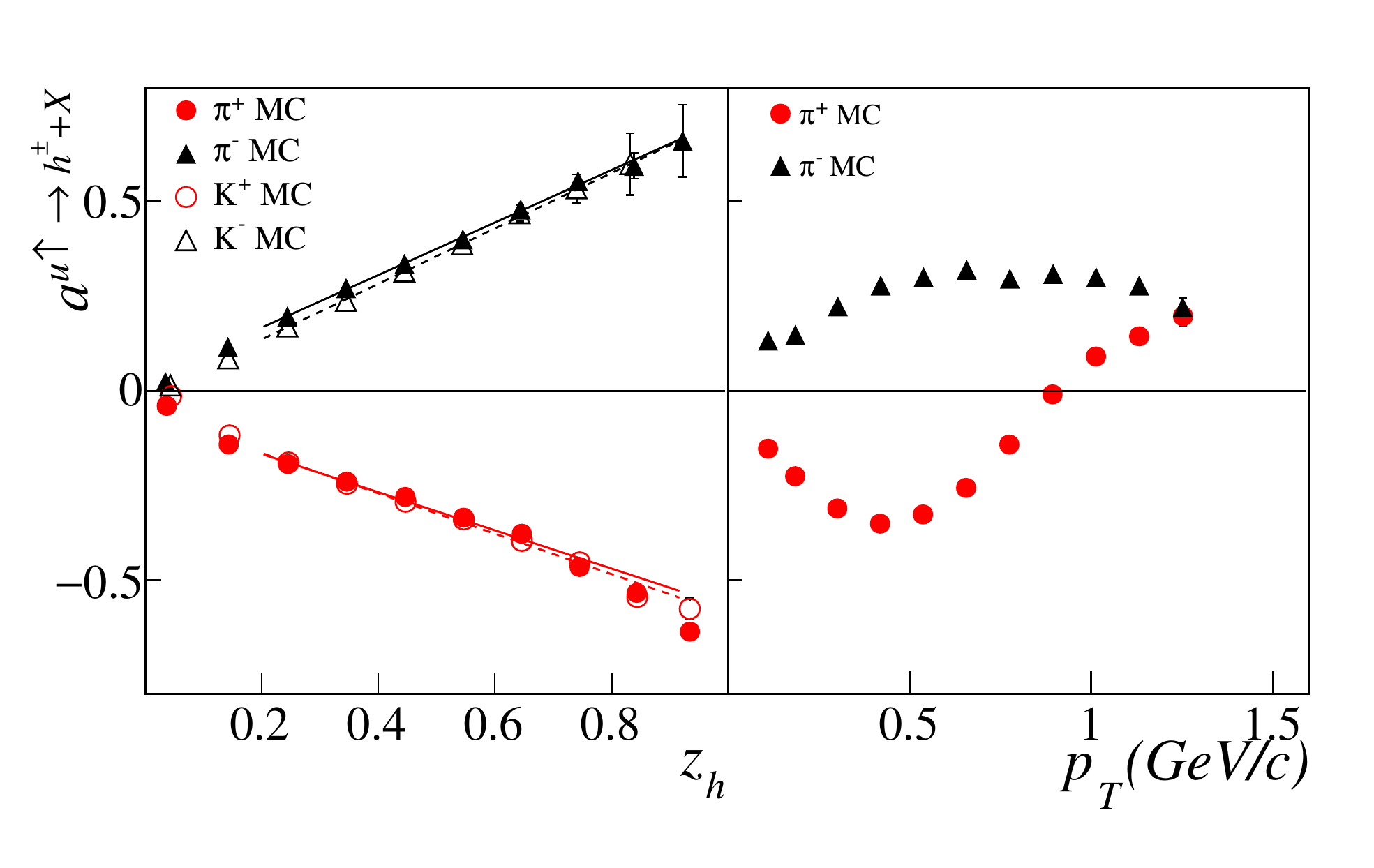}
\end{minipage}%
\caption{\small{Left panel: Collins analyzing power as function of $z_h$ for charged pions produced in simulations of transversely polarized $u$ quark jets. The lines are fits to the different analysing powers. Right panel: simulated Collins asymmetry as function of $p_{\rm T}$.}}\label{fig:mc asymm}
\end{figure*}

The Collins analyzing power $a^{u\uparrow \rightarrow h + X}$ obtained from the simulated events is shown in Fig. \ref{fig:mc asymm} as function of $z_h$ for charged pions and kaons (left panel) and as function of $p_{\rm T}$ for charged pions (right panel). The typical experimental cuts $p_{\rm T}>0.1\rm{GeV}/c$ and $z_h>0.2$ have been applied when looking at the other variable. As expected from the classical string+${}^3P_0$ model, the analyzing power has opposite sign and almost equal magnitude for oppositely charged mesons. The mean values for the different hadrons are given in Tab. \ref{tab:1h mean analyz power}. These values correspond to $2\langle\sin\phi_C\rangle$ averaged on $z_h$ and $p\T$ in the observed phase space. This is the simplest \textit{estimator} of the quark transverse polarization. Other estimators, which may have somewhat better efficiency, have been proposed and tested with simulations in Ref. \cite{Artru-Belghobsi-estimators}. 

\begin{table}
\caption{\label{tab:1h mean analyz power}\small{Mean values of the analyzing power shown in Fig. \ref{fig:mc asymm} for positive and negative charges. The cuts $z_h>0.2$ and $p_{\rm T}>0.1\,\rm{GeV}/c$ have been applied.}}
\centering
\begin{tabular}{l*{6}{c}r}
$\langle a^{u\uparrow\rightarrow h+X}\rangle$         & $h^+$ & $h^-$   \\
\hline
$\pi$& $-0.260\pm 0.002$ & $0.268\pm 0.002$   \\
$K$ & $-0.270\pm 0.003$ & $0.234\pm 0.004$\\
\end{tabular}
\end{table}

 The analyzing power vanishes for small $z_h$ and is almost linear in the range $0.2<z_h<0.8$.  A linear dependence on $z_h$ is also suggested by the BELLE data \cite{M.B.B} when the analysing power for the favoured fragmentation is assumed to be opposite to that for unfavoured fragmentation.

\begin{figure}[tbh]
\centering
\begin{minipage}{.8\textwidth}
\centering
\includegraphics[width=0.6\textwidth]{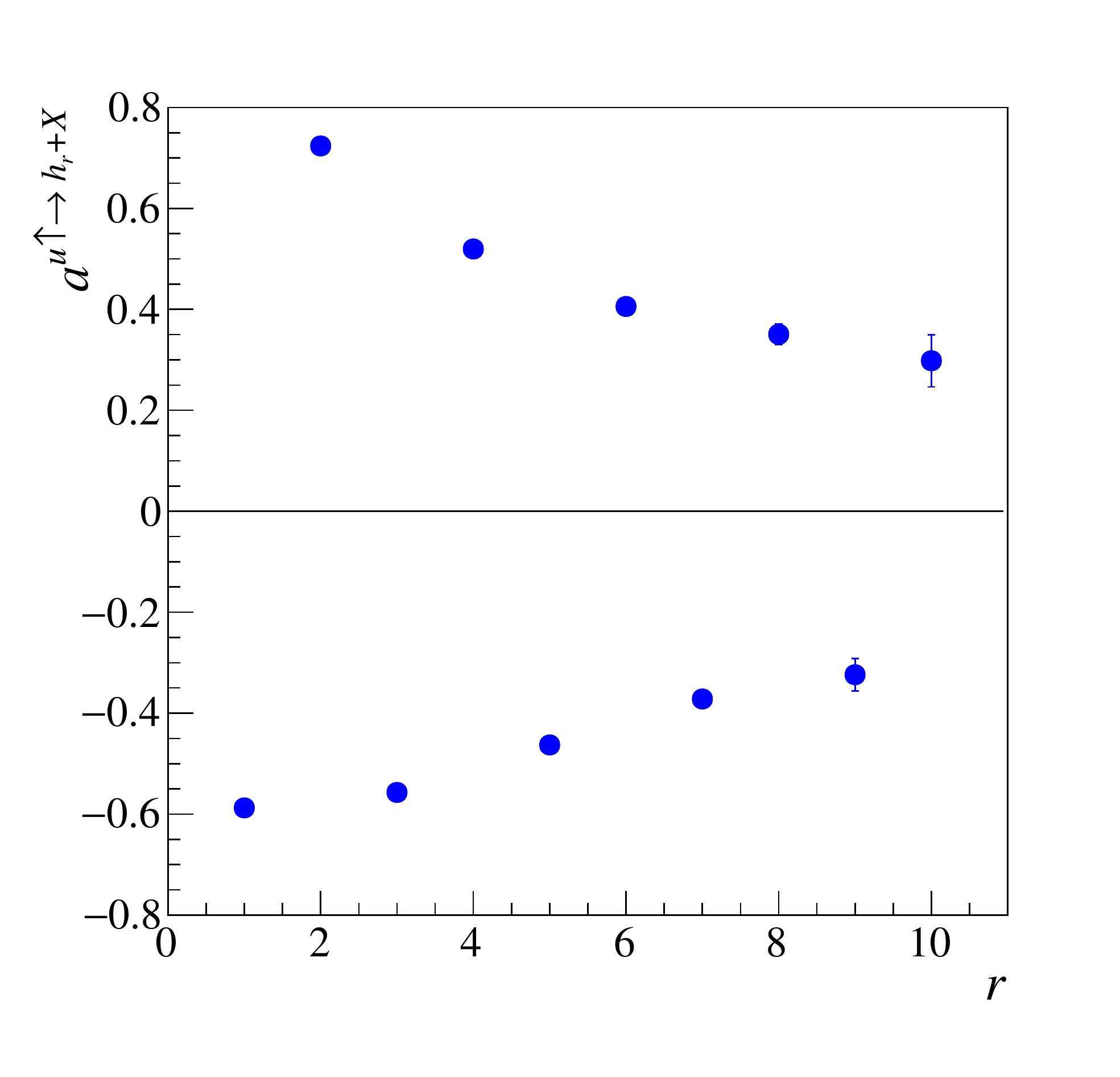}
\end{minipage}
\caption{Collins analysing power for positive pions as function of their rank. The cuts $z_h>0.1$ and $p_{\rm T}>0.1\,\rm{GeV}/c$ have been applied.}\label{fig:ap rank}
\end{figure}

The sign and the monotonic dependence of the analysing power on $z_h$ can be understood by writing $a^{u\uparrow \rightarrow h + X}$ as the sum of different rank hadron contributions weighted by the number of hadrons of that rank. The analysing power can be written as
\begin{equation}
a^{u\uparrow \rightarrow h + X}(t)=\frac{\sum_r N_{h_r}(t)a^{u\uparrow\rightarrow h_r+X}(t)}{\sum_r N_{h_r}(t)}
\end{equation}
where the variable $"t"$ can be either $z_h$ or $p_{\rm T}$. $N_{h_r}$ is the number of hadrons of type $h$ and of rank $r$ and $a^{u\uparrow\rightarrow h_r + X}$ is the analysing power associated with rank $r$, both calculated at the same value $t$. The analysing power for the different rank hadrons is shown in Fig. \ref{fig:ap rank}. It has opposite sign for even and odd ranks, as suggested by classical string+$^{3}P_0$ mechanism, and decreases with the rank. Such decrease is due to the depolarization of the recurrent quark, which turns out to be a weak effect with the current choice of parameters. Indeed in each splitting roughly $10\%$ of the recurrent quark transverse polarization is lost.
The main cause of decay of the analysing power at small $z_h$ is the mixture of contributions from even and odd ranks. The fact that the $z_h$ dependence is roughly linear, not another power law, is a priori accidental.

Concerning the sign of the analysing power, for an initial $u$ quark, a fast positive pion can be produced at first rank. On the contrary a negative pion can never be produced at first rank but from $r=2$ only. Since the contribution of larger rank hadrons is smaller because $N_{h_r}(r)$ decreases with rank due to the finite $W$, the signs of the $\pi^+$ and $\pi^-$ analysing powers are essentially those of the first and second ranks respectively. The same considerations hold for charged kaons.

From the left panel of Fig. \ref{fig:mc asymm}, we notice also that the slope for negative mesons, which are unfavoured in $u$ chains, is slightly larger than the slope for positive ones. This effect is easily explained by the fact that the absolute value of the analysing power for a rank two hadron is somewhat larger than the analysing power for a rank one, as can be seen from Fig. \ref{fig:ap rank}.
Finally we can see that the slope for $\pi^-$ and $K^-$ are similar, as expected because both start to be produced from rank two.

Concerning the analysing power as function of $p_{\rm T}$, shown in the right panel of Fig. \ref{fig:mc asymm}, there are clearly different behaviours for positive and negative mesons.
An interesting feature is the change of sign of the analyzing power for positive pions at $p_{\rm T}\simeq 0.9\,\rm{GeV}/c$.
The rank analysis at this value of $p_{\rm T}$ shows that the number of $\pi^+$ of rank $1$ and $3$ is roughly the same as the number of $\pi^+$ of rank $2$ and $4$. This is due to the fact that positive pions with large $p_{\rm T}$ are more likely produced as rank two, following a rank one $\pi^0$ or $\eta$, than as rank one, at least when the intrinsic transverse momentum is zero. This effect combines with the opposite sign of the analysing power for rank 1 and rank 2 hadrons giving $a^{u\uparrow\rightarrow \pi^+ + X}(p_{\rm T}=0.9\,\rm{GeV}/c)\simeq 0$.
For $r\geq 3$ the number of pions decreases quickly with the rank and they give only a small contribution to the asymmetry. The value of $p_{\rm{T}}$ at which the analyzing power changes sign depends on the choice of the parameters, in particular of $\bt$ and of $\bl$.

\begin{figure*}[tb]\centering
\begin{minipage}{0.8\textwidth}
  \includegraphics[width=1.0\linewidth]{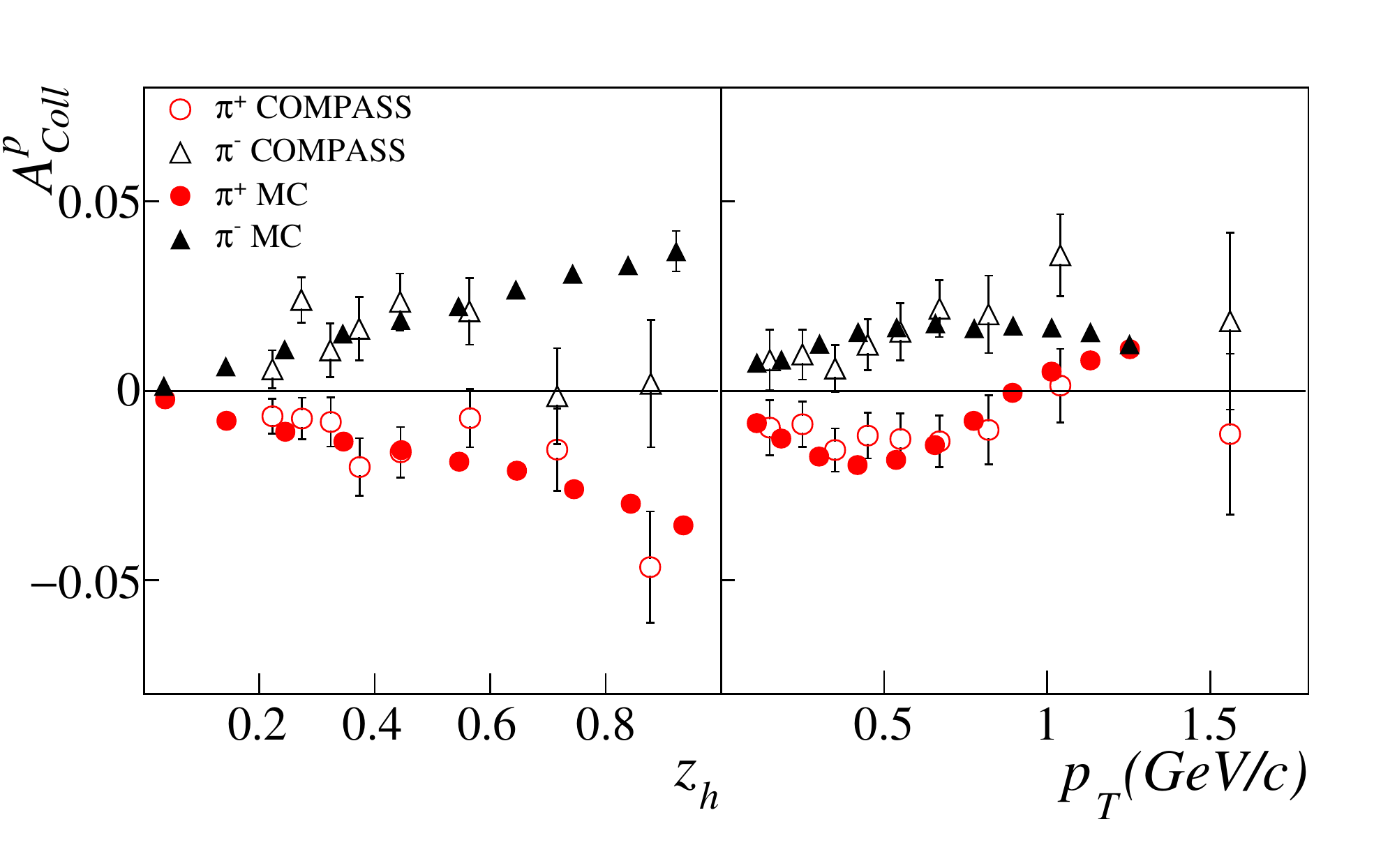}
\end{minipage}%
   \caption{\small{Comparison between the Collins asymmetry $A_{Coll}^p$ measured by COMPASS \cite{COMPASS-collins-sivers} (open points) and the Monte Carlo analysing power $a^{u\uparrow\rightarrow \pi^{\pm}+X}$ scaled by $\lambda_1$ (full points): as function of $z_h$ (left panel) and of $p_{\rm T}$ (right panel). The cuts $p_{\rm T}>0.1\,\rm{GeV}/c$ and $z_h>0.2$ have been applied in both cases.}}\label{fig:comp asymm}
\end{figure*}

These trends are compatible with the Collins asymmetry for charged pions produced in SIDIS off transversely polarized protons measured by COMPASS \cite{COMPASS-collins-sivers}. The comparison is shown in Fig. \ref{fig:comp asymm} as function of $z_h$ (left plot) and as function of $p_{\rm T}$ (right plot).
The Monte Carlo values in both panels are those of Fig. \ref{fig:mc asymm} multiplied by an overall scale factor $\lambda_{1}=0.055 \pm 0.010$ obtained from the minimization of
\begin{equation}\label{eq:chi^2 1h}
    \chi^2=\sum_{p_{\rm{T}i}} \left(A_{Coll}^{p,\pi^-}(p_{\rm{T}i})-\lambda_1 a^{u\uparrow\rightarrow \pi^-+X}(p_{\rm{T}i})\right)^2/\sigma_i^2,
\end{equation}
where $A_{Coll}^{p,\pi^-}(p_{\rm{T}i})$ is the experimental Collins asymmetry for $\pi^-$ in the bin centered in $p_{\rm{T}i}$ and $\sigma_i$ the associated statistical error. In the $u$-dominance hypothesis for a proton target and neglecting the effect of the intrinsic transverse momentum, $\lambda_1$ is the ratio of the $x_B$-integrated $u$-quark transversity and the $x_B$-integrated unpolarized $u$ quark density, multiplied by the depolarization factor $D_{\rm NN}$ of lepton-quark scattering.

As apparent from the right panel of Fig. \ref{fig:comp asymm}, the MC describes quite well the $p_{\rm T}$ dependence of the experimental points, in particular for negative pions. Also, within the statistical uncertainties the data does not exclude a change of the $\pi^+$ asymmetry sign for $p_{\rm T}>0.9\,\rm{GeV}/c$.
As function of $z_h$ (left panel of Fig. \ref{fig:comp asymm}), although the large statistical uncertainties, the experimental data suggest a different trend for the negative pions. The measured Collins asymmetry is in agreement with the MC for $z_h<0.6$, whereas at larger values of $z_h$ the experimental data fall off whereas the MC maintains the increasing trend.
All in all, the agreement is satisfactory in spite of simulating only the pseudoscalar meson production.



\subsection{Dihadron transverse spin asymmetries}\label{sec: 2h asymm M18}
As already anticipated, the string+${}^3P_0$ model for the fragmentation of a transversely polarized quark produces also a dihadron asymmetry. It originates from the combination of a (single hadron) Collins effect and the local compensation of transverse momentum. In the following the positive and the negative hadron of the pair, both produced in the same jet, are labelled with $1$ and $2$. They have fractional energies $z_{1}$ and $z_{2}$ and transverse momenta $\textbf{p}_{1\rm{T}}$ and $\textbf{p}_{2\rm{T}}$. As seen in Chapter 1, they are distributed according to
\begin{equation}\label{eq:2h distribution}
\frac{d^3N_{h_1h_2}}{dz\,dM_{inv}\,d\phi_R}\propto 1+a^{q\A\uparrow\rightarrow h_1h_2+X}(z,M_{inv})S_{\rm AT}\sin(\phi_R-\phi_{\textbf{S}_{\rm A}}),
\end{equation}
where $z=z_{1}+z_{2}$ is the total fractional energy and $M_{inv}$ the invariant mass of the pair. $\phi_R$ is the azimuthal angle of the vector characterizing the pair introduced in Chapter 1.
The analysing power $a^{q\A\uparrow\rightarrow h_1h_2+X}(t)$ is given in Eq. (\ref{eq:dihadron ap}) and is evaluated as $2\langle \sin(\phi_R-\phi_{S\A})\rangle$ taking into account all possible pairs of the jets and shown as function of $t=z,M_{inv}$. The same sample of simulated events used here is the same as that used for the study of the single transverse spin asymmetries in the previous section.

\subsection*{Comparison with BELLE data}
In order to compare the simulation results with the asymmetry of pairs of dihadrons in the $e^+e^-$ annihilation data, the quantity $\epsilon(M_{inv})\equiv\langle a^{u\uparrow\rightarrow \pi^+\pi^-+X}\rangle a^{u\uparrow\rightarrow \pi^+\pi^-+X}(M_{inv})$ turned out to be the most suitable. It has been evaluated from the BELLE data \cite{belle-spin-asymmetries} using the values of the $a_{12}$ asymmetry as function of $z$ and $M_{inv}$. The asymmetry has been averaged over $z$, and $\epsilon(M_{inv})$ has then been evaluated taking into account charge conjugation and isospin invariance, and neglecting the contribution of the polarized dihadron fragmentation function of strange quarks $H_{1s}^{\sphericalangle \pi^+\pi^-}$.
In the expression of $\epsilon(M_{inv})$ the quantity $\langle a^{u\uparrow\rightarrow \pi^+\pi^-+X}\rangle$ indicates the analyzing power averaged over all the kinematic variables, including $M_{inv}$.


For this comparison the analyzing power $a^{u\uparrow\rightarrow \pi^+\pi^-+X}$ of the simulated events has been estimated replacing the angle $\phi_R$ with the azimuthal dihadron angle used by the BELLE collaboration, namely the azimuthal angle of the vector $\textbf{p}_{1\rm T}-\textbf{p}_{2\rm T}$\,\footnote{Transverse vectors are defined in the $e^+e^-$ c.m. frame with respect to the thrust axis, which is an approximation of the string axis. The corresponding azimuthal angles are measured with respect to the plane defined by the $e^+e^-$ axis and the thrust axis.} which can be written as
\begin{equation}\label{eq:PT}
\textbf{p}_{1\rm T}-\textbf{p}_{2\rm T}=2\textbf{R}_{\rm T}+(z_{h_1}-z_{h_2})\textbf{P}_{\rm T}/z.
\end{equation}
$\textbf{P}_{\rm T}=\textbf{p}_{1\rm T}+\textbf{p}_{2\rm T}$ is the transverse momentum of the pair. Defining as "pure" dihadron asymmetry the one defined with respect to the vector $\textbf{R}_{\rm T}$, the asymmetry extracted from the BELLE data is a combination of the "pure" dihadron asymmetry and of the Collins effect of the pair.
 
\begin{figure}[tb]\centering
\begin{minipage}{.8\textwidth}
\centering
  \includegraphics[width=0.6\textwidth]{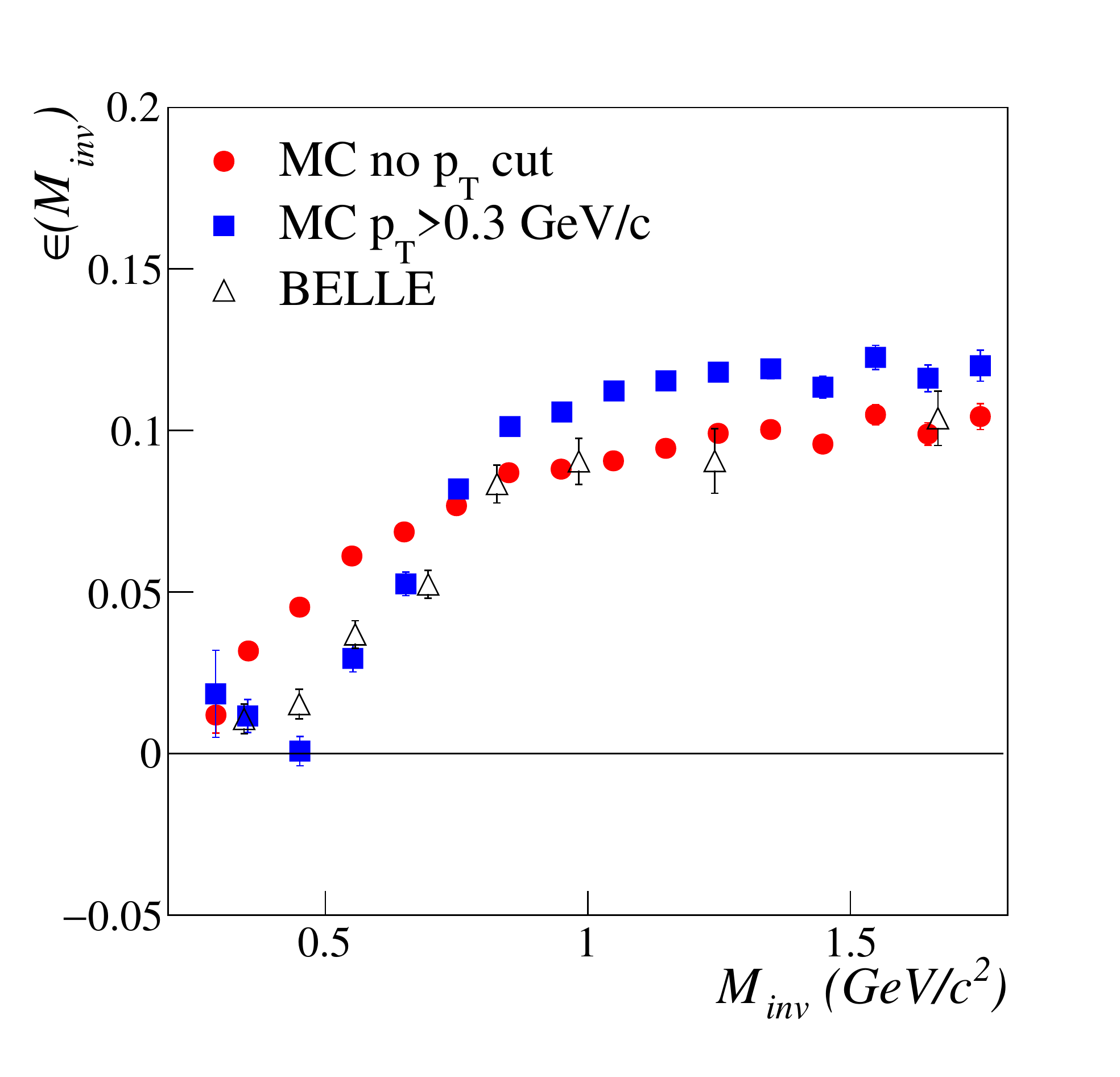}
\end{minipage} 
\caption{\small{Monte Carlo calculation of $\epsilon(M_{inv})$ for pions pairs produced in transversely polarized $u$ jets asking for each pion of the pair $z_h>0.1$ (circles) and also $p_{\rm T}>0.3\, \rm{GeV}/c$ (squares). The black open triangles are the values of $\epsilon(M_{inv})$ obtained from BELLE data \cite{belle-spin-asymmetries}.}}\label{fig:minv pt2>0.1}
\end{figure}

Figure \ref{fig:minv pt2>0.1} shows the results for $\epsilon(M_{inv})$ from the simulation when $z_{1,2}>0.1$ with no cut in $p_{\rm T}$ (circles) and for $p_{\rm T}>0.3\,\rm{GeV}/c$ (squares). The open triangles show the values of $\epsilon$ as measured by BELLE \cite{belle-spin-asymmetries}.
The qualitative agreement is satisfactory. Both in the simulation and in the data the analyzing power shows a saturation for large values of the invariant mass while for small values it falls to zero. However, the trend at small invariant mass, in the MC, depends on the $p_{\rm{T}}$ cut.
Also, the MC data sample has a different invariant mass spectrum with respect to BELLE data because of the absence of resonances, e.g. vector mesons. Anyhow, both in BELLE and in simulation results, no structure can be seen in $\epsilon(M_{inv})$. 

\subsection*{Comparison with COMPASS data}
\begin{figure*}[tb]
\centering
\begin{minipage}{.8\textwidth}
  \includegraphics[width=1.0\linewidth]{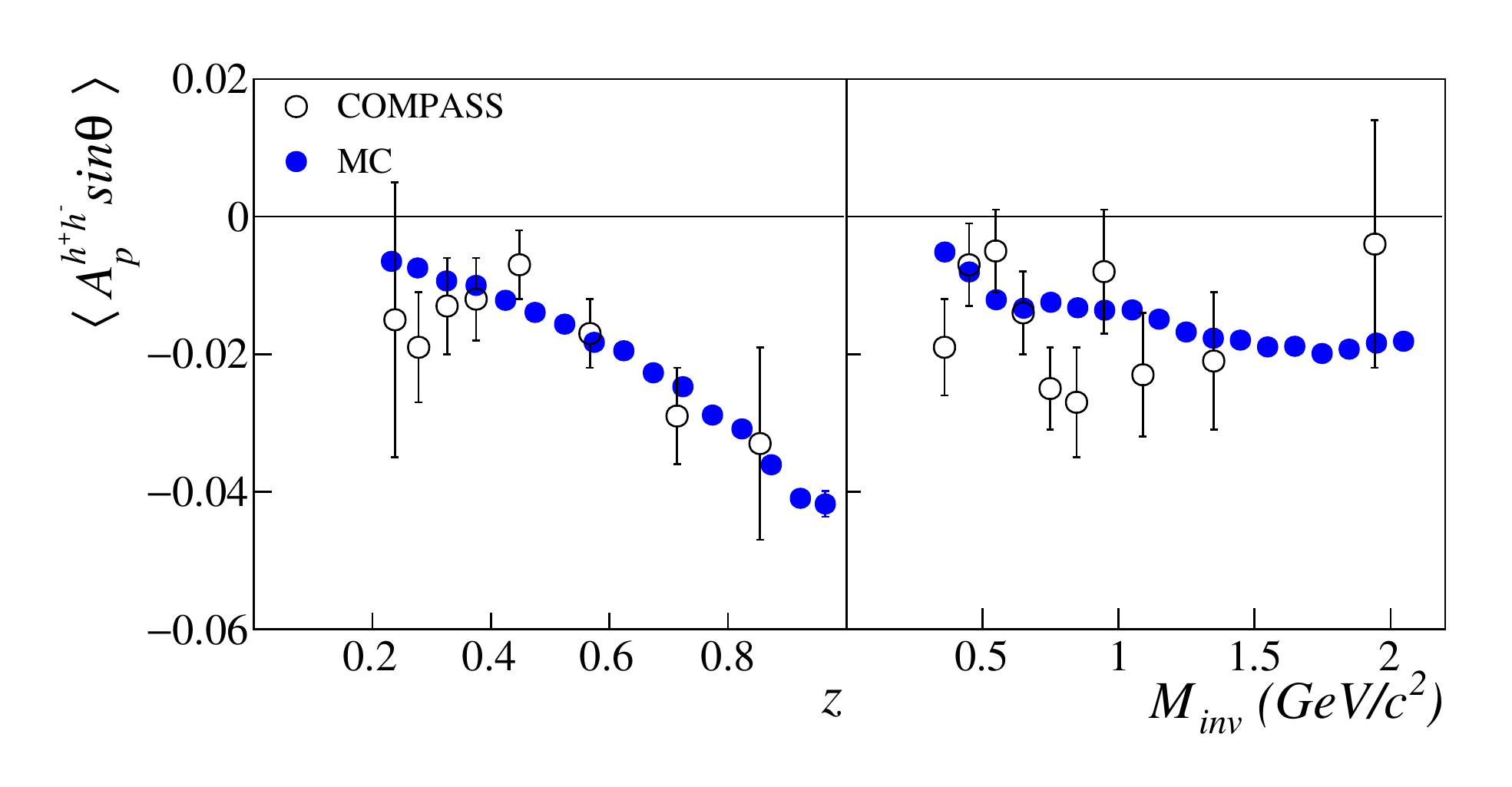}
\end{minipage}
 \caption{\small{Comparison between the dihadron asymmetry $\langle A_{p}^{\rm h^+h^-}\rm \sin\theta\rangle$ measured by COMPASS \cite{compass-dihadron} (open points) and the MC dihadron analysing power $a^{u\uparrow\rightarrow h^+h^-+X}$ scaled by $\lambda_2$ (full points): as function of $z=z_{1}+z_{2}$ (left panel) and as function of $M_{inv}$ (right panel)}.}
  \label{fig:zh compass}
\end{figure*}

Figure \ref{fig:zh compass} shows the comparison between the MC and the COMPASS dihadron asymmetry for $h^+h^-$ pairs measured in SIDIS off transversely polarized protons as function of $z$ (left) and $M_{inv}$ (right). Both in COMPASS data and in simulations the cuts $z_{1,2}>0.1$, $x_{F}>0.1$, $R_{\rm T}>0.07\,\rm{GeV}/c$ and $|\textbf{p}_{1,2}|>3\,\rm{GeV}/c$ have been applied.
The left plot of Fig. \ref{fig:zh compass} shows the dependence on $z$. The Monte Carlo points are scaled by a factor $\lambda_2$ estimated by a $\chi^2$ minimization procedure as in Eq. (\ref{eq:chi^2 1h}) comparing in this case with the COMPASS dihadron asymmetry as function of $z$. The result is $\lambda_2=0.055\pm 0.008$, in perfect agreement with the value of $\lambda_1$ obtained in the single hadron asymmetry case, as it should be. The results from the MC are in good agreement with the experimental data within the statistical uncertainties.
The right plot of Fig. \ref{fig:zh compass} shows the dependence of the analysing power on $M_{inv}$. After scaling by the same parameter $\lambda_2$, the MC points describe quite well the trend of the data.

\subsection{Comparison between single hadron and dihadron asymmetries}
The dihadron analyzing power $a^{u\uparrow \rightarrow h^+h^-+X}$ has been found to be related to $a^{u\uparrow \rightarrow h^{\pm}+X}$ in a recent experimental work of the COMPASS collaboration \cite{interplay}.
Following that work, the relationship between the Collins and the dihadron analyzing powers for hadron pairs in the same $u$ quark jet, as function of the relative azimuthal angle $\Delta\phi=\phi_1-\phi_2$, has been studied. In that analysis using only the events with at least one $h^+$ and one $h^-$ two kinds of asymmetries had been extracted: the "Collins Like" (CL) asymmetries $A_{CL1(2)}^{\sin\phi_C}$ for positive (negative) hadrons and the dihadron asymmetry for oppositely charged hadron pairs $A_{CL,2h}^{\sin\phi_{2h,S}}$. In each bin of $\Delta\phi$, the CL asymmetry is the Collins asymmetry of $h^+$ ($h^-$) of the pair.

As in Ref. \cite{interplay}, $a^{u\uparrow\rightarrow h^+h^-+X}$ has been calculated using $\Phi=\phi_{2h}$, where $\phi_{2h}$ is the azimuthal angle of the vector $\hat{\textbf{p}}_{1\rm T}-\hat{\textbf{p}}_{2\rm T}$ and $\hat{\textbf{p}}_{\rm T}\equiv \textbf{p}_{\rm T}/|\textbf{p}_{\rm T}|$. 
Due to the relation
\begin{eqnarray}\label{eq:pT compass}
\nonumber \hat{\textbf{p}}_{1\rm T}-\hat{\textbf{p}}_{2\rm T}&=&\textbf{R}_{\rm T}(1/|\textbf{p}_{1\rm T}|+1/|\textbf{p}_{2\rm T}|)\\
&+&\textbf{P}_{\rm T}\frac{z_{h_1}/|\textbf{p}_{1\rm T}|-z_{h_2}/|\textbf{p}_{2\rm T}|}{z},
\end{eqnarray}
the considered asymmetry is a combination of the "pure" dihadron asymmetry and of the global Collins asymmetry of the hadron pair. However, as discussed in Ref. \cite{interplay}, the azimuthal angle $\phi_R$ is strongly correlated with $\phi_{2h}$, and the dihadron asymmetry measured from $2\langle \sin\phi_{2h,S}\rangle$ with $\phi_{2h,S}=\phi_{2h}-\phi_{S\A}$, is essentially the same as the "pure" dihadron asymmetry. This has been verified in the simulations as well.

\begin{figure*}[tb]\centering
 \renewcommand{\thesubfigure}{a}
\begin{subfigure}[b]{.4\textwidth}
  \includegraphics[width=0.9\textwidth]{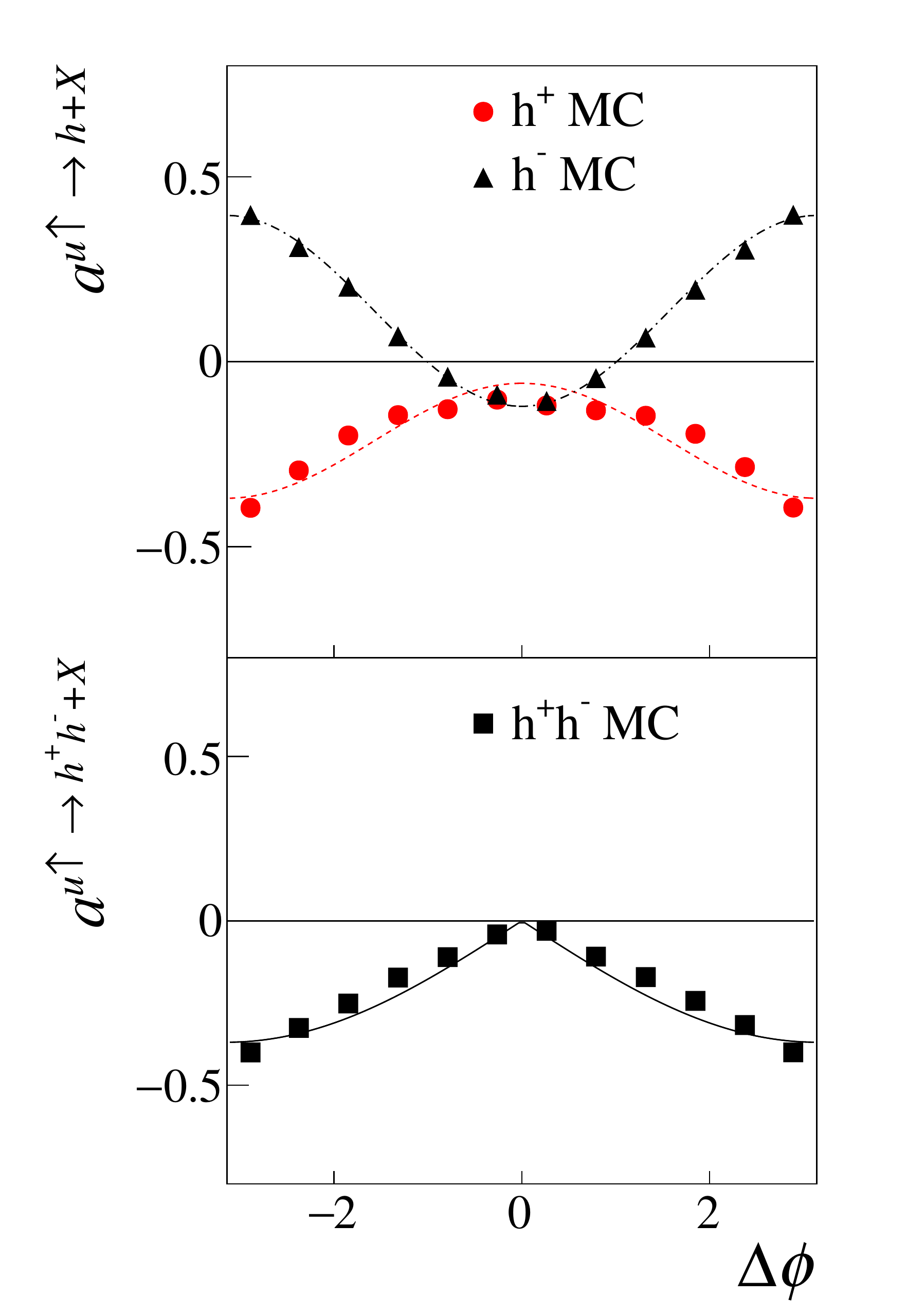}
  \caption{}
\end{subfigure}%
 \renewcommand{\thesubfigure}{b}
\begin{subfigure}[b]{.4\textwidth}
  \includegraphics[width=0.9\textwidth]{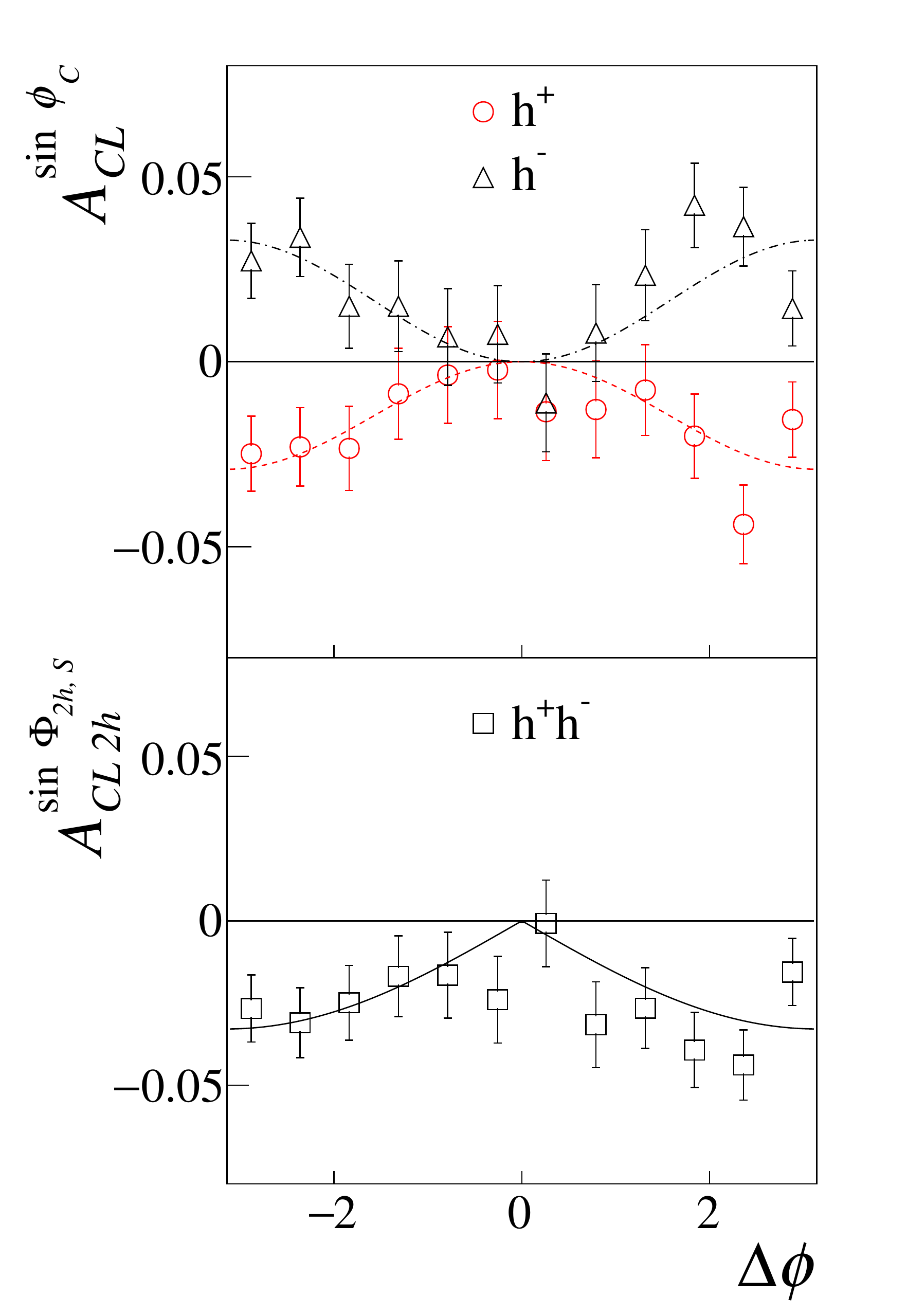}
  \caption{}
\end{subfigure}%

\caption{\small{(a): MC results for the analyzing power in the case of "Collins Like" asymmetries (top) and dihadron asymmetries (bottom) as function of $\Delta\phi$. (b): the corresponding asymmetries measured by COMPASS \cite{interplay}.}}\label{fig:delta phi}
\end{figure*}

The CL analysing power $A_{CL1(2)}^{\sin\phi_C}$ for $h^+$ (circles) and for $h^-$ (triangles) are shown in the top plot of Fig.\ref{fig:delta phi} (a). The corresponding COMPASS data are shown in top plot of Fig. \ref{fig:delta phi} (b). The trend is very similar. The MC points are fitted with functions of the type $\delta_{1(2)}+c_{1(2)}\cos\Delta\phi$, as suggested in Ref. \cite{interplay}, and the results are represented by the red and the black dashed lines.
The slight up-down disymmetry for $h^+$ and $h^-$ in the simulated results is due to the different values of the analyzing power for $h^+$ and $h^-$. The red and the black dashed lines in Fig.\ref{fig:delta phi} (b) represent the fits to the experimental CL asymmetries as shown in Ref. \cite{interplay}, which are consistent with $\delta_{1(2)}=-c_{1(2)}$.

The blue squares in the bottom plot of Fig.\ref{fig:delta phi} (a) show the dihadron analyzing power $a^{u\uparrow\rightarrow \pi^+\pi^-+X}$ calculated in the MC as function of $\Delta\phi$. The blue curve is the result of the fit with the function $c\sqrt{2(1-\cos\Delta\phi)}$ suggested in Ref. \cite{interplay}.
The bottom plot in Fig. \ref{fig:delta phi} (b) shows the asymmetry $A_{CL,2h}^{\sin\phi_{2h,S}}$ as measured in COMPASS. As can be seen, the agreement is good. The $A_{CL,2h}^{\sin\phi_{2h,S}}$ asymmetry is smaller than $a^{u\uparrow\rightarrow \pi^+\pi^-+X}$ by a factor of $0.1$ analogous to $\lambda_2$ but for the higher range $x_B>0.032$ (i.e. highest signal for the transversity distribution) used in the COMPASS analysis.

\subsection{Effect of the intrinsic transverse momentum}\label{sec:kt}

\begin{figure}[h]
\centering
\begin{minipage}{.8\textwidth}
\centering
  \includegraphics[width=0.6\linewidth]{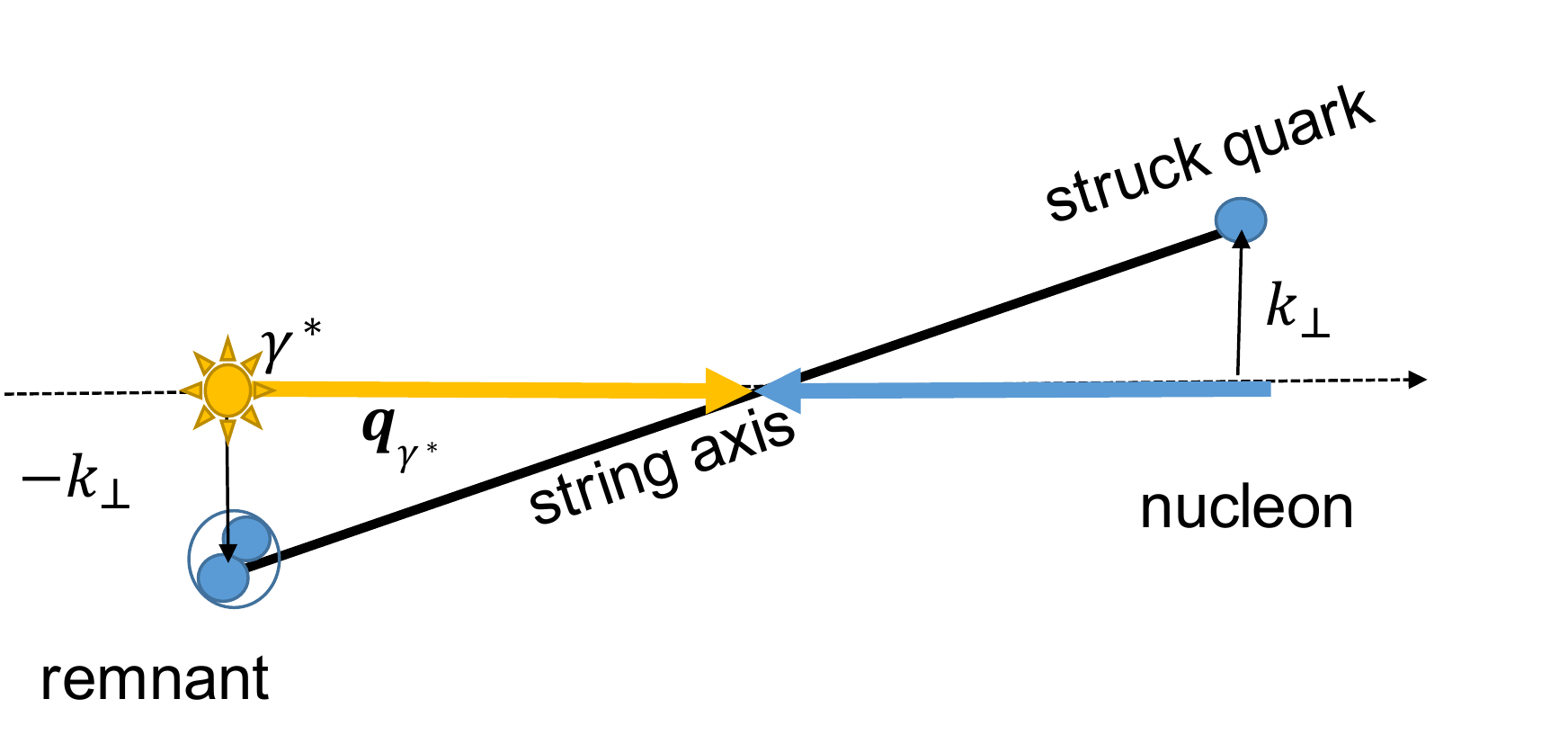}
  \end{minipage}
 \caption{\small{Rotation of the string axis in the string center of mass frame.}}
  \label{fig:rotation}
\end{figure}

In the previous sections the intrinsic transverse momentum $\kperp$ of the initial quark was not considered. This section shows the results when the initial quark $q_A$ has a non zero $\kperp$, which for a DIS event in the GNS tilts the string axis with respect to the $\gamma^*$-nucleon axis, as shown in Fig. \ref{fig:rotation}. The target remnant has the opposite $-\kperp$.
The effects of a random $\kperp$ are the broadening of the spectra of hadrons transverse momenta and a partial smearing of the single hadron asymmetry.

The intrinsic transverse momentum is generated according to the probability
\begin{equation}
d^2\kperp\, \pi^{-1} \langle \kperpkperp\rangle^{-1}\,\exp(-\kperpkperp/\langle \kperpkperp \rangle)
\end{equation}
where $\langle k^2_{\perp}\rangle$ is a free parameter.
The fragmentation of the initial transversely polarized quark $q_A$ is performed using the rotated string axis as $\hat{\textbf{z}}$ axis and then rotating the produced hadrons back to the GNS. 

In the small angle approximation, the rotation in the string center of mass frame is practically equivalent to make the following shift in $\pt$ (which is relative to the string axis)
\begin{equation}\label{eq:shift}
\textbf{P}_{\perp}=x_F\,\kperp+\pt
\end{equation}
where $\textbf{P}_{\perp}$ is the hadron transverse momentum with respect to the $\gamma^*$ axis and $x_F=(2p_z/W)_{c.m.}$ is the Feynman scaling variable\footnote{The hadron four-momentum $P_h$ can be expanded in the basis formed by $k\A$, $k_{\bar{B}}$ and $p\T$, where $p\T=(0,\pt,0)$ is the hadron transverse momentum with respect to the string axis generalized to a four-vector. Taking into account that $q\A$ and $\bar{q}\B$ travel along the forward and backward lightcones, the expression for the hadron momentum is $P_h=Z_+k_A+Z_-k_{\bar{B}}+p\T$ with $x_F=Z_+-Z_-$. Then, if $q\A$ and $q_{\bar{B}}$ have intrinsic transverse momenta $\kperp$ and $-\kperp$, one gets Eq. (\ref{eq:shift}).}. The shift is zero at $x_F=0$ and opposite to $\kperp$ in the backward hemisphere as can be guessed from Fig. \ref{fig:rotation}. Since $x_F=z_h-\epsilon_h^2/(z_hW^2)$, Eq. (\ref{eq:shift}) almost coincides at large $x_F$ with the often used relation $\textbf{P}_{\perp}=z_h\,\kperp+\pt$.

\begin{figure}[h]\centering
\begin{minipage}{1.0\textwidth}
\centering
 \includegraphics[width=0.4\linewidth]{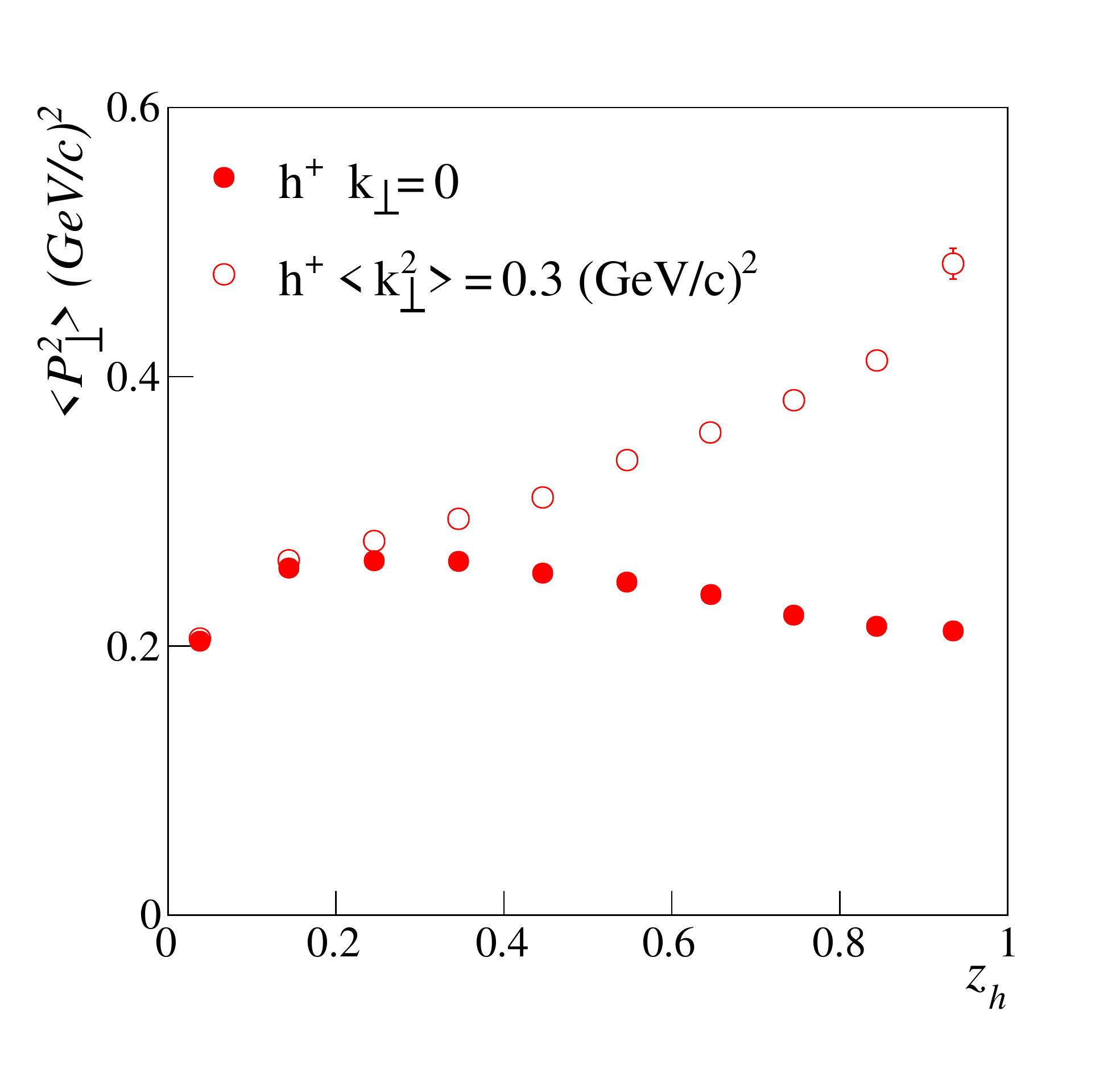}
 \end{minipage}
 \caption{$\langle \textbf{P}^2_{\perp}\rangle$ as function of $z_h$ for positive hadrons without (closed points) and with (open points) the intrinsic transverse momentum.} \label{fig:pt2 zh kt}
 \end{figure}

From Eq. (\ref{eq:shift}) at fixed $x_F$ it is
\begin{equation}
\langle \textbf{P}_{\perp}^2\rangle = x_F^2\langle \textbf{k}^2_{\perp}\rangle +\langle \ptpt\rangle.
\end{equation}
The effect of $\kperp$ is clearly seen in Fig. \ref{fig:pt2 zh kt} which shows the $\langle \textbf{P}^2_{\perp}\rangle$ as function of $z_h$ for positive hadrons when the fragmenting quark has $\langle k^2_{\perp} \rangle =0.3\,(\rm{GeV}/c)^2$. The large $z_h$ region, where $z_h\simeq x_F$, is more sensitive to the effect of the intrinsic transverse momentum. The effect decays then with $z_h$. It turns out that the difference between $\langle \ptpt\rangle$ for positive and negative hadrons shown in Fig. \ref{fig:pt2 zh} is somewhat reduced due to the $x_F^2\langle \textbf{k}^2_{\perp}\rangle$ term but still the negative hadrons are produced with larger transverse momenta.
  
\begin{figure*}[tb]\centering
\begin{minipage}{.7\textwidth}
  \includegraphics[width=0.9\textwidth]{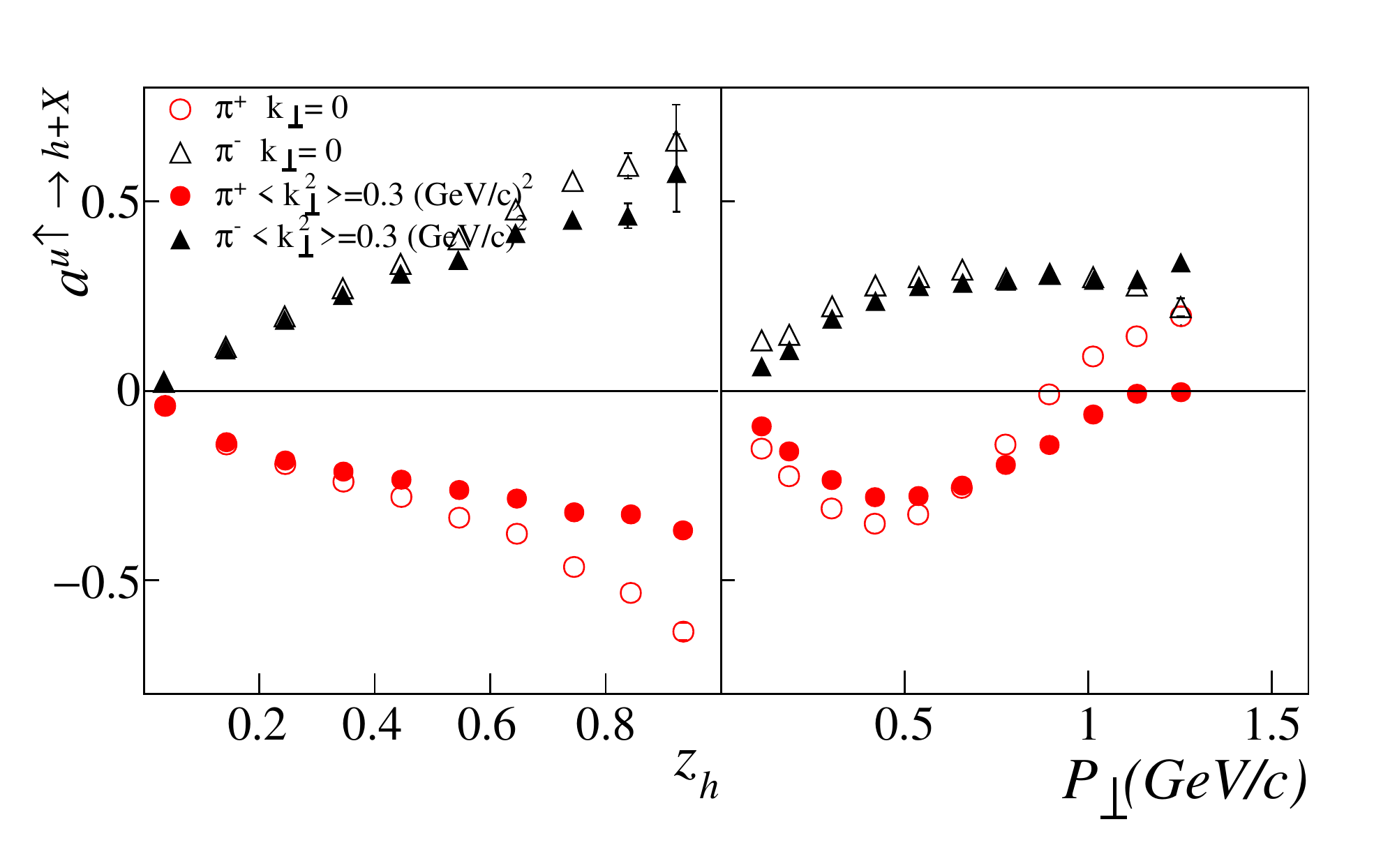}
\end{minipage}%
\caption{\small{Collins analyzing power for charged pions as function of $z_{h}$ (left) and $P_{\perp}$ (right) for $\langle \kperpkperp \rangle = 0$ (open points) and for $\langle \kperpkperp \rangle = 0.3\,(\rm{GeV}/c)^2$ (closed points).}}\label{fig:kteffect}
\end{figure*}

Figure \ref{fig:kteffect} shows the effect of the intrinsic transverse momentum on the Collins analyzing power from MC as function of $z_{h}$ (left plot) and as function of $P_{\perp}$ (right plot) for positive and negative pions. The analysing power for $\langle k^2_{\perp}\rangle=0.3\,(\rm{GeV}/c)^2$ (full points) is compared to that obtained without intrinsic transverse momentum (open points). The reduction of the analysing power is visible at large $z_h$ (left plot) and at low $P_{\perp}$ (right plot). A further consequence of the introduction of $\kperp$ is that the change of sign of the analysing power for positive pions as function of $P_{\perp}$ is no more there.
Similar effects are also observed for charged kaons.


Table \ref{tab:kteffect} shows the mean values of the single hadron and dihadron analysing powers for charged pions for different values of $\langle \textbf{k}^2_{\perp}\rangle$. At variance with the Collins asymmetry for single hadrons, the asymmetry for pairs of oppositely charged hadrons is practically not affected by the noise introduced by $\kperp$.

\begin{table}[tb]
\centering
\caption{\label{tab:kteffect}\small{Mean value of the analyzing powers shown in Fig.\ref{fig:kteffect} (left) for positive and negative pions with cuts $z_h>0.1$ and $P_{\perp}>0.1\,(\rm{GeV}/c)$ have been applied. Also shown are the mean values of the asymmetry for $\pi^+\pi^-$ pairs with the same cuts.}}
\begin{tabular}{l*{6}{c}r}
$\langle k^2_{\perp}\rangle$         & $\langle a^{u\uparrow\rightarrow \pi^+ +X}\rangle$ & $\langle a^{u\uparrow\rightarrow \pi^- +X}\rangle$  &$\langle a^{u\uparrow\rightarrow \pi^+\pi^- +X}\rangle$ \\
\hline
no $\textbf{k}_{\perp}$ & $-0.208\pm 0.001$ & $0.188\pm 0.002$  & $-0.276\pm0.002$\\
$0.1\,(\rm{GeV}/c)^2$ & $-0.197\pm 0.001$ & $0.181\pm 0.002$ &$-0.271\pm0.002$ \\
$0.3\,(\rm{GeV}/c)^2$ & $-0.183\pm 0.001$ & $0.175\pm 0.002$ &$ -0.269\pm0.002$ \\
$0.5\,(\rm{GeV}/c)^2$ & $-0.172\pm 0.001$ & $0.169\pm 0.002$ & $-0.266\pm0.002$ \\
\end{tabular}
\end{table}

\section{Results on the jet handedness}\label{sec:handedness}
As already stressed, the present model can treat at the same time both longitudinal and transverse polarizations. In particular it can predict \textit{jet handedness} which was introduced in Refs. \cite{Nachtmann,Efremov} as a tool to access the longitudinal polarizations of quarks and gluons in polarized processes. The jet handedness is a parity odd correlation of the type $S_{\rm A L}\zu\cdot(\textbf{p}_1\times\textbf{p}_2) = S_{\rm A L}\zu\cdot(\textbf{p}_{\rm{T}1}\times\textbf{p}_{\rm{T}2})$ in the distribution of a $h_1h_2$ pair produced in the fragmentation of a longitudinally polarized quark $q\A$. $\textbf{p}_1$ and $\textbf{p}_2$ are the momenta of $h_1$ and $h_2$, chosen according to some prescription, e.g. $h_1$ is the positive and $h_2$ the negative hadron. The jet handedness thus correlates the longitudinal polarization of the fragmenting quark with the transverse momenta of the hadrons of the observed pair. The distribution of the pair can be parameterized in the form
\begin{equation}\label{eq:handedness}
\frac{d^6N_{h_1h_2}}{d^3\textbf{p}_1d^3\textbf{p}_2}\propto 1+a_{\rm JH}^{\vec{q_A}\rightarrow h_1h_2+X}S_{\rm A L}\sin(\phi_2-\phi_1),
\end{equation}
where $a_{\rm JH}^{\vec{q_A}\rightarrow h_1h_2+X}$ is the jet-handedness analyzing power. This distribution is similar to Eq. (\ref{eq:2h distribution}) but obtained for fragmentations of longitudinally polarized quarks and using as relevant angle the relative azimuthal angle between the transverse momenta of $h_2$ and $h_1$ with respect to the string axis.
The toy model of Ref. \cite{DS09} predicts a jet-handedness effect with an analysing power proportional to $\RE(\mu)\IM(\mu^2)$. The same factor appears in the present model.

\begin{figure}[!h]
\begin{minipage}{1.0\textwidth}
\centering
  \includegraphics[width=0.6\textwidth]{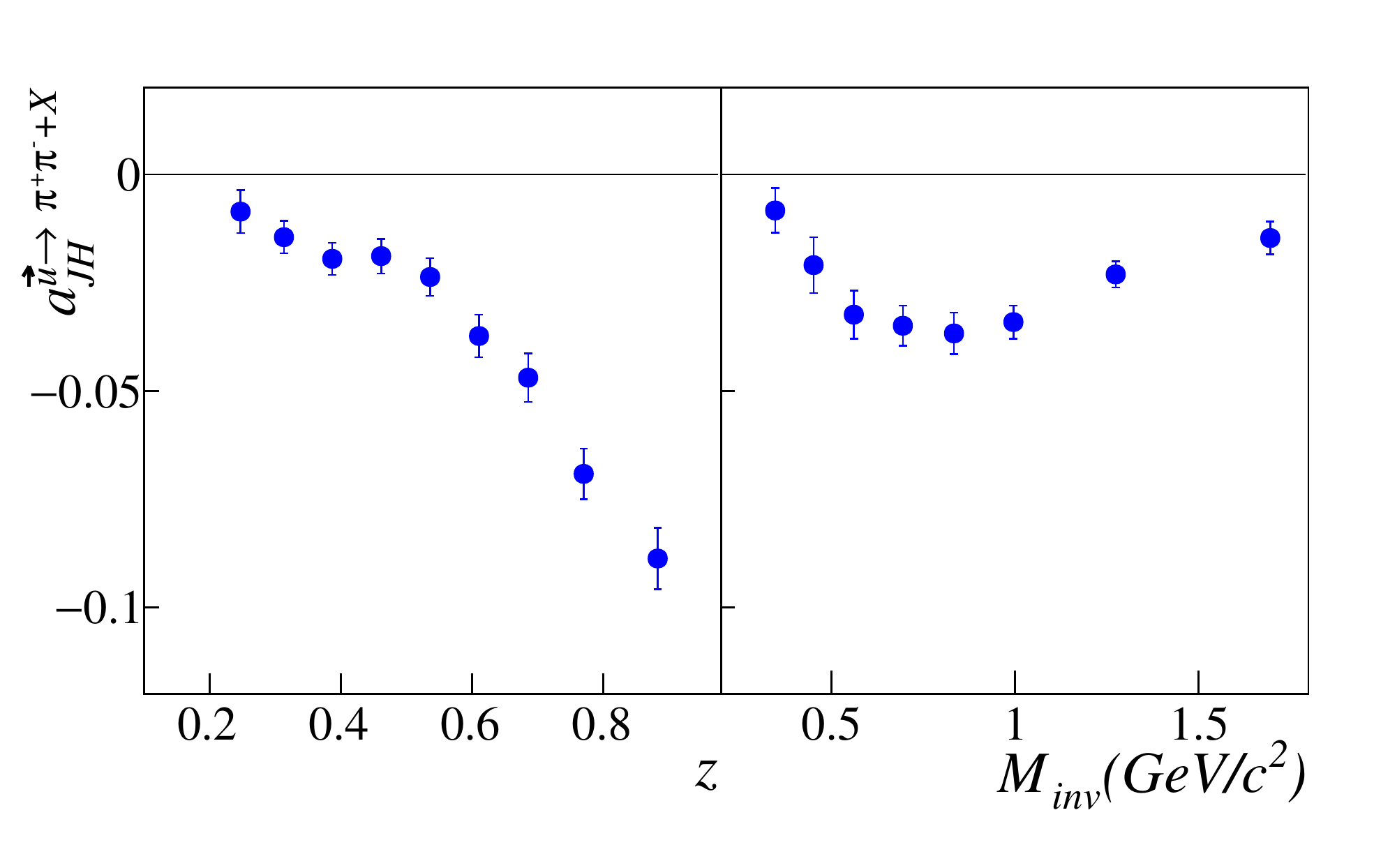}
\end{minipage}%
\caption{\small{The analysing power for the jet handedness effect in the fragmentation $\vec{u}\rightarrow\pi^+\pi^-+X$, defined in Eq. (\ref{eq:handedness}), as function of $z=z_{1}+z_{2}$ (left plot) and as function of the invariant mass of the pion pair (right plot). The cuts $z_{1,2}>0.1$ and $p_{1,2\rm T}>0.1\,\rm{GeV}/c$ have been applied.}}
\label{fig:handedness}
\end{figure}

The handedness analysing power $a_{\rm JH}^{\vec{u}\rightarrow\pi^+\pi^-+X}$ has been calculated for $\pi^+\pi^-$ pairs produced in fragmentations of initial longitudinally polarized $u$ quarks as $2\langle\sin(\phi_2-\phi_1)\rangle$.
Figure \ref{fig:handedness} shows the dependencies of $a_{\rm JH}^{\vec{u}\rightarrow\pi^+\pi^-+X}$ on the invariant mass $M_{inv}$ of the pion pair (left plot) and on the sum of their fractional energies $z_1+z_2$ (right plot). 
The handedness analysing power increases with $z_1+z_2$, which is expected since at large $z_1+z_2$ both hadrons have nearly fixed ranks (rank $1$ for $\pi^+$ and rank 2 for $\pi^-$). No strong dependence is observed as function of $M_{inv}$.
From the comparison between Fig. \ref{fig:handedness} and Fig. \ref{fig:zh compass}, keeping in mind that in the latter case the MC analysing power is scaled by the factor $\lambda_2$, an effect smaller by one order of magnitude than the dihadron asymmetry can be observed for the jet handedness.
Note that an opposite asymmetry can be obtained by reversing the sign of $\RE(\mu)$.

Up to now, attempts to observe jet handedness were not conclusive, see $e.g.$ Ref. \cite{Abe-handedness}. Several reasons can explain this failure: 
\begin{itemize}
\item[-] the sign of the asymmetry may vary too much with the charges, the rapidity ordering or the invariant mass of the $h_1-h_2$ pair. 
\item[-] the observable $\sin(\phi_2 - \phi_1)$ is very sensitive to a redefinition of the jet axis. It can be easily blurred by a too large experimental uncertainty on the orientation of the jet axis or by gluon radiation. 
\end{itemize}
Like for the Collins effect, the blurring effect can be eliminated by involving one more particle. Indeed, for three particles $h_1$, $h_2$ and $h_3$ of the jet, the pseudoscalar quantity
\begin{equation}
 J = (\textbf{p}_1 \times \textbf{p}_2)  \cdot \textbf{p}_3  = (\textbf{p}_{1,\perp P} \times \textbf{p}_{2,\perp P})  \cdot \textbf{P},
\end{equation}
where $\textbf{P}=\textbf{p}_1+\textbf{p}_2+\textbf{p}_3$, is independent of the jet axis and one may take $\langle J \rangle $ as helicity-sensitive  estimator (the estimator $\langle \rm{sign}(J) \rangle $ was proposed in Ref. \cite{ Efremov}). However it requires the clean measurement of three particle momenta and its amplitude depends on six kinematic variables, $e.g.$, $z_1$, $z_2$, $z_3$, $ |\textbf{p}_{1,\perp P}|$ , $ |\textbf{p}_{2,\perp P}|$  and $ |\textbf{p}_{3,\perp P}|$. 

\section{Comparison between M18 and M19}\label{sec:comparison M18 and M19}
In this section the MC results obtained with M18 described in so far are compared with M19. As a reminder, at variance with M18, M19 does not contain spin-independent $\kt$-$\kpt$ correlations which simplifies the description of the spin transfer from $q$ to $q'$. In particular the preliminary task of M18 is not needed and the quark intermediate spin density matrix coincides with its true spin density matrix. The simulation programs for the two variants are the same except for the points described in Sec. \ref{sec:structure of stand alone MC Chapter 3 subsection}. The parameters used for this comparison are the same as given in Sec. \ref{sec: free parameters and kinematical distributions M18 M19} for M18, except for $\bt$ which has been increased by a factor $1.63$ (e.g. $\bt=8.43\,(\rm{GeV}/c)^{-2}$ in M19) in order to compensate the exponential growth of the free input function of M19 at large $\ptpt$ and thus to generate $\ptpt$ distributions similar to those obtained with M18.

In the following results on the kinematical distributions, on the Collins and on the dihadron analyzing powers are compared. They are obtained from fragmentations of transversely polarized $u$ quarks whose momenta are calculated from the same sample of $x_B$ and $Q^2$ values of SIDIS events used with M18.

The rank dependence of the kinematical distributions is a typical feature of recursive models and is about the same for M19 and M18. In particular, in both cases the $Z$ and $\ptpt$ distributions do not depend on the rank $r$ for $r\geq 2$ as in Fig. \ref{fig:Z distribution} and in Fig. \ref{fig:pt distribution}.

\begin{figure*}[tb]
    \centering
    \begin{subfigure}[b]{0.5\textwidth}
        \includegraphics[width=1.1\textwidth]{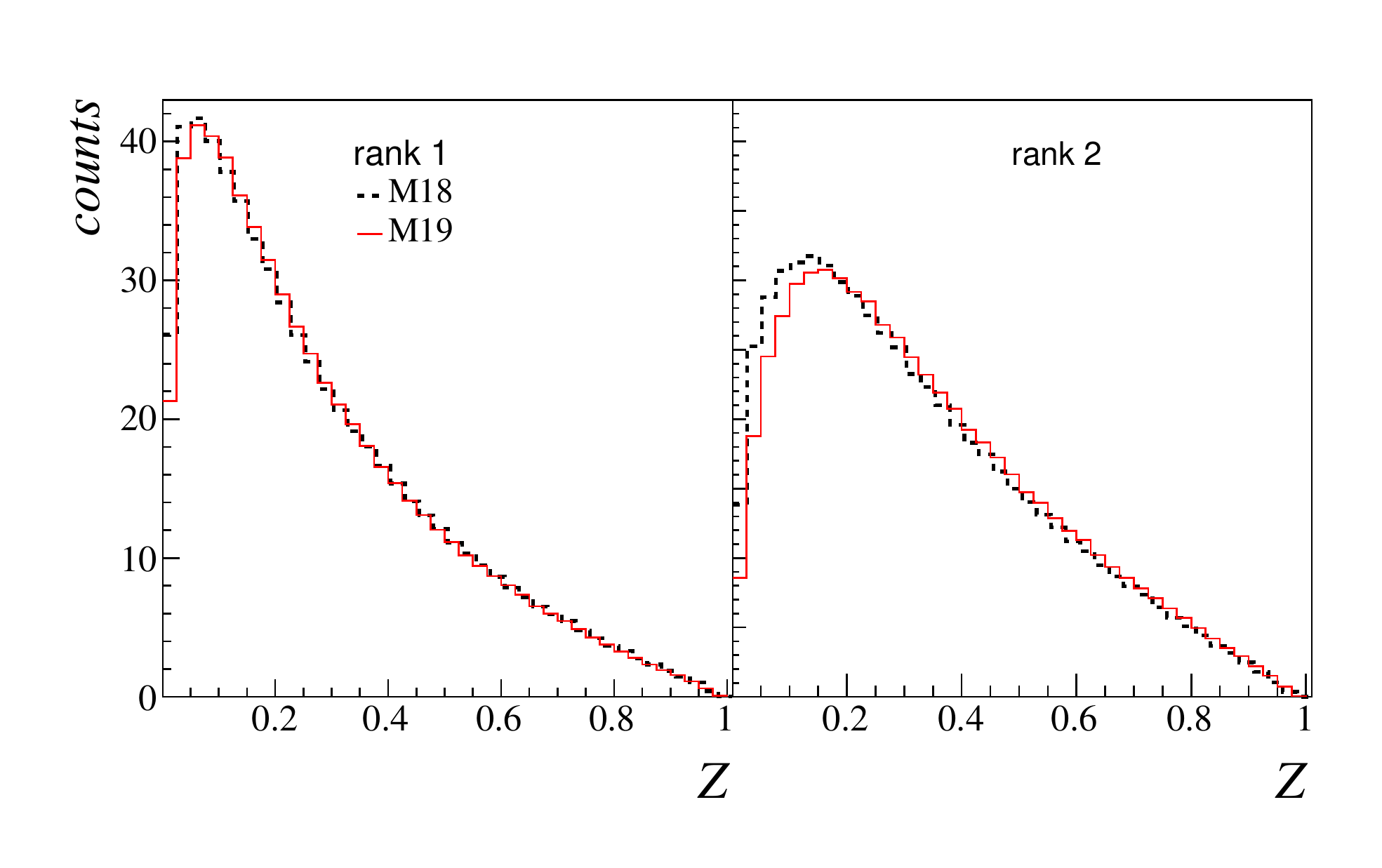}
       \caption{}
       \label{fig:M19 Z_r12}
    \end{subfigure}
    \centering
     \begin{subfigure}[b]{0.5\textwidth}
        \includegraphics[width=1.1\textwidth]{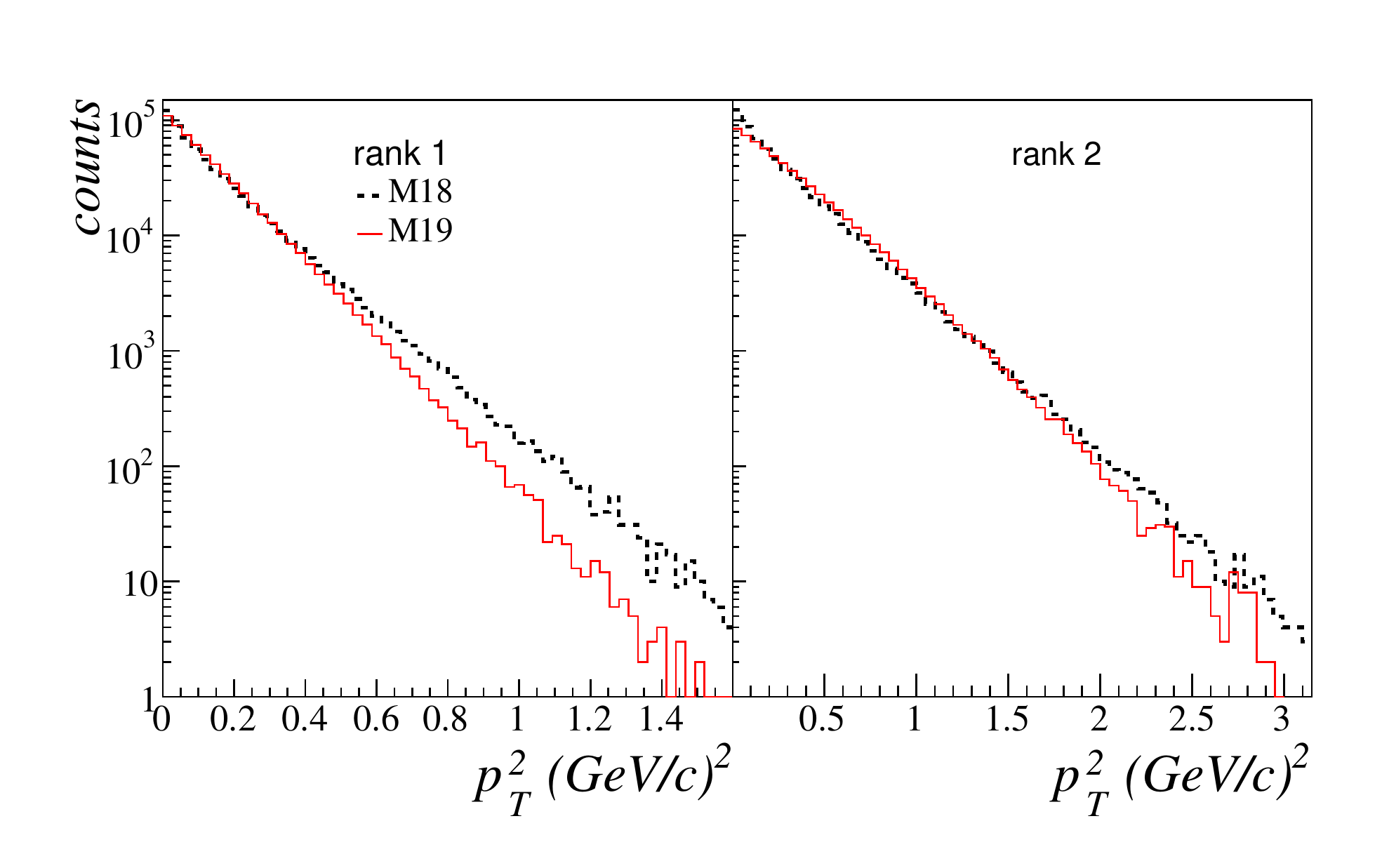}
        \caption{}
        \label{fig:M19 pT2_r12}
    \end{subfigure}
 \caption{Comparison between M18 (dotted black histogram) and M19 (continuous red histogram) for: (a) $Z$ distribution for rank 1 hadrons, (b) $Z$ distribution for rank 2 hadrons, (c) $\ptpt$ distribution for rank 1 hadrons and (d) $\ptpt$ distribution for rank 2 hadrons. Their ratios are shown in the bottom plots. Note the different horizontal scales in plots (c) and (d).}\label{fig:Z_pT2_rank}   
\end{figure*}

In Fig. \ref{fig:M19 Z_r12} and in Fig. \ref{fig:M19 pT2_r12} the $Z$ and $\ptpt$ distributions for the rank $r=1$ (left) and $r=2$ (right) hadrons as obtained with M19 (continuous histograms) and with M18 (dotted histograms) are compared. The two models produce almost the same $Z$ distribution for rank 1 (plot (a)). For rank 2 (plot (b)) the $Z$ distribution in M19 is slightly shifted towards greater values of $Z$. This is correlated to the somewhat larger $\langle \ptpt \rangle$, as can be seen from plot (d).
From plot (c) it is also clear that the $\ptpt$ distribution for rank $1$ of M18 has two slopes, at variance with M19. Indeed the $\ptpt$ distribution of M18 is a sum of contributions of different slopes, one for each $Z$, due to the factor $\exp(-\bl\varepsilon_h^2/Z)$. In M19 also there is a different slope for each $Z$, but effect is reduced by the factor $1/N_a(\varepsilon_h^2)$.

\begin{figure}[tb]
\begin{subfigure}[b]{1.0\textwidth}
\centering
   \includegraphics[width=0.78\linewidth]{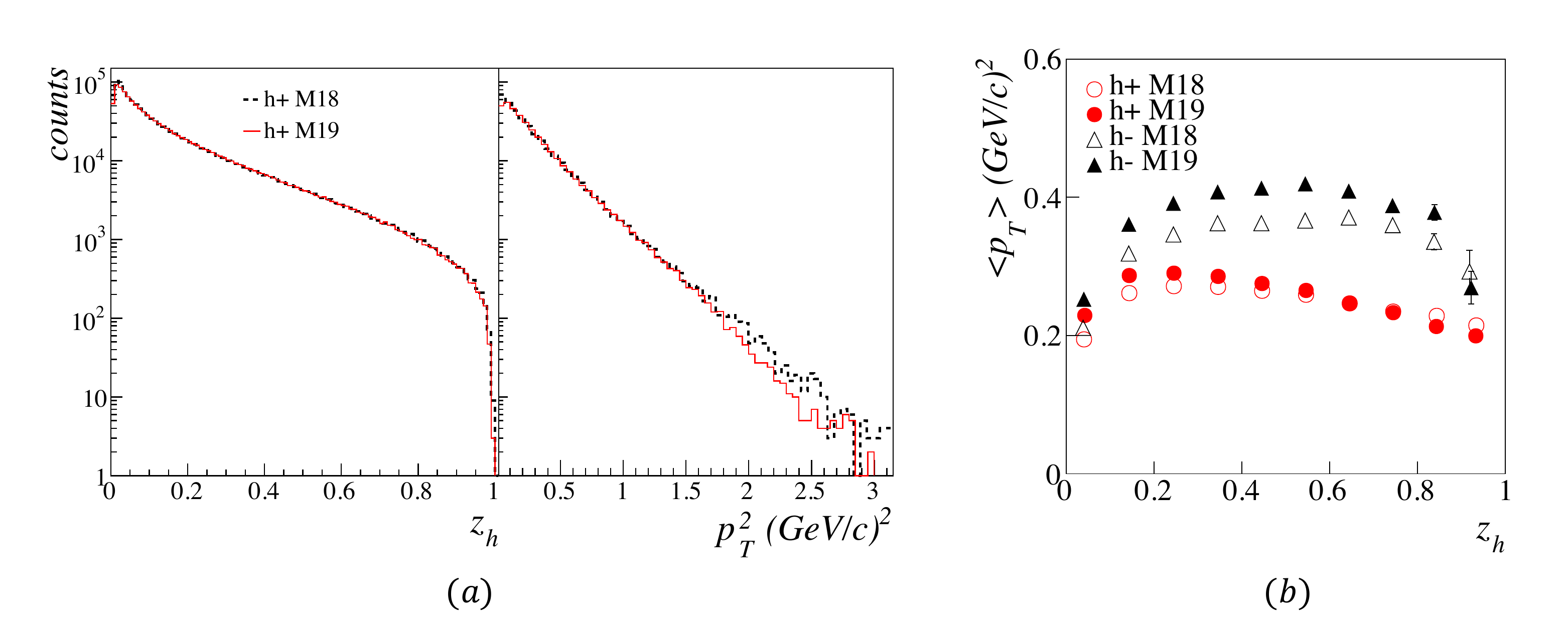}
\end{subfigure}
\caption{(a): Comparison between M18 (dotted black histogram) and M19 (continuous red histogram) for: $z_h$ (left) and $\ptpt$ (right) distributions of positively charged hadrons. The cuts $p\T>0.1\,\rm{GeV}/c$ and $z_h>0.2$ have been applied. (b): Comparison between the $z_h$ dependence of $\langle \ptpt \rangle$ in M19 (full points) and in M18 (open points).}
 \label{fig:M19 zh pT2 distributions}   
\end{figure}

The differences are even smaller when looking the distributions of the fraction $z_h$ of the fragmenting quark energy carried by hadrons. The $z_h$ distributions for positive hadrons with $p_{\rm{T}}>0.1\,\rm{GeV}/c$ from the two models are shown in Fig. \ref{fig:M19 zh pT2 distributions}a. The region of very small $z_h$ is less populated in M19.
The $\ptpt$ distribution for positive hadrons with $z_h>0.2$ is almost the same in both models as shown in the same figure.

Figure \ref{fig:M19 zh pT2 distributions}b compares the $z_h$ dependence of $\langle \ptpt \rangle$ of charged hadrons in the two models. M19 gives a larger difference between the $\langle \ptpt \rangle$ for positive hadrons and the $\langle \ptpt \rangle$ for negative hadrons than M18, which already was not in agreement with experiments (see Fig. \ref{fig:fig 3.3} and the related discussion in sec. \ref{sec: free parameters and kinematical distributions M18 M19}). Indeed, due to the pure spin correlations which gives $\langle \kt\cdot\kpt\rangle <0$, now at ranks larger than one it is $\langle \ptpt \rangle > 2\langle \ktkt\rangle$. In M18, on the other hand, the spin-independent correlation, if taken alone, would give the opposite correlation $\langle \kt\cdot\kpt\rangle >0$, therefore $\langle \ptpt \rangle<2\langle \ktkt\rangle$.

\begin{figure}[tb]
\centering
\begin{subfigure}[b]{0.80\textwidth}
   \includegraphics[width=1\linewidth]{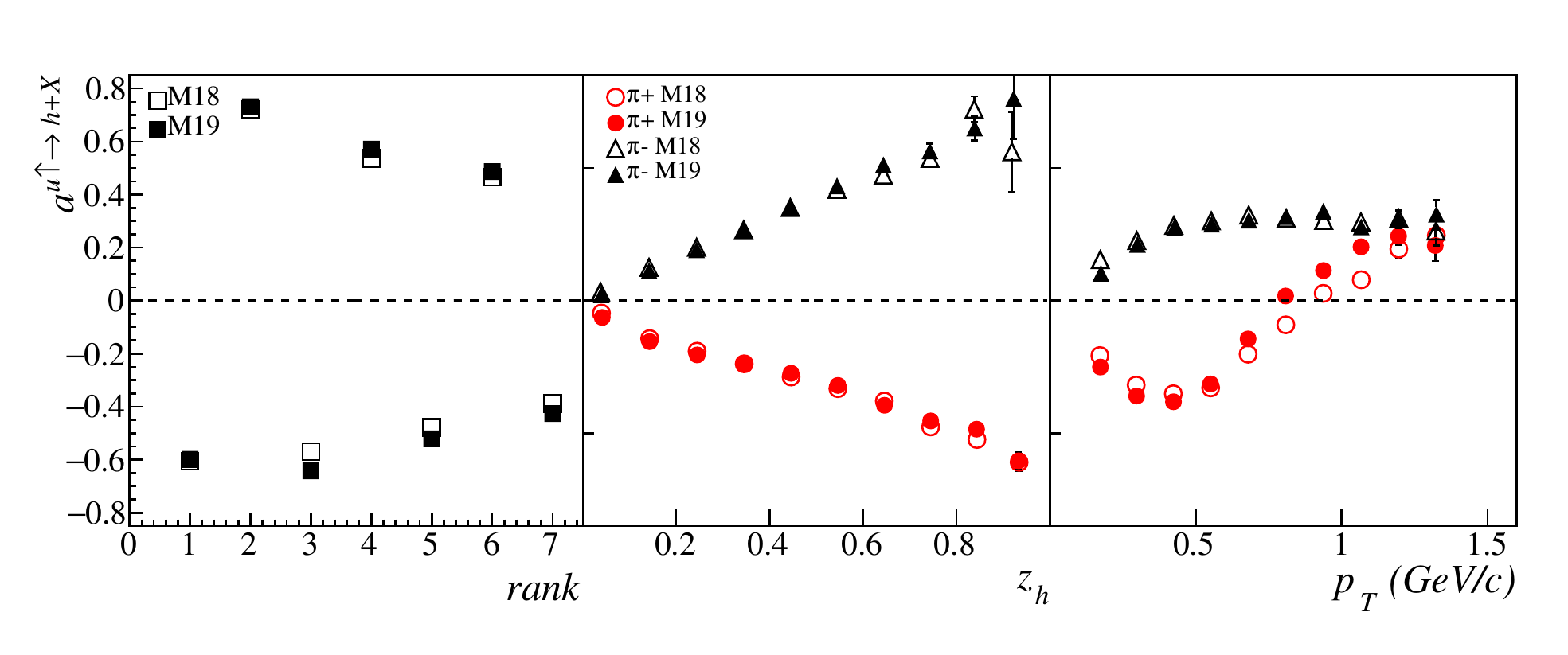}
   \caption{}
   \label{fig:M19 asymmetries-collins} 
\end{subfigure}
\begin{subfigure}[b]{0.65\textwidth}
   \includegraphics[width=1\linewidth]{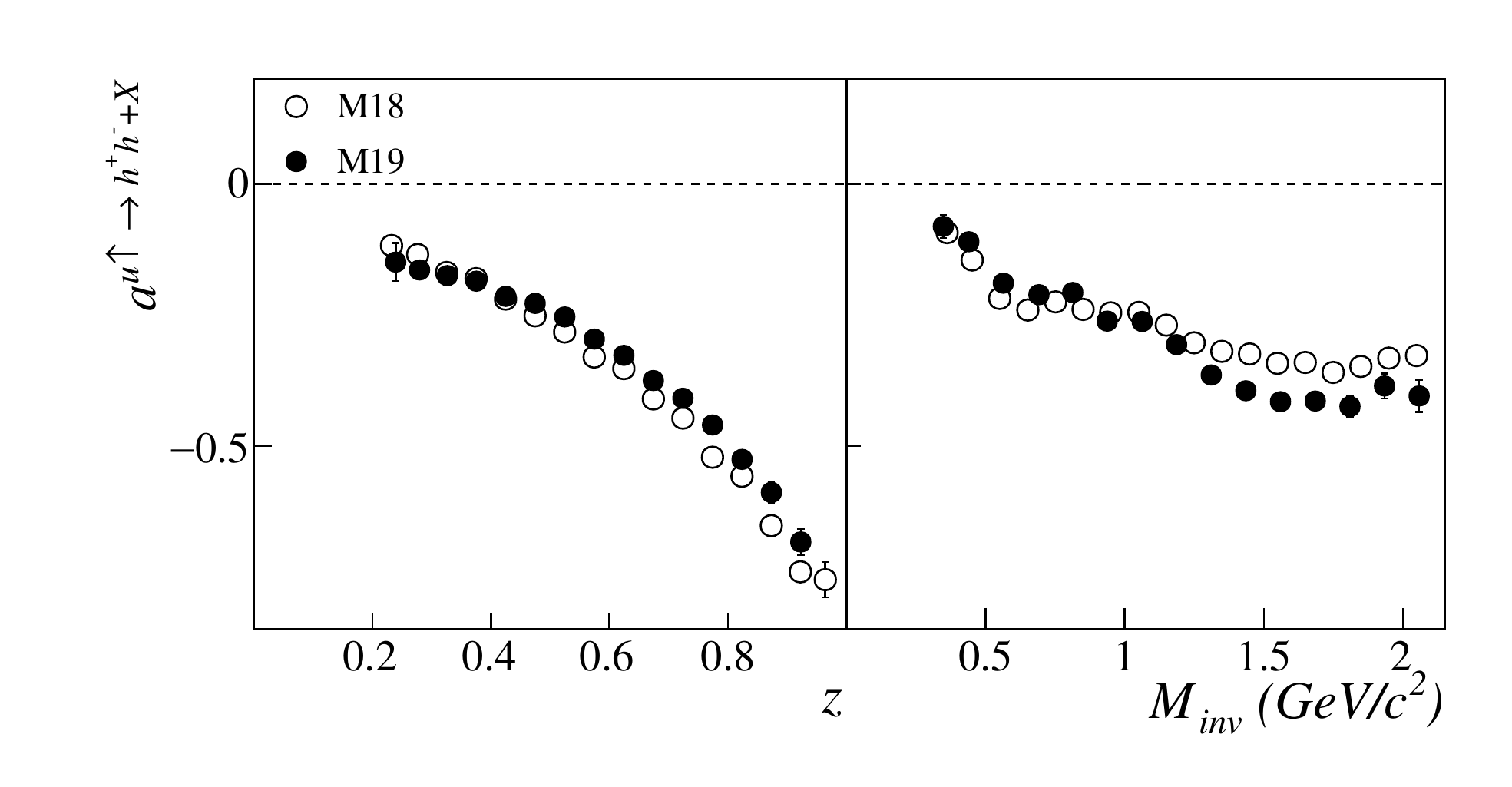}
   \caption{}
   \label{fig:M19 asymmetries-dihadron}
\end{subfigure}
\caption{(a): Collins analysing power for positive charge (red circles) and negative charge (black triangles) pions as function of rank (left panel), $z_h$ (middle panel) and $p_{T}$ (right panel) as obtained with M19 (full points) and with M18 (open points). (b): dihadron asymmetry for unidentified hadrons as function of $z$ (left panel) and $M_{inv}$ (right panel).}
\label{fig:M19 asymmetries}
\end{figure}

Figure \ref{fig:M19 asymmetries-collins} shows the Collins analysing power for charged pions produced in jets of transversely polarized $u$ quarks obtained with M19 (full points) compared with the results of M18 (open points). The analysing power is shown as function of rank (left panel), $z_h$ (middle panel) and as function of $p_T$ (right panel). The cuts $z_h>0.2$ and $p_{\rm{T}}>0.1\, \rm{GeV}/c$ have been applied to evaluate the analysing power as function of $p_{\rm{T}}$ and $z_h$ respectively. Both cuts are applied to evaluate the analysing power as function of rank. The models produce almost the same analysing power. As function of rank, the polarization decays almost with the same speed. The slight differences in the analysing powers as function of $p_T$ for $\pi^+$ are due to the different $\kptkpt$ dependencies of the splitting functions.

Figure \ref{fig:M19 asymmetries-dihadron} compares the dihadron $h^+h^-$ analysing power as function of $z$ (left panel) and $M_{inv}$ (right panel) as obtained with M19 (full points) and with M18 (open points). The cuts $z_{1,2}>0.1$, $R\T>0.07\,\rm{GeV}/c$ and $|\textbf{p}_{1,2}|>3\,\rm{GeV}/c$ have been applied.
The overall trends are the same in both models and only some slight differences can be seen. In particular the result of M19 is somewhat larger in absolute value at large $M_{inv}$. All in all, the main features of the results obtained from the two implementations are the same.

To summarize, the simulations using the M18 and the M19 models for the fragmentation of polarized quarks give very similar results. In spite of the fact that only pseudoscalar mesons are generated, the analysing powers for transversely polarized quarks turn out to be clearly different from zero and in statisfactory agreement with experimental data. This encouraging result has been the main reason for the implementation in \verb|PYTHIA| described in the next chapter. Since the results of M18 and M19 are very close, the interface with \verb|PYTHIA| has been developed for M19, which code is simpler and more flexible. M19 has been extended including vector meson production and the resulting model, called M20, is described in Chapter 5.



\chapter{Implementation of the model in PYTHIA}
The stand alone MC program simulating the fragmentation of a polarized quark based on the string+${}^3P_0$ model (M18 and M19) described in Chapter 2, allows one to satisfactorily describe the existing transverse spin data. In particular, as shown in Chapter 3, the Collins and the dihadron analysing powers obtained from simulations reproduce the main features of the asymmetries measured in SIDIS processes.

A complete generation of the physics events is required in order to make a more quantitative comparison with data and to deliver a more realistic tool. Monte Carlo event generators are a powerful tool in this respect. They provide a full event description in all the allowed kinematic region, have a high level of sophistication in the simulating the physics processes and allow to study correlations between particles produced in the same event. For these reasons, the model M19 has been interfaced with the \verb|PYTHIA 8.2| event generator allowing for the first time to simulate the spin effects in the hadronization process of complete events.

\verb|PYTHIA| is one of the currently most used event generators. It is a standard tool for the generation of events in high energy collisions. In spite of the fact that the largest user community obviously come from the LHC experiments, \verb|PYTHIA| allows to simulate in a complete and detailed way other complex processes like DIS and $e^+e^-$ annihilation into hadrons which are the basic processes to investigate the nucleon transverse spin structure.

The choice of \verb|PYTHIA| as event generator for the implementation of spin effects in the hadronization process is the most natural since the hadronization process in \verb|PYTHIA| is based on the Lund string model. In particular, the splitting function of the M19 for unpolarized quarks is very similar to that implemented in \verb|PYTHIA|.

This chapter describes the work done to interface M19 to \verb|PYTHIA| in order to simulate spin effects in SIDIS. A summary of the first part can also be found in Ref. \cite{kerbizi-lonnblad}. Section \ref{sec:SIDIS in Pythia} is dedicated to a short description of the main features of the simulation of the SIDIS process in \verb|PYTHIA|. The strategy for the implementation of the spin effects in the hadronization process of \verb|PYTHIA|, the description, and the validation of the interface are presented in section \ref{sec:spin in fragmentation process}. The use of the transversity PDF and the results for the simulated Collins and dihadron asymmetries in SIDIS off transversely polarized proton and deuteron targets are presented in section \ref{sec:results for transversely pol nucleon}. Finally, in section \ref{sec:adding other TMDs} a general recipe for the implementation of other TMD PDFs is discussed.

\section{The SIDIS process in PYTHIA}\label{sec:SIDIS in Pythia}




\verb|PYTHIA| is a very general, detailed and complex event generator. Since the version \verb|PYTHIA 8| it is written in \verb|C++| and uses object oriented methods \cite{pythia8}. Here only a short description, focused on the interface with M19, is given.

The simulation is done by using a main program which communicates with the top level \verb|Pythia| class. The main program consists of an initialization phase where the initial conditions for the event generation are specified. They concern the beam type and momentum, the target type and momentum, the selection of the interaction process, definition of the phase space cuts, the initialization of the free parameters and of the various flags which regulate the different processes.

\begin{figure}[tb]
\begin{minipage}[t]{1.0\textwidth}
  \centering
    \includegraphics[width=0.8\textwidth]{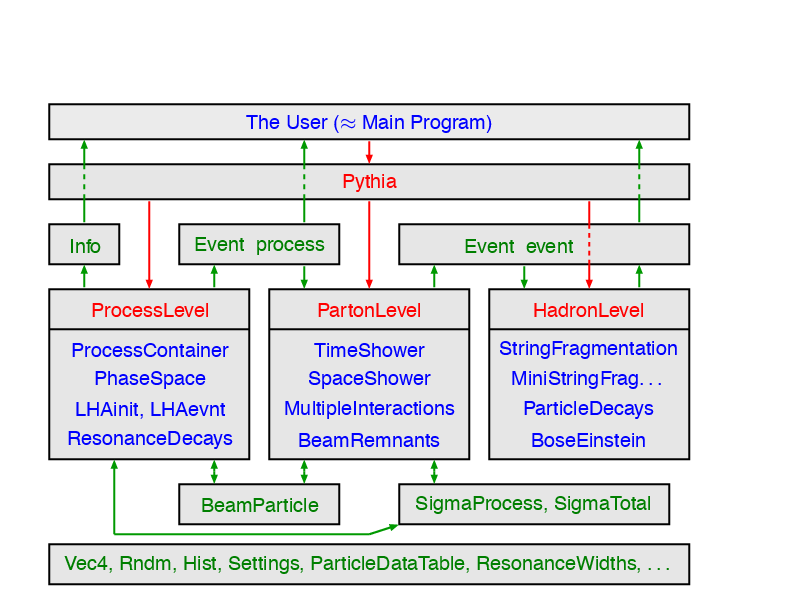}
\end{minipage}
\cprotect\caption{Structure of the \verb|PYTHIA 8| code taken from Ref. \cite{Sjostrand-Pythia-history}.}
\label{fig:pythia-structure}
\end{figure}

After the initialization phase a loop over the events has to be started. Each event is handled by the \verb|Pythia| class which coordinates the communication between three main classes, namely the \verb|ProcessLevel|, the \verb|PartonLevel| and the \verb|HadronLevel| classes. These classes are shown in Fig. \ref{fig:pythia-structure} \cite{Sjostrand-Pythia-history}, which describes the general structure of the \verb|PYTHIA 8| code.

The event generation consists of different phases. First the simulation of the interaction process. Different processes are allowed: hard QCD (elementary partonic interactions like $qq\rightarrow qq$, $qg\rightarrow qg$, $q\bar{q}\rightarrow gg$), soft QCD (interactions like elastic and diffractive scatterings), electroweak processes (for instance $ff\rightarrow ff$ with $\gamma^*$, $Z^0$, $W^{\pm}$ exchanges) and many other processes.
The simulation of the interaction process is handled by the \verb|ProcessLevel| class, which in addition takes care also of other tasks as for instance the decays of intermediate resonances like $Z^0$ or $W^{\pm}$. At the \verb|ProcessLevel| usually a small number of partons is produced.

\begin{figure}[tb]
\begin{minipage}[t]{1.0\textwidth}
\centering
    \includegraphics[width=1.0\textwidth]{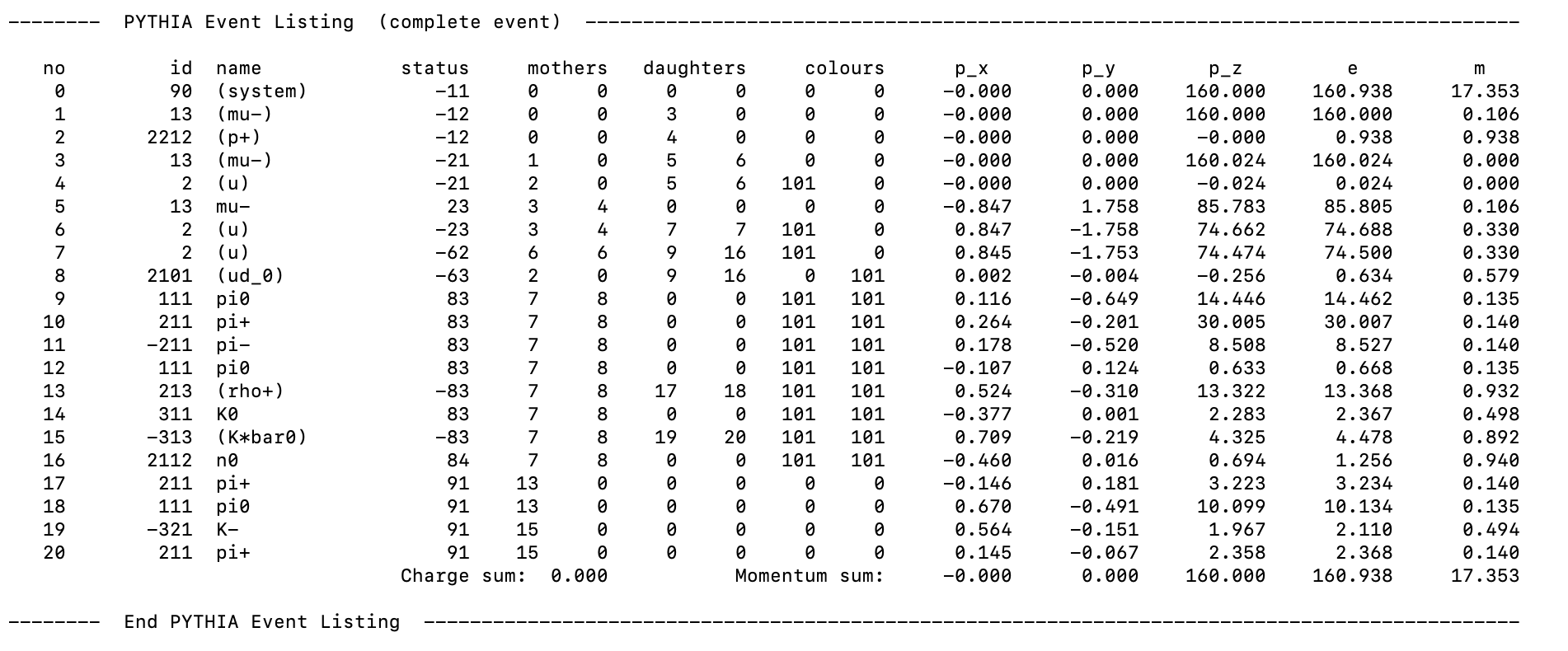}
\end{minipage}
  \cprotect\caption{Example of a SIDIS event generated with \verb|PYTHIA 8|.}\label{fig:pythia-event-record}
\end{figure}

Being interested in the SIDIS process, in this work only the elastic scattering of fermions 
$ff\rightarrow ff$ with $\gamma^*$ exchange is selected as interaction process. The beam lepton is taken to be a muon with four-momentum $l$ and the target is a proton (or neutron) at rest with four-momentum $P_N$. This corresponds to the typical configuration of a fixed target experiment and defines the laboratory system, indicated in the following with the superscript \textit{lab}. The momenta of the beam lepton and of the target are written at lines 1 and 2 of the event record (an example is shown in the Fig. \ref{fig:pythia-event-record}).

To simulate the DIS hard scattering, \verb|PYTHIA| samples the variables $x_B$ and $Q^2$ according to the inclusive DIS cross section. The interacting quark, referred here to as the \textit{initial quark} and indicated with $q_A$, has a flavor selected on the basis of PDFs. It is the quark that participates in the hard elastic scattering $f+q_A\rightarrow f + q_A$. After the hard scattering kinematics has been simulated, the initial quark has four-momentum $k_a$
whereas the scattered quark, called also \textit{final} or \textit{fragmenting quark}, has four-momentum $k_A=k_a+q_{\gamma^*}$, $q_{\gamma^*}=l-l'$ being the four-momentum of the exchanged virtual photon and $l'$ indicates the momentum of the scattered lepton. In addition to the values of $x_B$ and $Q^2$, \verb|PYTHIA| samples also the intrinsic transverse momentum of the initial quark\footnote{The modulus squared of the intrinisc transverse momentum $\textbf{k}^2_{\perp}$ is generated according to an exponential distribution with a width $\langle \textbf{k}^2_{\perp}\rangle$ dependent on $Q$ and on the invariant mass of the lepton-quark system. The azimuthal angle $\phi_{\perp}$ instead is sampled from a flat distribution in $[0,2\pi]$. This latter choice does not take into account the Cahn effect, which would require a $\phi_{\perp}$ distribution of the type $1+a_{1}\cos\phi_{\perp}+a_2\cos\phi_{\perp}$ with $a_1$ and $a_2$ expressed in powers of $|\textbf{k}_{\perp}|/Q$ and obtained from the non coplanar $l+q_A\rightarrow l+q_A$ hard scattering.}. The momenta $k_a$, $l'$ and $k_A$ are written in the event record (lines 4, 5 and 6 of the event record in the example in Fig. \ref{fig:pythia-event-record}).

Both the initial and the scattered quark may undergo parton showering, namely gluons are attached to the initial (\textit{initial state radiation}, ISR) and to scattered quark legs (\textit{final state radiation}, FSR). The parton showering is performed by the \verb|PartonLevel| class. The ISR and FSR evolve the quark virtualities until some lower cut in virtuality is reached. If ISR is switched on, the momentum $k_a$ is the momentum of the initial quark after ISR has been completed. Analogusly, if FSR is switched on, then the fragmenting quark four-momentum $k_A$ is the result of the evolution of the four-momentum that the scattered quark had right after the virtual photon vertex.
As a consequence of these processes, a large number of partons may be generated and the overall event picture becomes more complex. Since ISR and FSR produce complicated string topologies that are presently not treated with spin, they are disabled when using the interface with M19.
After the showering, the same class takes care of possible secondary scatterings among the partons, which are called \textit{multiple parton interactions} (MPI). The MPI machinery, however, does not apply to DIS or to $e^+e^-$ annihilation to hadrons.
The final result of the \verb|PartonLevel| class is a collection of partons with given momenta and colors.

\begin{figure}[!h]
  \centering
    \includegraphics[width=0.9\textwidth]{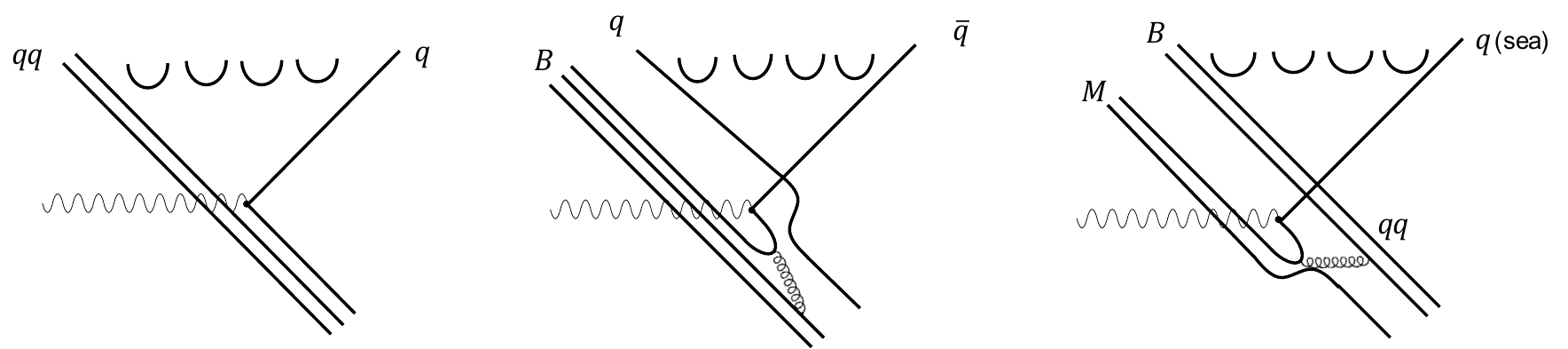}
  \cprotect\caption{Different string topologies in \verb|PYTHIA|.}
  \label{fig:remnant_pythia}
\end{figure}

To complete the list of partons that will then fragment into hadrons, the flavor and the momentum of the \textit{beam remnants} (in the \verb|PYTHIA| language) are decided.
The selection of the remnant depends on the flavor of $q_A$. If $q_A$ has a flavor found in the valence wave function of the proton, then it is considered to be a valence quark or a sea quark depending on the $x_B$ and $Q^2$ values of the event, by using the corresponding values of the valence and sea PDFs. This is relevant for the introduction of the quark transversity distribution, which is measured to be different from zero only in the valence region. As shown in the left diagram in Fig. \ref{fig:remnant_pythia}, if $q_A$ is a valence quark then the target proton remnant is a diquark $(q_fq_{f'})_{s_{tot}s_z}$, where $s_{tot}$ is the sum of the spin of $q_f$ and $q_{f'}$, and $s_z$ is the component of $s_{tot}$ along the quantization axis. The diquark can be either scalar, indicated also with $(q_fq_{f'})_0$, or vector, indicated also generically with $(q_fq_{f'})_1$, according to the $SU(6)$ wave function. For instance, if $q_A=u$, then
\begin{equation}\label{eq:proton wave function}
    |p^{\uparrow}\rangle = \frac{1}{\sqrt{18}}\left[u^{\uparrow}\left(3(ud)_{0}+(ud)_{10}\right)-\sqrt{2}u^{\downarrow}(ud)_{11}-\sqrt{2}d^{\uparrow}(uu)_{10}+2d^{\downarrow}(uu)_{11}\right]
\end{equation}
gives the weights $9/12$ ($0.75$) for the scalar diquark $(ud)_0$ and $3/12$ ($0.25$) for the vector diquark $(ud)_1$. These weights are hard coded in \verb|PYTHIA|.
If $q_A$ comes from the sea, the remnant is treated differently depending wether $q_A$ is a sea quark or a sea antiquark. In particular if it is $\bar{q}_A=\bar{u},\bar{d},\dots$, then $q_A$ comes always from the sea, and it is assumed to come from the splitting $g\rightarrow q_A\bar{q}_A$ which implies that the proton valence $q(qq)$ state is a color octet state, as in the middle diagram in Fig. \ref{fig:remnant_pythia}. In this case the proton is split in a quark $q$ and a diquark $(qq)$ with probabilities taken from Eq. (\ref{eq:proton wave function}). The diquark is paired with $q_A$ to form the baryon $q_A(qq)$ and eventually a string is stretched from $q$ to $\bar{q}_A$.
If $q_A$ is a sea quark, it is again assumed to come from the splitting $g\rightarrow q_A\bar{q}_A$ and therefore the $q(qq)$ system is again in a color octet state, as in the right diagram in Fig. \ref{fig:remnant_pythia}. In this case $\bar{q}_A$ is paired with a quark $q$, which is selected by splitting the proton in a quark and a diquark using the wave function in Eq. (\ref{eq:proton wave function}), and the meson $q\bar{s}$ is formed. Eventually a string is stretched from $q_A$ to the remaining diquark $(qq)$.

Once the nucleon remnant treatment is finished, \verb|PYTHIA| searches for color singlets among the produced partons. They can be open strings, closed strings or junctions. However, since gluon radiation is disabled, in this work, for each event only one open string with no gluons in between can be formed. The endpoints of the string are then written in the event record. In Fig. \ref{fig:pythia-event-record} they correspond to lines 7 and 8.

\begin{figure}[tb]
  \centering
    \includegraphics[width=0.6\textwidth]{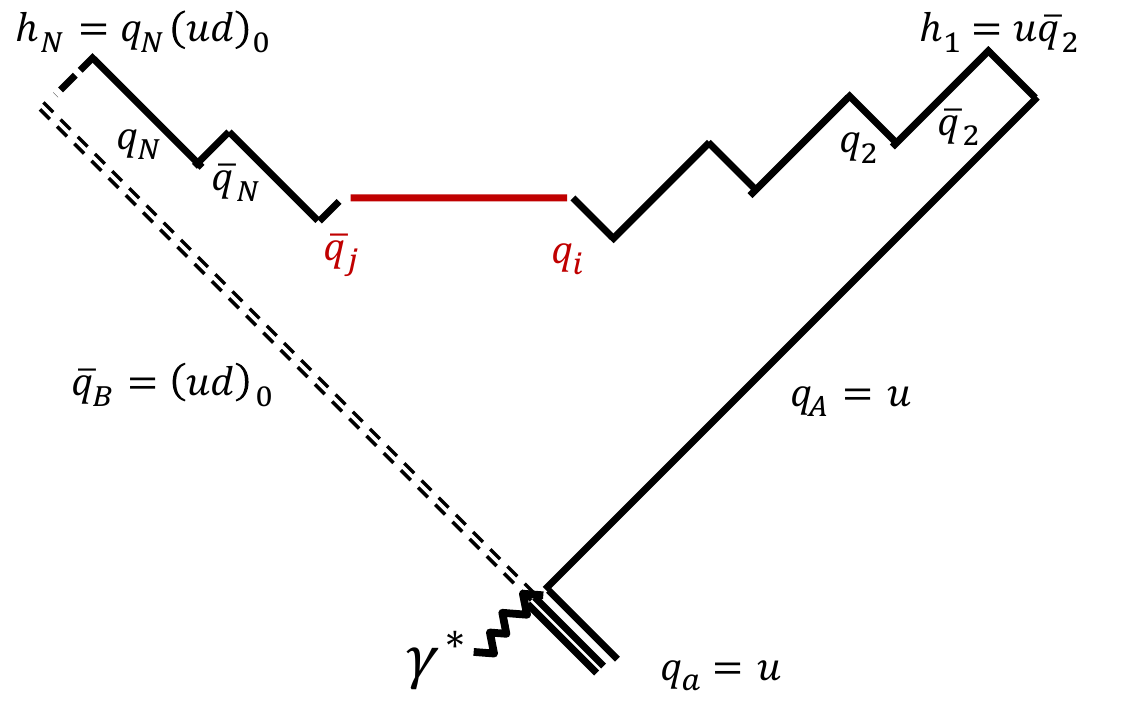}
  \cprotect\caption{String fragmentation in \verb|PYTHIA|: $i$-breaking points from the $q_A$ side and $j$ breaking points from the $\bar{q}_B$ side. The remaining string piece is stretched between $q_i$ and $\bar{q}_j$.}\label{fig:DIS_pythia}
\end{figure}

The string is now ready to hadronize. This process and the following are handled by the \verb|HadronLevel| class. The string is fragmented according to the Lund Model by repeating recursively the elementary splitting $q\rightarrow h + q'$, as shown in Fig. \ref{fig:DIS_pythia} for the case of a $u$ fragmenting quark and a $(ud)_0$ remnant. To start the recursive algorithm, \verb|PYTHIA| choses randomly with equal probability between the $q_A$ side (\textit{quark side}) and the \textit{remnant side}. Hence $q$ can be either $q_A$ or the remnant. Supposing that the first splitting is taken from the $q_A$ side, hence it is $q_A\rightarrow h_1+q_2$, the flavor of $q_2$ of the produced $q_2\bar{q}_2$ pair is generated. The $q_2$ can be a quark or, in order to allow for baryon production, an anti-diquark. For the quark production, $s$ quarks are suppressed by the factor \verb|StringFlav:probStoUD| (set by default to 0.19) with respect to $u$ and $d$ quarks. Diquark production is suppressed with respect to the quark production by the factor \verb|StringFlav:probQQtoQ| (set by default to 0.09). For diquark production, there are also parameters for the suppression of strange with respect to non strange diquarks and for the suppression of spin 1 diquarks with respect to spin 0 diquarks.

If $q_2$ is a quark, then \verb|PYTHIA| choses if the the hadron $h_1=q_A\bar{q}_2$ is a pseudoscalar or a vector meson. The suppression factor vector/pseudo-scalar is given by the parameter \verb|StringFlav:mesonUDvector| (set by default to 0.62) if $h$ does not contain a strange quark and by the parameter \verb|StringFlav:mesonSvector| (set by default to 0.725) if $h_1$ contains at least one strange quark. There are similar parameters also for mesons containing $c$ and $b$ quarks. The type of $h_1$ is finally decided according to the isospin wave functions and by some parameters for the treatment of the mixing between iso-scalar states ($\eta$ and $\eta'$ or $\omega$ and $\phi$) and of supressions of these states with respect to their normal production rates, as for instance $\eta$ with respect to $\pi^0$.
The recipe for the pseudoscalar meson production in \verb|PYTHIA| is similar to the recipe implemented in M19 but it is extended to five quark flavors and it is more sophisticated in the mixing between isoscalar states.


If $q_2$ is an anti-diquark, two models are presently implemented in PYTHIA for baryon production. They are the \textit{diquark model} \cite{Pythia6} and the \textit{pop-corn model} \cite{simple-popcorn,advanced-popcorn}. 

After the determination of the identity of $h_1$, its mass is defined (either fixed or drawn according to a Breit-Wigner distribution depending whether $h_1$ is stable or a resonance). The four-momentum of $h_1$ is then obtained by generating first $\textbf{k}_{2\rm{T}}$, which gives $\textbf{p}_{1\rm{T}}=\textbf{k}_{A\rm{T}}-\textbf{k}_{2T}$ (with $\textbf{k}_{A\rm{T}}=\textbf{0}$) as in M19, and then $Z$ using the splitting function in Eq. (\ref{eq:F Pythia}). The meson type and its four momentum are then added to the event record (line 9 of Fig. \ref{fig:pythia-event-record}).

Now the remaining string piece to be hadronized at the next step is stretched between the quark $q_2$ and the remnant. \verb|PYTHIA| chooses again from which side to take the next splitting, whether from the side of $q_2$ or from the remnant and so on. 

This procedure is repeated and hadrons constituting the quark and the remnant jets are emitted from any of the two sides. After the emission of some hadrons\footnote{The average number of hadrons produced in the string fragmentation is proportional to $\log W$, $W$ being the mass of the final hadronic system.}, the mass of the remaining string piece $\bar{q}_jq_i$ can fall off below a given threshold\footnote{The threshold value is not constant like in M18 or M19. It is rather generated uniformly in an interval which center and length are tunable parameters.}. If this is the case \verb|PYTHIA| calls the exit method and the string piece is split into the two final hadrons. This allows to join the quark and the remnant jets. The total four-momentum, the charge, baryon number and other quantum numbers are automatically conserved. In the string rest frame, this joining takes place on average in the central rapidity region. The produced hadrons are stored in the event record (lines 9-16 of Fig. \ref{fig:pythia-event-record}).

After string fragmentation, the short lived particles like $\rho$ mesons are decayed (lines 17-20 of Fig. \ref{fig:pythia-event-record}). Then Bose-Einstein correlations can be taken into account and finally the long lived hadrons are decayed.

At the end of these stages the event generation is terminated and all produced particles are listed in the event record and are ready to be analysed.

\section{Spin in the fragmentation process}\label{sec:spin in fragmentation process}
The simulation of a polarized SIDIS process requires, in general, to take into account both beam and target polarizations. The lepton beam can be logitudinally polarized, for instance as a consequence of the production mechanism (a $\mu^+$ is produced with helicity $+1/2$ in the decay $\pi^{+}\rightarrow \mu^+ + \nu_{\mu}$, and a $\mu^-$ is produced with helicity $-1/2$ in the decay $\pi^-\rightarrow \mu^-+\bar{\nu}_{\mu}$).
As can be seen from the SIDIS cross section in Eq. (\ref{eq:pol SIDIS x section}), a helicity polarized beam is coupled to some structure functions related to convolutions of worm-gear TMDs with the Collins FF.
The direction of the target polarization, depends on the experimental set-up and can be either longitudinal or transverse with respect to the lepton beam.
This work is focused on the Collins and dihadron asymmetries, which are measured with unpolarized lepton beams and transversely polarized targets and, in the following, only this configuration is considered. Spin effects in the fragmentation are however interesting also for unpolarized targets, for instance because of the Boer-Mulders function. 

\subsection{Strategy for the introduction of spin effects}
The strategy for inclusion of the spin effects in the simulation of a SIDIS event with \verb|PYTHIA| is to accept the hadrons produced during string fragmentation according to rules based on the ${}^3P_0$ mechanism as implemented in M19. 
In particular, each hadron generated by the standard \verb|PYTHIA| procedure is rejected if \begin{itemize}[noitemsep,topsep=0pt,parsep=0pt,partopsep=0pt]
    \item[-] it is produced after a string breaking taken from the remnant side
    \item[-] it is not a pseudoscalar meson
    \item[-] it does not respect a weight that takes into account the quark polarization (see Eq. (\ref{eq:w})).
\end{itemize}

If the hadron proposed by \verb|PYTHIA| passes the selection rules, it is stored in the event record, otherwise a new one is tried. If accepted, the previous quark polarization is transferred to the next quark and the hadron produced at the next splitting undergoes the same selection procedure.

The \verb|PYTHIA| fragmentation chain is thus modified and emulates the polarized splittings of M19. Namely, the fragmentation chain evolves from the $q_A$ side towards the remnant side producing only pseudo-scalar mesons.
The chain is terminated according to the \verb|PYTHIA| exit condition without changes.

To apply this strategy, it is necessary to communicate with the generator. This is done using the so called \verb|UserHooks| class, a \verb|PYTHIA| class that allows the external user to step-in at particular points during the normal execution of the event generation, to read information on the current event status and to make changes. In practice, it consists in implementing a derived \verb|UserHooks| class which estabilishes the communication between \verb|PYTHIA|, a \verb|Fortran| module containing the modified routines of M19 for the calculation of the spin dependent quantities, and a \verb|C++| file which contains the parameterization of the quark transversity PDF.

In the next sections, the interface is described in more detail and the comparison of the results of a simplified simulation are compared with M19 to test the correct implementation of the spin effects in \verb|PYTHIA|.


\subsection{The code}
The DIS events in the laboratory frame are generated using a \verb|PYTHIA| main program which includes the header file with the developed \verb|UserHooks| class.
As the event generation starts, \verb|PYTHIA| selects the variables $x_B$ and $Q^2$, picks up the flavor of the initial quark $q_A$, performs the hard scattering and sets up a string between the scattered $q_A$ and the target remnant. At this point, by calling a boolean function, \verb|PYTHIA| asks if user intervention during the fragmentation process has to be activated. Normally this function returns false. By modifying it to return true, for each hadron generated during string fragmentation, in \verb|Hadronlevel|, the user is allowed to decide to veto it or not. The procedure described in the following is that presented in Sec. \ref{sec:SIDIS in Pythia}, but with the modifications necessary for the incorporation of the spin effects.

The next step is to define the polarization vector $\textbf{S}_A$ of the scattered quark $q_A$ in the string rest frame. At leading order it is the same as that in the GNS. It can either be chosen freely, i.e. fully polarized along some given axis, or calculated using a parameterization of the transversity PDF, as will be described below. This vector defines the spin density matrix $\rho(q_A)=(\textbf{1}+\boldsymbol{\sigma}\cdot\textbf{S}_A)/2$. At this point the polarized splittings of M19 can be applied.

Reading from the event record the momentum $k_A^{lab}$ of $q_A$ and the momentum $k_B^{lab}$ of the remnant $q_B$, a matrix is set-up to transform momenta from this system to the string rest system, with the string axis that defines the $\zu$ axis and points toward $\textbf{k}_A$.

\verb|PYTHIA| starts the string fragmentation by choosing if the first splitting is on the $q_A$ side or on the remnant side, generates a first $q'\bar{q}'$ pair, and forms the hadron $h$ according to its standard procedure. As soon as $h$ is generated, \verb|PYTHIA| calls an other boolean function to ask if $h$ has to be accepted. The function has been modified to reject the hadron if it is on the remnant side, if $h$ is not a pseudoscalar meson or, if these are not the cases, with probability $1-w$, where\cprotect\footnote{The $\kptkpt$ distribution in \verb|PYTHIA| is different with respect to M19. However since the chosen values of the parameters are such that $|\mu|^2\gg \bt^{-1}$, no big difference is expected.}
\begin{equation}\label{eq:w}
    w(\kpt,\textbf{S}_{q})=\frac{1}{2}\left(1-\frac{2\IM(\mu)\textbf{S}_{q}\cdot(\zu\times\kpt)}{|\mu|^2+\kptkpt}\right).
\end{equation}
This weight is obtained from the splitting function of M19 given in Eq. (\ref{eq:F_explicit simple 3P0}) and can be thought as the probability for the current splitting to be a ${}^3P_0$ splitting. It depends on the transverse momentum $\kpt$ of $q'$ and on the polarization vector $\textbf{S}_q\equiv\textbf{S}_{A}$. The four momentum of $q'$ is $k'=k_A-P_h$, where the hadron momentum $P_h$ is obtained transforming $P_h^{lab}$, as read from the event record, to the string rest frame.

If the hadron is rejected, a new one is generated by \verb|PYTHIA| and tested again, until a hadron is accepted. When this happens, the spin density matrix of $q'$ is calculated according to Eq. (\ref{eq:rho'=TrhoT}), namely
\begin{equation}
    \rho(q')=\frac{(\mu+\sigma_z\boldsymbol{\sigma}\cdot\kpt)\,\sigma_z\,\rho(q_A)\,\sigma_z\,(\mu^*-\sigma_z\boldsymbol{\sigma}\cdot\kpt)}{\rm{tr}\left[ (\mu+\sigma_z\boldsymbol{\sigma}\cdot\kpt)\,\sigma_z\,\rho(q_A)\,\sigma_z\,(\mu^*-\sigma_z\boldsymbol{\sigma}\cdot\kpt)\right]},
\end{equation}
$q_A$ is replaced with $q'$, $\textbf{S}_q$ with $\textbf{S}_{q'}$ extracted from $\rho(q')$, and the procedure is repeated recursively until the \verb|PYTHIA| exit condition is verified. The spin effects are switched off and the last two hadrons are generated by the default procedure of \verb|PYTHIA|.

In principle this recipe is applicable when the target remnant is a scalar diquark, as for instance in Fig. \ref{fig:DIS_pythia}. In that case the spin information flows along the quark line, namely from the initial quark to the scattered quark, and then from the scattered quark towards the target remnant along the fragmentation chain. If the target remnant were instead a spin $1$ diquark, one should in principle consider a joint quark-diquark spin density matrix to take into account their correlated spin state. This would produce space-like correlations between the momenta of hadrons emitted from the quark side with those emitted from the remnant side, which would thus manifest when analyzing hadron pairs of which one is selected from the quark jet and the other from the remnant jet. These correlations do not play a role when looking at the current fragmentation region. Since for the evaluation of the Collins and dihadron asymmetries one considers single hadrons or pairs of hadrons coming both from the quark side, this recipe is appropriate. In $\verb|PYTHIA|$+${}^3P_0$ the target remnant is thus treated as unpolarized.

An improved version of the interface capable of treating properly the case of polarized target remnants would require similar techniques to those needed for the implementation of spin-dependent effects in $e^+e^-$ annihilation to hadrons, where the spin effects are manifested through azimuthal correlations between hadrons produced in the quark and anti-quark jets \cite{Boer:1997-e+e-,Artru-Collins}.

\subsection{Test of the implementation}\label{sec:test of implementation}
To make sure that the spin effects are correctly implemented in \verb|PYTHIA|, some relevant observables generated in the same conditions with M19 and with \verb|PYTHIA| interfaced to the ${}^3P_0$ model (in the following indicated as \verb|PYTHIA|+${}^3P_0$), have been compared.
The results of M19 shown in the previous chapter have been obtained from fragmentations of fully transversely polarized $u$ quarks which momentum has been calculated using a sample of $x_B$ and $Q^2$ values from real COMPASS events. 
The same conditions are obtained with \verb|PYTHIA| simulating SIDIS events where muons with $160\,\rm{GeV}/c$ momentum are scattered off a transversely polarized proton target at rest. The relevant parameters that enter the LSSF in \verb|PYTHIA| have been set to the values used in M19, given in Tab. \ref{tab:pythia parameters}. In the same table, the correspondence with the parameters of M19 and the values of the default \verb|PYTHIA| setting are also given.
By setting to zero \verb|StringPT:enhancedFraction| forces \verb|PYTHIA| to generate the quark transverse momentum at string breaking using a single exponential which average value is \verb|StringPT:sigma|.

\begin{table}[h!]
\centering
\begin{tabular}{|c|c|c|c|} 
 \hline
 \verb|PYTHIA| & M19 & value & default\\ [0.5ex] 
 \hline\hline
 \verb|StringZ:aLund| & $a$ & $0.9$ & $0.3$\\ 
 \verb|StringZ:bLund| & $\bl$ & $0.5\,(\rm{GeV}/c^2)^{-2}$ & $0.8\,(\rm{GeV}/c^2)^{-2}$ \\
 \verb|StringPT:sigma| & $\bt^{-1/2}$ & $0.34\,(\rm{GeV}/c)$ & $0.304\,(\rm{GeV}/c)$ \\
 \verb|StringPT:enhancedFraction| &  & $0$ & $0.01$\\
 \hline\hline
 \multicolumn{1}{|c}{\UseVerb{verbPythia}+${}^3P_0$} & \multicolumn{1}{|c|}{M19}  & \multicolumn{2}{c|}{value} \\ [0.5ex]
 \hline\hline
  \verb|complexMass| & $\mu$ & \multicolumn{2}{c|}{$(0.42+i0.76)\,\rm{GeV}/c^2$} \\[1ex] 
 \hline
\end{tabular}
\cprotect\caption{Relation between the parameters of the fragmentation process in \verb|PYTHIA| and in M19, and the corresponding values used in simulations.}
\label{tab:pythia parameters}
\end{table}

To be in the kinematics of the COMPASS experiment \cite{COMPASS-collins-sivers}, the phase space cuts on the DIS variables $W>5\,\rm{GeV}/c^2$, $0.2<y<0.9$ and $Q^2>1.0\,(\rm{GeV}/c)^2$ have been applied on the generated events. 
Finally, among these events only those where the string is stretched between a struck $u$ quark and a remnant $(ud)_0$ diquark have been selected. As in M19, the struck quark is taken fully transversely polarized in the string rest frame, i.e. $\textbf{S}_A=(0,1,0)$.

The results shown in the present section are obtained in the string rest frame looking at hadrons with $p_{\rm{T}}>0.1\,\rm{GeV}/c$ and $z_h>0.2$. The $z_h$ cut is lowered to $z_h>0.1$ when considering hadron pairs.

\begin{figure}[tb]
  \centering
    \includegraphics[width=0.6\textwidth]{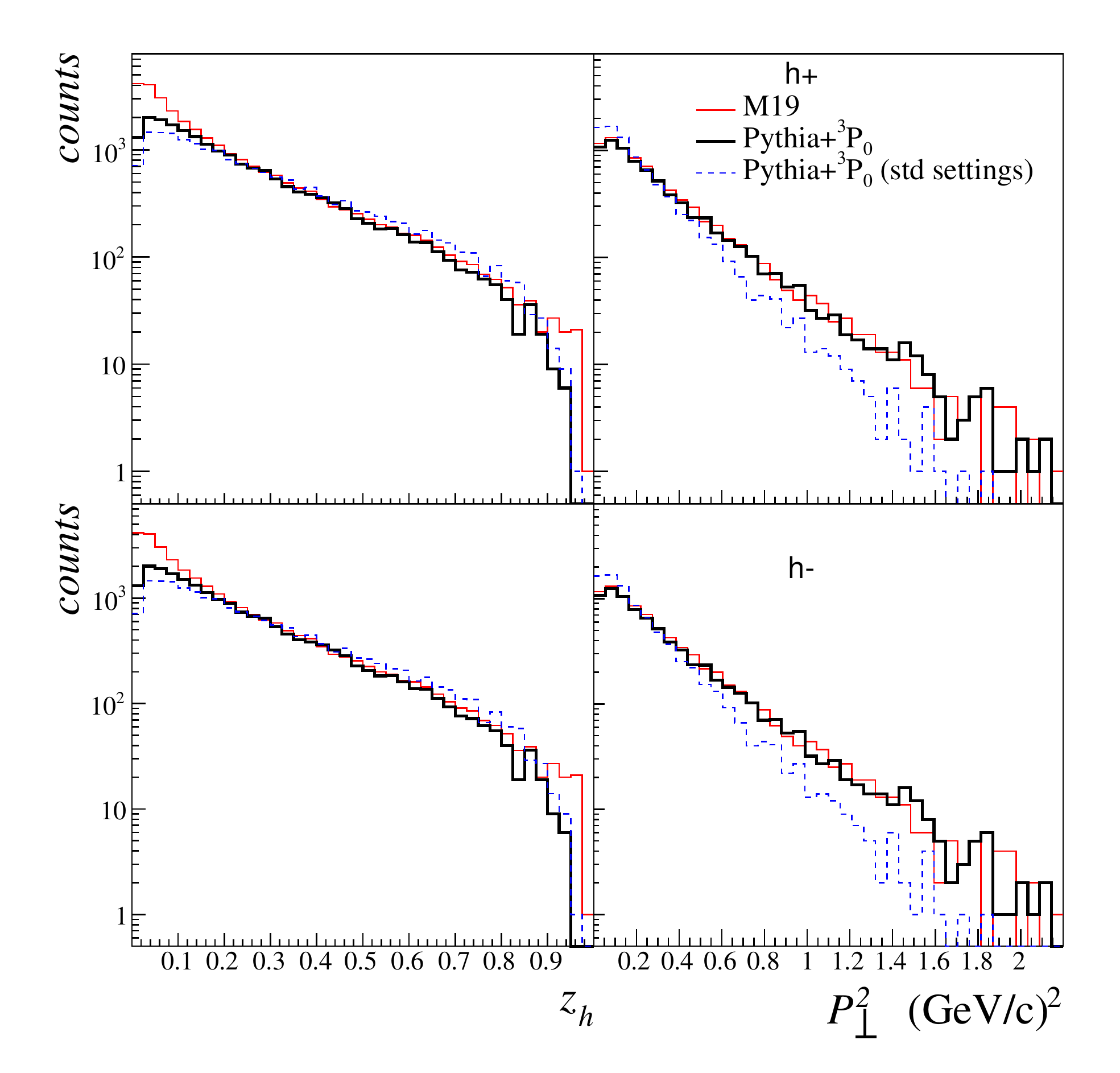}
  \cprotect\caption{Distributions of $z_h$ (left plots) and $p^2_{\rm{T}}$ (right plots) for positive hadrons (upper row) and negative hadrons (lower row). The red histogram is obtained with M19, the solid black histogram is obtained with \verb|PYTHIA|+${}^3P_0$ with the $^{}3P_0$ settings, and the dashed blue histogram is obtained with \verb|PYTHIA|+${}^3P_0$ with standard \verb|PYTHIA| settings.}\label{fig:PY+3P0 distrib}
\end{figure}

Figure \ref{fig:PY+3P0 distrib} compares the $z_h$ (left plots) and $p^2_{\rm{T}}$ distributions (right plots) for positive (upper row) and negative (lower row) hadrons, as obtained from simulations with M19 (red histogram), with \verb|PYTHIA|+${}^3P_0$ with the parameter values of Tab. \ref{tab:pythia parameters} (solid black histogram) and with the default setting (dashed blue histogram). It can be noticed that the $z_h$ distributions have different trends for $z_h<0.1$. This is due to the different exit conditions in M19 and in \verb|PYTHIA|. In M19 the recursive algorithm is terminated at smaller remaining string energies which produces more hadrons with small fractional energies. This difference is, however, not relevant for the observables considered in the following due to the $z_h$ cut. Looking at the positive hadrons, some difference can be noticed also at large $z_h$. This is due to the presence of the heavy $(ud)_0$ diquark in \verb|PYTHIA| which prevents first rank hadrons to take large amount of fractional energies. This is not seen in the $z_h$ distribution for negative hadrons since they are produced starting with the second rank.
These distributions are also similar to those obtained with \verb|PYTHIA|+${}^3P_0$ but with the standard \verb|PYTHIA| parameters.
The $p^2_{\rm{T}}$ distributions obtained with stand alone M19 program and with \verb|PYTHIA|+${}^3P_0$ are also very similar both for positive and negative hadrons. They have different slopes with respect to the distributions obtained with \verb|PYTHIA|+${}^3P_0$ with the default \verb|PYTHIA| settings.

\begin{figure}[!h]
  \centering
 \begin{minipage}{0.48\textwidth}
    \includegraphics[width=1.0\textwidth]{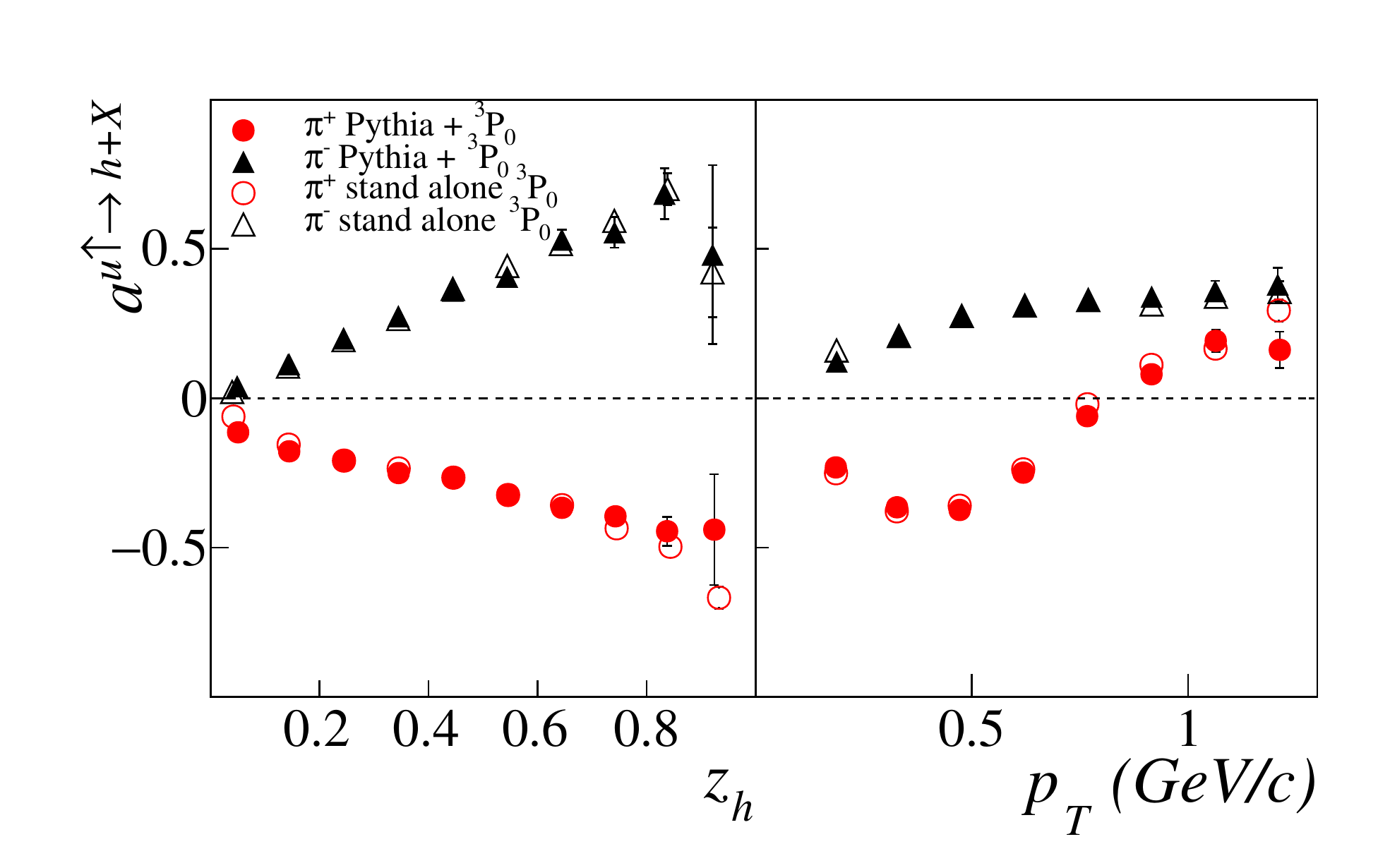}
\end{minipage}
 \begin{minipage}{0.48\textwidth}
    \includegraphics[width=1.0\textwidth]{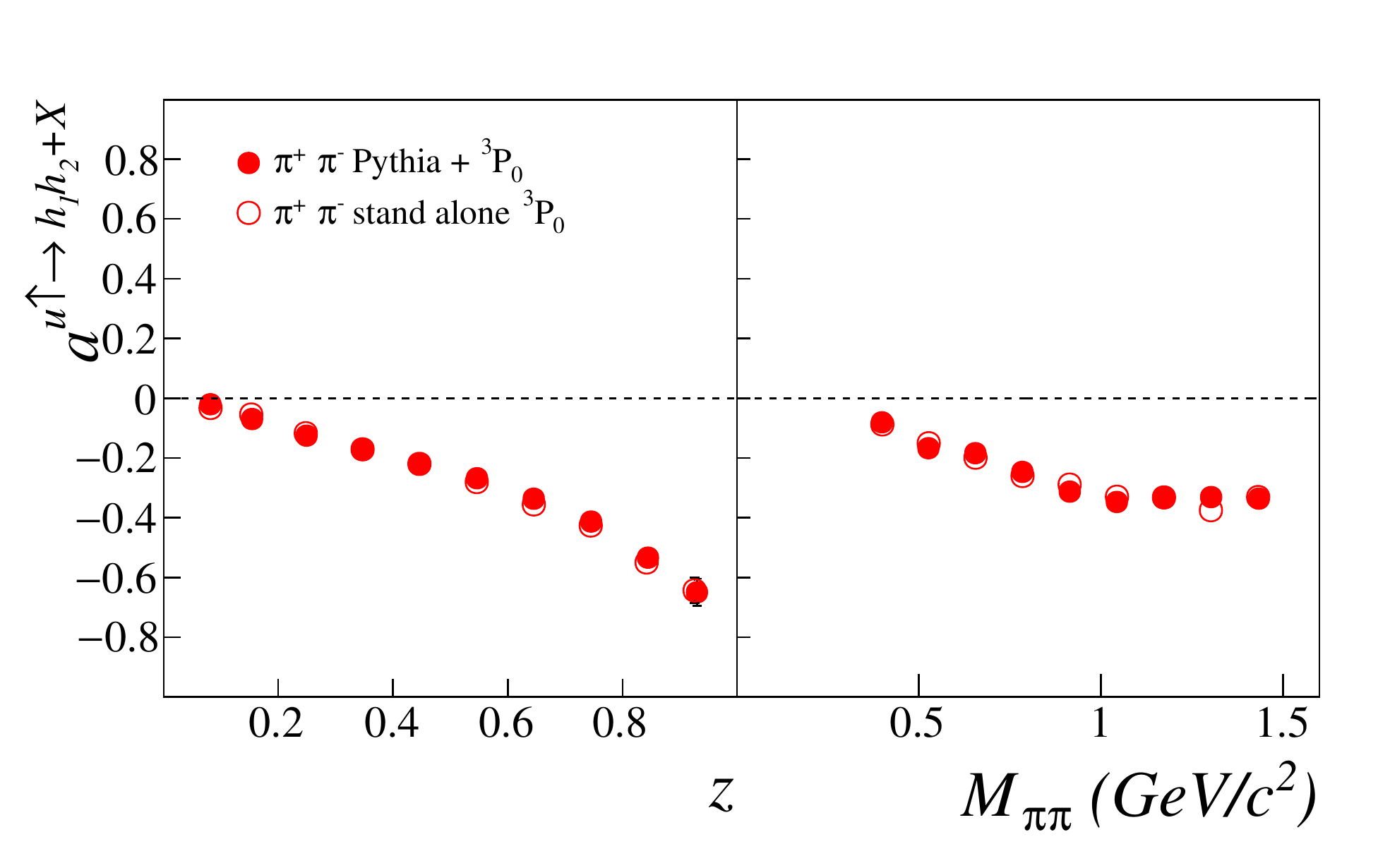}
 \end{minipage}
  \cprotect\caption{Left: comparison between the Collins analysing power as function of $z_h$ (left plot) and $p\T$ (right plot) for $\pi^{\pm}$, as obtained with M19 (open points) and with \verb|PYTHIA|+${}^3P_0$ (closed points). Right: comparison of the dihadron analysing power.}\label{fig:PY+3P0 collins ap}
\end{figure}

Figure \ref{fig:PY+3P0 collins ap} compares the Collins analyzing power for charged pions as obtained with M19 (open points) and with \verb|PYTHIA|+${}^3P_0$ (closed points). The analysing power is given as function of $z_h$ in the left plot and as function of $p\T$ in the right plot. The comparison of the dihadron analysing power as function of $z$ and of the invariant mass is shown in the right panel of Fig. \ref{fig:PY+3P0 collins ap}. As expected if the spin effects are correctly introduced in \verb|PYTHIA| the results of the two simulations are very close.

\section{Results for transversely polarized nucleon}\label{sec:results for transversely pol nucleon}
To fully exploit the predictive power of \verb|PYTHIA|+${}^3P_0$ model for the Collins and di-hadron asymmetries the quark transversity PDF has to be used to calculate the polarization of the fragmenting quark $q_A$ starting from the nucleon polarization. This is done in the interface as explained in detail in the following section. Then the results from simulations of SIDIS events with transversely polarized protons and deuterons targets are discussed.

\subsection{The transversity PDF and struck quark polarization}


For each event, the transverse polarization vector of the initial quark $\textbf{S}_{a\perp}^{gns}$ in the GNS\footnote{The longitudinal $\zu^{gns}$ axis is defined by the direction of the exchanged $\gamma^*$ momentum. The normal to the scattering plane defines the $\hat{\textbf{y}}^{gns}$ axis and the $\hat{\textbf{x}}^{gns}$ axis is such that $(\hat{\textbf{x}}^{gns},\hat{\textbf{y}}^{gns},\hat{\textbf{z}}^{gns})$ forms a right handed frame.} is calculated in the interface implementing the \verb|UserHooks| class as
\begin{equation}\label{eq: S_a perp collinear}
    \textbf{S}_{a\perp}^{gns} = \frac{h_1^{q_A}(x_B,Q^2)}{f_1^{q_A}(x_B,Q^2)}\textbf{S}_{\perp}^{gns},
\end{equation}
where $\textbf{S}_{\perp}^{gns}$ is the target nucleon transverse polarization vector. This relation implies that for $h_1>0$ the quark is polarized along the same direction as the nucleon, whereas for $h_1<0$ it is polarized along the opposite direction. The degree of quark transverse polarization is $h_1/f_1$ as from the definition of the transversity PDF. The quark transversity PDF $h_1$ and the unpolarized PDF $f_1$ are assumed to have the same dependence on the intrinsic quark transverse momentum.

In order to evaluate $\textbf{S}_{a\perp}^{gns}$ it is necessary to introduce a parameterization for $h_1^{q_A}$. Also, the polarization vector of the target nucleon, which in experiments is given in the laboratory system, has to be calculated in the GNS. These two steps are described in the following.

\begin{figure}[tb]
\centering
 \begin{subfigure}[b]{0.4\textwidth}
    \includegraphics[width=\textwidth]{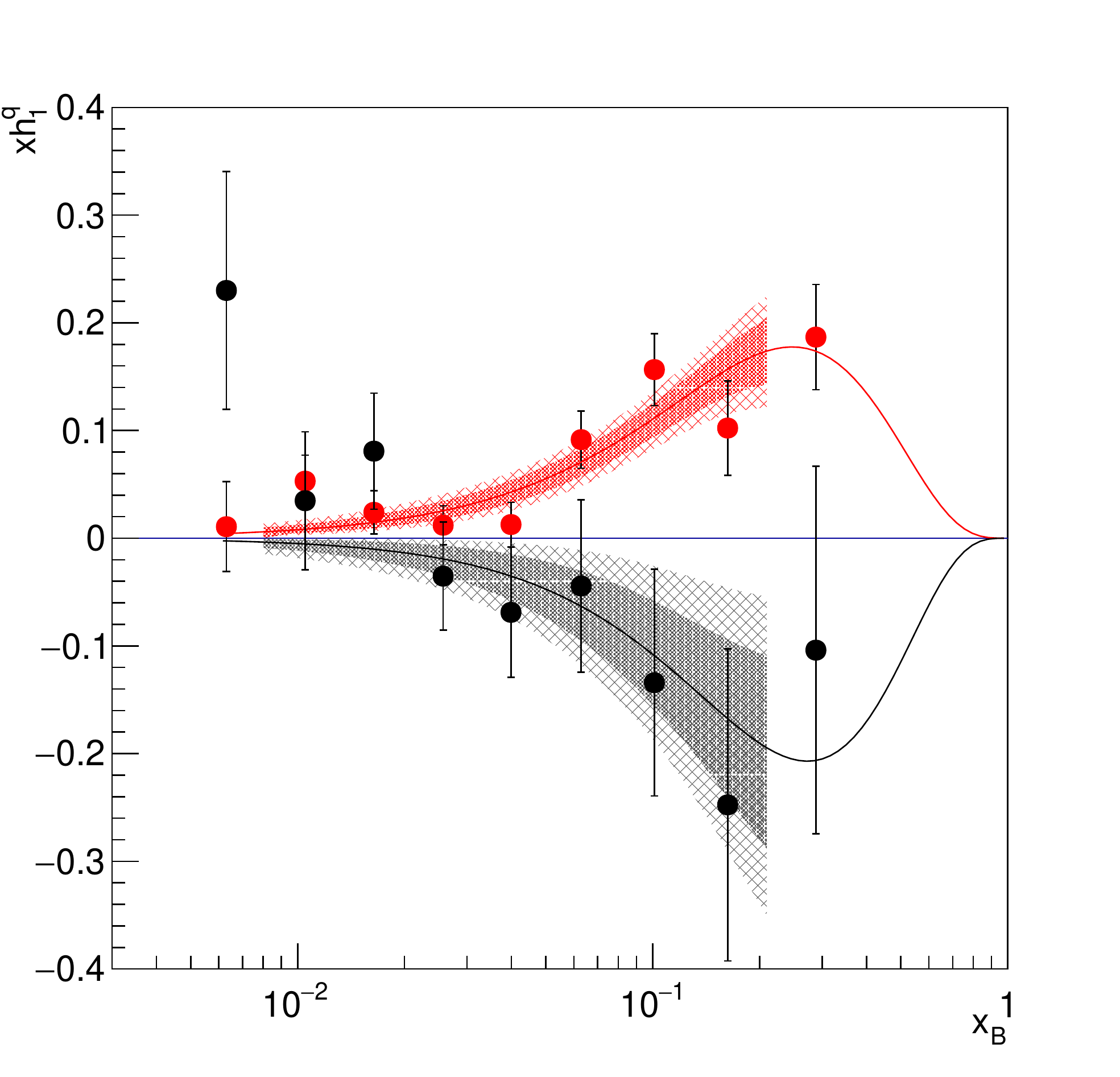}
\end{subfigure}
  \caption{Values of $h_1^{u_v}$ (red points) and of $h_1^{d_v}$ (black points) from Ref. \cite{M.B.B}. The continuous curves are the results of the fits. The bands give the $68\%$ and $90\%$ confidence levels \cite{deuteron-proposal}.}\label{fig:h1 parameterization}
\end{figure}

According to the most recent extractions \cite{Anselmino-transversity-extraction,M.B.B}, $h_1^q$ is non vanishing for the valence $u_v$ and $d_v$ quarks. The parameterizations used here
\begin{eqnarray}\label{eq:h1q parametrization}
    x_Bh_1^{u}(x_B) = 3.2\,x_B^{1.28}\,(1-x_B)^4, & x_Bh_1^{d}(x_B)=-4.6\,x_B^{1.44}\,(1-x_B)^4,
\end{eqnarray}
are obtained fitting the point by point extraction of $h_1$ performed in Ref. \cite{M.B.B} and are shown in Fig. \ref{fig:h1 parameterization}. This extraction uses the Collins asymmetries for proton and deuteron targets as measured by the COMPASS collaboration at the average scale $\langle Q^2\rangle\simeq 10\,\rm{GeV}^2$ and the BELLE $e^+e^-$ data. The Soffer bound has not been applied. Also, no evolution for $h_1$ has been used, thus the parameterizations give $h_1$ at the $Q^2$ values of $x_B$ in the COMPASS kinematics.

The quark polarizations $h_1/f_1$ have been obtained using for $f_1$ the CTEQ5L LO parameterization, which is the default setting in \verb|PYTHIA|. With this procedure the transverse polarization of the $d^v$ quark is close to one at large $x_B$. The parameterization has however large statistical errors due to the poor existing deuteron data set, which does not allow to constrain $h_1^{d^v}$ at large $x_B$. This is the motivation for the future deuteron run of COMPASS which will take place in 2021 and will provide more precise data.

In the present implementation of M19 in \verb|PYTHIA| the parameterization of $h_1^{q_A}$ is defined in an external file and it can be easily changed.

The target nucleon polarization is usually measured in a reference system defined by the direction of nucleon transverse polarization ($\yu^{lab}$) and by the direction of the incoming lepton beam momentum ($\zu^{lab}$). The $\xu^{lab}$ axis is such that $\hat{\textbf{x}}^{lab}$,$\hat{\textbf{y}}^{lab}$ and $\hat{\textbf{z}}^{lab}$ form a right handed system.
In this system, the polarization of a transversely polarized nucleon is therefore $\textbf{S}^{lab}=(0,1,0)$. It is rotated to GNS giving $\textbf{S}^{gns}$.

\begin{figure}[tb]
  \centering
    \includegraphics[width=0.5\textwidth]{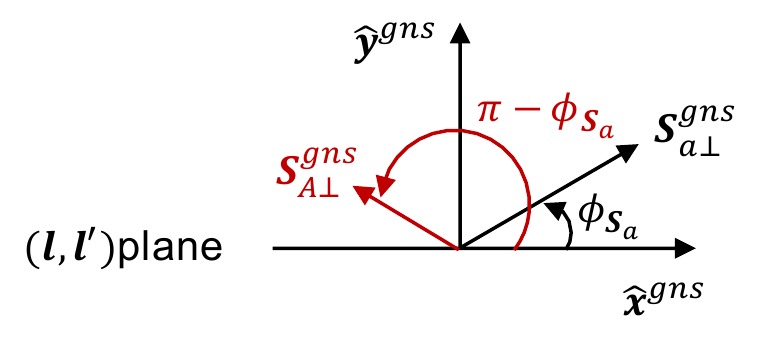}
  \caption{Relation between the polarization vector of the initial and scattered quarks in the GNS.}\label{fig:DNN}
\end{figure}

The polarization vector of the scattered quark in the same event is different from that of the initial quark because of the interaction with the virtual photon. In the GNS reference system the polarization vector of the struck quark $q_A$ is given by
\begin{eqnarray}\label{eq:S_A reflected}
    \textbf{S}^{gns}_{A\perp} = D_{\rm{NN}}\left[\textbf{S}^{gns}_{a\perp}-2\,(\textbf{S}^{gns}_{a\perp}\cdot\hat{\textbf{x}}^{gns})\,\hat{\textbf{x}}^{gns}\right], & S^{gns}_{A\parallel} = -S^{gns}_{a\parallel},
\end{eqnarray}
where the subscripts "$\perp$" and "$\parallel$" indicate the components perpendicular and parallel to the $\gamma^*$ momentum. After the hard scattering, the initial quark polarization is reduced by the \textit{depolarization factor} $D_{\rm{NN}}=2(1-y)/(1+(1-y)^2)$ \cite{Collins:1993kq} and it is reflected with respect to the normal to the scattering plane. The explicit relation between the azimuthal angle of $\textbf{S}^{gns}_{a\perp}$ and of $\textbf{S}^{gns}_{A\perp}$ is
\begin{equation}\label{eq:phiS_A = pi - phiS_a}
    \phi_{\textbf{S}^{gns}_{A\perp}} = \pi - \phi_{\textbf{S}^{gns}_{a\perp}},
\end{equation}
and is represented graphically in Fig. \ref{fig:DNN}.
The longitudinal component, instead, is reflected after the hard scattering due to helicity conservation. Equation (\ref{eq:S_A reflected}) is valid when terms of order $M_N/Q$ and $k_{\perp}/Q$ are neglected. At this order it is $\textbf{S}_A\equiv \textbf{S}_{A}^{gns}$ and this vector is finally used to calculate the spin density matrix of the fragmenting quark in the string rest frame.

\subsection{Transverse spin asymmetries for a proton target}
The results shown in the present section are obtained from simulations of the SIDIS process for a transversely polarized proton target at rest, in the kinematical regime of the COMPASS experiment as in Sec. \ref{sec:test of implementation}.

The azimuthal distribution of the produced hadrons in the GNS is
\begin{equation}
    \frac{d^2N_h}{d\phi_C\,dX} = N_h^0(X)\left[1+D_{\rm{NN}}\,|\textbf{S}_{\perp}^{gns}|\,A_{Coll}^p(X)\,\sin\phi_C\right]
\end{equation}
where $\phi_C$ is the Collins angle. The Collins asymmetry $A_{Coll}^p$ is extracted as $2\langle \sin\phi_C\rangle$ $/(\langle D_{\rm{NN}}\rangle \,\langle |\textbf{S}^{gns}_{\perp}|\rangle)$ as function of the kinematic variable $X$ which can be either $x_B$, $z_h$ or $P_{\perp}$.

\begin{figure}[tb]
  \centering
    \includegraphics[width=0.8\textwidth]{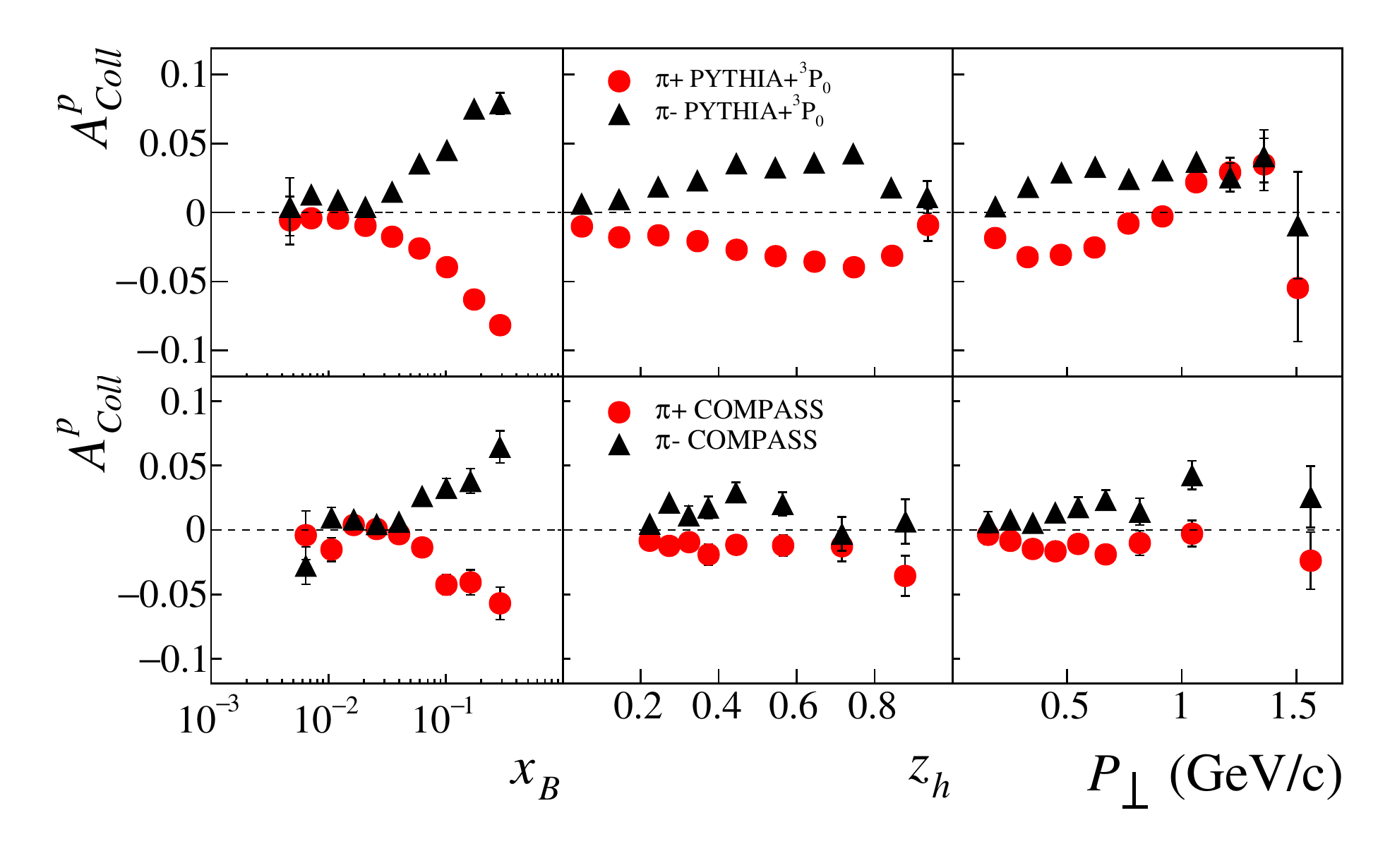}
  \cprotect\caption{Collins proton asymmetry as function of $x_B$, $z_h$ and $P_{\perp}$ for positive pions (circles) and negative pions (triangles) as obtained from \verb|PYTHIA|+${}^3P_0$ (upper row) compared to COMPASS data \cite{COMPASS-collins-sivers} (lower row).}\label{fig:PY+3P0 collins asymm}
\end{figure}

The proton Collins asymmetry as obtained from \verb|PYTHIA|+${}^3P_0$ is shown in the upper row of Fig. \ref{fig:PY+3P0 collins asymm} for positive pions (circles) and negative pions (triangles) as function of $x_B$, $z_h$ and $P_{\perp}$. The lower row shows the corresponding asymmetry measured by the COMPASS Collaboration \cite{COMPASS-collins-sivers}.

The dependence on $x_B$ and $P_{\perp}$ of the Collins asymmetry is very similar in the two cases. In particular the mirror symmetry of the $\pi^+$ and $\pi^-$ asymmetries as function of $x_B$ is very well reproduced by the MC. The $x_B$-dependence reflects that of the ratio $h_1/f_1$.

As function of $z_h$, the characteristic linear behavior of the simulated Collins analysing power seen in $u$-jets (shown for instance in Fig. \ref{fig:mc asymm}) is modified by the realistic mixture of quarks generated by \verb|PYTHIA|. The trend is still different with respect to data for $0.4<z_h<0.7$. In this region, the experimental asymmetry decreases, at variance with the simulation results. This difference could be due to the contribution of pions produced in the decay of vector mesons.

Also, the values of the asymmetry obtained from PYTHIA+${}^3P_0$ are larger than the measured ones. For these simulations the same value of $\IM(\mu)$ as in M19 has been used. This parameter was tuned by comparing the stand alone Collins analysing power as obtained from simulations with that obtained from BELLE data. The comparison with the measured Collins asymmetries in Chapter 3 was done rescaling the simulation results since the transversity PDF was not used. The results in Fig. \ref{fig:PY+3P0 collins asymm} indicate that a retuning of $\IM(\mu)$ is needed. It has not been done here in order to keep the same parameter values throughout all this work and in particular for the comparison with the new developments presented in Chapter 5.

\begin{figure}[tb]
  \centering
    \includegraphics[width=0.8\textwidth]{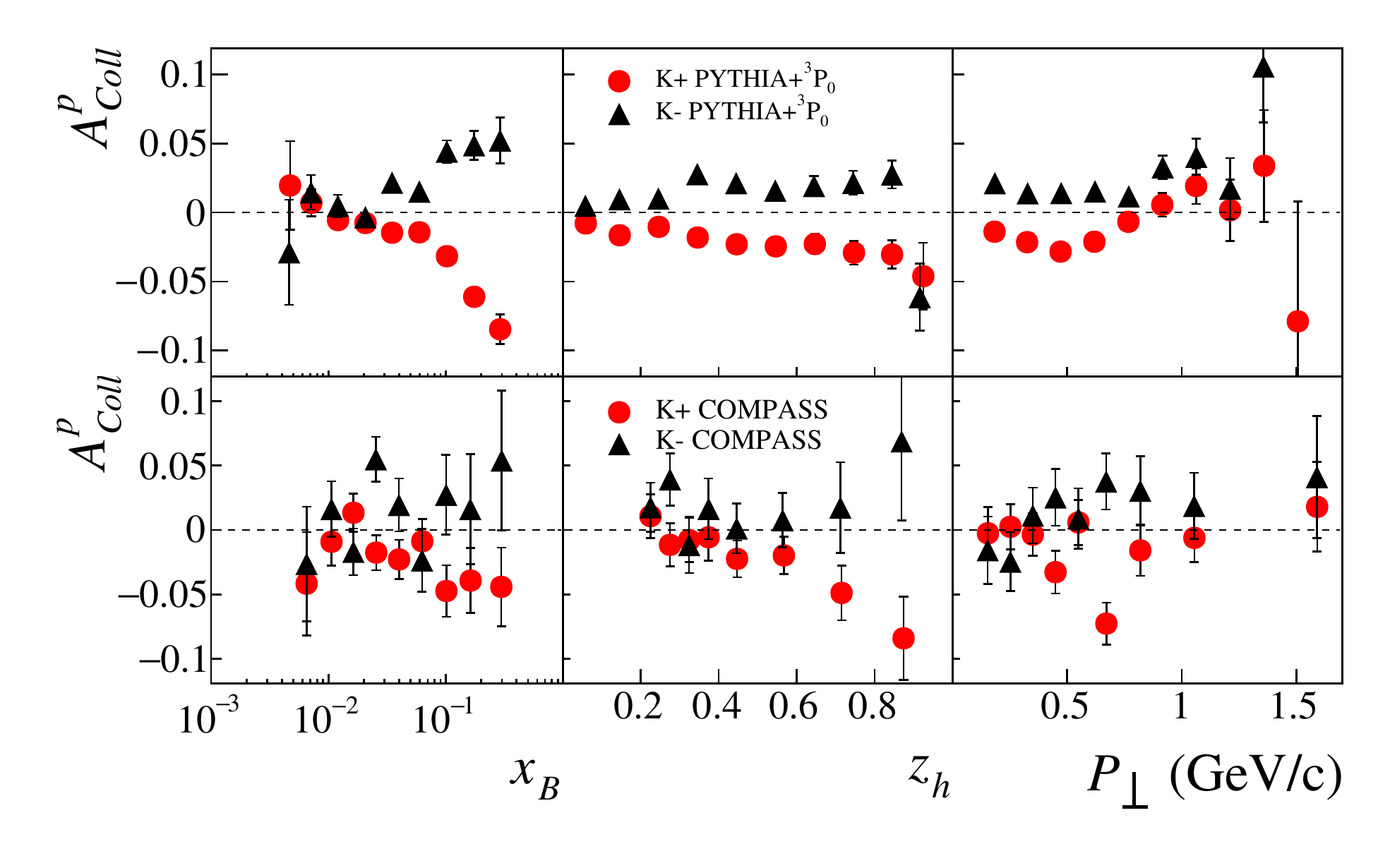}
  \cprotect\caption{Collins proton asymmetry as function of $x_B$, $z_h$ and $P_{\perp}$ for positive kaons (circles) and negative kaons (triangles) as obtained from \verb|PYTHIA|+${}^3P_0$ (upper row) compared to COMPASS data \cite{COMPASS-collins-sivers} (lower row).}\label{fig:PY+3P0 collins asymm K}
\end{figure}

Figure \ref{fig:PY+3P0 collins asymm K} shows the comparison between the Collins asymmetry for $K^+$ (circles) and $K^-$ (triangles) as obtained from simulations (upper row) with the COMPASS data \cite{COMPASS-collins-sivers} (lower row). The similarity is striking even if more precise data are needed for a quantitative comparison.

\begin{figure}[tb]
  \centering
    \includegraphics[width=0.8\textwidth]{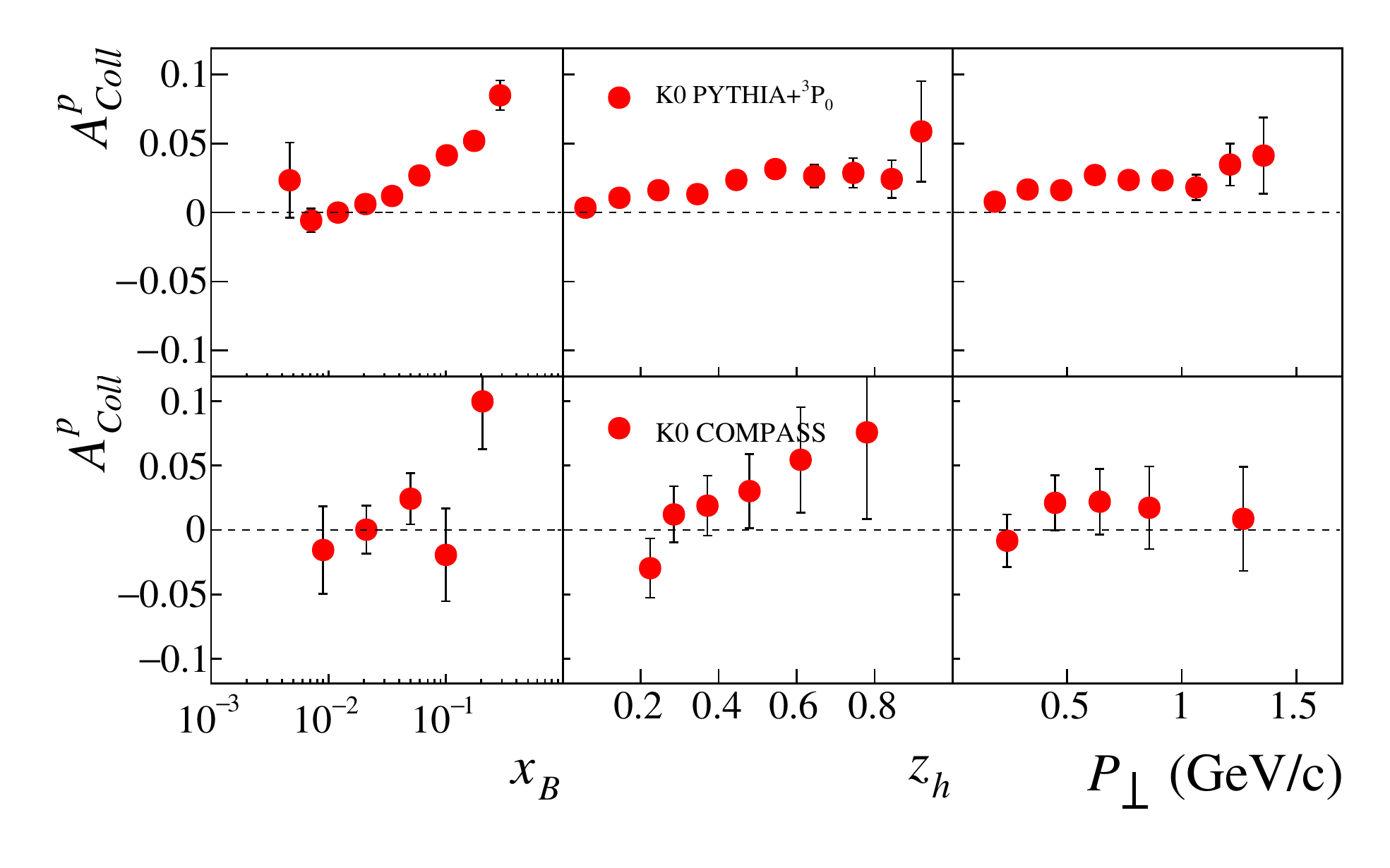}
  \cprotect\caption{Collins proton asymmetry as function of $x_B$, $z_h$ and $P_{\perp}$ for $K^0$ as obtained from \verb|PYTHIA|+${}^3P_0$ (upper row) compared to COMPASS data \cite{COMPASS-collins-sivers} (lower row).}\label{fig:PY+3P0 collins asymm K0}
\end{figure}

In Fig. \ref{fig:PY+3P0 collins asymm K0} is shown the Collins asymmetry for $K^0$ mesons as obtained from \verb|PYTHIA|+${}^3P_0$ (upper row) compared to the corresponding asymmetry measured by the COMPASS experiment \cite{COMPASS-collins-sivers} (lower row). Also in this case the comparison is satisfactory and again more precise data are needed.

The same sample of simulated data has been used to calculate also the dihadron asymmetry for $h^+h^-$ pairs in the same jet, which in GNS are distributed according to
\begin{equation}
    \frac{d^2N_{h^+h^-}}{d\phi_{RS}\,dX} = N_{h^+h^-}^0(X)\left[1+D_{\rm{NN}}\,\sin\theta\,A^{h^+h^-}(X)\,|\textbf{S}^{gns}_{\perp}|\,\sin\phi_{RS}\right]
\end{equation}
where $\phi_{RS}=\phi_R+\phi_{S_{\perp}}-\pi$ and $\theta$ is the polar angle of one of the hadrons (e.g. the positive one) in the dihadron rest frame with respect to the dihadron boost axis. The dihadron asymmetry $\langle \sin\theta A_p^{h^+h^-}\rangle$ is given in Eq. (\ref{eq: dihadron asymmetry}) and is extracted from the simulated data as $2\langle \sin\phi_{RS}^{gns}\rangle / (D_{\rm{NN}}\,|\textbf{S}_{\perp}^{gns}|)$ as function of the kinematic variable $X$ which can be either $z=z_1+z_2$ or the invariant mass $M_{h^+h^-}$.

\begin{figure}[tb]
  \centering
    \includegraphics[width=0.8\textwidth]{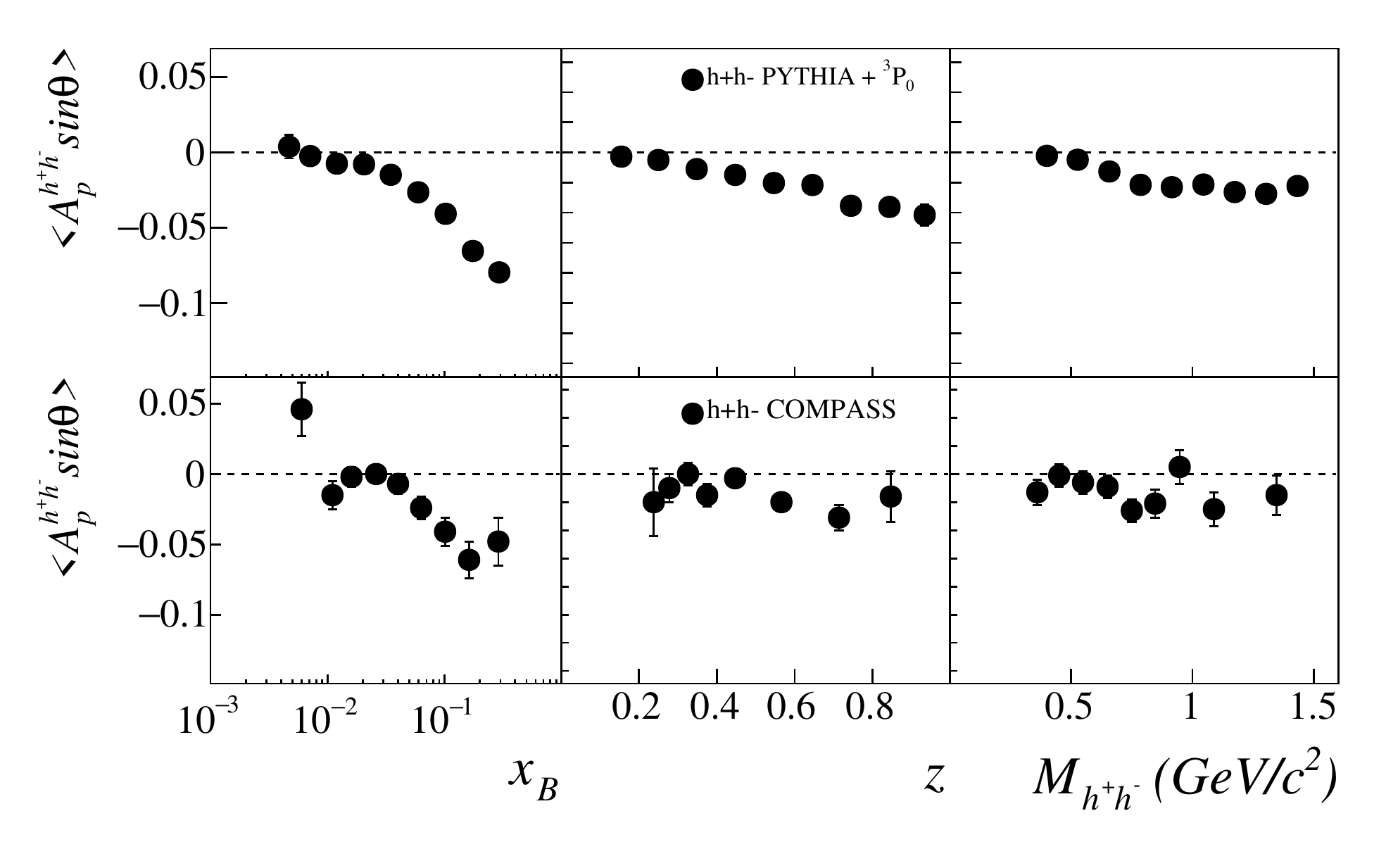}
  \cprotect\caption{Dihadron proton asymmetry as function of $x_B$, $z$ and invariant mass for $h^+h^-$ pairs as obtained from \verb|PYTHIA|+${}^3P_0$ (upper row) compared to COMPASS data \cite{compass-dihadron} (lower row).}\label{fig:PY+3P0 dihadron asymm}
\end{figure}

The results are shown in the first row of Fig. \ref{fig:PY+3P0 dihadron asymm} as function of $z=z_1+z_2$ and of the invariant mass. In the lower row the corresponding COMPASS asymmetries \cite{compass-dihadron} are shown. The same cuts have been applied in both analyses. The simulation results have trends similar to those of the measured asymmetries and again they have larger values. 

\subsection{Deuteron target}
The SIDIS process for a deuteron target is simulated at first order by merging two simulation samples, one obtained for a proton target and the other for a neutron target.
For the case of a neutron target, the transversity distributions for $u$ and $d$ quarks are exchanged as required by the isospin symmetry. Namely it is
\begin{eqnarray}
    h_1^{u/p}(x_B) = h_1^{d/n}(x_B), & h_1^{d/p}(x_B) = h_1^{u/n}(x_B).
\end{eqnarray}

\begin{figure}[h]
    \centering
    \includegraphics[width=0.8\textwidth]{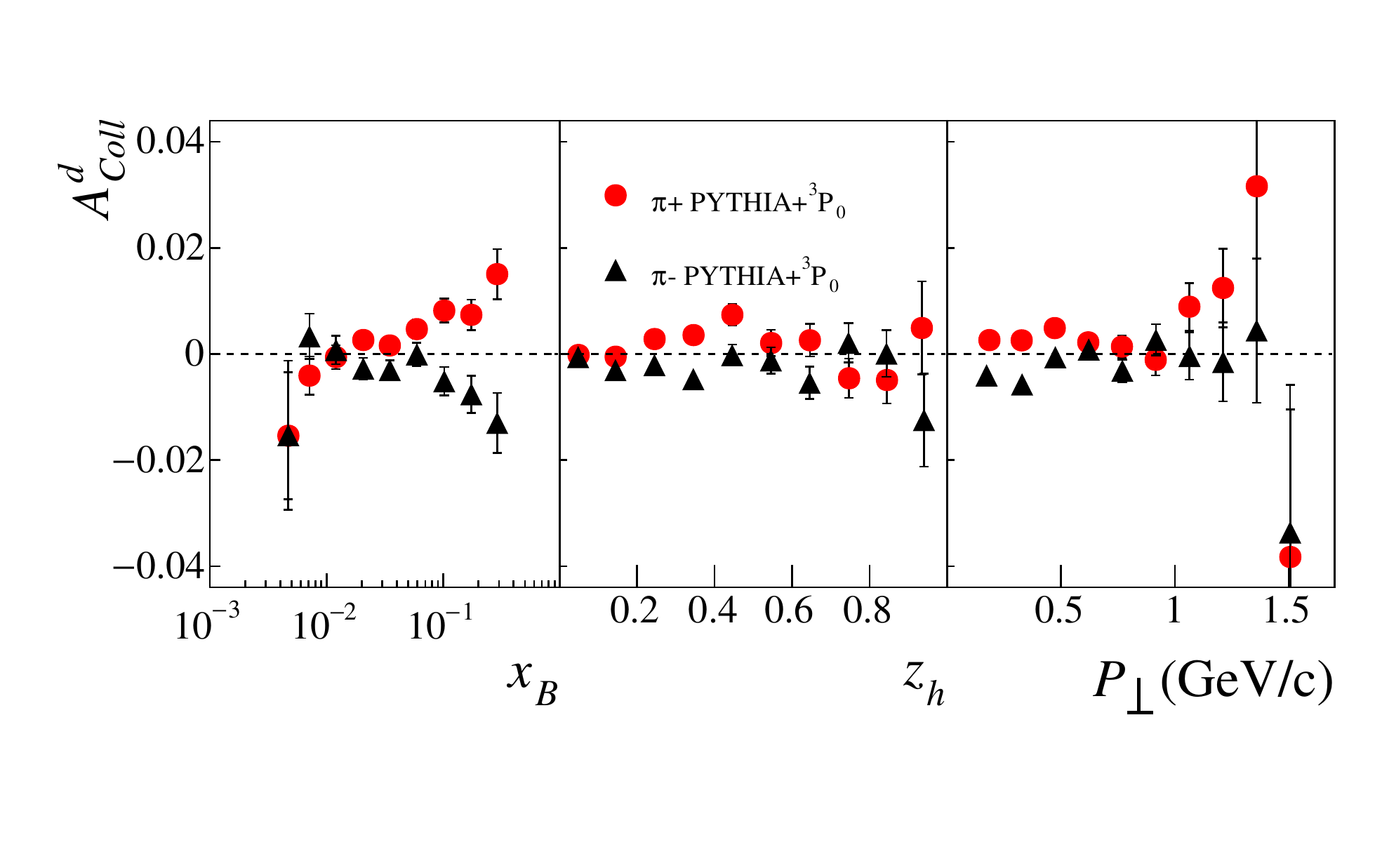}
    \cprotect \caption{Collins asymmetry as function of $x_B$, $z_h$ and $P_{\perp}$ for positive pions (circles) and negative pions (triangles) as obtained from \verb|PYTHIA|+${}^3P_0$ for a deuteron target.}\label{fig:PY+3P0 collins asymm deuteron}
\end{figure}

The Collins asymmetry obtained from simulations of the transversely polarized SIDIS process in the same kinematic conditions as above is shown in Fig. \ref{fig:PY+3P0 collins asymm deuteron} as function of $x_B$, $z_h$ and $P_{\perp}$. The asymmetry is very small, below $2\%$, as in the experimental data \cite{COMPASS-2006}. This happens due to cancellations between $h_1^u$ and $h_1^d$ which in the Collins asymmetry for a deuteron target enter with the same weight, namely it is $A_{Coll}^{d}\simeq \left[(h_1^u+h_1^d)/(f_1^u+f_1^d)\right] \left[(4H_{1u}^{\perp h}+H_{1d}^{\perp h})/(4D_{1u}^h+D_{1d}^h)\right]$. As function of $x_B$, for $x_B \gtrsim 0.1$ the asymmetry is slightly positive (negative) for $\pi^+$ ($\pi^-$). This is due to the fact that for the parameterization used in this work, in the region $x_B \gtrsim 0.1$, it is $h_1^u+h_1^d<0$, as can be seen in Fig. \ref{fig:h1 parameterization}.

\begin{figure}[h]
\centering
\includegraphics[width=0.8\textwidth]{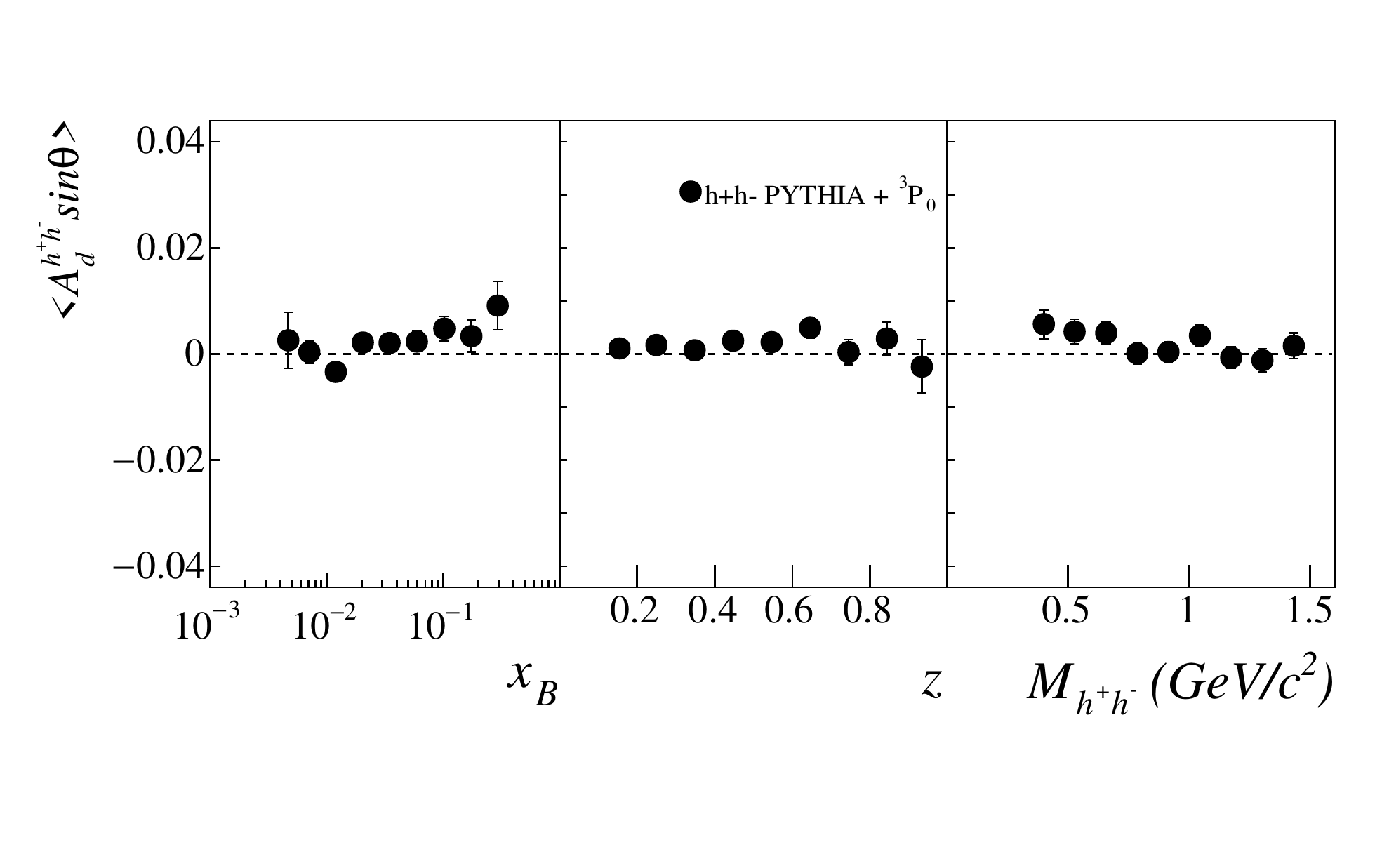}
\cprotect\caption{Dihadron asymmetry for $h^+h^-$ pairs as function of $x_B$, $z$, and invariant mass as obtained from \verb|PYTHIA|+${}^3P_0$ for a deuteron target.}\label{fig:PY+3P0 2h asymm deuteron}
\end{figure}

The dihadron asymmetry, obtained from the same simulated sample, is given in Fig. \ref{fig:PY+3P0 2h asymm deuteron}. Also this asymmetry is below the $2\%$, as in the experimental data \cite{compass-dihadron}. At large $x_B$, it is slightly positive, reflecting the trend of the Collins asymmetry for $\pi^+$ shown in Fig. \ref{fig:PY+3P0 collins asymm deuteron}.


\section{Adding other TMD PDFs}\label{sec:adding other TMDs}
The recipe for the implementation of the quark transversity distribution presented in the previous sections can be generalized to include other TMD PDFs which can be used to calculate the polarization vector of the initial quark and the orientation of its intrinsic transverse momentum.

\begin{figure}[tb]
  \centering
    \includegraphics[width=0.5\textwidth]{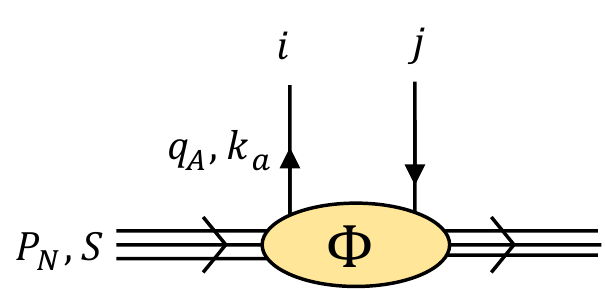}
  \caption{Diagramatic representation of the quark-quark correlator.}\label{fig:quark-quark correlatior}
\end{figure}

The starting point is the calculation of the general quark distribution in a polarized nucleon which can be derived from the quark-quark correlator \cite{Boer-Mulders_corelator}
\begin{eqnarray}
\Phi^{(q_A)}_{ij}(x,\kperp;\textbf{S})=\int \frac{d\xi^-d^2\boldsymbol{\xi}_{\perp}}{(2\pi)^3} e^{ik\cdot\xi} \langle P_N,S|\bar{\psi}_j(0)\psi_i(\xi)|P_N,S\rangle|_{\xi^+=0}.
\end{eqnarray}
The genuine gauge invariant definition of the quark-quark correlator includes a Wilson line connecting the quark fields. Alternatively, it can be defined by making a fixed gauge choice. The expression given here holds in the light-cone gauge $A^+=0$, $A$ being the gauge field vector. It is a $4\times 4$ hermitian matrix defined in the quark Dirac spin space and it describes the possible correlations, as allowed by parity conservation, between the quark momentum and the nucleons polarization. It is defined in terms of the expectation value of the quark field $\psi$ on the nucleons state $|P_N,S\rangle$ and it is represented by the diagram of Fig. \ref{fig:quark-quark correlatior}.
The index $i=1,..,4$ refers to the $i$-th component of the quark field. For an extensive description of the quark-quark correlator definition and use see for instance Ref. \cite{Collins-libro}.

The parameterization of the quark-quark correlator is usually given in the so-called \textit{infinite momentum frame}, namely in the reference frame where the nucleon travels with infinite momentum along some direction which is taken as the longitudinal direction, for instance the direction of the virtual photon momentum in the GNS.
In the infinite momentum frame the nucleon momentum can be parameterized as $P_N^{\mu}=P_N^+ n^{\mu}_+/2+M_N^2 n^{\mu}_-/(2P_N^+)$, where $n_{\pm}^{\mu}=(1,0,0,\pm1)$ are light-like vectors and $P_N^+$ is the component of the nucleon momentum along $n_+^{\mu}$. The quark momentum in the same frame can be parametererized as $k_a^{\mu}=xP_N^+ n_+^{\mu}/2+\kperpkperp\,n_-^{\mu}/(2xP_N^+)+k_{\perp}^{\mu}$ where $x=k_a^+/P_N^+$ and $k_{\perp}^{\mu}=(0,\kperp,0)$.

At leading order in the nucleons forward momentum $P_N^+$, or equivalently at leading order in the hard scale $Q$, the parameterization of the quark-quark correlator is \cite{Boer-Mulders_corelator}
\begin{eqnarray}\label{eq:correlator parameterisation}
\nonumber \Phi^{(q_A)}(x,\kperp;\textbf{S}) &=& \frac{1}{2}\bigg\lbrace f_1^{q_A}\, \slashed n_+ - f_{1\rm{T}}^{q_A\perp}\,\frac{\textbf{S}_{\perp}\cdot(\hat{\textbf{P}}_N\times \kperp)}{M_N}\slashed n_+ +\left(S_{\parallel}g_{1\rm{L}}^{q_A}+\,g_{1\rm{T}}^{q_A\perp}\,\frac{\kperp\cdot \textbf{S}_{\perp}}{M_N}\right)\\
\nonumber &\times& \gamma_5\slashed n_+ + h_{1\rm{T}}^{q_A}\frac{\left[\slashed S_{\perp},\slashed n_+\right]}{2}\gamma_5 + \left(S_{\parallel} h_{1\rm{L}}^{q_A\perp}+\frac{\kperp\cdot\textbf{S}_{\perp}}{M_N}h_{1\rm{T}}^{\perp}\right)\,\frac{\left[\slashed k_{\perp},\slashed n_+\right]}{2M_N}\gamma_5 \\
&+& i\,h_1^{q_A\perp}\frac{\left[\slashed k_{\perp},\slashed n_+\right]}{2M_N} \bigg\rbrace,
\end{eqnarray}
where the Feynman slash notation, defined as $\slashed k = \gamma^{\mu}k_{\mu }$ for a generic four-vector $k$, has been used. The nucleons transverse polarization four-vector $S_{\perp}^{\mu}$ is defined as $S_{\perp}^{\mu}=(0,\textbf{S}_{\perp},0)$. The functions $f_{1T}^{\perp}$, $g_{1T}$, $h_{1L}^{\perp}$, $h_1^{\perp}$ and $h_{1T}^{q_A\perp}$ are the other TMD PDFs already introduced in Chapter 1 and reported in Tab. \ref{tab:tmd}. 

The quantity $\Phi^{(q_A)}\gamma^0$ can be interpreted as the spin density matrix of $q_A$ inside the nucleon. Hence, supposing to observe $q_A$ by a gedanken experiment, the probability of observing $q_A$ with forward momentum fraction\footnote{The variables $x$ and $x_B$ are related by $x=\frac{x_B}{2}\left(1+\sqrt{1+4\kperpkperp/Q^2}\right)$ when the nucleon mass is neglected. Hence at leading order it is $x\simeq x_B$.} $x=k_a^+/P_N^+\simeq x_B$, with intrinsic transverse momentum $\kperp$ and with polarization vector $\check{\textbf{S}}_a$, is given by
\begin{equation}\label{eq:Pq def}
    \mathcal{P}_{q_A}(x,\kperp;\check{\textbf{S}}_{a}|\textbf{S})= \frac{1}{2xP_N^+}\trace{\left(\Phi^{(q_A)}\check{\rho}(q_A)\right)}.
\end{equation}
The $4\times 4$ spin matrix
\begin{eqnarray}
\check{\rho}(q_A)=\frac{xP_N^+}{2}\,\gamma^+\left(1+\check{S}_{a\parallel}\gamma_5-\gamma_5\boldsymbol{\gamma}_{\perp}\cdot\check{\textbf{S}}_{a\perp}\right)
\end{eqnarray}
is the expression, at leading order in $P_N^+$, of the matrix $\gamma^0\rho^{rel.}(q_A)\gamma^0$, where $\rho^{rel.}(q_A)=\slashed k (1-\check{S}_{a\parallel}\gamma_5-\gamma_5\boldsymbol{\gamma}_{\perp}\cdot\textbf{S}_{a\perp})/2$ is the relativistic spin density matrix for a massless on-shell spin $1/2$ fermion \cite{landau-book} with imposed polarization vector $\check{\textbf{S}}_a$. This vector does not depend on $x$, $\kperp$ or on the nucleon polarization. The matrix $\gamma^+=\gamma^0(1+\alpha_z)$ is a projector onto the subspace of Dirac spinors describing a mass-shell fast quark moving along the $\zu$ axis, for a mass-shell particle being $v^z=\langle\alpha_z\rangle$.

The general distribution of quark $q_A$ in the nucleon can now be obtained explicitly by using the parametrization of the quark-quark correlator in Eq. (\ref{eq:correlator parameterisation}) and Eq. (\ref{eq:Pq def}). It gives
\begin{eqnarray}\label{eq:q(x,kT,S)}
\nonumber \mathcal{P}_{q_A}(x,\kperp;\check{\textbf{S}}_{q_a}|\textbf{S})&=& f_{1}^{q_A}(x,\kperpkperp) - \,\frac{\textbf{S}_{\perp}\cdot(\hat{\textbf{P}}_N\times \kperp)}{M_N}\,f_{1T}^{q_A \perp}(x,\kperpkperp)\\
\nonumber &+& g_{1\rm{L}}^{q_A}(x,\kperpkperp) \check{S}_{q\parallel} S_{\parallel}-\frac{\kperp\cdot \textbf{S}_{\perp}}{M_N}\,\check{S}_{q\parallel}\,g_{1\rm{T}}^{q_A\perp}(x,\kperpkperp)\\
 \nonumber &+& h_1^{q_A}(x,\kperpkperp)\,\textbf{S}_{\perp}\cdot\check{\textbf{S}}_{q\perp}+\frac{\kperp\cdot \check{\textbf{S}}_{q\perp}}{M_N}\,S_{\parallel}\,h_{1\rm{L}}^{q_A \perp}(x,\kperpkperp)\\
&-&\frac{\check{\textbf{S}}_{q\perp}\cdot(\hat{\textbf{P}}_N\times\kperp)}{M_N}\,h_{1}^{q_A \perp}(x,\kperpkperp)\\
\nonumber &-&\left[\frac{(\kperp\cdot\check{\textbf{S}}_{q\perp})(\kperp\cdot \textbf{S}_{\perp})-\kperpkperp(\textbf{S}_{\perp}\cdot\check{\textbf{S}}_{q\perp})/2}{M_N^2}\right]\,h_{1T}^{q_A \perp}(x,\kperpkperp).
\end{eqnarray}

This function allows to calculate the polarization vector of the initial quark using the relation
\begin{equation}
    \textbf{S}_a = \frac{\nabla_{\check{\textbf{S}}_a}\mathcal{P}_{q_A}(x,\kperp;\check{\textbf{S}}_{a}|\textbf{S})}{\mathcal{P}_{q_A}(x,\kperp;\check{\textbf{S}}_{q_a}=0|\textbf{S})},
\end{equation}
which is analogus to Eq. (\ref{eq: Sq' = grad C}) and the symbol $\nabla_{\check{\textbf{S}}_a}$ indicates the gradient vector with respect to the components of $\check{\textbf{S}}_a$.
It gives
\begin{eqnarray}
\label{eq:Sq parallel} S_{a\parallel} &=& \frac{g_{1\rm{L}}^{q_A}(x,\kperpkperp) S_{\parallel}-\kperp\cdot \textbf{S}_{\perp}\,g_{1\rm{T}}^{q_A\perp}(x,\kperpkperp)/M_N}{\mathcal{P}_{q_a}(x,\kperp;\check{\textbf{S}}_{a}=0|\textbf{S})} \\
\label{eq:Sq perp} \textbf{S}_{a\perp} &=& \frac{1}{\mathcal{P}_{q_a}(x,\kperp;\check{\textbf{S}}_{a}=0|\textbf{S})}\Big\lbrace h_1^{q_A}(x,\kperpkperp)\,\textbf{S}_{\perp} + \kperp\,S_{\parallel}\,h_{1\rm{L}}^{q_A \perp}(x,\kperpkperp)/M_N\\
\nonumber &-&(\hat{\textbf{P}}_N\times\kperp)\,h_{1}^{q_A \perp}(x,\kperpkperp)/M_N\\
\nonumber &-&\left[(\kperp\cdot \textbf{S}_{\perp})\kperp-\kperpkperp\textbf{S}_{\perp}/2\right]\,h_{1T}^{q_A \perp}(x,\kperpkperp)/M_N^2 \Big\rbrace,
\end{eqnarray}
where the unpolarized quark distribution is
\begin{equation}\label{eq:P_q at Sq=0}
    \mathcal{P}_{q_A}(x,\kperp;\check{\textbf{S}}_{a}=0|\textbf{S}) = f_{1}^{q_A}(x,\kperpkperp) - \,\textbf{S}_{\perp}\cdot(\hat{\textbf{P}}_N\times \kperp)\,f_{1T}^{q_A \perp}(x,\kperpkperp)/M_N.
\end{equation}

If parameterizations for all TMD PDFs were available, one could in principle calculate the polarization vector of the initial quark using Eqs.(\ref{eq:Sq parallel}-\ref{eq:Sq perp}).
Then the simulations can be performed by following the same procedure as done in the case of the transversity PDF in the previous section. Namely, the polarization vector $\textbf{S}_A$ of the final quark is obtained by reducing $\textbf{S}_{a\perp}$ by $D_{\rm{NN}}$ and reflecting it about the normal to the lepton scattering plane. The longitudinal component $S_{A\parallel}$ is the opposite of $S_{a\parallel}$.

In this more general case, however, one further step at the level of lepton quark hard scattering has to be performed. That is, the intrinsic quark transverse momentum should be generated according to Eq. (\ref{eq:P_q at Sq=0}). This means that the distribution of the modulus squared $\kperpkperp$ is given by
\begin{equation}
    d\kperpkperp\,\int \frac{d\phi_{\perp}}{2\pi}\mathcal{P}_{q_a}(x,\kperp;\check{\textbf{S}}_{a}=0|\textbf{S}) = f_1^{q_A}(x,\kperpkperp) d\kperpkperp
\end{equation}
evaluated at the generated values of $x_B$ and $Q^2$.
The distribution of the azimuthal angle of $\textbf{k}_{\perp}$, instead, is given by
\begin{equation}\label{eq:phi_perp Sivers}
    d\phi_{\perp}\,\frac{1}{2\pi}\left[1 - \frac{k_{\perp}}{M_N}\frac{f_{1T}^{q_A \perp}(x,\kperpkperp)}{f_1^{q_A}(x,\kperpkperp)}\,S_{\perp}\,\sin\left(\phi_{\perp}-\phi_S\right) \right].
\end{equation}
This recipe could be implemented in \verb|PYTHIA| by extending the interface presented above to enable user intervention also at the \verb|PartonLevel|. This would allow to bias the azimuthal angle of the intrinsic quark transverse momentum according to Eq. (\ref{eq:phi_perp Sivers}) and to simulate the Sivers asymmetry.
\chapter{Vector meson production in polarized quark fragmentation}\label{chapter5}
In the previous chapters the treatment of the fragmentation process of polarized quarks was focused on the production of pseudoscalar mesons. The comparison of the results of the MC simulations with the experimental data was already satisfactory.
However, the production of other particles has to be taken into account in order to build a more complete model of the polarized quark fragmentation. In this chapter the introduction of the vector meson octet in the string+${}^3P_0$ model is described. Experimental data show for instance that $\rho$ and $K^*$ mesons are produced in the fragmentation process roughly as many as $\pi$ and $K$ mesons \cite{Pei:1996kq}, hence vector meson production is not negligible if compared to pseudoscalar meson production.

The introduction of vector mesons makes a distinction between two classes of observable hadrons: the \textit{primary} and the \textit{secondary} hadrons. To the first class belong the directly produced pseudoscalar mesons and the directly produced vector mesons. To the second one belong the hadrons produced in the decay of the primary vector mesons. The notion of rank in this case applies only to the primary hadrons.
The decays of vector mesons change the kinematical distributions and the spin asymmetries of the observable hadrons, which now are a mixture of primary and secondary hadrons.
Therefore, a complete
simulation includes genererating the momenta of the decay products. 

The main difficulties are due to the vector meson polarization, which has to be taken into account in their production and decay, as well as in the spin propagation along the fragmentation chain. In particular, this requires the calculation of the spin density matrix of the emitted vector meson which governs the angular distribution of the decay products. Then, to continue the recursive process, the spin density matrix of the leftover quark is calculated introducing a \textit{decay matrix}. This matrix carries back to the vector meson birth the information about the quantum mechanical correlation between the polarization of the leftover quark and the momenta of the decay products.

All that has been done introducing two new parameters and the formalism is described in section \ref{sec:new string+3P0 model}. The new model (M20) has been implemented in a stand alone MC program which has the same structure of that used for M19 and is summarized in section \ref{sec:M20 MC implementation}. The results of the simulations have been deeply studied. The kinematic distributions, the analysing powers and the sensitivity to the values of the free parameters are discussed in section \ref{sec:M20 results from simulations}. The last section of this chapter contains the comparison with the experimental data.

Other effects like the interference between the amplitude for direct production of pseudoscalar mesons and the amplitude for pseudoscalar meson production via a vector meson decay are presently not treated in the string$+{}^{3}P_0$ model.

 \section{The new string+${}^3P_0$ model}\label{sec:new string+3P0 model}
 
 As for the pseudoscalar mesons, the core of the theoretical description of vector meson production in polarized quark jets is the polarized splitting function which characterises the elementary splitting $q\rightarrow h+q'$. It is obtained from the corresponding splitting matrix which can be written starting from that for the pseudoscalar meson case and introducing a new coupling matrix which describes the interaction of quarks with a spin 1 vector meson. The splitting matrix allows also to study the spin correlations between the vector meson and the leftover quark and to calculate the spin density matrix of the vector meson, as explained in Sec. \ref{sec:h-q' spin correlations}. Section \ref{sec:decay of a vector meson} is instead dedicated to the decay process of a vector meson, which requires as input the vector meson spin density matrix and gives as output the decay matrix. This matrix is then used for the calculation of the spin density matrix of $q'$ in order to take into account the spin correlations between $h$ and $q'$. The knowledge of the spin density matrix of $q'$ allows then to continue recursively the fragmentation chain.

 \subsection{Elementary splitting for vector meson production}\label{sec:elementary splitting for vm}
 As for pseudoscalar mesons, the first step for the study of vector meson production in polarized quark jets is the calculation of the splitting function which, as already seen, describes the energy sharing between the emitted vector meson $h$ and the leftover quark $q'$ in the elementary splitting $q\rightarrow h+q'$. It is used in simulations for the generation of the four-momentum of $h$. The splitting function is derived from the splitting matrix for vector meson production which can be written as
\begin{eqnarray}\label{eq: T final vm}
\nonumber    T_{q',h,q}&=& \sum_{\alpha}T_{q',h,q}^{(\alpha)}V_{\alpha}^*\\
\nonumber &=& C_{q',h,q}\,D_h(M^2)\,\check{g}(\varepsilon_h^2)\left[(1-Z)/\varepsilon_h^2\right]^{a/2}\,\exp{\left[-\bl\varepsilon_h^2/(2Z)\right]}\\
    &\times& \Delta_{q'}(\kpt)\Gamma_{h,V}\hat{u}_q^{-1/2}(\kt),
\end{eqnarray}
and is the analogue of the splitting matrix for pseudoscalar meson emission given in Eq. (\ref{eq: T final}).
The complex spectral function $D_h(M^2)$ implements the finite width of the resonance and here it is taken to be
\begin{equation}\label{eq:BW distribution}
    D_h(M^2)=\frac{1}{M^2-m_h^2+i\,m_h\gamma_{h}}.
\end{equation}
The modulus square gives the distribution of the resonance mass $M$.
The factor $\gamma_{h}=\sum_i \gamma_h^{(i)}$ is the total width of the resonance and $\gamma_{h}^{(i)}$ is the partial width corresponding to the $i$-th decay channel. The ratio $\gamma_h^{(i)}/\gamma_h$ is then the branching ratio for the $i$-th channel. 
The introduction of the spectral function requires also to modify the phase space factor which, in the case of a resonance, is\footnote{It can be obtained from the phase space factor of a stable particle as
\begin{eqnarray}\label{eq:phase space stable hadron}
\nonumber       \frac{dZ}{Z}d^2\pt = \frac{dp^+}{p^+}d^2\pt= dp^+dp^- \delta(p^2-m_h^2)d^2\pt,
\end{eqnarray}
where $\delta(p^2-m_h^2)$ is replaced with $|D_h(M^2)|^2$ and $p^-$ is traded for $M^2$ using the relation $dp^+dp^-=dM^2dp^+/p^+$.}
\begin{eqnarray}\label{eq:phase space vm}
dM^2\frac{dZ}{Z}d^2\pt.
\end{eqnarray}

The matrix $\Gamma_{h,V}$ describes the coupling of a quark with a vector meson. It can be written as \cite{DS09}
 \begin{eqnarray}\label{eq: Gamma vm}
 \Gamma_{h,V} &=& G_T\,\boldsymbol{\sigma}\T\sigma_z\cdot \textbf{V}\T^* + G_L\, 1_{2\times 2}\,V_z^* = \Gamma_{h,\alpha}V_\alpha^*
 \end{eqnarray}
and replaces the matrix $\Gamma_h=\sigma_z$ which appears in the splitting matrix of the pseudoscalar meson emission. $\textbf{V}=(\textbf{V}_{\rm{T}},V_z)$ is the polarization vector of the vector meson in the vector meson rest frame normalized such that $\textbf{V}^*\cdot\textbf{V}=1$ (the "$*$" here indicates the complex conjugate). It is obtained boosting the corresponding polarization four-vector first by a boost along $\zu$ which brings the meson at $p_z=0$ then by a boost along $\pt$ which brings the meson at rest. This composition boosts preserves the LR symmetry. $G_L$ and $G_T$ are two complex constants for the coupling to a vector meson with linear polarization longitudinal and transverse with respect to the string axis. The label $\alpha=\rm{m},\rm{n},\rm{l}$ refers to the polarization state of the vector meson in the $(\Mx,\Nx,\Lx)$ basis. The term proportional to $G_T$ comes from the reduction to Pauli spinors of the $4\times 4$ matrix $\gamma^{\mu}V_{\mu}^*$ whereas the term proportional to $G_L$ comes from the reduction of the chiral-odd matrix $\sigma^{\mu\nu}V^*_{\mu}p_{v}$. The coupling to longitudinally polarized vector mesons should in principle be included to reproduce the non vanishing alignment of vector mesons seen in experimental data \cite{OPAL-alignment-rho,OPAL-alignment-K,DELPHI-alignment}. The expression in Eq. (\ref{eq: Gamma vm}) satisfies the symmetries $S1$-$S3$ introduced in Sec. \ref{sec: string decay in mult form} and it is a minimal coupling: other terms involving the transverse momenta $\kt$ and $\kpt$ in combination with Pauli matrices and $\textbf{V}$ may appear without spoiling symmetries $S1$-$S3$ and the LR symmetry.

The reduced single quark density matrix $\hat{u}_q$, introduced in Eq. (\ref{eq: hat uq}) for the pseudoscalar mesons, in the case of vector meson emission becomes
\begin{eqnarray}\label{eq: hat uq vm}
 \nonumber   \hat{u}_q(\kt) &=&\sum_{h}\,|C_{q',h,q}|^2\int dM^2\,|D_h(M^2)|^2\, \int dZZ^{-1} d^2\kpt\,\left(\frac{1-Z}{\varepsilon_h^2}\right)^a\,e^{-\bl\varepsilon_h^2/Z}\,\check{g}^2(\varepsilon_h^2)\\
    &\times& \sum_{\alpha}\Gamma^{\dagger}_{h,\alpha}\Delta^{\dagger}_{q'}(\kpt)\Delta_{q'}(\kpt)\Gamma_{h,\alpha}.
    \end{eqnarray}
With the choice $C_1$ for the function $\check{g}$ it is
\begin{equation}\label{eq: uq M20}
    \hat{u}_q= \textbf{1}\,(2|G_T|^2+|G_L|^2)\,\sum_h |C_{q',h,q}|^2 \left(|\mu|^2+\langle \ktkt\rangle_{f\T}\right),
\end{equation}
to be compared with Eq. (\ref{eq:hat uq ps explicit}).
Also in this case $\hat{u}_q$ is proportional to the unit matrix and the model has the same simplifying features of M19: the intermediate spin density matrix $\hat{\rho}_{int}(q)$ coincides with the true spin density matrix $\rho(q)$ of $q$, and the elementary splitting $q\rightarrow h + q'$ in flavor space is described only by the coefficient $C_{q',h,q}$ (see Eq. (\ref{eq:p_q->h simple 3P0})). As for the pseudoscalar meson emission, $C_{q',h,q}$ takes into account the suppression of $s$ quarks with respect to $u$ and $d$ quarks and the weights coming from the isospin wave function, which now are those of the spin-1 mesons. In this case there is no suppression among the flavor neutral states, for instance the $\omega$ meson is not suppressed with respect to the $\rho^0$ meson.

The splitting function for non analysed vector meson production with the choice $C_1$ is obtained using Eq. (\ref{eq:F=TrhoT pol}) and the splitting matrix in Eq. (\ref{eq: T final vm}) and Eq. (\ref{eq: uq M20}). It is given by
\begin{eqnarray}\label{eq:F_explicit vm}
\nonumber F_{q',h,q}(M^2,Z,\pt;\kt,\textbf{S}_q)&=& \frac{|C_{q',h,q}|^2}{\sum_H |C_{q',H,q}|^2}\,|D_h(M^2)|^2\, \left(\frac{1-Z}{\varepsilon_h^2}\right)^a \frac{\exp{(-\bl \varepsilon_h^2/Z)}}{N_a(\varepsilon_h^2)} \\
\nonumber &\times& \frac{|\mu|^2+\kptkpt}{|\mu|^2+\langle \ktkt\rangle_{f\T}} f_{\T}^2(\kptkpt) \\
&\times& \left[ 1+\frac{|G_L|^2}{2|G_T|^2+|G_L|^2}\frac{2\IM(\mu)\,\rm{k'}_{\rm{T}}}{|\mu|^2+\kptkpt}\textbf{S}_{q}\cdot\tilde{\textbf{n}}(\kpt) \right],
\end{eqnarray}
where the tilde symbol, already introduced in Eq. (\ref{eq:U(q)=U0+U1 + tilde definition}), indicates the cross product $\ntil=\zu\times\textbf{n}$ and the unit vector $\textbf{n}$ is defined as $\textbf{n}(\kt)=\kt/|\kt|$. This function is the analogue of Eq. (\ref{eq:F_explicit simple 3P0}) for pseudoscalar mesons and will be used in simulations to generate $\kpt$ and $Z$.
The last line of the splitting function gives the Collins effect for the vector meson. By comparing Eq. (\ref{eq:F_explicit vm}) and Eq. (\ref{eq:F_explicit simple 3P0}), one can see that, for rank $r=1$, the analysing power ratio is
\begin{equation}\label{eq: ap vm / ap ps}
    \frac{a^{q_A\uparrow\rightarrow (VM)_r +X}|_{r=1}}{a^{q_A\uparrow\rightarrow (PS)_r+X}|_{r=1}} = - \frac{|G_L|^2}{2|G_T|^2+|G_L|^2},
\end{equation}
namely the Collins effect for a leading vector meson is opposite to the Collins effect for a leading pseudoscalar meson. It is also scaled by the factor $1/(1+2|G_T|/|G_L|)$ which for $|G_L|/|G_T|=1$ is $1/3$, in agreement with the Czyzewski prediction \cite{Czyzewski-vm} obtained in the non relativistic quark model. However, in general the factor $|G_L|/|G_T|$ can be different from unity.

For a deeper investigation of the relation in Eq. (\ref{eq: ap vm / ap ps}), it is useful to decompose the splitting function for vector meson production along the different linear polarization states, namely
\begin{eqnarray}
F_{q',h,q}(M^2,Z,\pt;\kt,\textbf{S}_q)=\sum_{\alpha} F_{q',h,q}^{(\alpha)}(M^2,Z,\pt;\kt,\textbf{S}_q).
\end{eqnarray}
where $F_{q',h,q}^{\alpha}$ indicates the splitting function for the emission of a vector meson with linear polarization along the axis $\alpha=\rm{m},\rm{n},\rm{l}$.
Carrying out explicitly the calculations, it is
\begin{eqnarray}
\label{eq: F vm M} F_{q',h,q}^{(\rm{m})}&=& \dots \,\frac{|G_T|^2}{2|G_T|^2+|G_L|^2}\,\left[ 1+\frac{2\IM(\mu)\,\rm{k'}_{\rm{T}}}{|\mu|^2+\kptkpt}\textbf{S}_{q}\cdot\tilde{\textbf{n}}(\kpt) \right] \\
\label{eq: F vm N} F_{q',h,q}^{(\rm{n})}&=& \dots \,\frac{|G_T|^2}{2|G_T|^2+|G_L|^2}\,\left[ 1-\frac{2\IM(\mu)\,\rm{k'}_{\rm{T}}}{|\mu|^2+\kptkpt}\textbf{S}_{q}\cdot\tilde{\textbf{n}}(\kpt) \right] \\
\label{eq: F vm L} F_{q',h,q}^{(\rm{l})}&=& \dots \,\frac{|G_L|^2}{2|G_T|^2+|G_L|^2}\,\left[ 1+\frac{2\IM(\mu)\,\rm{k'}_{\rm{T}}}{|\mu|^2+\kptkpt}\textbf{S}_{q}\cdot\tilde{\textbf{n}}(\kpt) \right]
\end{eqnarray}
where the dots represent the first and second lines of Eq. (\ref{eq:F_explicit vm}) and are the same in all three cases.  From these equations one can see that the Collins effect for vector mesons with longitudinal linear polarization along $\zu$ or transverse linear polarization along $\hat{\textbf{k}}'\T$ is opposite to that of pseudoscalar mesons. Whereas, vector mesons with transverse linear polarization along $\zu\times\hat{\textbf{k}}'\T$ have the same Collins effect as pseudoscalar mesons. The ratio of Eq. (\ref{eq: ap vm / ap ps}) is obtained after summing over the vector meson polarization. In the summation the contributions coming from transversely polarized vector mesons cancel out and remains only the contribution of the coupling with longitudinally polarized vector mesons.

Finally, looking at Eq. (\ref{eq:F_explicit vm}), it turns out that also for the generation of a vector meson in simulations, it is easier to generate first the splitting in flavor space, then the mass of the vector meson according to the spectral function $|D_h(M^2)|^2$, afterwards $\kpt$ and eventually $Z$.

\subsection{The spin correlations between $h$ and $q'$}\label{sec:h-q' spin correlations}
In the elementary splitting $q\rightarrow h+q'$ with emission of a pseudoscalar meson $h$, the spin information of $q$ is transferred directly to $q'$. In fact, $h$ does not carry spin information, at least in the form of a spin density matrix.

The production of a vector meson complicates the description of the spin flow in the splitting process since, being a spin 1 object, the vector meson carries spin information. Its different polarization states can be encoded in a hermitian $3\times 3$ spin density matrix $\hat{\rho}(h)$ which, for a pure spin state, is defined as $\hat{\rho}_{\alpha\alpha'}(h)=V_{\alpha}V_{\alpha'}^*$, $\textbf{V}$ being the polarization vector in the rest frame of the vector meson.
The condition of hermiticity, expressed as $\hat{\rho}_{\alpha\alpha'}=\hat{\rho}^*_{\alpha'\alpha}$, implies that the real part $\RE(\hat{\rho})$ is a symmetric matrix whereas the pure imaginary part $\IM(\hat{\rho})$ is an antisymmetric matrix.
Hence, $\hat{\rho}(h)$ can be decomposed as \cite{X.A_et_al_spin_observables}
\begin{equation}
    \hat{\rho}_{\alpha \alpha'}(h) = \frac{1}{3} \delta_{\alpha\alpha'} + T_{\alpha\alpha'}-\frac{i}{2}\varepsilon_{\alpha\alpha'\beta}\,P_{\beta}.
\end{equation}
The first term describes an unpolarized vector meson, or equivalently an equal mixture of states with linear polarization along $\Mx$, $\Nx$ and $\Lx$, and it has unit trace. The matrix $\textbf{T}$ is a real and symmetric matrix with vanishing trace. It encodes the \textit{tensor polarization} of the vector meson. The term involving the fully anti-symmetric tensor $\varepsilon_{\alpha\beta\gamma}$ encodes the \textit{axial polarization} $\textbf{P}=(P_{\rm{m}},P_{\rm{n}},P_{\rm{l}})$ of the vector meson. This term does not play a role in the decay processes considered here and it is neglected in the following. The spin density matrix of the vector meson is identified thus with its real part. 

Since the spin information coming from $q$ is shared among the left-over quark $q'$ and the vector meson $h$ at each elementary splitting, the polarization of the vector meson is in general correlated to that of $q'$ and this fact deserves a closer investigation.

\subsection*{Correlation matrix of $h$ and $q'$}
\begin{figure}[tb]
\centering
  \includegraphics[width=0.30\linewidth]{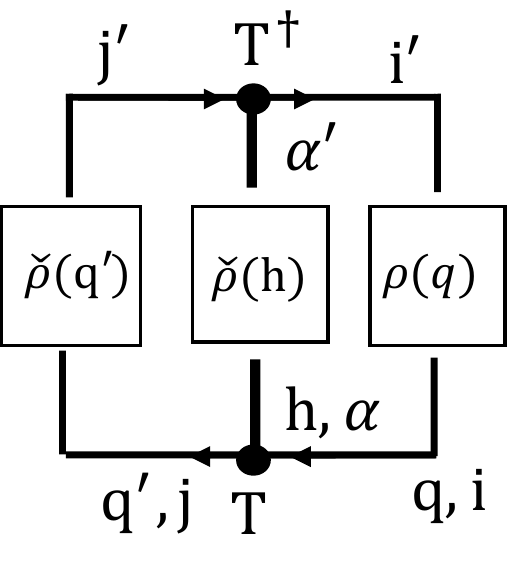}
  \caption{Unitary diagram corresponding to the splitting function for vector meson production when the polarization states of $h$ and $q'$ are not summed over. See second line of Eq. (\ref{eq:all pol F vm}).}\label{fig:unitary joint h q'}
\end{figure}

All possible spin correlations between the vector meson and the leftover quark are encoded in the \textit{correlation matrix}\footnote{An introduction to this formalism can be found for instance in Ref. \cite{X.A_et_al_spin_observables}.} of $h$ and $q'$. It can be obtained by generalizing the splitting function for vector meson emission in Eq. (\ref{eq:F_explicit vm}), which is summed over the polarization states of $h$ and $q'$, to the case where the polarizations of both $h$ and $q'$ are analysed. This generalization of the procedure followed in the study of the positivity conditions for M19 (see sec. \ref{sec:simple splitt function}) requires the introduction of the acceptance spin density matrix $\check{\rho}(h)$ of $h$ and of the acceptance spin density matrix $\check{\rho}(q')$ matrix of $q'$ which do not dependent neither on the kinematical variables $M^2$, $Z$, $\pt$ nor on the polarization vector of $q$. The splitting function is schematically represented in the unitary diagram of Fig. \ref{fig:unitary joint h q'} to which corresponds the formal expression
\begin{eqnarray}\label{eq:all pol F vm}
\nonumber F_{q',h,q}(M^2,Z,\pt,\check{\rho}(q'),\check{\rho}(h);\textbf{S}_{q}) &=&\langle q', h|T_{q',h,q}|q\rangle \langle q|T^{\dagger}_{q',h,q}|q',h\rangle \\
\nonumber &=& \sum_{ii',jj',\alpha\alpha'}\nonumber \langle j\alpha |T_{q',h,q}|i\rangle \rho_{ii'}(q)\langle i'|T^{\dagger}_{q',h,q}|j'\alpha'\rangle \\
\nonumber &\times& \check{\rho}_{j'j}(q')\check{\rho}_{\alpha'\alpha}(h) \\
&=& \sum_{ii',jj',\alpha\alpha'}R_{j\alpha;j'\alpha'}\, \check{\rho}_{j'j}(q')\check{\rho}_{\alpha'\alpha}(h),
\end{eqnarray}
where
\begin{eqnarray}
\nonumber \langle j\alpha|T_{q',h,q}|i\rangle &=& C_{q',h,q}\,D_h(M^2)\,\check{g}(\varepsilon_h^2)\left[(1-Z/\varepsilon_h^2\right]^{a/2}\,\exp{\left[-\bl\varepsilon_h^2/(2Z)\right]} \\
&\times& \left(\Deltaqp\,\Gamma_{h,\alpha}\hat{u}_q^{-1/2}\right)_{ij}
\end{eqnarray}
The indices $i$,$j$ and $\alpha$ refer to the spin states of $q$, $q'$ and $h$ respectively. The primed indices $i'$, $j'$ and $\alpha'$ label the spin states of $q$, $q'$ and $h$ in the complex conjugated amplitude, i.e. the upper part of the unitary diagram in Fig. \ref{fig:unitary joint h q'}. The elements of the correlation matrix $R$ are
\begin{eqnarray}\label{eq:R definition}
R_{j\alpha;j'\alpha'} &=& \langle j\alpha |T_{q',h,q}|i\rangle \rho_{ii'}(q)\langle i'|T^{\dagger}_{q',h,q}|j'\alpha'\rangle
\end{eqnarray}
and they describe all possible spin correlations between $q'$ and $h$ allowed by parity invariance. The complete expression of $R$ is given in Appendix \ref{appendix:R}. The full trace of $R$ over the spin indices of $q'$ and $h$ gives the splitting function for vector meson emission introduced in Eq. (\ref{eq:F_explicit vm}). In particular it is
\begin{eqnarray}
\nonumber \textrm{Tr}_{q',h}R(q',h) &\equiv& \sum_{j,\alpha}R_{j\alpha;j\alpha}(q',h) \\
&=& F_{q',h,q}(M^2,Z,\pt;\kt,\textbf{S}_q),
\end{eqnarray}
This is also equivalent to Eq. (\ref{eq:all pol F vm}) when the polarizations of $h$ and $q'$ are not analysed, namely their acceptance spin density matrices are replaced by the respective unit matrices. The ratio of the correlation matrix and its trace gives then the joint spin density matrix of $h$ and $q'$, which takes into account the fact that they are produced in a correlated spin state. Such correlation is taken into account in the stand alone MC implementation of M20 as explained in sec. \ref{sec:M20 MC implementation}.

\subsection{Spin density matrix of the vector meson}
The spin density matrix of the vector meson is obtained from the correlation matrix in Eq. (\ref{eq:R definition}) summing over the polarization states of $q'$. It gives
\begin{eqnarray}\label{eq:rho vm matrix elements}
\hat{\rho}_{\alpha\alpha'}(h) = \frac{\sum_{j}R_{j\alpha;j\alpha'}}{\sum_{i,\beta}R_{i\beta;i\beta}} = \frac{\textrm{tr}\left(\Deltaqp\,\Gamma_{h,\alpha}\hat{u}_q^{-1/2}\,\rho(q)\,\hat{u}_q^{-1/2}\,\Gamma_{h,\alpha'}^{\dagger}\,\Deltaqpdag\right)}{\sum_{\beta}\textrm{tr}\left(\Deltaqp\,\Gamma_{h,\beta}\hat{u}_q^{-1/2}\,\rho(q)\,\hat{u}_q^{-1/2}\,\Gamma_{h,\beta}^{\dagger}\,\Deltaqpdag\right)}.
\end{eqnarray}
The summation over the spin indices of $q'$ means that the polarization of $q'$ is not analysed. The division by the full trace of $R(q',h)$ guarantees that the spin density matrix of the vector meson has unit trace.

The expressions of the vector meson matrix elements in the $(\Mx,\Nx,\Lx)$ basis are
\begin{eqnarray}\label{eq:rho vm MNL decomposition}
\nonumber \hat{\rho}_{\rm{mm}}(h) &=& \frac{|G\T|^2}{N_h}\left(|\mu|^2+\kptkpt+2\IM(\mu)k'\T\,S_{q\rm{n}}\right) \\
\nonumber \hat{\rho}_{\rm{nn}}(h) &=& \frac{|G\T|^2}{N_h}\left(|\mu|^2+\kptkpt-2\IM(\mu)k'\T\,S_{q\rm{n}}\right) \\
\nonumber \hat{\rho}_{\rm{ll}}(h) &=& \frac{|G_{\rm{L}}|^2}{N_h}\left(|\mu|^2+\kptkpt+2\IM(\mu)k'\T\,S_{q\rm{n}}\right) \\
\nonumber \RE{\hat{\rho}}_{\rm{ml}}(h) &=& -\frac{|G\T||G_{\rm{L}}|\sin\theta_{\rm{LT}}}{N_h}\big[\left(|\mu|^2+\kptkpt\right)\,S_{q\rm{n}}+2\IM(\mu)k'\T\big] = \RE{\hat{\rho}}_{\rm{lm}}(h) \\
\nonumber \RE{\hat{\rho}}_{\rm{mn}}(h) &=& -\frac{|G\T|^2}{N_h}\,2\IM(\mu)k'\T\,S_{q\rm{m}} = \RE{\hat{\rho}}_{\rm{nm}}(h) \\
\nonumber \RE{\hat{\rho}}_{\rm{nl}}(h) &=& \frac{|G\T||G_{\rm{L}}|}{N_h}\big[\left(|\mu|^2+\kptkpt\right)\,\sin\theta_{\rm{LT}}\,S_{q\rm{m}}+2\IM(\mu)\,\cos\theta_{\rm{LT}}\,k'\T\,S_{q\rm{l}}\big] = \RE{\hat{\rho}}_{\rm{ln}}(h)\\
\,
\end{eqnarray}
where $\theta_{LT}=\arg(G_L/G_T)$ and
\begin{equation}
    N_h = \left(|\mu|^2+\kptkpt\right)\,(2|G\T|^2+|G_{\rm{L}}|^2)+2\IM(\mu)k'\T|G_L|^2\,S_{q\rm{n}}.
\end{equation}
is the normalization factor.
These matrix elements are the cartesian components of the vector meson tensor polarization. The non diagonal elements $\RE{\hat{\rho}}_{\rm{mn}}(h)$, $\RE{\hat{\rho}}_{\rm{ml}}(h)$ and $\RE{\hat{\rho}}_{\rm{nl}}(h)$ give the \textit{oblique} polarization in the $(\rm{\textbf{m}},\rm{\textbf{n}})$, $(\rm{\textbf{m}},\rm{\textbf{l}})$ and $(\rm{\textbf{n}},\rm{\textbf{l}})$ planes respectively. Concerning the diagonal elements, $\hat{\rho}_{\rm{mm}}(h)-\hat{\rho}_{\rm{nn}}(h)$ is the \textit{straight linear polarization} and $\hat{\rho}_{\rm{ll}}(h)-(\hat{\rho}_{\rm{mm}}(h)+\hat{\rho}_{\rm{nn}}(h))/2$ is the \textit{alignment} along the $\zu$ axis \cite{X.A_et_al_spin_observables}.
\begin{figure}[tb]
\centering
\begin{minipage}[t]{.7\textwidth}
  \includegraphics[width=1.0\linewidth]{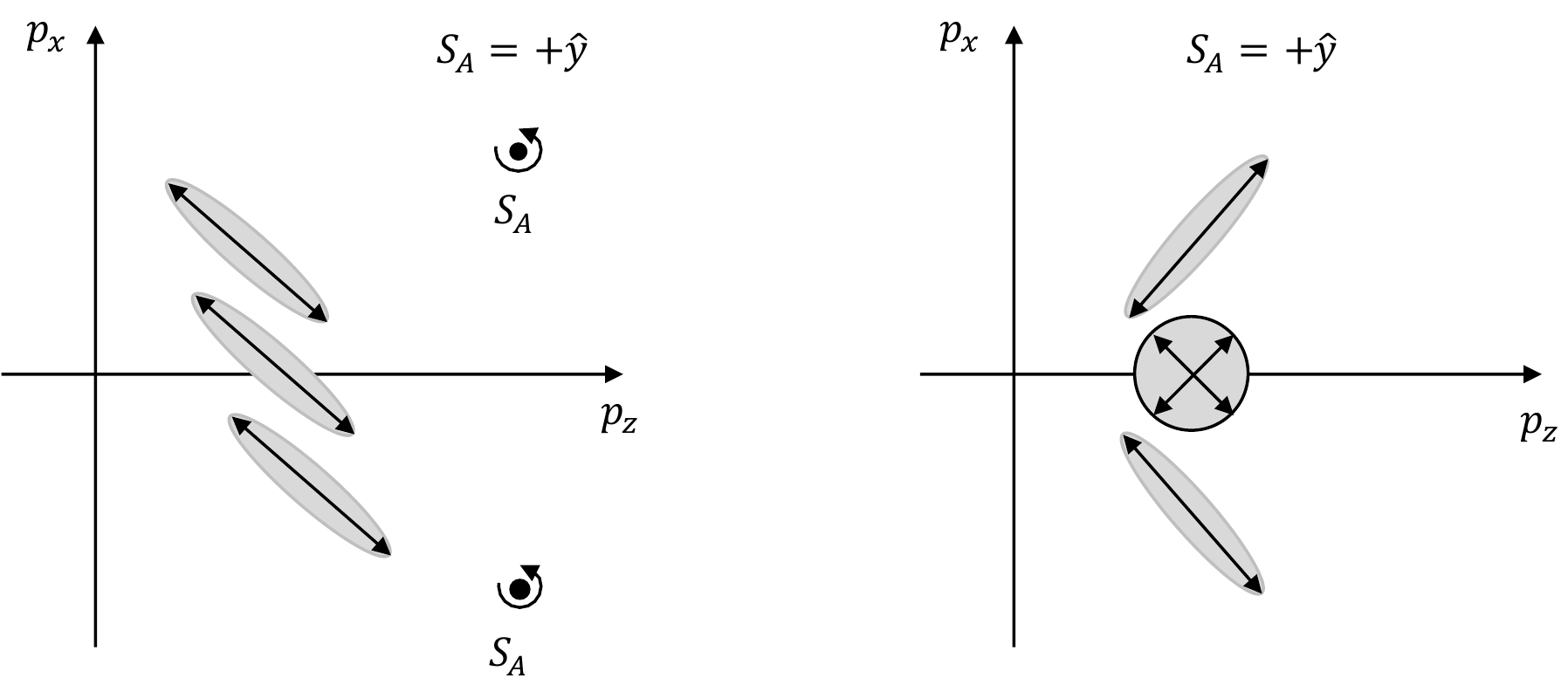}
  \end{minipage}
  \caption{Vector meson oblique polarization in the $(\Mx,\Lx)$ plane with only the first term of $\RE\hat{\rho}_{\rm{ml}}$ in Eq. (\ref{eq:rho vm MNL decomposition}) (left) and with only the second term of $\RE\hat{\rho}_{\rm{ml}}$ in Eq. (\ref{eq:rho vm MNL decomposition}) (right).}\label{fig:oblique polarization}
\end{figure}

Particularly important is the oblique polarization in the $(\rm{\textbf{m}},\rm{\textbf{l}})$ plane, namely in the plane perpendicular to $\zu\times\kpt$, described by $\RE{\hat{\rho}}_{ml}(h)$. The effect of this matrix element can be understood considering the two-body decay of a rank 1 vector meson, e.g. $\rho\rightarrow \pi\pi$, produced in the fragmentation of a quark polarized along $\yu$ axis, i.e. $\textbf{S}_A=+\yu$. To simplify, the decay is assumed to occur in the $(\xu,\zu)$ plane, $\zu$ being the string axis. Also, the parameters $\theta_{LT}$ and $\IM(\mu)$ are assumed to be both positive. Then, the first term of $\RE\hat{\rho}_{\rm{ml}}$ produces the oblique polarization (represented by a double arrow inside an ellipse) in the left picture of Fig. \ref{fig:oblique polarization} whereas the right picture in the same figure shows the oblique polarization produced by the second term of $\RE\hat{\rho}_{\rm{ml}}$. In both pictures, the oblique polarization is shown as function of $p_x$ in the plane where the horizontal axis is $p_z$ and the vertical axis is $p_x$. For a rank 1 vector meson, the lower ellipses in both pictures are favored.
Then, the first part of $\rho_{ml}$ (left picture) favors, for the fastest (with largest $z_h$) pion, a Collins effect opposite to the one of a direct rank 1 pion. Taking only one of the two decays pions and averaging over $z_h$ washes out the contribution of the oblique polarization. The second part of $\rho_{ml}$ attenuates $\langle \ptpt\rangle$ for the fastest pion. These effects are non-zero if $\theta_{LT}\neq 0,\pi$.

Among the other matrix elements, $\RE{\hat{\rho}}_{nl}$ gives rise to a jet handedness effect. In particular, it can be seen that the term proportional to $S_{q\rm{l}}$ produces a correlation of the type $S_{q\rm{l}}\left[\pt^{fast}\times \pt^{slow}\right]\cdot\zu$, namely a handedness effect, similar to the one introduced in sec. \ref{sec:handedness}, where the decay pions are distinguished according to their fractional energies and not according to their charges.

For the implementation of vector meson production in MC simulations, it is more convenient to consider the cartesian basis for the polarization states defined by the axes $(\xu,\yu,\zu)$ where $\zu$ is the string axis and $\yu$ is defined by the polarization vector of the initial quark $q_A$. The spin density matrix in this case has components $\hat{\rho}_{ab}$ with $a,b=x,y,z$ and can be calculated using again Eq. (\ref{eq:rho vm matrix elements}). The relation with the matrix elements in Eq. (\ref{eq:rho vm MNL decomposition}) is
\begin{eqnarray}
    \hat{\rho}_{ab}(h) = \sum_{\alpha\alpha'}\left(R_{\zu}(\phi_{\kpt})\right)_{a\alpha}\,\hat{\rho}_{\alpha\alpha'}(h)\left(R^{-1}_{\zu}(\phi_{\kpt})\right)_{\alpha'b}
\end{eqnarray}
where the matrix $R_{\zu}(\phi_{\kpt})$ is the rotation matrix about the $\zu$ axis and $\phi_{\kpt}$ is the azimuthal angle of $\kpt$.

\subsection{The decay of a polarized vector meson}\label{sec:decay of a vector meson}

The hadronic decays of vector mesons can be two-body or three-body. The decay channels and the related branching ratios are listed in Tab. \ref{tab:decay channels}. $\rho$ mesons have only one hadronic 2-body decay channel whereas $K^*$ mesons have two channels which branching ratios are obtained from the Clebsch-Gordan coefficients involved in the corresponding decay amplitudes in isospin space.
The $\omega$ and $\phi$ mesons, instead, are characterized by both two-body and three-body hadronic decay channels.
In this case the branching ratios are calculated adjusting slightly the PDG values \cite{PDG2019} so that for each meson they sum to one. The decays of the $\eta$ and $\eta'$ mesons, which can be both two-body and three-body, have been neglected. Indeed, taking into account the production rates of such mesons in the fragmentation process as obtained from simulations and the branching ratios of the decay channels involving pions, the effect of such decays on kinematic variables and transverse spin asymmetries is expected to be small. This has also been verified using the interface of M19 with \verb|PYTHIA| described in the previous chapter.

\begin{table}[h]
\centering
\begin{tabular}{ |p{2.5cm}|p{2.5cm}|p{3.0cm}| }
 \hline
 meson          &   decay channel                           & branching ratio  \\
 \hline
 \hline
 $\rho^+$        &  $\pi^+  \pi^0$        & 1                 \\
 $\rho^-$        &  $\pi^-  \pi^0$        & 1                 \\
 $\rho^0$        &  $\pi^+  \pi^-$        & 1                 \\
 \hline
 $K^{*+}$        &  $K^+  \pi^0$          & $1/3$               \\
                 &  $K^0  \pi^+$          & $2/3$               \\
\hline
$K^{*-}$        &  $K^-  \pi^0$          & $1/3$               \\
                 &  $\bar{K}^0  \pi^-$          & $2/3$               \\
\hline
$K^{*0}$        &  $K^+  \pi^-$          & $2/3$               \\
                 &  $K^0  \pi^0$          & $1/3$               \\
\hline
$\bar{K}^{*0}$        &  $\bar{K}^0  \pi^0$          & $1/3$               \\
                 &  $K^-  \pi^+$          & $2/3$               \\
\hline
$\omega$        & $\pi^+\pi^-\pi^0$ &       $90.1\,10^{-2}$ \\
                & $\pi^0\gamma$ &   $8.4\,10^{-2}$ \\
                & $\pi^+\pi^-$   &   $1.5\,10^{-2}$ \\
\hline
$\phi$          & $K^+K^-$  &   $49.0\,10^{-2}$ \\
                & $K^0_S K^0_L$ &   $34.2\,10^{-2}$ \\
                & $\pi^+\pi^-\pi^0$ & $15.3\,10^{-2}$ \\
                & $\eta \gamma$ & $1.3\,10^{-2}$ \\
                & $\pi^0\gamma$ & $0.1\,10^{-2}$ \\
                \hline
\end{tabular}
\caption{Decay channels and corresponding branching rations of vector mesons used in the present work.}
\label{tab:decay channels}
\end{table}

The kinematics of the two-body and three-body decays processes are rather different. They are both studied in the rest frame of the vector meson where the momenta are indicated with $*$, i.e. the momentum of the vector meson in this frame is $p^*=(M,\textbf{0})$. The momenta of the decay products are labelled with $p^*_{h1}$ and $p^*_{h2}$ for a two-body decay and with $p^*_{+}$, $p^*_{0}$ and $p^*_{-}$ for the momenta of $\pi^+$, $\pi^0$ and $\pi^-$ respectively in a three-body decay.

The other quantity needed to specify the decay process is the spin density matrix of the vector meson in the meson rest frame along the cartesian basis $(\xu,\yu,\zu)$. 

\subsection*{Two body decay}
Defining the relative momentum between $h_1$ and $h_2$ in the rest frame of the vector meson $\textbf{r}^*=(\textbf{p}^*_{h1}-\textbf{p}^*_{h2})/2$ and using the energy-momentum conservation, the momenta of the mesons are $p_{h1}^*=(E_1^*,\textbf{r}^*)$ and $p_{h2}^*=(E_2^*,-\textbf{r}^*)$
 where the relative momentum is parameterized as
 \begin{eqnarray}\label{eq:r*}
 \nonumber \textbf{r}^*&=&|\textbf{r}^*|\left(\cos\phi^*\sin\theta^*,\sin\phi^*\sin\theta^*,\cos\theta^*\right)\\
 |\textbf{r}^*|&=&\frac{1}{2M}\sqrt{\left[M^2-\left(m_{h1}+m_{h2}\right)^2\right]\left[M^2-\left(m_{h1}-m_{h2}\right)^2\right]},
 \end{eqnarray}
and $\theta^*$ and $\phi^*$ are the polar and the azimuthal angles of $\textbf{r}^*$.

The joint probability distribution of these angles depends on the spin density matrix of the vector meson and on the type of coupling with the decay mesons encoded in the matrix element $\mathcal{M}^{(2)}_h$. It is obtained from the differential decay width for a two-body decay which is
\begin{equation}\label{eq: dGamma 2-body}
    d\Gamma^{(2)}_h=\frac{1}{2M}|\mathcal{M}^{(2)}_h|^2\,d\Phi_h^{(2)}(p^*;p_{h1}^*,p_{h2}^*).
\end{equation}
The two-body phase space can be written as
\begin{eqnarray}\label{eq:dPhi^(2)}
\nonumber    d\Phi^{(2)}(p^*;p_{h1}^*,p_{h2}^*)&=&(2\pi)^4\delta^{(4)}(p^*-p_{h1}^*-p_{h2}^*)
    \frac{d^3\textbf{p}_{h1}^*}{(2\pi)^3 2E_1^*}\frac{d^3\textbf{p}_{h2}^*}{(2\pi)^3 2E_2^*}\\
    &=& \frac{|\textbf{r}^*|}{M}\frac{1}{16\pi^2}d\cos\theta^*\,d\phi^*.
\end{eqnarray}
The matrix element $\mathcal{M}^{(2)}_h$ depends on the two-body decay process and it is different for the decay of a vector meson in two pseudo-scalar mesons $\textrm{vm}\rightarrow \textrm{ps}_1 + \textrm{ps}_2$ and for the decay of a vector meson in a photon and a pseudo-scalar meson $\textrm{vm}\rightarrow \gamma +\textrm{ps}$.

For the case $\textrm{vm}\rightarrow \textrm{ps}_1 + \textrm{ps}_2$ the matrix element has the form
\begin{equation}\label{eq:M^(2)}
    \mathcal{M}^{(2)}_h=g_{hh_1h_2}\,V^{\mu}{r^*_{\mu}}=-g_{hh_1h_2}\textbf{V}\cdot\textbf{r}^*.
\end{equation}
The normalized angular distribution, obtained using Eqs. (\ref{eq:r*})-(\ref{eq:M^(2)}), is
\begin{eqnarray}\label{eq: dN/dphi dcostheta}
  \frac{1}{\Gamma_h^{(2)}}\frac{d\Gamma_h^{(2)}}{d\phi^*d\cos\theta^{*}}&=&\frac{3}{4\pi}\sum_{ab}\hat{\textbf{r}}^*_{a}\,\RE{\hat{\rho}}_{ab}\,\hat{\textbf{r}}^*_{b}\\
    \nonumber &=&\frac{3}{4\pi}\big(\hat{\rho}_{xx}\cos^2\phi^*\sin^2\theta^*+\hat{\rho}_{yy}\sin^2\phi^*\sin^2\theta^*+\hat{\rho}_{zz}\cos^2\theta^*\\
 \nonumber   &+& \RE{\hat{\rho}}_{xz}\cos\phi^*\sin2\theta^*+\RE{\hat{\rho}}_{xy}\sin 2\phi^*\sin^2\theta^*+\RE{\hat{\rho}}_{yz}\sin\phi^*\sin 2\theta^*\big)
\end{eqnarray}
which is the joint probability distribution for $\theta^*$ and $\phi^*$. It depends quadratically on $\textbf{r}^*$, hence it is invariant under the parity transformation $\textbf{r}^*\rightarrow -\textbf{r}^*$. This means that both $h_1$ and $h_2$ have the same angular distributions.

Integrating Eq. (\ref{eq: dN/dphi dcostheta}) over $\phi^*$ gives the distribution of $\cos\theta^*$ which is
\begin{equation}\label{eq: dN dcostheta}
    \frac{1}{\Gamma_h^{(2)}}\frac{d\Gamma_h^{(2)}}{d\cos\theta^*}=\frac{3}{4}\big[1-\hat{\rho}_{zz}-(1-3\hat{\rho}_{zz})\cos^2\theta^*\big].
\end{equation}
It is a parabola concave upward or downward depending whether $\hat{\rho}_{zz}$ is smaller or larger than $1/3$. For $\hat{\rho}_{zz}=1/3$ the $\cos\theta^*$ distribution is instead flat. In MC simulations this distribution is used to generate the angle $\hat{\theta}^*$.

Evaluating Eq. (\ref{eq: dN/dphi dcostheta}) at $\hat{\theta}^*$ gives the distribution of the azimuthal angle $\phi^*$
\begin{eqnarray}\label{eq:dN dphi}
\nonumber \frac{1}{\Gamma_h^{(2)}}\frac{d\Gamma_h^{(2)}}{d\phi^*}&=&\frac{3}{4\pi}\big[ \hat{\rho}_{yy}\sin^2\hat{\theta}^*+\hat{\rho}_{zz}\cos^2\hat{\theta}^*+(\hat{\rho}_{xx}-\hat{\rho}_{yy})\sin^2\hat{\theta}^*\cos^2\phi^*\\
\nonumber  &+&\RE{\hat{\rho}}_{xz}\sin2\hat{\theta}^*\cos\phi^*+\RE{\hat{\rho}}_{yz}\sin 2\hat{\theta}^*\sin\phi^*\\
  &+&\RE{\hat{\rho}}_{xy}\sin^2\hat{\theta}^*\sin 2\phi^*\big].
\end{eqnarray}

Finally combining Eqs. (\ref{eq: dN dcostheta})-(\ref{eq:dN dphi}) and Eq. (\ref{eq:r*}) allows to construct the momenta of the two decay mesons.

The same steps are followed also for the decay $\textrm{vm}\rightarrow \gamma+\textrm{ps}$, with the difference that the matrix element in the vector meson rest frame is $\mathcal{M}^{(2)}_h\propto\textbf{V}\cdot(\textbf{V}_{\gamma}\times \textbf{p}_{h_2})$ which leads to the normalized angular distribution
\begin{eqnarray}\label{eq: dN/dphi dcostheta gamma}
  \frac{1}{\Gamma_h^{(2)}}\frac{d\Gamma_h^{(2)}}{d\phi^*d\cos\theta^{*}}&=&\frac{3}{4\pi}\sum_{ab}\RE{\hat{\rho}}_{ab}\,\left[ \delta_{ab}-(\hat{p}_1^*)_{a}(\hat{p}_1^*)_{b}\right].
\end{eqnarray}
In this case the polarization vector $\textbf{V}_{\gamma}$ of the photon is not analysed, namely it is summed over.

\subsection*{Three body decay}
\begin{figure}[tb]
\centering
\begin{minipage}[t]{.6\textwidth}
  \includegraphics[width=1.0\linewidth]{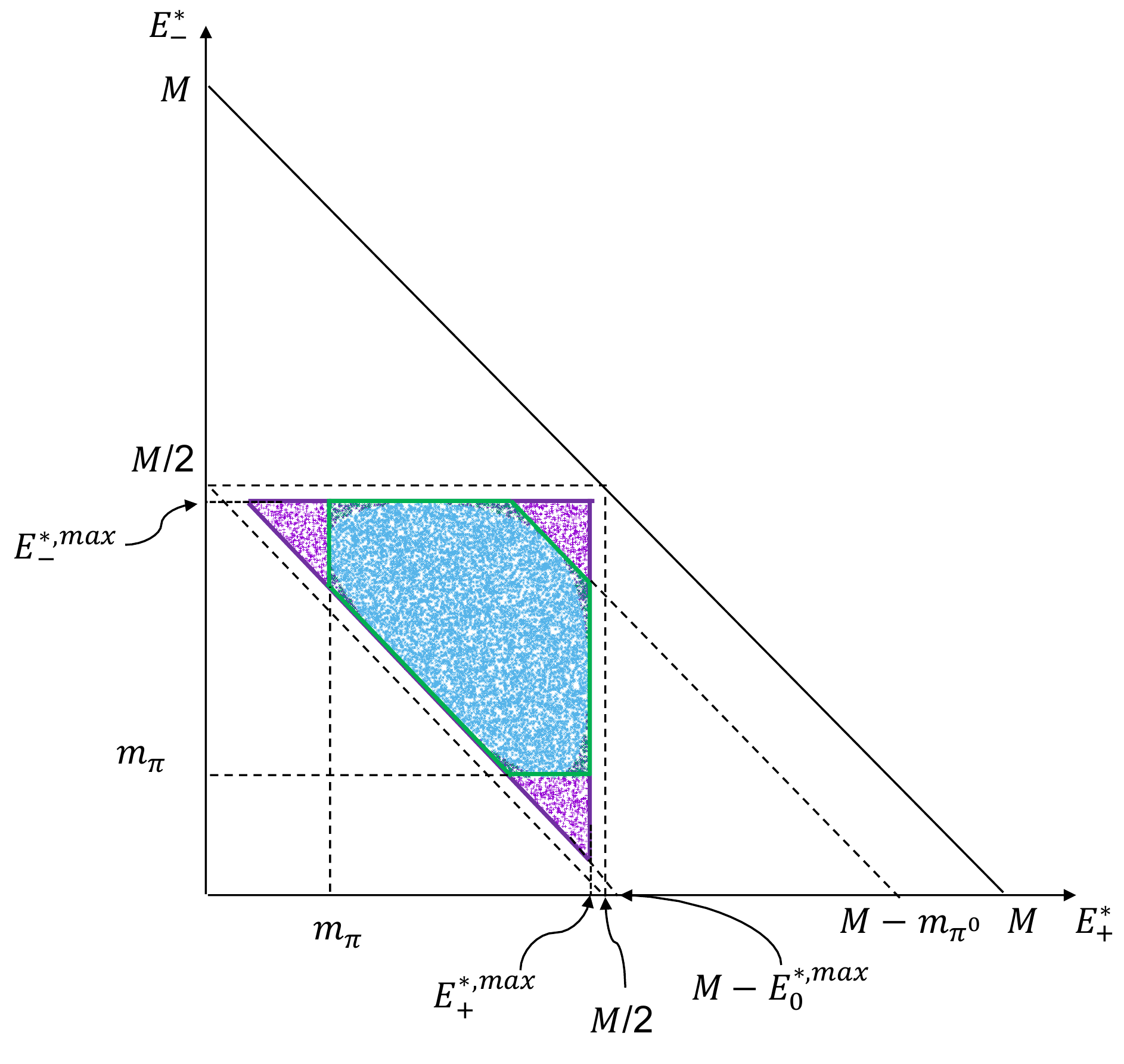}
  \end{minipage}
  \caption{Representation of the allowed phase space in the $\omega/\phi\rightarrow \pi^+\pi^-\pi^0$ decay.}\label{fig:3-body triangle E}
\end{figure}

In the case of three-body decay, the energies of the produced pions do not have fixed values but are rather distributed within the available kinematic phase space. The allowed kinematic region can be studied starting with the conservation of energy
\begin{equation}\label{eq:E+ + E- + E0 = mh}
    E_+^* +E_-^* + E_0^* = M.
\end{equation}
This relation can be represented as a triangle in a plane where the horizontal axis is $E_+^*$ and the vertical axis is $E_-^*$, as shown in Fig. \ref{fig:3-body triangle E}.
The pion energies have to be larger than the pion masses, i.e. $E_i^*>m_{\pi^i}$, and smaller than the maximum energy $E_i^{*,max}$ allowed for each pion, i.e. $E_i^*<E_i^{*,max}$, for $i=+,-,0$. Considering $i$, $j$ and $k$ to be cyclic permutations of $+$, $-$ and $0$, then the maximum energy allowed for $\pi^i$ is $E_i^{*,max}=(M^2+m_{\pi^i}^2-(m_{\pi^j}+m_{\pi^k})^2)/(2M)$, reached when $\pi^j$ and $\pi^k$ are at rest with respect to each other. 
Also, it is $E_i^{*}\leq M/2$, namely the energy of each decay pion can not exceed half of the vector meson mass.
The combination of the constraints $E_i^*>m_{\pi^i}$ and $E_i^*<E_i^{*,max}$ gives the region enclosed by the thick green polygon in Fig. \ref{fig:3-body triangle E}. Each point in such region corresponds to certain values of $E_+^*$, $E_-^*$ and $E_0^*$ which fulfill Eq. (\ref{eq:E+ + E- + E0 = mh}). These values, however, do not necessarily obey momentum conservation, which has then to be imposed as a further constraint.

\begin{figure}[tb]
\centering
\begin{minipage}[t]{.5\textwidth}
  \includegraphics[width=1.0\linewidth]{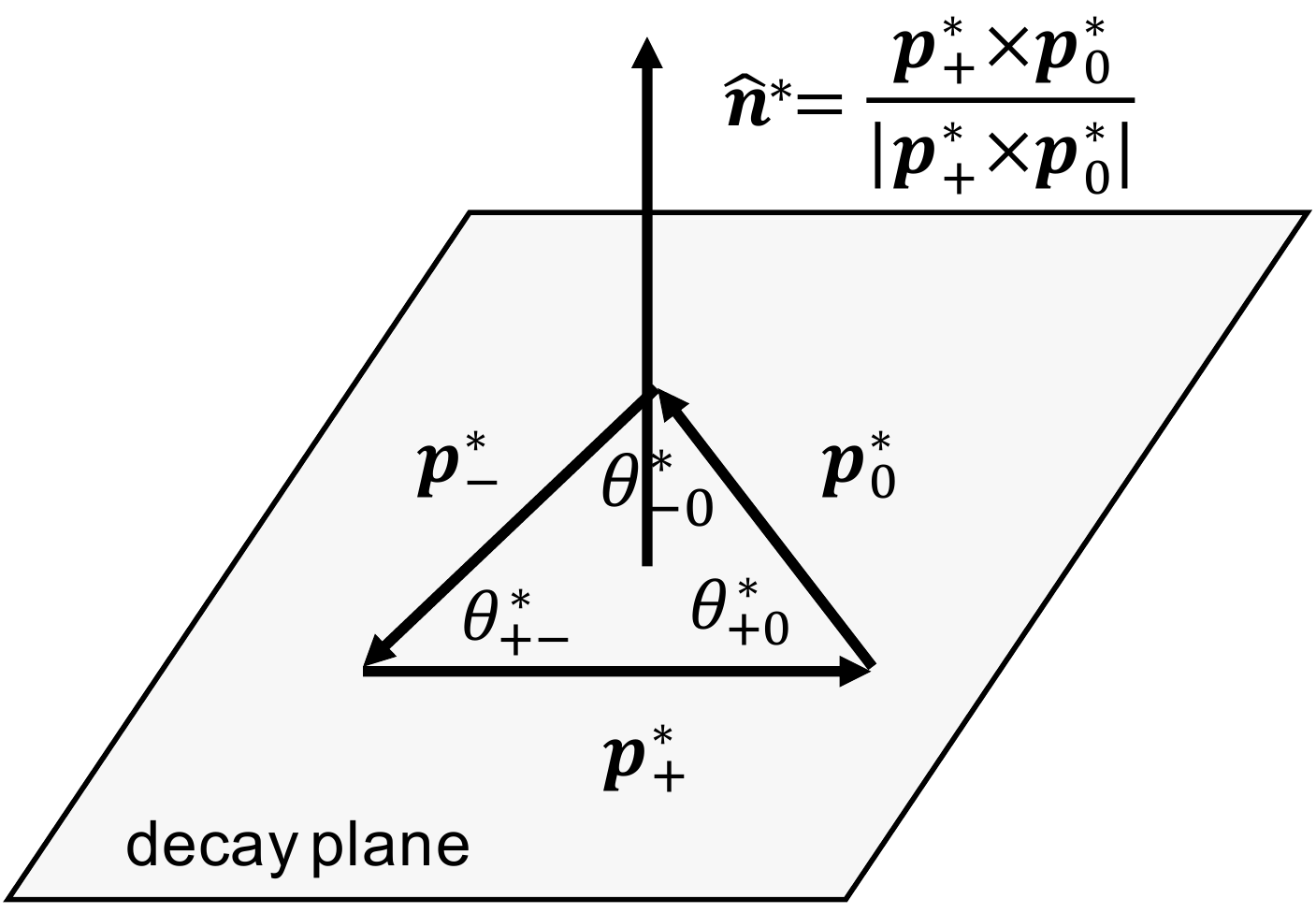}
  \end{minipage}
  \caption{Plane containing the momenta of the pions produced in the $\omega/\phi\rightarrow \pi^+\pi^-\pi^0$ decay.}\label{fig:3-body triangle momenta}
\end{figure}

The momenta $\textbf{p}^*_+$, $\textbf{p}^*_-$ and $\textbf{p}^*_0$, lie in the decay plane with normal vector $\nunit=\textbf{p}_+^*\times\textbf{p}_0^*/|\textbf{p}_+^*\times\textbf{p}_0^*|$, as shown in Fig. \ref{fig:3-body triangle momenta}. The modulus of each momentum is fixed by the mass-shell condition whereas their relative orientations in the decay plane are fixed by
\begin{eqnarray}\label{eq:cos theta+-}
\cos\theta_{ij}^*=\frac{p_k^{*2}-p_i^{*2}-p_j^{*2}}{2p_i^*p_j^*},
\end{eqnarray}
where $\theta_{ij}^*$ is the angle between $\textbf{p}_i^*$ and $\textbf{p}_j^*$. This relation is a direct consequence of momentum conservation $\textbf{p}_+^*+\textbf{p}_-^*+\textbf{p}_0^{*}=\textbf{0}$.
The pion $\pi^i$ has maximum momentum when $\pi^j$ and $\pi^k$ have parallel momenta opposite to $\pi^i$, and are at rest in the $\pi^j\pi^k$ center of mass frame.
For instance, the maximum allowed squared momenta for $\pi^+$ and $\pi^0$ are given by
\begin{eqnarray}
p_{+,max}^{*2} = \left[M^2-(2m_{\pi}+m_{\pi^0})^2\right]\left[M^2-m^2_{\pi^0}\right]/(4M^2) \\
p_{0,max}^{*2} = \frac{\left[M^2-(2m_{\pi}+m_{\pi^0})^2\right]\left[M^2-(m_{\pi^0}-2m_{\pi})^2\right]}{4M^2}.
\end{eqnarray}

The distribution of the final pions in the rest frame of the vector meson is given by differential decay width for the three body decay
\begin{equation}\label{eq: dGamma 3-body}
    d\Gamma^{(3)}_h=\frac{1}{2M}|\mathcal{M}^{(3)}_h|^2|\mathcal{F}_h|^2d\Phi_h^{(3)}(p^*;p_+^*,p_-^*,p_0^*),
\end{equation}
where $\mathcal{M}_h^{(3)}$ is the matrix element for the decay $vm\rightarrow\rm{ps}_1\rm{ps}_2\rm{ps}_3$, $\mathcal{F}_h$ is a form factor and $d\Phi_{h}^{(3)}$ is the three-body invariant phase space.

For this decay the matrix element is
\begin{equation}\label{eq:matrix element M for 3-body decay}
    \mathcal{M}_h^{(3)}=\varepsilon_{\alpha\beta\gamma\delta}V^{\alpha}p_h^{*\beta}p_+^{*\gamma}p_0^{*\delta}=-M\textbf{V}\cdot(\textbf{p}_0^*\times\textbf{p}_+^*)=-M\textbf{V}\cdot\hat{\textbf{n}}^*\,|\textbf{p}_+^*\times\textbf{p}_0^*|.
\end{equation}
It couples the polarization vector $\textbf{V}$ with the normal $\nunit^*$ to the decay plane. The totally anti-symmetric tensor $\varepsilon_{\alpha\beta\gamma\delta}$ is needed because of the coupling with pseudoscalar mesons.

\begin{figure}[tb]
\centering
\begin{minipage}[t]{.8\textwidth}
  \includegraphics[width=1.0\linewidth]{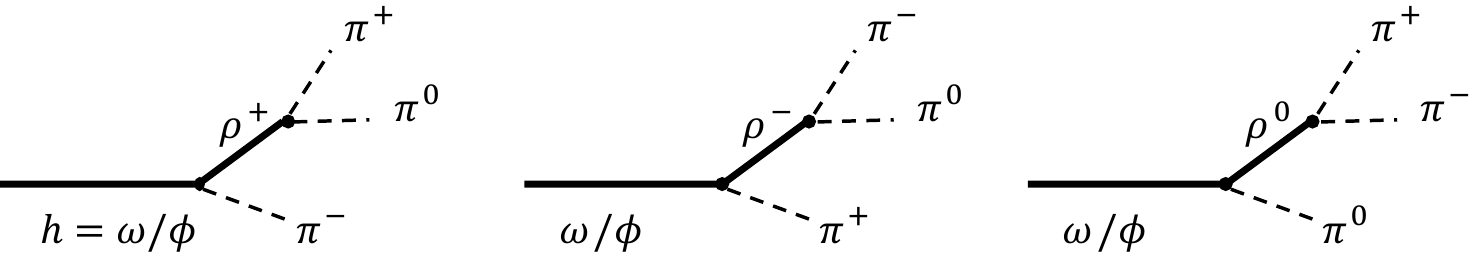}
  \end{minipage}
  \caption{Diagrams for the $\omega/\phi\rightarrow \pi^+\pi^-\pi^0$ decay via intermediate $\rho$-mesons. }\label{fig:3-body decay}
\end{figure}

The form factor $\mathcal{F}_h$ is obtained using the isobar model where e.g. the $\omega/\phi\rightarrow \pi^+\pi^-\pi^0$ decay occurs via intermediate $\rho$ mesons as schematically represented in Fig. \ref{fig:3-body decay}. In this model the form factor can be written as
\begin{equation}\label{eq:form factor F}
    \mathcal{F}_h(m_{+-}^2,m_{+0}^2) = \sum_{ij=+-,+0,-0} \frac{g_{h\rho^k\pi}\,g_{\rho^{k}\pi\pi}}{m^2_{ij}-\hat{m}^2_{\rho^k}+i\hat{m}_{\rho^k}\hat{\gamma}_{\rho^{k}}},
\end{equation}
where the variables $m_{ij}^2$ are defined as
\begin{eqnarray}\label{eq:m0 m+ m-}
    m_{ij}^2=(p_i+p_j)^2=M^2+m_{\pi^k}^2-2ME^*_k.
\end{eqnarray}
The variables $m_{+-}$, $m_{+0}$ and $m_{-0}$ are thus the invariant masses of the $\pi^+\pi^-$, $\pi^+\pi^0$ and $\pi^-\pi^0$ pairs respectively. Also, they obey the constraint
\begin{equation}
    m_{+-}^2+m_{+0}^2+m_{-0}^2=M^2+2m^2_{\pi}+m^2_{\pi^0},
\end{equation}
meaning that only two of them, e.g. $m_{+-}$ and $m_{+0}$, can be taken as independent variables.
The factors $g_{h\rho\pi}$ and $g_{\rho\pi\pi}$ are coupling constants. By isospin symmetry they do not depend on the charge of the intermediate $\rho$ meson, namely it is $g_{h\rho^+\pi^-}=g_{h\rho^-\pi^+}=g_{h\rho^0\pi^0}$ and $g_{\rho^+\pi^+\pi^0}=g_{\rho^-\pi^-\pi^0}=g_{\rho^0\pi^+\pi^-}$.


The last ingredient needed to write down the differential decay width in Eq. (\ref{eq: dGamma 3-body}) is the three-body invariant phase space
\begin{eqnarray}
\nonumber    d\Phi_3(p^*;p_+^*,p_-^*,p_0^*)&=&(2\pi)^4\delta^{(4)}(p_h^*-p_+^*-p_-^*-p_0^*)\\
    &\times& \frac{d^3\textbf{p}_+^*}{(2\pi)^3 2E_+^*}\frac{d^3\textbf{p}_-^*}{(2\pi)^3 2E_-^*}\frac{d^3\textbf{p}_0^*}{(2\pi)^3 2E_0^*},
\end{eqnarray}
which, after some algebra, can then be re-written as
\begin{equation}\label{eq: dPhi_3 uniform}
    d\Phi_3(M;m_{+-}^2,m_{+0}^2)=\frac{1}{8(2\pi)^5} dm_{+-}^2 dm_{+0}^2 d\phi_+^* d\cos\theta_+^* d\phi_{+0}^*.
\end{equation}
The angles $\phi_+^*$ and $\theta_+^*$ are the azimuthal and polar angles of $\pi^+$, and $\phi_{+0}$ is the azimuthal angle of $\pi^0$ with respect to $\pi^+$. It is useful to note that the three-body phase space is flat as function of $m_{+-}^2$ and $m_{+0}^2$, and that $dm_{ij}^2$ can be replaced by $dE_k^*$ since $m_{ij}^2$ and $E_k^*$ are linearly related via Eq. (\ref{eq:m0 m+ m-}).

Finally, inserting Eqs. (\ref{eq:matrix element M for 3-body decay}-\ref{eq:form factor F}) and Eq. (\ref{eq: dPhi_3 uniform}) in Eq. (\ref{eq: dGamma 3-body}), the differential decay width can be written as
\begin{eqnarray}\label{eq:dGamma^3 final}
\nonumber    \frac{d\Gamma_h^{(3)}}{dm_{+-}^2dm_{+0}^2d\phi_+^* d\cos\theta_+^* d\phi_{+0}^*}&\propto& p_+^{*2} p_0^{*2}\,\sin^2\theta_{+0}^*\,|\mathcal{F}(m_{+-}^2,m_{+0}^2)|^2 \\
&\times& \hat{\textbf{n}}^*_a\,\RE{\hat{\rho}}_{ab}\,\hat{\textbf{n}}^*_b.
\end{eqnarray}
For the MC simulation of the three-body decay this equation is used as the starting point for the generation of the energies of the decay pions and for the orientation of their momenta. This task is more complicated compared to the two-body case, where only the direction of one decay product had to be generated,
and the algorithm is the following.


\begin{itemize}
\item[(1)] Generate the energies $E^*_+$ and $E^*_-$ with uniform distribution within the thick green polygon of Fig. \ref{fig:3-body triangle E} by applying the proper constraints. $E_0^*$ is then obtained from Eq. (\ref{eq:E+ + E- + E0 = mh}).
\item[(2)] Using the energies generated in the previous step and imposing the mass shell constraints, evaluate the momenta $p_+^*$, $p_-^*$ and $p_0^*$. In order to respect momentum conservation accept the values of $E_+^*$, $E_-^*$ and $E_0^*$ only if $|\cos\theta_{+-}^*|<1$, where $\cos\theta_{+-}^*$ is given in Eq. (\ref{eq:cos theta+-}).
This expression ensures that the momenta are in the physical kinematic region and restricts the phase space to the smooth blue area within the green polygon in Fig. \ref{fig:3-body triangle E}.
\item[(3)] Using Eq. (\ref{eq:m0 m+ m-}) and the values of $E_+^*$, $E_-^*$ and $E_0^*$ accepted at the second step, calculate the invariant masses $m_{+-}^2$ and $m_{+0}^2$, and accept or reject them according to the probability
\begin{equation}\label{eq:w(m+-,m+0)}
    w(m_{+-}^2,m_{+0}^2) = \frac{(p_+^* p_0^*)^2 \sin^2\theta^*_{+-}}{(p_{+,max}^*p_{0,max}^*)^2} \frac{|\mathcal{F}(m_0^2,m_+^2)|^2}{|\mathcal{F}|_{max}^2}.
\end{equation}
This quantity is obtained by integrating the differential decay width in Eq. (\ref{eq:dGamma^3 final}) over the azimuthal and polar angles, and by summing over the polarization states of the vector meson. It takes into account the fact that the invariant mass distributions are not uniform within the allowed phase space region due to the not constant matrix element and form factor. The denominators of the first and second factor in Eq. (\ref{eq:w(m+-,m+0)}) are such that $w(m_{+-}^2,m_{+0}^2)<1$.
The majorant of the form factor squared
\begin{equation}
    |\mathcal{F}|_{max}^2=\left(\sum_a \frac{g_{h\rho^a\pi}\,g_{\rho^a\pi\pi}}{\hat{m}_a\gamma_{\rho^a}}\right)^2,
\end{equation}
is obtained noting that $|z_1+z_2+z_3|^2\leq (|z_1|+|z_2|+|z_3|)^2$ for any arbitrary complex numbers $z_1,z_2,z_3$.

At the end of this step the final values of the invariant masses and therefore of the pion energies are available.
\item[(4)] For the generation of the directions of the pion momenta, draw first the normal $\nunit$ to the pion decay plane. Namely draw its polar angle $\theta^*_{\nunit}$ and its azimuthal $\phi^*_{\nunit}$ defined in the $(\xu,\yu,\zu)$ reference frame using Eq. (\ref{eq: dN/dphi dcostheta}) with the replacement $\hat{\textbf{r}}^*\rightarrow \nunit^*$, motivated by the second line of Eq. (\ref{eq:dGamma^3 final}). 


\item[(5)] Consider the plane orthogonal to $\nunit$ which contains the pion momenta, shown in Fig. \ref{fig:3-body triangle momenta}. To draw the pion momenta in this plane, generate first the orientation of $\textbf{p}^*_+$ with uniform distribution, namely generate an azimuthal angle $\varphi$ with uniform distribution in $[0,2\pi]$. Then, the pion momenta in the vector meson rest frame are given by
\begin{eqnarray}
    \textbf{p}_+^*=R_{\zu\times \nunit}(\theta^*_{\nunit})R_{\zu}(\varphi)(p_+,0,0)^{\rm{T}}\\
    \textbf{p}_-^*=R_{\zu\times \nunit}(\theta^*_{\nunit})R_{\zu}(\varphi)(p_-\cos\theta_{+-},-p_-\sin\theta_{+-})^{\rm{T}}\\
    \textbf{p}_0^*=R_{\zu\times \nunit}(\theta^*_{\nunit})R_{\zu}(\varphi)(p_0\cos\theta_{+0},p_0\sin\theta_{+0})^{\rm{T}},
\end{eqnarray}
where $\cos\theta_{+0}$ is calculated according to Eq. (\ref{eq:cos theta+-}).
\end{itemize}
After this last step, the pion momenta are properly generated according to the differential three-body decay width.

\subsection*{Boosting the decay mesons}
The momenta of the particles produced in the vector meson decay are finally obtained in the vector meson rest frame. To transform them to the string rest frame, a composition of two boosts is applied. First the transverse boost (along $\pt$) $\Lambda_T(\pt)$ defined as
\begin{eqnarray}\label{eq:Lambda T boost}
    \Lambda_T(\pt) =
    \begin{pmatrix}
    \varepsilon_h/M   &   p_x/M                             & p_y/M &                         0 \\
    p_x/M              &   1+p_x^2/[M(\varepsilon_h+M)]   & p_xp_y/[M(\varepsilon_h+M)]   & 0  \\
    p_y/M              &   p_yp_x/[M(\varepsilon_h+M)]    & 1+p_y^2/[M(\varepsilon_h+M)]   & 0  \\
    0                    & 0                                    & 0                                     & 1
    \end{pmatrix},
\end{eqnarray}
and afterwards the longitudinal boost (along the string axis) $\Lambda_L(p_z)$ defined as
\begin{eqnarray}\label{eq:Lambda L boost}
    \Lambda_L(p_z) =
    \begin{pmatrix}
    E/\varepsilon_h   & 0   &   0   & p_z/\varepsilon_h \\
    0   & 1 & 0 & 0 \\
    0 & 0 & 1 & 0 \\
    p_z/\varepsilon_h & 0 & 0 & E/\varepsilon_h
    \end{pmatrix}.
\end{eqnarray}
$\Lambda_T(\pt)$ is the transformation that brings the vector meson from its rest to the frame where it has transverse momentum $\pt$ with respect to the string axis. $\Lambda_L(p_z)$ is the transformation which gives to the vector meson the longitudinal momentum $p_z$. Hence the composition of the two boosts brings $p^*$ to $p$, namely it is $p = \Lambda_L(p_z)\Lambda_T(\pt)\,p^*$.

\subsection{Spin density matrix of $q'$}
After the production of a vector meson, the spin density matrix of the remaining quark $q'$ can be obtained from Eq. (\ref{eq:R definition}) taking into account the information coming from the decay process. Its expression is
\begin{eqnarray}\label{eq: rho(q') after V decay}
\nonumber    \hat{\rho}_{jj'}(q';\check{D})&=&\frac{\sum_{\alpha\alpha'}R_{j\alpha;j'\alpha'}\check{D}_{\alpha'\alpha}}{\sum_{i,\beta\beta'}R_{i\beta;i\beta'}\check{D}_{\beta'\beta}}\\
&=& \frac{\left(\sum_{\alpha\alpha'}\check{D}_{\alpha\alpha'} \,\Deltaqp \Gamma_{h,\alpha}\,\rho(q)\,\Gamma^{\dagger}_{h,\alpha'}\Deltaqpdag\right)_{jj'}}{\sum_{\beta\beta'}\check{D}_{\beta\beta'} \,\rm{tr}\left(\Deltaqp \Gamma_{h,\beta}\,\rho(q)\,\Gamma^{\dagger}_{h,\beta'}\Deltaqpdag\right)}.
\end{eqnarray}
The matrix $\check{D}$ is called \textit{decay matrix} \cite{collins-corr,knowles-corr} and encodes the information coming from the decay process of the produced vector meson. It allows to propagate the quantum mechanical spin correlations between the orientation of the decay products and the spin of $q'$. It satisfies the unitarity condition \cite{collins-corr}
\begin{equation}
    \sum_{states} \check{D}_{\alpha\alpha'} = \delta_{\alpha\alpha'},
\end{equation}
where the summation runs over all possible states (in momentum and spin space) of the decay mesons.
If the decay of the vector meson is not analysed, i.e. the decay products of the vector meson are not detected, then $\check{D}_{\alpha\alpha'}=\delta_{\alpha\alpha'}$ which is equivalent to summing over the polarization states of the vector meson.
If the decay of the vector meson is analysed, then the formula for $\check{D}$ depends on the decay process. It can be obtained from the angular distribution of the decay products, which is of the form $(1/\Gamma)d\Gamma/d\cos\theta^*\,d\phi^*=\RE\hat{\rho}_{\alpha\alpha'}\,\check{D}_{\alpha'\alpha}$ \cite{collins-corr}.
From Eq. (\ref{eq: dN/dphi dcostheta}) it follows that for the decay $vm\rightarrow \rm{ps}_1\rm{ps}_2$ the decay matrix is
\begin{equation}\label{eq: D V -> PS PS}
    \check{D}_{\alpha\alpha'}=\frac{3}{4\pi}\hat{\textbf{r}}^*_{\alpha}\hat{\textbf{r}}^*_{\alpha'}
\end{equation}
evaluated at the generated value of $\hat{\textbf{r}}^*$.

To obtain the decay matrix for $\omega/\phi\rightarrow \pi^+\pi^-\pi^0$ decay, it is sufficient to make the substitution $\hat{\textbf{r}}^*\rightarrow \nunit$, which gives
\begin{eqnarray}\label{eq: D V -> PS PS pS}
\check{D}_{\alpha\alpha'}=\frac{3}{4\pi}\hat{\textbf{n}}^*_{\alpha}\hat{\textbf{n}}^*_{\alpha'}.
\end{eqnarray}

Finally the decay matrix for the process $\rm{V}\rightarrow \gamma \rm{PS}$ can be obtained from Eq. (\ref{eq: dN/dphi dcostheta gamma}) and it is
\begin{equation}\label{eq: D V -> gamma PS}
    \check{D}_{\alpha\alpha'}=\frac{3}{8\pi}\left(\delta_{\alpha\alpha'}-\hat{\textbf{r}}^*_{\alpha}\hat{\textbf{r}}^*_{\alpha'}\right).
\end{equation}
Eqs. (\ref{eq: D V -> PS PS})-(\ref{eq: D V -> gamma PS}) are used in the MC simulations when the vector meson decay is analysed.

The introduction of vector mesons changes the dynamics of the polarization transfer from $q$ to $q'$ along the fragmentation chain. In particular the "depolarization coefficients" $D_{\rm{TT}}^{vm}$ and $D_{\rm{LL}}^{vm}$ are different with respect to those of M19 given in Eq. (\ref{eq: DTT DLL}).
These coefficients can be obtained from the spin density matrix of $q'$ calculated using Eq. (\ref{eq: rho(q') after V decay}). The calculation involves an integration over the momenta of the decay products. Considering, for instance, the two-body decay $vm\rightarrow ps_1+ps_2$ and integrating Eq. (\ref{eq: rho(q') after V decay}) over $d\Omega^*=d\phi^* d\cos\theta^*$ yields
\begin{eqnarray}\label{eq:rho(q') after vm decay averaged}
\nonumber    \hat{\rho}(q') &=& \int d\Omega^*\,\frac{d\Gamma_h^{(2)}}{\Gamma_h^{(2)}d\Omega^*} \,\frac{\check{D}_{\alpha\alpha'} \,\Deltaqp \Gamma_{h,\alpha}\,\rho(q)\,\Gamma^{\dagger}_{h,\alpha'}\Deltaqpdag}{\check{D}_{\beta\beta'} \,\rm{tr}\left(\Deltaqp \Gamma_{h,\beta}\,\rho(q)\,\Gamma^{\dagger}_{h,\beta'}\Deltaqpdag\right)}\\
\nonumber    &=&\frac{3}{4\pi} \int d\Omega^*\,\hat{\textbf{r}}^*_{\sigma}\hat{\textbf{r}}^*_{\sigma'}\RE{\hat{\rho}}_{\sigma\sigma'}\,\frac{\hat{\textbf{r}}^*_{\alpha}\hat{\textbf{r}}^*_{\alpha'} \,\Deltaqp \Gamma_{h,\alpha}\,\rho(q)\,\Gamma^{\dagger}_{h,\alpha'}\Deltaqpdag}{\hat{\textbf{r}}^*_{\beta}\hat{\textbf{r}}^*_{\beta'} \,\rm{tr}\left(\Deltaqp \Gamma_{h,\beta}\,\rho(q)\,\Gamma^{\dagger}_{h,\beta'}\Deltaqpdag\right)} \\
\nonumber    &=& \left(\frac{3}{4\pi}\,\int d\Omega^*\,\hat{\textbf{r}}^*_{\alpha}\hat{\textbf{r}}^*_{\alpha'}\right)\,\frac{\Deltaqp \Gamma_{h,\alpha}\,\rho(q)\,\Gamma^{\dagger}_{h,\alpha'}\Deltaqpdag}{\rm{tr}\left(\Deltaqp \Gamma_{h,\beta}\,\rho(q)\,\Gamma^{\dagger}_{h,\beta'}\Deltaqpdag\right)}\\
    &=&\delta_{\alpha\alpha'}\,\frac{\Deltaqp \Gamma_{h,\alpha}\,\rho(q)\,\Gamma^{\dagger}_{h,\alpha'}\Deltaqpdag}{\rm{tr}\left(\Deltaqp \Gamma_{h,\beta}\,\rho(q)\,\Gamma^{\dagger}_{h,\beta'}\Deltaqpdag\right)},
\end{eqnarray}
where Eq. (\ref{eq:rho vm matrix elements}) and Eq. (\ref{eq: dN/dphi dcostheta}) have been used, and the summation over repeated indices has been understood. Clearly, such integration is equivalent to take $\check{D}=1$ in Eq. (\ref{eq: rho(q') after V decay}), namely the resulting spin density matrix of $q'$ is the same as if $h$ were not decayed. Thus, the decay of the vector meson turns out to have no effect on the mechanism of the quark spin transfer along the fragmentation chain. The depolarization coefficients obtained from Eq. (\ref{eq:rho(q') after vm decay averaged}) (see sec. \ref{sec:simple splitt function}) are
\begin{eqnarray}\label{eq:DTT and DLL for vm emission}
    D_{\rm{TT}}^{\rm{vm}} = \frac{|G_L|^2}{2|G_T|^2+|G_L|^2}|D_{\rm{TT}}^{ps}| && D_{\rm{LL}}^{vm} = \frac{|G_L|^2-2|G_T|^2}{2|G_T|^2+|G_L|^2}\,D_{\rm{LL}}^{ps}.
\end{eqnarray}
The factors $D_{\rm{TT}}^{ps}$ and $D_{\rm{LL}}^{ps}$ are the depolarization coefficients of M19. From this result it is clear that the emission of a vector meson does not flip the quark transverse polarization as in the case of pseudoscalar meson production. Since the vector meson carries spin information, the depolarization coefficients are smaller than in the pseudoscalar case depending on the ratio $|G_L|^2/|G_T|^2$. In particular, if quarks couple only to vector mesons with transverse liner polarization, i.e. for $G_L=0$, there is no transverse spin transfer from $q$ to $q'$ and the transverse polarization of $q$ is transferred to the vector meson. If quarks couple only to vector mesons with linear longitudinal polarization, i.e. for $G_T=0$, the transverse polarization of $q$ is transferred to $q'$ with the same fraction as in M19.
Concerning the longitudinal polarization of $q$, it is transferred to $q'$ as in M19 without flip for $G_L=0$ and with flip if $G_T=0$.
The overall effect on the fragmentation chain with both vector and pseudoscalar meson production is that the initial spin information decays faster. A more quantitative study of this effect is given in Appendix \ref{appendix:polarization transfer}.

\section{Monte Carlo implementation}\label{sec:M20 MC implementation}
The basic structure of the stand alone MC implementation of M20 is the same as that of M19 described in section \ref{sec:structure of stand alone MC Chapter 3}. Namely, initially the flavor, the momentum and the spin density matrix of the fragmenting quark $q_A$ are defined. Then the elementary splitting $q\rightarrow h + q'$ is repeated recursively. As already mentioned, also in the case of M20 the matrix $\hat{u}_q$ is proportional to the identity matrix and does not need to be tabulated as preliminary task. Consequently $\hat{\rho}_{int}(q)\equiv \rho(q)$ also for M20.

The iteration procedure follows the steps:
\begin{itemize}
    \item [(1)] generate a new $q'\bar{q}'$ taking into account the suppression of s quarks
    \item [(2)] form the pair $h=q\bar{q}'$ and then: choose if it represents a vector meson with probability $f_{vm}$ or a pseudoscalar meson with probability $f_{ps}$, afterwards choose the hadron type and assign its mass $m_h$. For a vector $h$, generate its mass $M$ according to the distribution $|D_h(M^2)|^2$.
    \item [(3)] if $h$ is pseudoscalar, generate $\kpt$ according to the $Z$-integrated splitting function in Eq. (\ref{eq:F_explicit simple 3P0}). If $h$ is a vector meson, generate $\kpt$ according to the $Z$-integrated splitting function of M19 in Eq. (\ref{eq:F_explicit vm}) at the given value of $m_h$ drawn at step (2). In both cases construct $\pt=\kt-\kpt$ with $\kt=0$ for $q_A$.
    \item [(4)] for pseudoscalar meson $h$, generate $Z$ according to the splitting function of M19 in Eq. (\ref{eq:F_explicit simple 3P0}) evaluated at the generated value of $\pt$. For a vector meson $h$, use instead the splitting function in Eq. (\ref{eq:F_explicit vm}).
    \begin{itemize}
    \item [(4.1)] calculate $p^-$ imposing the mass shell condition $p^+p^-=m_h^2+\ptpt$ for a pseudoscalar meson and $p^+p^-=M^2+\ptpt$ for a vector meson
    \item [(4.2)] test the exit condition and if it is not satisfied continue with the next step, otherwise the current hadron is removed and the fragmentation chain ends
    \item [(4.3)] calculate the hadron momentum $p=(E_h,\pt,p_L)$, where $E_h=(p^++p^-)/2$ and $p_L=(p^+-p^-)/2$, and store it in the event record
    \end{itemize}
    \item [(5)] for pseudoscalar $h$, calculate the spin density matrix of $q'$ using Eq. (\ref{eq:rho'=TrhoT}) and return to step 1. For a vector meson $h$, if its decay is switched off, calculate the spin density matrix of $q'$ using Eq. (\ref{eq: rho(q') after V decay}) with unit decay matrix $\check{D}$. Otherwise
    \begin{itemize}
        \item [(5.1)] calculate the spin density matrix $\hat{\rho}(h)$ of $h$ using Eq. (\ref{eq:rho vm matrix elements})
        \item [(5.2)] simulate the decay process in the vector meson rest frame using $\hat{\rho}(h)$ to generate the angular distribution of the decay products following the recipes explained above for the different decay types. If more than one decay channel is allowed, select it according to the branching ratios.
        \item [(5.3)] boost the decay products first along $\pt$ using Eq. (\ref{eq:Lambda T boost}) and then along the string axis using Eq. (\ref{eq:Lambda L boost}) to have their momenta in the string rest frame. Store the momenta in the event record.
        \item[(5.4)] return the decay matrix $\check{D}$ using the expressions of Eqs. (\ref{eq: D V -> PS PS}-\ref{eq: D V -> gamma PS} ) for the different decay types, and calculate the spin density matrix of $q'$ using Eq. (\ref{eq: rho(q') after V decay}).
    \end{itemize} 
    \item[(6)] follow steps (1)-(5.4) until the exit condition (same as in M18 and M19) is satisfied.
\end{itemize}

The parameter giving the suppression of $s$ quarks with respect to $u$ and $d$ quarks is taken the same as in M19.

The probabilities $f_{vm}$ and $f_{ps}$ that a given $q\bar{q}'$ pair is a vector or a pseudoscalar meson,  have been taken $\left(f_{vm}/f_{ps}\right)_{ud}=0.68$ for a non-strange $q\bar{q}'$ pair and $\left(f_{vm}/f_{ps}\right)=0.725$ if at least one quark of the pair is a strange quark. These values are the default values of \verb|PYTHIA 8| \cite{pythia8}.

The probabilities for the identification of a given spin-1 $q\bar{q}'$ pair with a vector meson type at step (2) are obtained from the isospin wave functions, as in the pseudoscalar mesons case. However, for vector meson production there is no suppression among flavor neutral states like $\rho^0$ and $\omega$. Namely a $u\bar{u}$ or a $d\bar{d}$ pair is identified either with a $\rho^0$ or with a $\omega$ with probability $0.5$. Hence, the factor $C_{q',h,q}$ for the case of vector meson production is obtained from Eq. (\ref{eq: Cq'hq matrix}) taking $p_{\eta}=0$.

\section{Results of simulations}\label{sec:M20 results from simulations}
This section is dedicated to the results obtained from simulations with the model M20.
In addition to the parameters $a$, $\bl$, $\bt$ and $\mu$, which were already present in the pseudoscalar case, this model introduces two new free parameters to describe the spin effects for vector meson production in polarized quark fragmentation. As can be seen from the corresponding splitting function in Eq. (\ref{eq:F_explicit vm}) and from the spin density matrix in Eq. (\ref{eq:rho vm matrix elements}), the new parameters are the ratio $|G_L|/|G_T|$ related to the probabilities that a quark couples to a vector meson with longitudinal or transverse linear polarizations, and the relative phase $\theta_{LT}$ of $G_L$ and $G_T$ which allows vector mesons to possess oblique linear polarization in the $(\xu,\zu)$ plane.
To study the kinematic distributions, the values $|G_L|/|G_T|=1$ and $\theta_{LT}=0$, in agreement with the model of Ref. \cite{Czyzewski-vm} have been used. Then the Collins and the dihadron analysing powers, and their sensitivity to the values of the new parameters are studied.

The results shown below are obtained from simulations of transversely and fully polarized $u$ quarks with initial energy obtained using the same sample of $x_B$ and $Q^2$ values used also in the simulations with M18 and M19. The primordial transverse momentum is switched off. And the other parameters $a$, $\bl$, $\bt$ and the complex mass $\mu$ are kept as in M19.
 The comparison with experimental data concludes this section.

\subsection{Kinematic distributions}
The relations between the $Z$ and $\ptpt$ distributions of vector mesons with different ranks are the same as those for the pseudoscalar mesons of M19 and will not be shown here. The $Z$ distribution of a first rank vector meson decreases faster than the $Z$ distributions for larger ranks.

\begin{table}[]
\centering
\begin{tabular}{|lc|l|l|}
\hline
\multicolumn{2}{|c|}{primary hadrons} &     \multicolumn{1}{c|}{final hadrons}                                                                     \\ \hline
\hline
\multicolumn{1}{|c|}{hadron}      & \multicolumn{1}{c|}{fraction}  & \multicolumn{1}{c|}{sec. / (prim. + sec.)}\\
\hline
\multicolumn{1}{|c|}{$\pi^+$}        & \multicolumn{1}{c|}{0.153}      &  \multicolumn{1}{c|}{0.464} \\
\multicolumn{1}{|c|}{$\pi^-$}        & \multicolumn{1}{c|}{0.072}       &  \multicolumn{1}{c|}{0.585} \\
\multicolumn{1}{|c|}{$\pi^0$}        & \multicolumn{1}{c|}{0.112}       &  \multicolumn{1}{c|}{0.559} \\
\multicolumn{1}{|c|}{$K^+$}          & \multicolumn{1}{c|}{0.027}       & \multicolumn{1}{c|}{0.256} \\
\multicolumn{1}{|c|}{$K^-$}          & \multicolumn{1}{c|}{0.010}       &  \multicolumn{1}{c|}{0.340}\\
\multicolumn{1}{|c|}{$K^0$}          & \multicolumn{1}{c|}{0.012}       &  \multicolumn{1}{c|}{0.513}\\
\multicolumn{1}{|c|}{$\bar{K}^{0}$}  & \multicolumn{1}{c|}{0.010}       &  \multicolumn{1}{c|}{0.321}\\
\cline{1-3}
\multicolumn{1}{|c|}{$\rho^{+}$}     & \multicolumn{1}{c|}{0.161}       &       \multicolumn{1}{c}{}                                   \\
\multicolumn{1}{|c|}{$\rho^{-}$}     & \multicolumn{1}{c|}{0.071}       &       \multicolumn{1}{c}{}                                      \\
\multicolumn{1}{|c|}{$\rho^{0}$}     & \multicolumn{1}{c|}{0.116}       & \multicolumn{1}{c}{}                                               \\
\multicolumn{1}{|c|}{$K^{*+}$}       & \multicolumn{1}{c|}{0.025}           & \multicolumn{1}{c}{}                                      \\
\multicolumn{1}{|c|}{$K^{*-}$}       & \multicolumn{1}{c|}{0.010}         & \multicolumn{1}{c}{}                                        \\
\multicolumn{1}{|c|}{$K^{*0}$}       & \multicolumn{1}{c|}{0.011}       & \multicolumn{1}{c}{}                                          \\
\multicolumn{1}{|c|}{$\bar{K}^{*0}$} & \multicolumn{1}{c|}{0.010}    & \multicolumn{1}{c}{}                                              \\
\multicolumn{1}{|c|}{$\omega$}       & \multicolumn{1}{c|}{0.115}      & \multicolumn{1}{c}{}                                           \\
\multicolumn{1}{|c|}{$\phi$}         & \multicolumn{1}{c|}{0.001}        & \multicolumn{1}{c}{}                                         \\ \cline{1-2}
\end{tabular}
\caption{Fractions of some hadron species in $u$ quark fragmentations. For each hadron the kinematical cuts $z_h>0.2$ and $p\T>0.1\,\rm{GeV}/c$ have been required. The third column shows the fraction of secondary pseudoscalar mesons among the final ones (primary and secondary).}\label{tab:fractions prim sec}
\end{table}

\begin{figure}[tb]
\centering
\begin{subfigure}{0.48\textwidth}
\centering
  \includegraphics[width=1.0\linewidth]{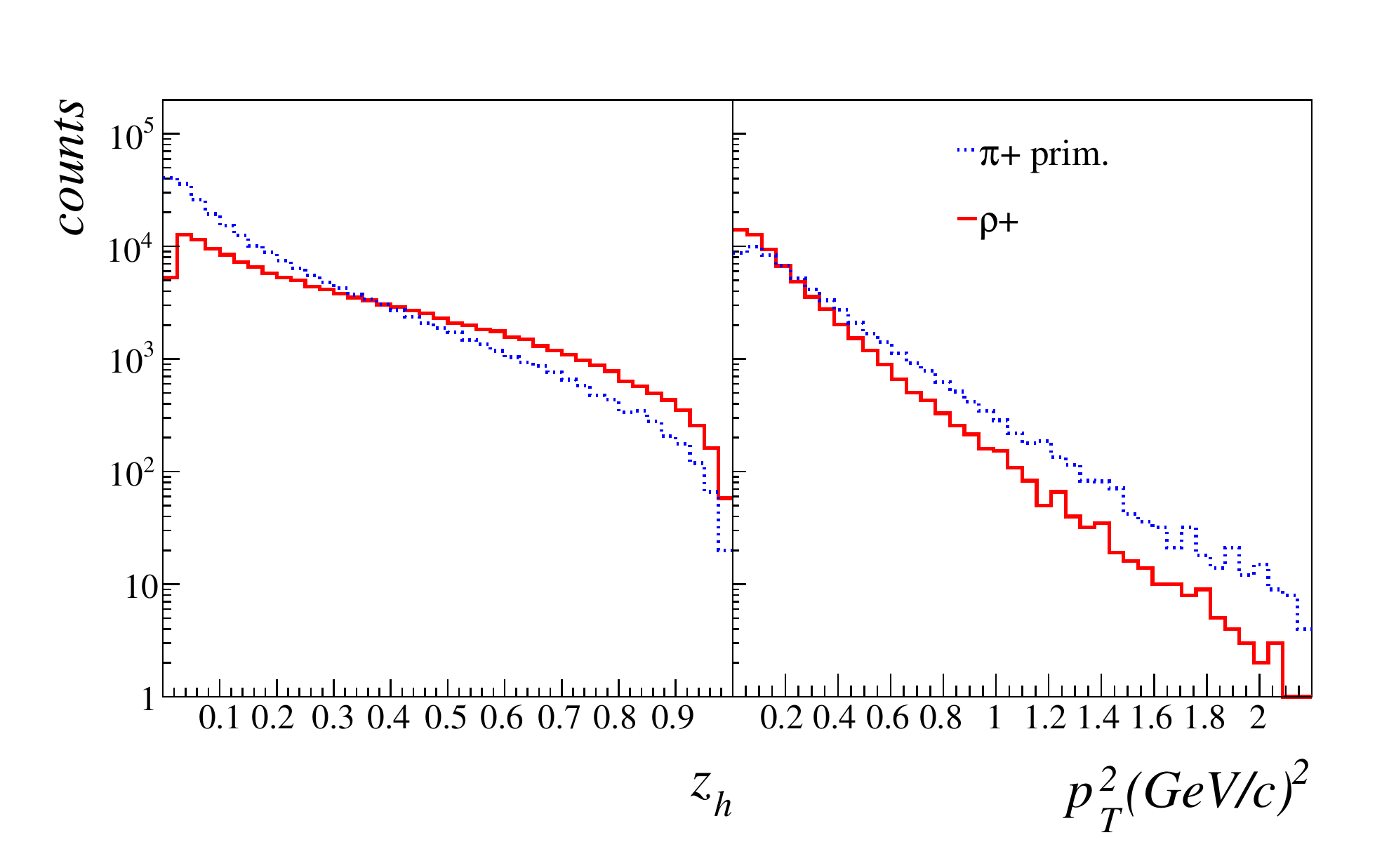}
  \end{subfigure}
\begin{subfigure}{0.48\textwidth}
\centering
	\includegraphics[width=1.0\linewidth]{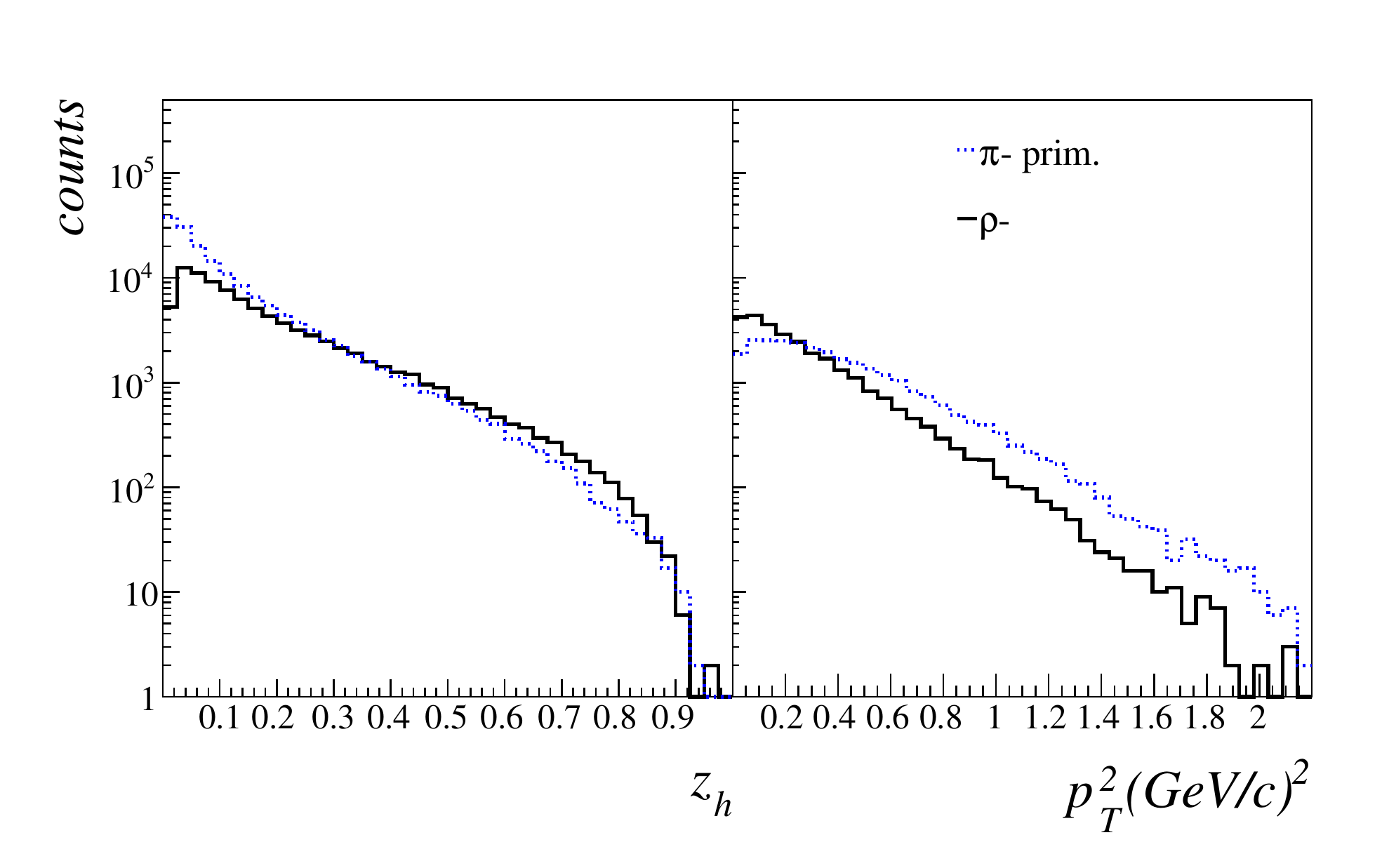}
\end{subfigure}
\caption{From left to right: distributions of $z_h$ and of $\ptpt$ for primary $\pi^+$ (red dotted histogram) and for $\rho^+$ (red continuous histogram), then the same for $\pi^-$ (blue dotted histogram) and for $\rho^-$ (black continuous histogram).}
\label{fig:zh pT prim pi+ vs rho+}
\end{figure}

\begin{figure}[tb]
\centering
\begin{subfigure}[t]{.48\textwidth}
  \includegraphics[width=1.0\linewidth]{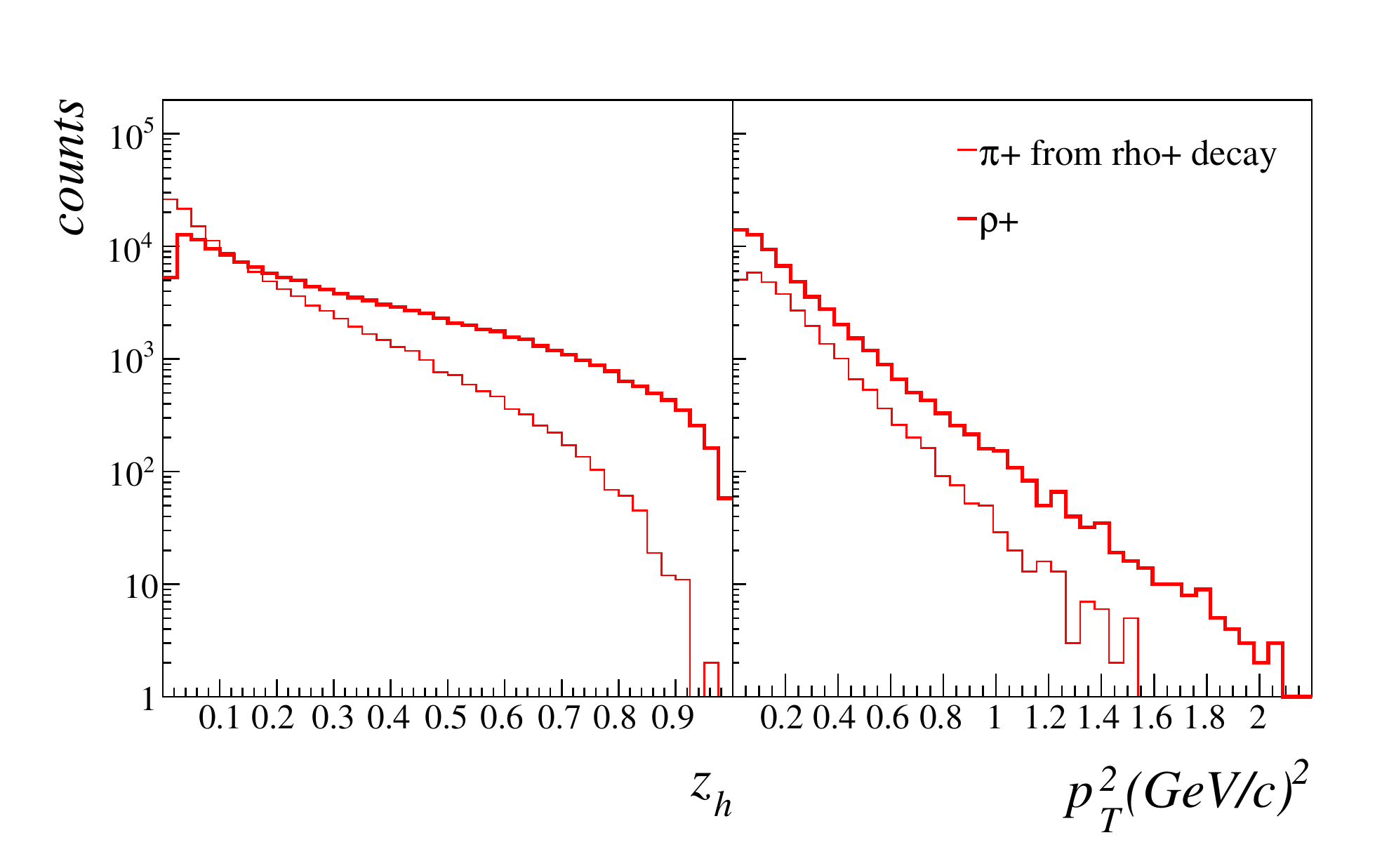}
  \end{subfigure}
\centering
\begin{subfigure}[t]{.48\textwidth}
	\includegraphics[width=1.0\linewidth]{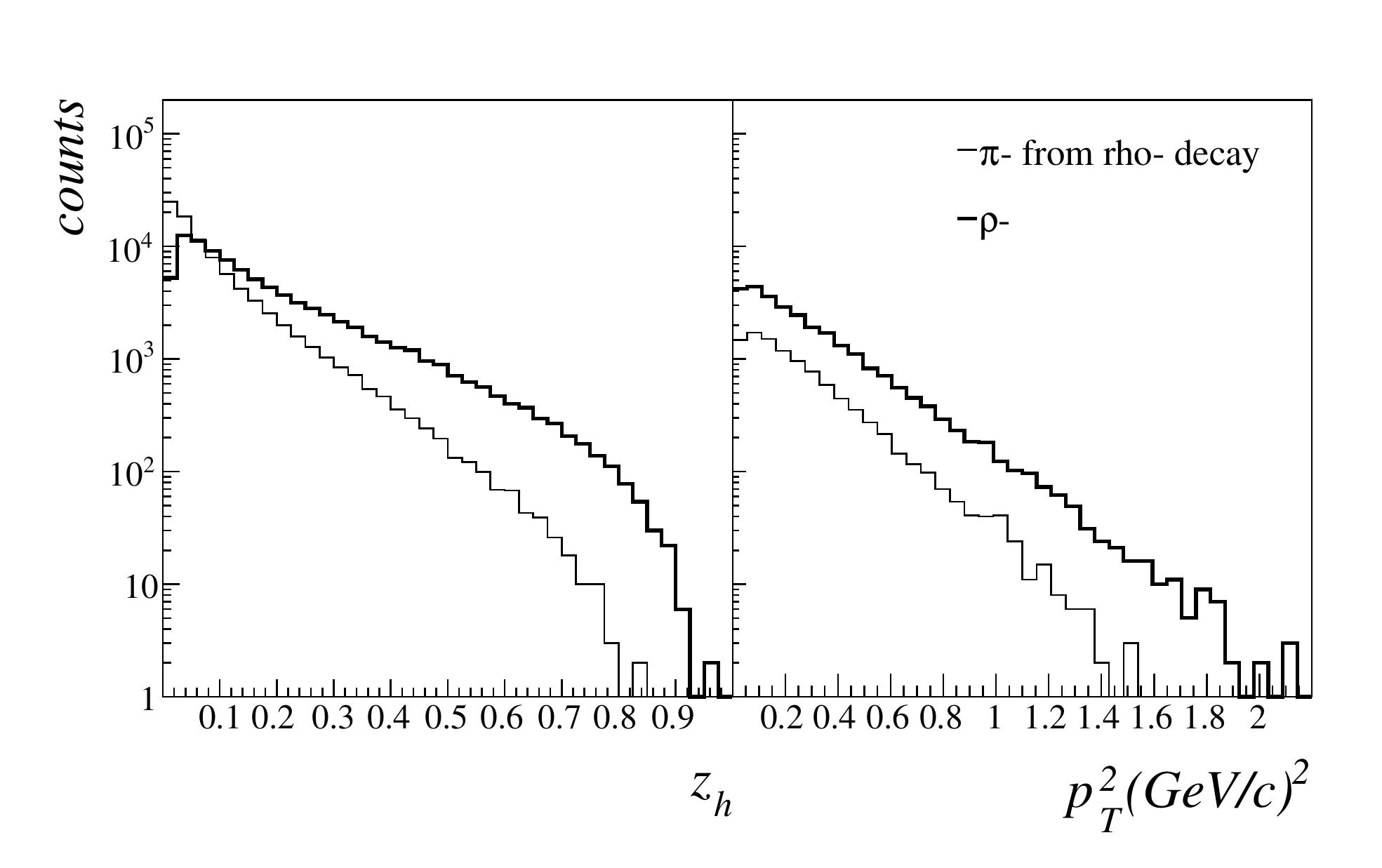}
\end{subfigure}
\caption{Left: distributions of $z_h$ and of $\ptpt$ for $\rho^+$ (red continous histogram) and for $\pi^+$ from $\rho^+$ decay (red dotted histogram). Right: the same for $\pi^-$ (black dotted histogram) and $\pi^-$ from $\rho^-$ decay (dotted black histogram).}
\label{fig:zh pT rho+ vs pi+ from rho+}
\end{figure}

\begin{figure}[tb]
\centering
\begin{minipage}[t]{.48\textwidth}
  \includegraphics[width=1.0\linewidth]{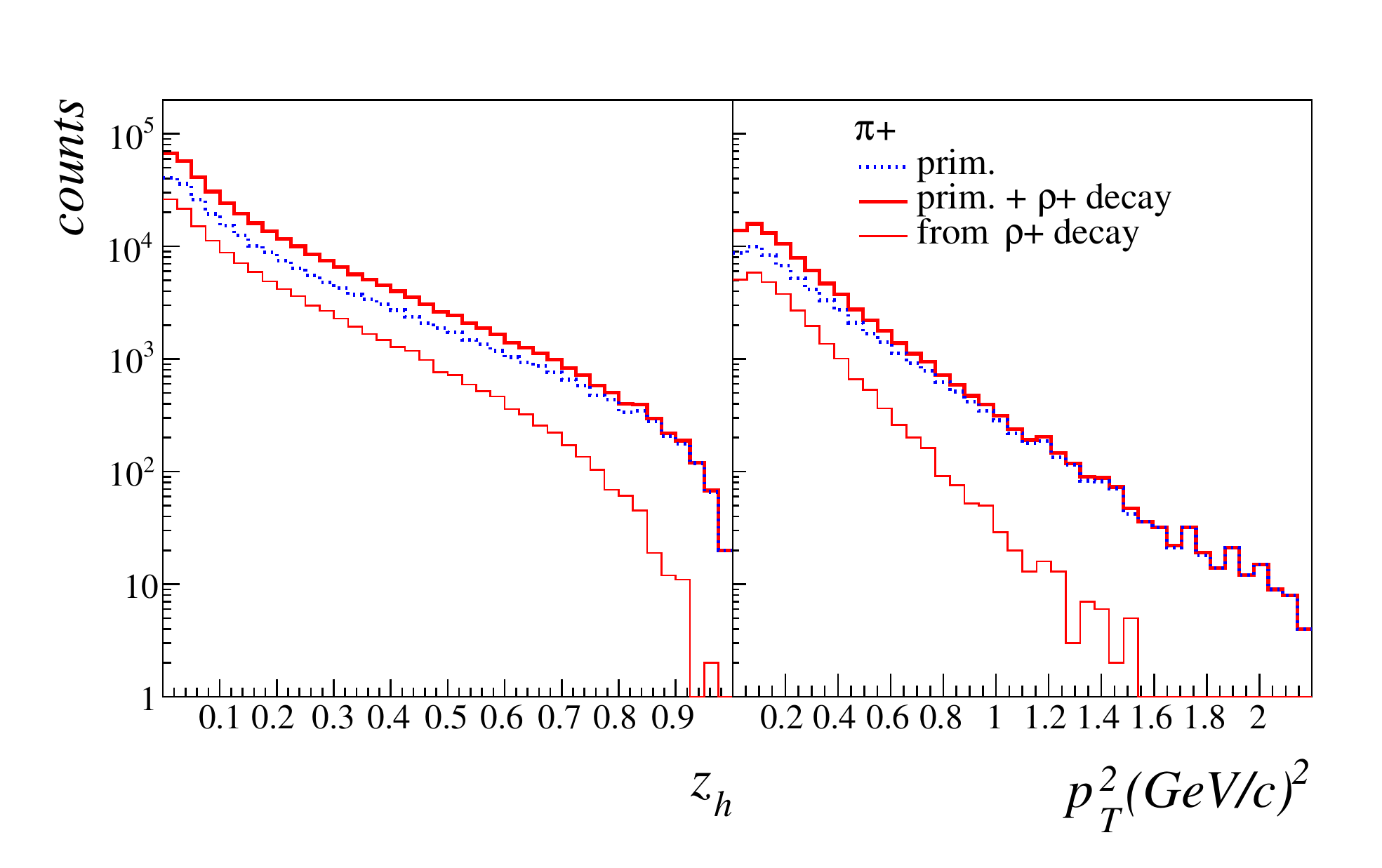}
  \end{minipage}
\begin{minipage}[t]{.48\textwidth}
	\includegraphics[width=1.0\linewidth]{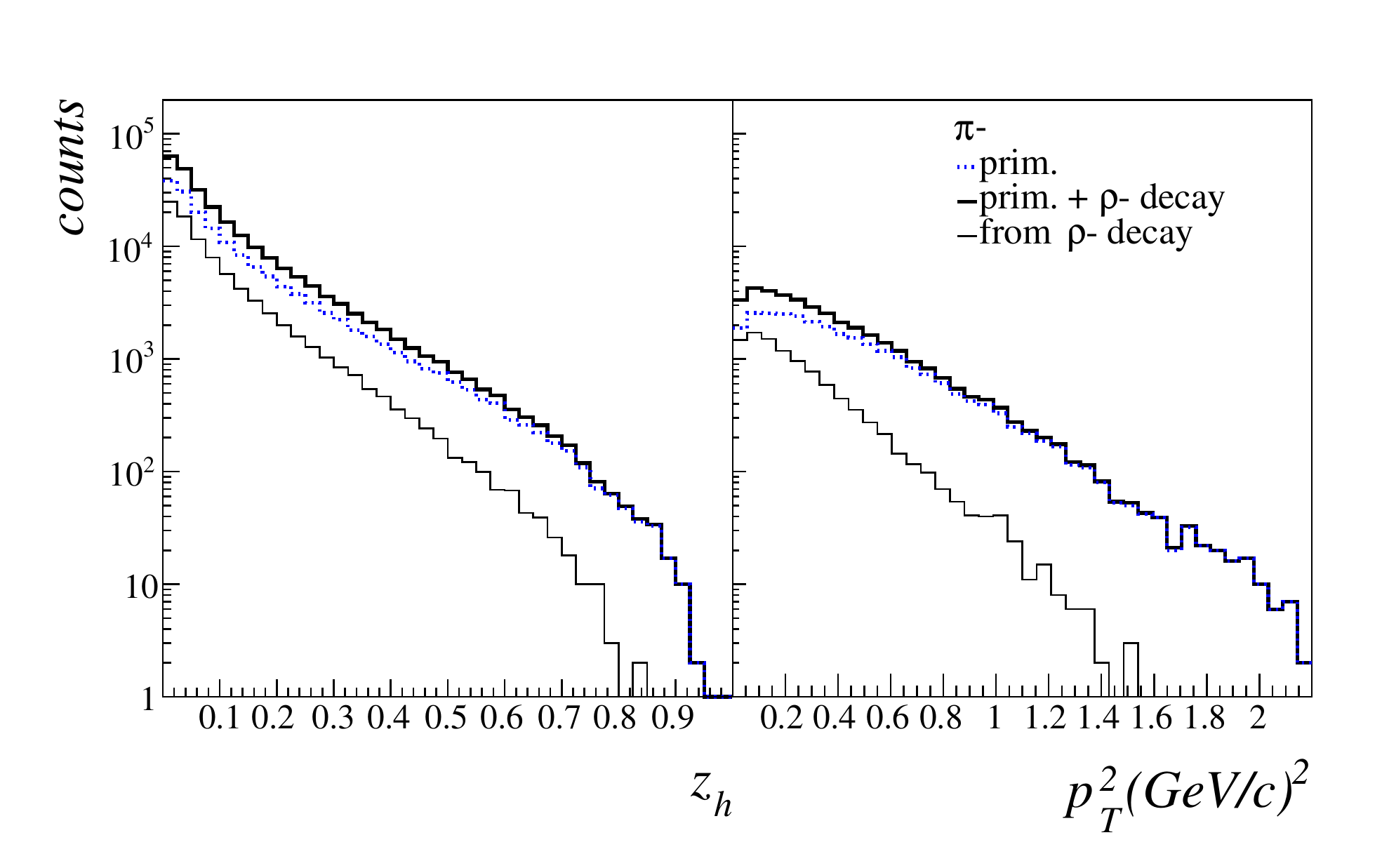}
\end{minipage}
\caption{Left: distributions of $z_h$ and of $\ptpt$ for primary $\pi^+$ (blue dotted histogram), for $\pi^+$ from $\rho^+$ decay (red continous thin histogram) and for the sum of the two contributions (red continous thick histogram). Right: the same for primary $\pi^-$ (blue dotted histogram), for $\pi^-$ from $\rho^-$ decay (black thin histogram) and all $\pi^-$ (black continous thick histogram).}
\label{fig:zh pT prim pi+ vs pi+ from rho+}
\end{figure}

\begin{figure}[tb]
\begin{minipage}[tb]{1.0\textwidth}
\centering
  \includegraphics[width=0.5\linewidth]{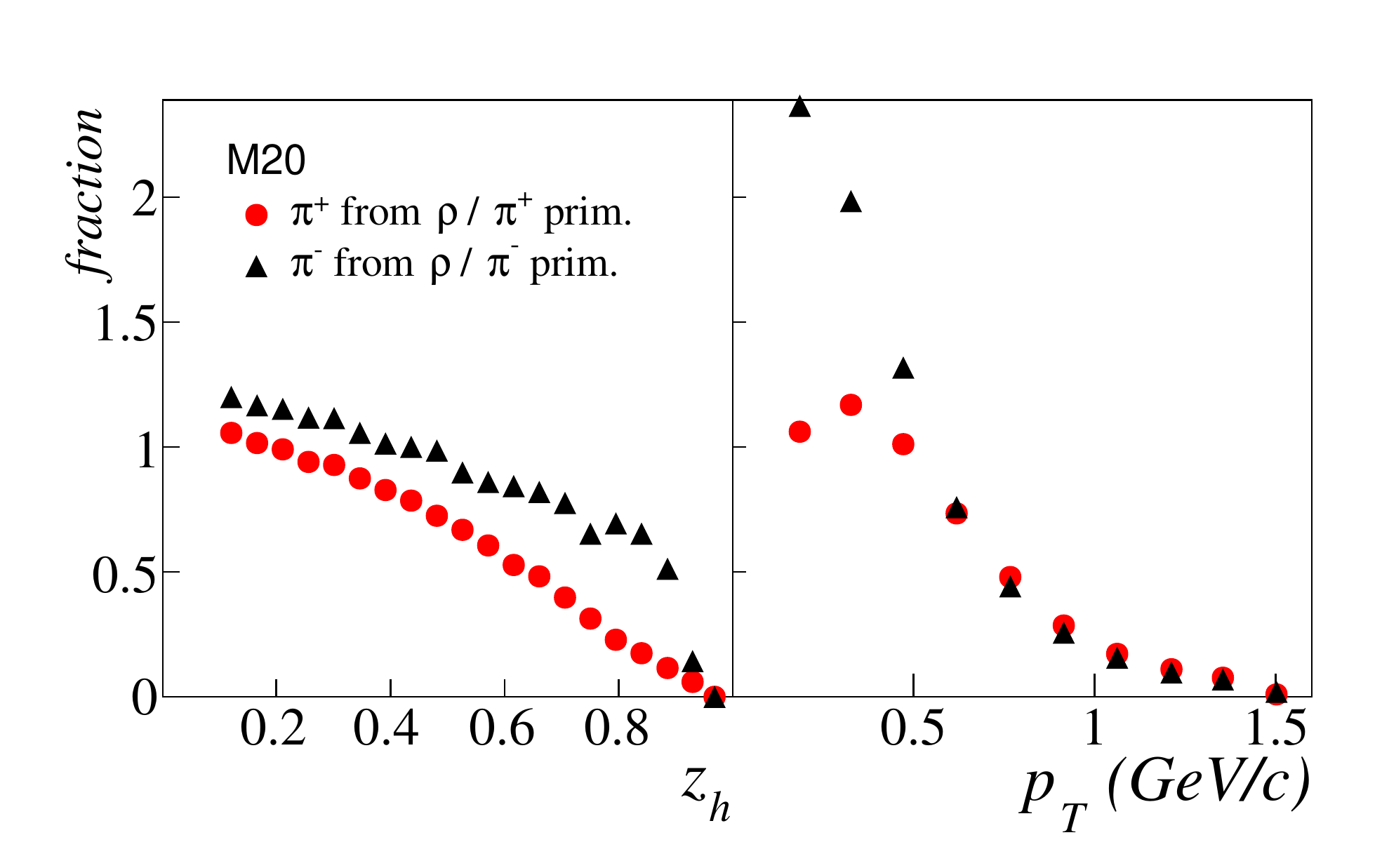}
  \end{minipage}
  \caption{Ratio between positive (circles) and negative pions (triangles) produced in decays of $\rho^{\pm,0}$ mesons and primary positive and negative pions, as function of $z_h$ (left panel) and as function of $p_{\rm{T}}$ (right plot).}\label{fig:pi+ from rho / prim. pi+}
\end{figure}

\begin{figure}[tb]
\centering
\begin{minipage}[t]{.48\textwidth}
  \includegraphics[width=1.0\linewidth]{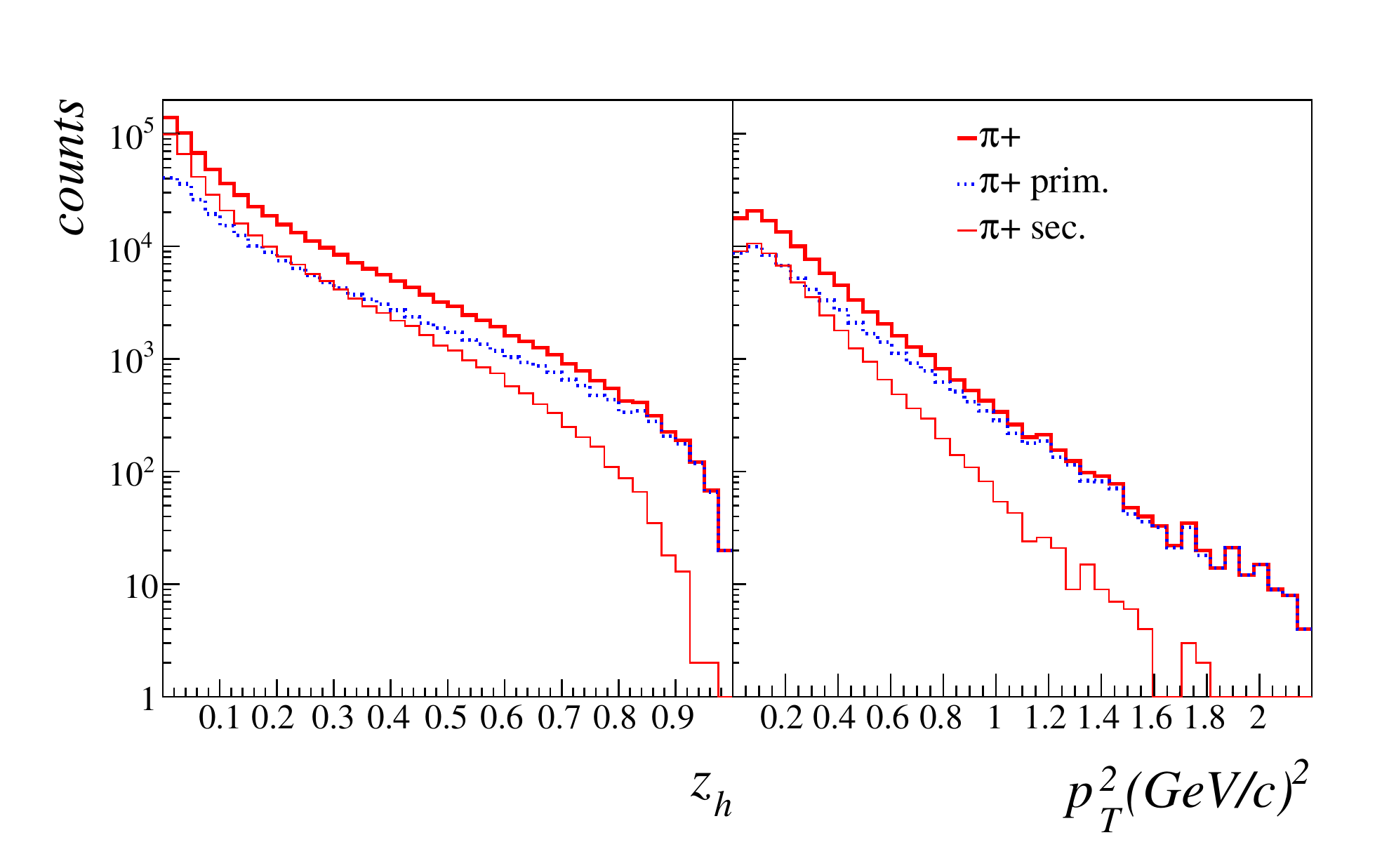}
  \end{minipage}
\begin{minipage}[t]{.48\textwidth}
	\includegraphics[width=1.0\linewidth]{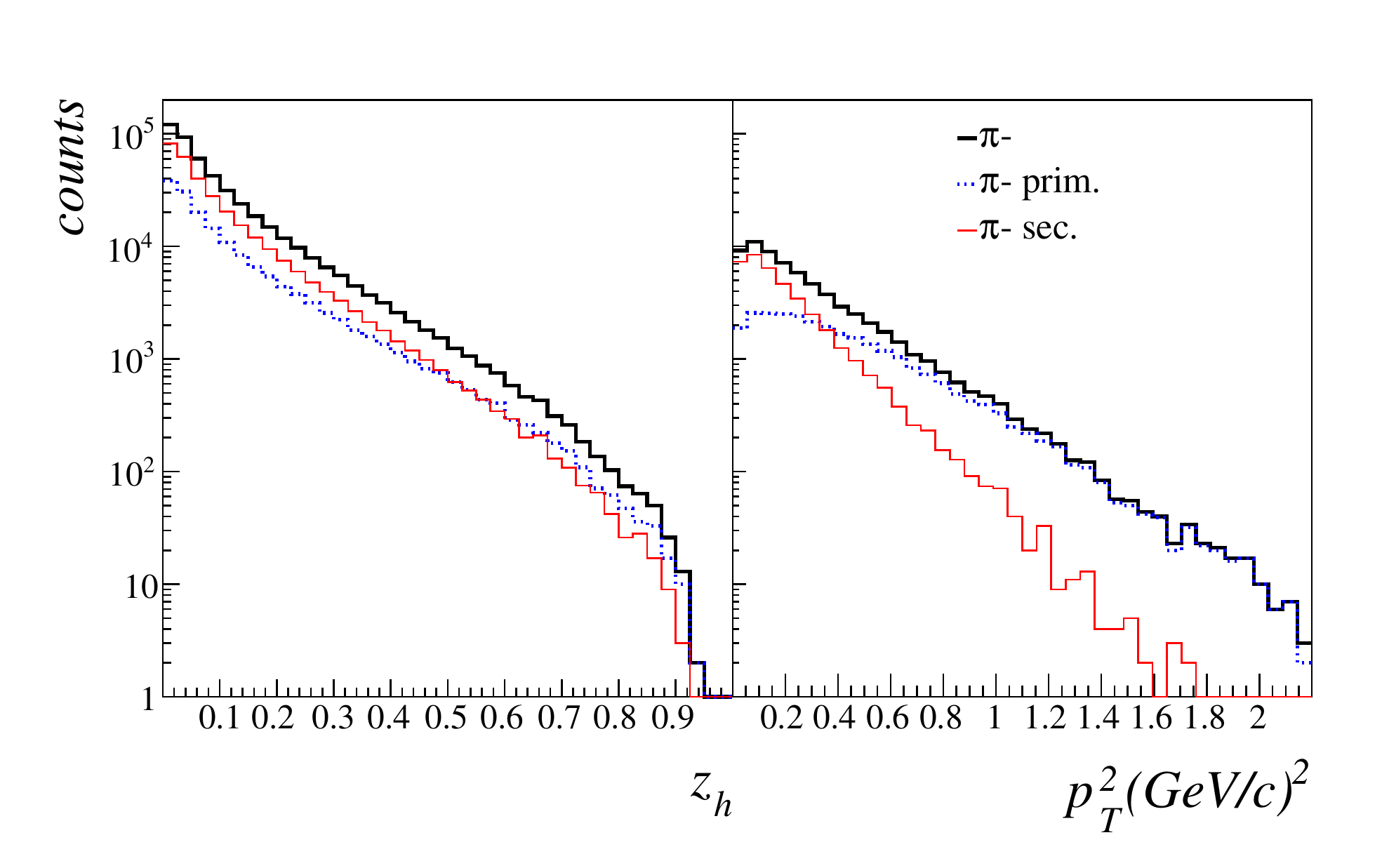}
\end{minipage}
\caption{Left: distributions of $z_h$ and of $\ptpt$ for primary $\pi^+$ (blue dotted histogram), secondary $\pi^+$ (red continous thin histogram) and for all $\pi^+$ (red continuous thick histogram) when decays of all vector mesons are considered. Right: the same for primary $\pi^-$ (dotted blue histogram), secondary $\pi^-$ (red thin continous histogram) and for all $\pi^-$ (red thick continous histogram).}\label{fig:zh pT prim pi+ vs all pi+}
\end{figure}

The comparison between the $z_h$ and $\ptpt$ distributions for vector mesons and primary pseudo-scalar mesons is shown in Figs. \ref{fig:zh pT prim pi+ vs rho+}-\ref{fig:zh pT prim pi+ vs all pi+}. The kinematic distributions are obtained requiring $p_{\rm{T}}>0.1\,\rm{GeV}/c$ when looking at the $z_h$ distribution and $z_h>0.2$ when looking at the $\ptpt$ distributions.

Figure \ref{fig:zh pT prim pi+ vs rho+} shows in the left panel the comparison between the $z_h$ and $\ptpt$ distributions for primary $\pi^+$ (dotted histogram) and $\rho^+$ mesons (continous histogram). The right panels show the corresponding distributions for $\pi^-$ and $\rho^-$. As can be seen the vector mesons carry typically larger fractions of the initial quark energy due to their larger mass. The squared hadron mass, indeed, enters in the exponential of the longitudinal splitting function, as can be seen from Eq. (\ref{eq:F_explicit vm}). When cutting on $z_h$, vector mesons turn out to be as many as the pseudoscalar mesons, in spite of $f_{vm}/f_{ps}$ being less than one. This can be seen from Tab. \ref{tab:fractions prim sec} which gives the fractions of the different hadron types produced in $u$ quark fragmentations. Concerning the $\ptpt$ distributions, vector mesons have typically smaller transverse momenta than pseudoscalar mesons. The reason for this difference is twofold. The genuine and largest contribution comes from the fact that in the string+${}^3P_0$ model the transverse momenta of the quarks that constitute the vector meson have on the average opposite directions whereas in the pseudoscalar case they lay on the average along the same direction. Concerning the other contribution, it is due to the fact that the largest proportion to the vector meson sample comes from rank 1 vector mesons, which have smaller $\ptpt$ with respect to larger ranks and for which the term $\exp\left[-\bl(m_h^2+\ptpt)/Z\right]/N_a(\varepsilon_h^2)$ in the splitting function, as a consequence of $m^2_{ps}<m^2_{vm}$, favors slightly smaller transverse momenta compared to pseudoscalar mesons when the cut $z_h>0.2$ is applied. The same features are seen for negative charged mesons. The genuine contribution is at variance with \verb|PYTHIA|, where the $Z$-integrated splitting function is the same for vector mesons and for primary pseudoscalar mesons. However, when a cut on $z_h$ is applied, \verb|PYTHIA| also produces primary pseudoscalar mesons with somewhat larger $\ptpt$ than vector mesons.

Figure \ref{fig:zh pT rho+ vs pi+ from rho+} shows the comparison of the $z_h$ and $\ptpt$ distributions of $\rho^{\pm}$ (continous histogram) with the corresponding distributions of $\pi^{\pm}$ (dotted histogram) produced in the $\rho^{\pm}$ decay. The decay pions have typically smaller fractional energies and smaller transverse momenta compared to those of their parent vector mesons. The tranverse momentum of a decay pion receives two contributions: it inherits part of the parent $\rho$ transverse momentum and a transverse kick with respect to the $\rho$ line of flight in the decay process. An approximate relation among the average transverse momentum of the decay pion $\langle \pt^2\rangle_{\pi}$, the transverse momentum $\langle \pt^2\rangle_{\rho}$ of the $\rho$ and the transverse momentum $\langle \pt^2\rangle_{\pi/\rho}\simeq M^2z_{\pi}(z_{\rho}-z_{\pi})/z_{\rho}^2$ that the pion receives in the decay process is $\langle \pt^2\rangle_{\pi}\simeq (z_{\pi}/z_{\rho})^2\langle \pt^2\rangle_{\rho}+ \langle\pt^2\rangle_{\pi/\rho}$.

The $z_h$ and $\ptpt$ distributions of secondary $\pi^{\pm}$ are shown in Fig. \ref{fig:zh pT prim pi+ vs pi+ from rho+}. The dotted histograms represent the distributions of primary pions, the thin continous histograms represent the the distributions of pions from the $\rho^{\pm}$ decay, and the continous ones represent the distributions of primary plus decay pions. In both cases the pions produced in the decay of vector mesons enhance the $z_h$ and the $\ptpt$ distributions at small values of fractional energies and transverse momenta.

The ratios between primary and primary plus secondary pseudoscalar mesons are given in Tab. \ref{tab:fractions prim sec} when the decay of all vector mesons is switched on. Among the charged pion sample, primary and secondary $\pi^+$ have roughly the same abundance. Secondary $\pi^-$, are on the contrary somewhat more than the primary ones due to the fact that the primary one are unfavored in $u$ quark jets. It is interesting to note that the kaon sample is dominated by primary mesons. Hence the properties of charged kaons, e.g. the Collins analysing power, arise mostly from direct production and are less affected by the decays of strange vector mesons.

Figure \ref{fig:pi+ from rho / prim. pi+} shows the ratio as function of $z_h$ and of $p_{\rm{T}}$ between secondary positive (red points) and negative (black triangles) pions coming from the decays of $\rho$ mesons and the primary ones. Circles represent positive pions whereas triangles represent negative pions.
Again, the contribution of $\rho$ meson decay increases with decreasing $z_h$. Also, the fraction of the secondary mesons is large at small transverse momenta and vanishes with increasing $p\T$.

Including the decays of all vector mesons, the $z_h$ and $\ptpt$ distributions of final pions are enhanced by a factor of two at small $z_h$ and $\ptpt$ with respect to the primary ones. The contribution of secondary pions is visible for $z_h\leq 0.6$ and for $\ptpt\leq 0.5\, (\rm{GeV}/c)^2$, as can be seen in Fig. \ref{fig:zh pT prim pi+ vs all pi+} for positive and negative pions.

\subsection{Analysing power}
\subsection*{Collins analysing power}
\begin{figure}[tb]
\centering
\begin{minipage}[t]{.6\textwidth}
  \includegraphics[width=0.8\linewidth]{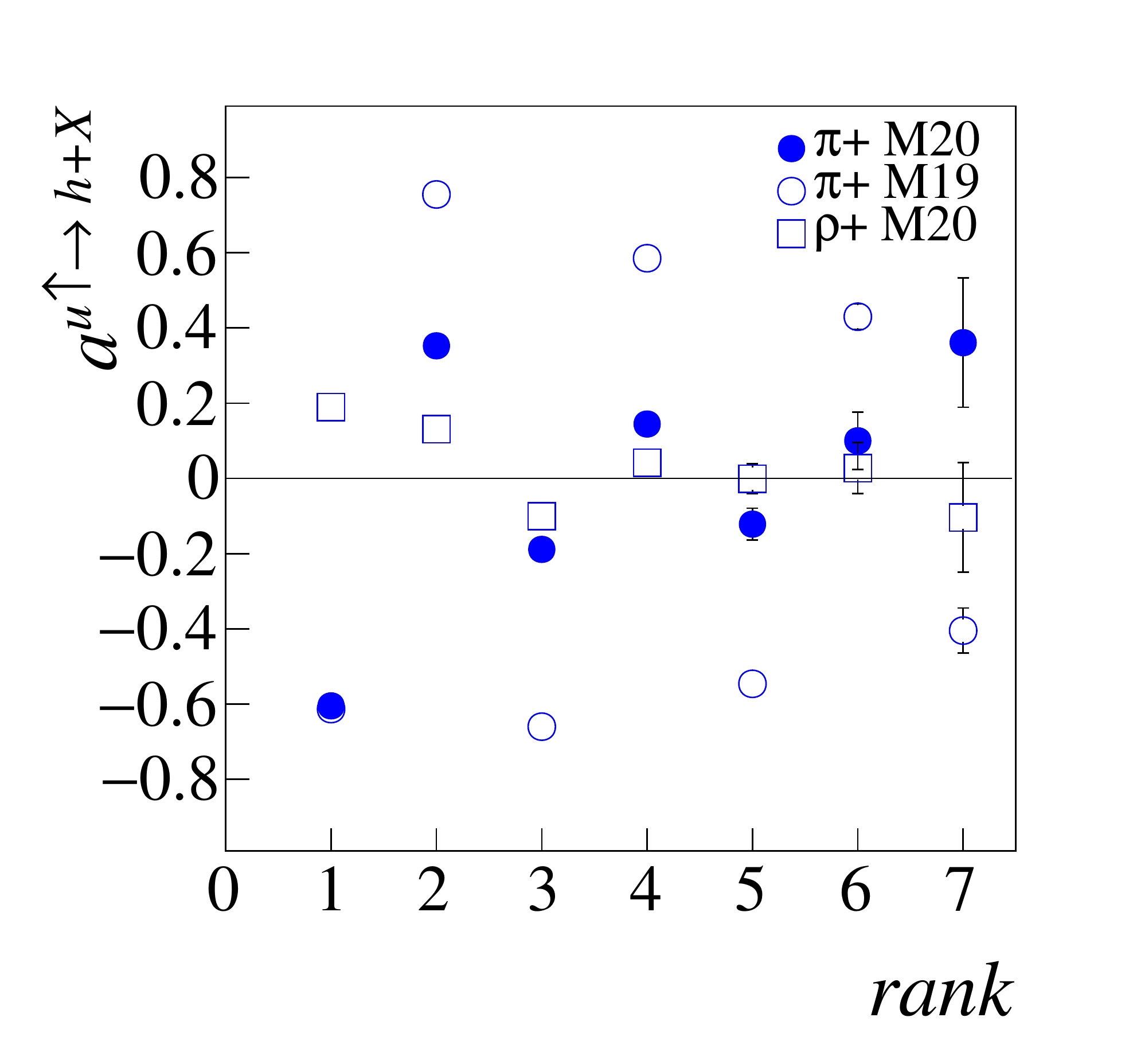}
  \end{minipage}
  \caption{Comparison between the Collins analysing power as function of rank for $\pi^+$ from M19 (open circles), directly produced $\pi^+$ from M20 (full circles) and $\rho^+$ (squares).}\label{fig:collins rank pi rho}
\end{figure}

From the calculation in Eq. (\ref{eq: ap vm / ap ps}) leading vector mesons are expected to have opposite Collins analysing power with respect to leading (directly produced) pions. This is also the result of the simulations. Figure \ref{fig:collins rank pi rho} shows the comparison between the Collins analysing power for positive pions as obtained with M19 (open circles) and for directly produced positive pions as obtained with M20 (full circles), with the analysing power for $\rho^+$ mesons (squares) as function of rank.
Clearly, first rank $\rho^+$ mesons have Collins analysing power opposite and reduced by a factor of $1/3$ with respect to first rank $\pi^+$ mesons. The factor $1/3$ is as expected from Eq. (\ref{eq: ap vm / ap ps}). Also, the analysing power of larger rank $\rho^+$ has the same sign as that of $\pi^+$ but smaller magnitude. The analysing power of $\pi^+$ mesons obtained with M20 decays much faster with rank. The reason is that a mixture of different particle types in the fragmentation chain leads to a faster decrease of the fragmenting quark polarization (see for instance Eq. (\ref{eq: polarization decay in M20})).

\begin{figure}[tb]
\centering
\begin{minipage}[t]{.6\textwidth}
  \includegraphics[width=1.0\linewidth]{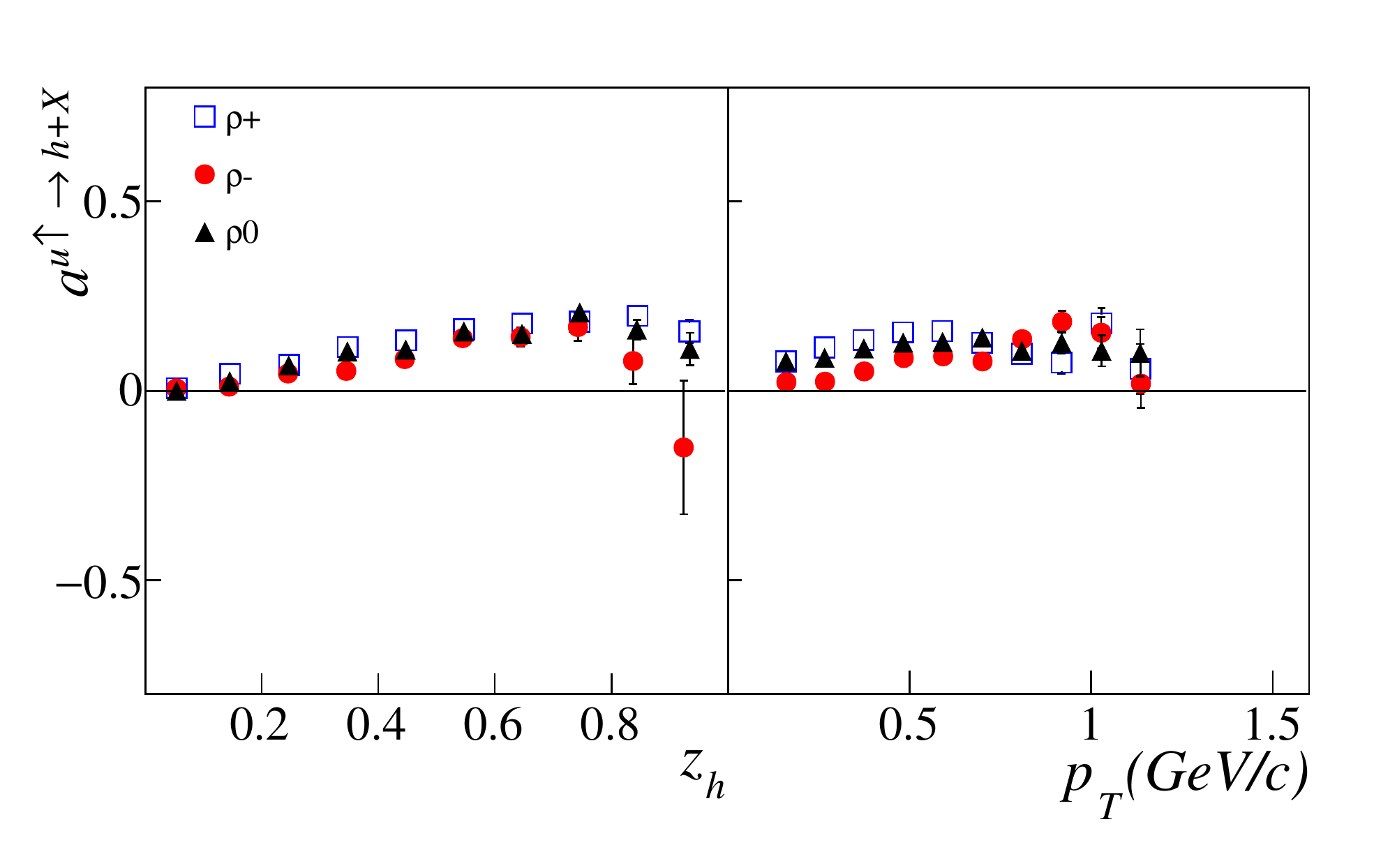}
  \end{minipage}
  \caption{Collins analysing power for $\rho^+$ (squares), $\rho^0$ (triangles) and $\rho^-$ (circles) as function of $z_h$ (left panel) and as function of $p\T$ (right panel).}\label{fig:collins for rho}
\end{figure}

All this can be seen more clearly in Fig. \ref{fig:collins for rho} which shows the Collins analysing power for $\rho$ mesons as function of $z_h$ (left panel) and as function of $p\T$ (right panel).
All $\rho$ mesons have positive analysing power, at variance with the primary pseudoscalar mesons. Indeed, $\rho^+$ and $\rho^0$ mesons are produced already at first rank which is characterized by a positive analysing power and this fixes the sign of the rank-integrated analysing power. On the contrary, $\rho^-$ mesons are produced from second rank either with negative analysing power after a first rank vector meson or with positive analysing power after a first rank pseudoscalar meson. Since the production of $\rho$ mesons is suppressed with respect to that of pions, the analysing power for $\rho^-$ is mostly determined from rank 2 $\rho^-$ mesons produced after a rank 1 pseudoscalar meson, and it is not very sensitive to the value of $|G_L|/|G_T|$.

The effect of vector meson production on the Collins analysing power for all hadrons (primary and secondary mesons) is the most important result of the simulation.
Vector mesons have a twofold contribution: they affect the Collins analysing power of primary pseudoscalar mesons because of the different propagation of the quark spin along the chain and they give a contribution to the analysing power of the secondary mesons because of their own Collins effect.

\begin{figure}[tb]
\centering
\begin{minipage}[t]{.48\textwidth}
  \includegraphics[width=1.0\linewidth]{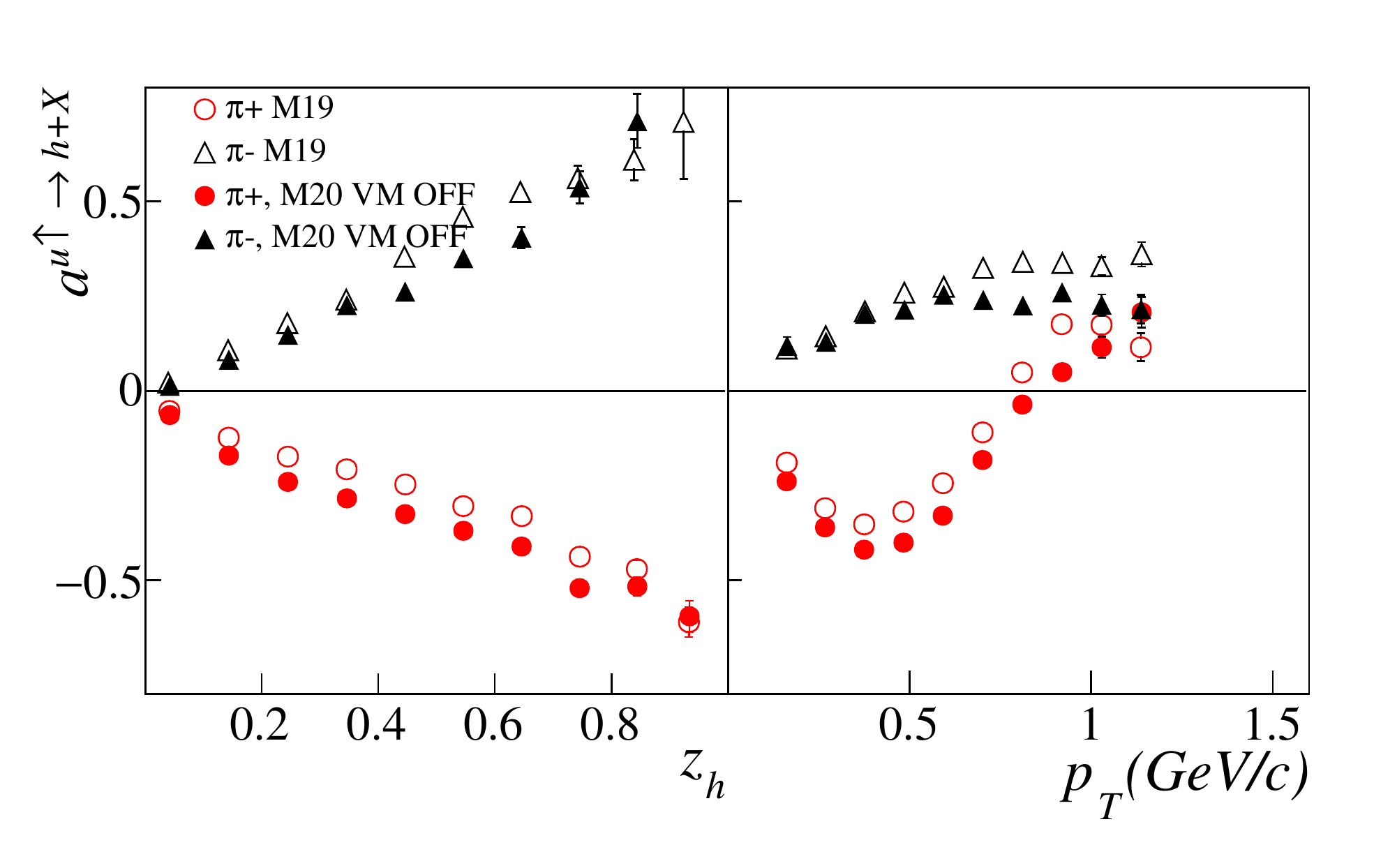}
  \end{minipage}
\begin{minipage}[t]{.48\textwidth}
  \includegraphics[width=1.0\linewidth]{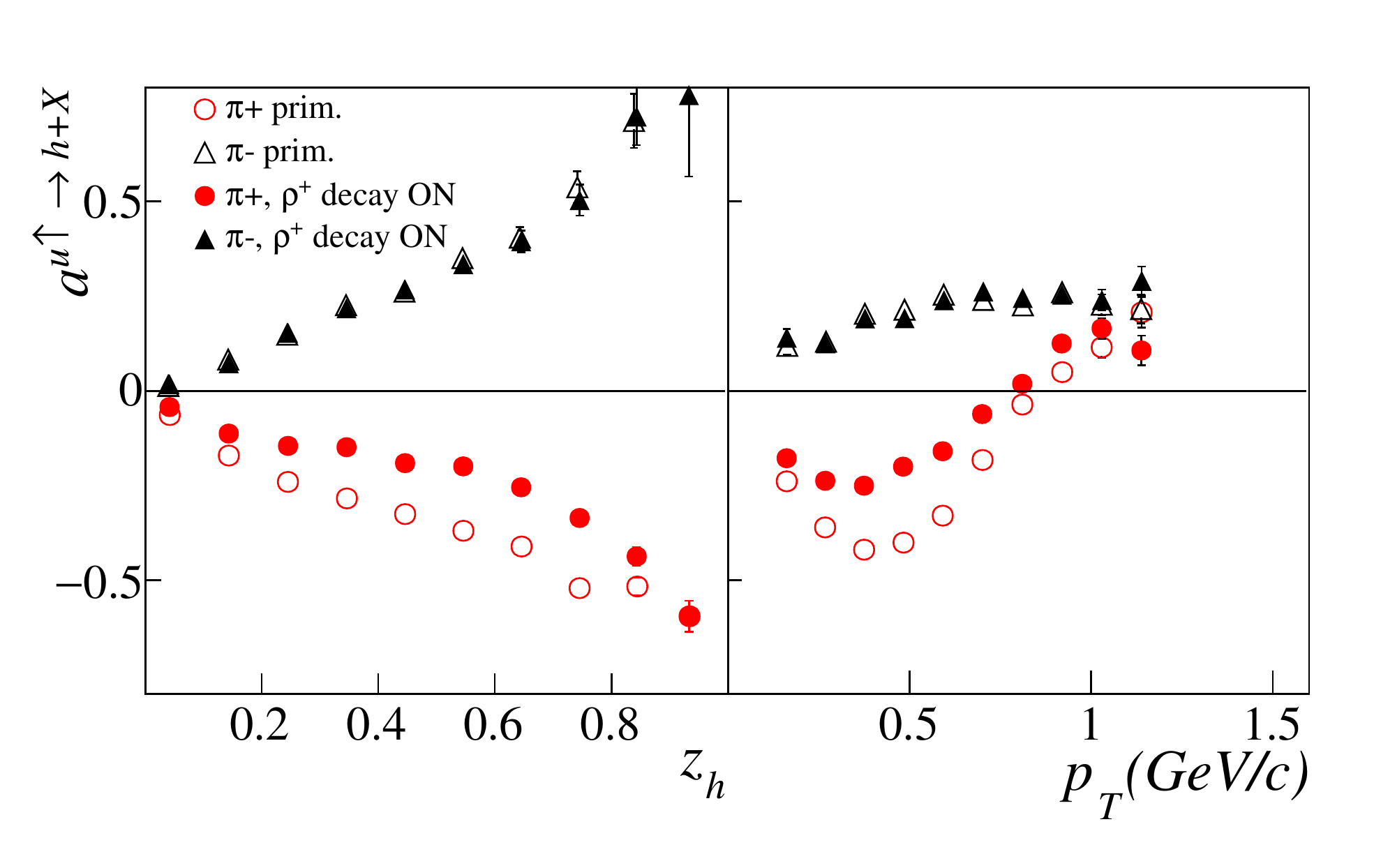}
  \end{minipage}
  \caption{Left: comparison between the Collins analysing power as function of $z_h$ and of $p\T$ for $\pi^+$ (circles) and $\pi^-$ (triangle) as obtained with M19 (open markers) and with M20 (full markers) when the vector meson decay is switched off. Right: effect of $\rho^+$ decay on the Collins analysing power.}
\label{fig:M19 vs M20 e effetto rhop pioni}
\end{figure}

\begin{table}
\centering
\begin{tabular}{ |p{3.5cm}|p{3.cm}|p{3.cm}| }
 \hline
 decay mode       &   $\pi^+$     & $\pi^-$             \\
 \hline
 \hline
 M19                                            &   $-0.251 \pm 0.004$       & $0.257 \pm 0.006$            \\
 M20 no decays                                     &   $-0.308 \pm 0.003$       & $0.218 \pm 0.005$             \\
 $\rho^+$                               &   $-0.178 \pm 0.003$       & $0.216 \pm 0.005$             \\
 $\rho^-$                               &   $-0.307 \pm 0.003$       & $0.172 \pm 0.004$             \\
 $\rho^0$                               &   $-0.210 \pm 0.003$       & $0.151 \pm 0.004$             \\
 $\rho^{\pm, 0}$                        &   $-0.136 \pm 0.003$       & $0.140 \pm 0.004$             \\
 all VM                                &   $-0.124 \pm 0.003$       & $0.124 \pm 0.003$             \\
 \hline
\end{tabular}
\caption{Average values of Collins analysing power for charged pions. For each hadron the cuts $z_h>0.2$ and $p\T>0.1\,\rm{GeV}/c$ have been applied.}\label{tab:collins ap averages}
\end{table}

The effect of vector meson production on the primary charged pions is shown in the left panel of Fig. \ref{fig:M19 vs M20 e effetto rhop pioni}. It compares the Collins analysing power for $\pi^+$ (circles) and $\pi^-$ (triangles) as obtained with M19 (open markers) and with M20 (full markers) when vector meson decays are switched off. The absolute value of the Collins effect for positive pions is slightly larger in M20 than in M19 whereas for negative pions the reverse is true. This is clear also from the average values given in Tab. \ref{tab:collins ap averages} and it can be understood considering the first two splittings in the fragmentation chain of M20. Rank 1 positive pions are produced with negative analysing power. At rank 2 they are produced either with positive analysing power after a pseudoscalar meson or with negative analysing power after a first rank vector meson. Since the first case is favored, the sign of rank 2 mesons is positive, and this enhances the overall analysing power. Similar considerations apply to negative pions which are produced from rank 2 with positive or negative analysing power depending whether the first rank meson is a pseudoscalar or a vector particle.

\begin{figure}[tb]
\centering
\begin{minipage}[t]{.6\textwidth}
  \includegraphics[width=1.0\linewidth]{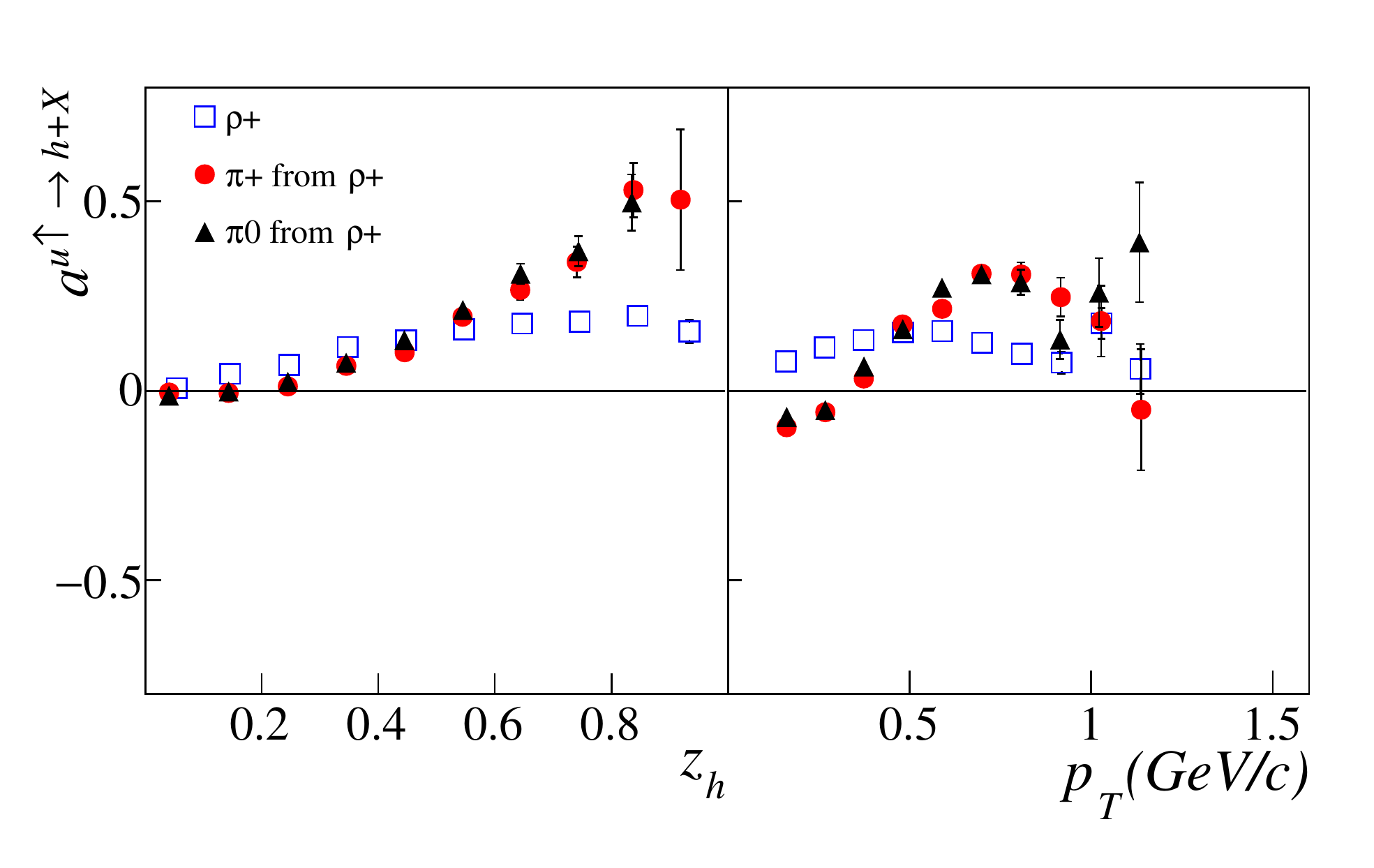}
  \end{minipage}
  \caption{Comparison between the Collins analysing power as function of $z_h$ (left panel) and as function of $p\T$ (right panel) for $\rho^+$ positive (squares) and the analysing power of the decay $\pi^+$ (circles) and $\pi^0$ (triangles).}\label{fig:collins rho+ vs decay pi+ pi0}
\end{figure}

In the decay process, the Collins effect of vector mesons is transferred to the decay products. This is shown in Fig. \ref{fig:collins rho+ vs decay pi+ pi0} for $\rho^+$ (rectangles) and the decay $\pi^+$ (circles) and $\pi^0$ (triangles) as function of $z_h$ (left panel) and as function of $p\T$ (right panel). The decay pions have the same analysing power since the decay process is invariant under parity, somewhat different with respect to the analysing power of the parent $\rho^+$.
Indeed requiring decay pions with large $z_h$ (note that for the pions it is $z_{\rho}>z_{\pi}$), selects preferably $\rho^+$ mesons with longitudinal linear polarization which, as already mentioned, have opposite Collins effect with respect to the leading pions. Instead, when looking at the Collins analysing power for $\rho^+$ their polarization states are summed over giving an analysing power of $\rho^+$ which is a factor of $3$ smaller than that of the $\pi^+$.
The differences between the $\rho^+$ and the decay pion analysing powers at large transverse momenta can be understood with similar considerations. Decay pions with large transverse momenta can be produced from a first rank $\rho^+$ linearly polarized along $\Mx$ (hence along its $\pt$) or from $\rho^+$ mesons with large transverse momenta. In the former case, the relative transverse momentum of the pion in the decay adds with the transverse momentum of $\rho^+$ meson, which also has positive analysing power as can be seen from Eq. (\ref{eq: F vm M}). In the latter case, it can be seen from the classical string+${}^3P_0$ picture that a $\rho^+$ with a large transverse momentum can be produced at second or higher rank when the transverse momenta of the constituent quarks are aligned along the same direction. This happens if the meson is linearly polarized along $\textbf{N}$, namely along the normal to its production plane. In this case, from Eq. (\ref{eq: F vm N}), we know that the $\rho^+$ has exactly the same Collins effect as a pseudoscalar meson, which for rank two is positive. This explains the larger analysing power of $\pi^+$ and of $\pi^0$ at large transverse momenta.
A further interesting feature is that the analysing power of decay pions becomes negative for $p\T<0.4\,\rm{GeV}/c$. This happens likely when the pion transverse momentum in the $\rho^+$ rest frame is larger than transverse momentum of the $\rho^+$ in the string rest frame. 
The same features are seen also in the decay of $\rho^-$ and $\rho^0$ mesons.

The effect of $\rho^+$, $\rho^-$ and $\rho^0$ decays have been studied separately and understood. To summarize, all $\rho$ decays reduce the analysing power of positive and negative pions, the most relevant reduction being that on the $\pi^+$ analysing power due to the $\rho^+$ decays, as can also be seen in Tab. \ref{tab:collins ap averages}.

\begin{figure}[tb]
\begin{minipage}[t]{1.0\textwidth}
\centering
  \includegraphics[width=0.6\linewidth]{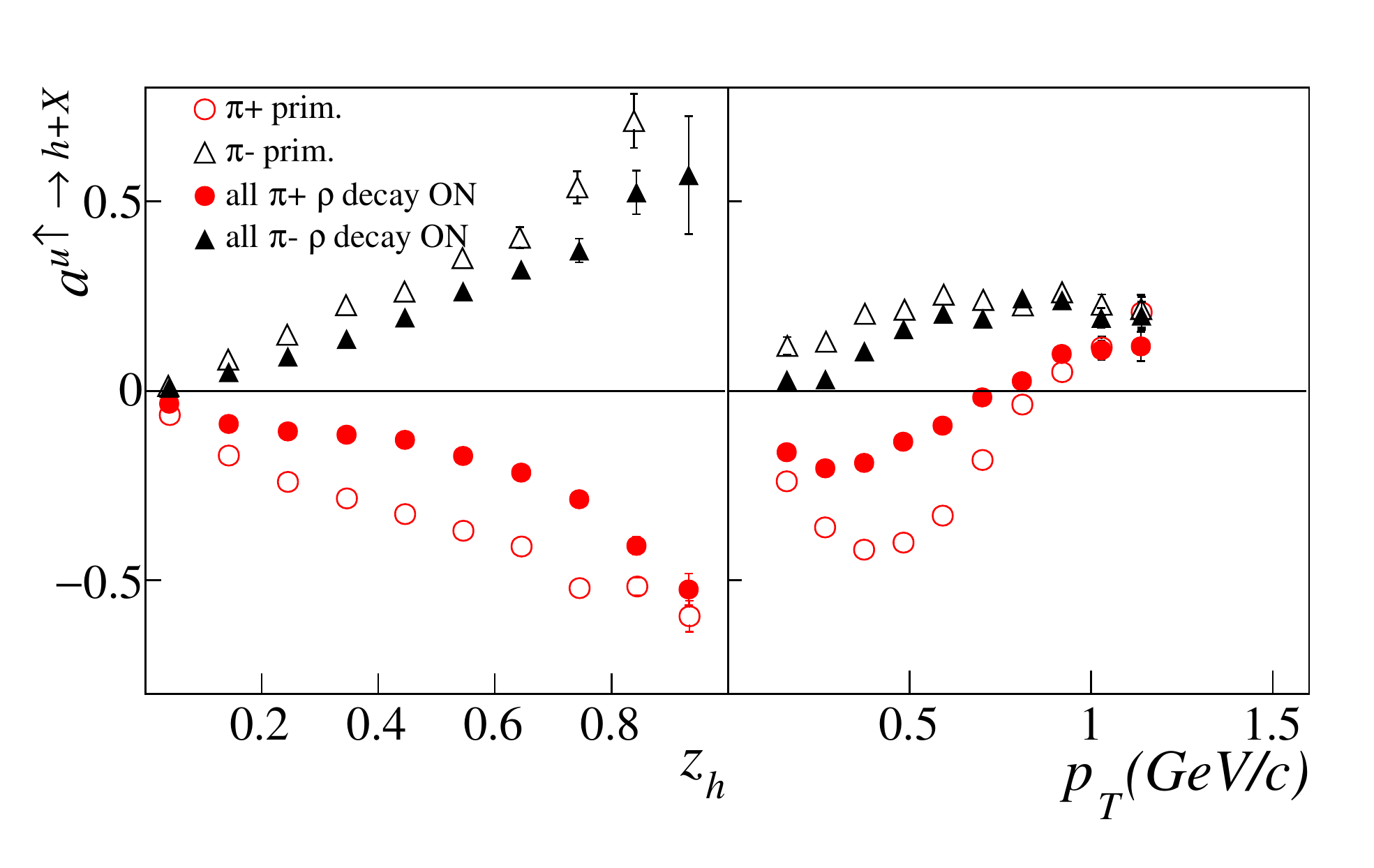}
  \end{minipage}
  \caption{The Collins analysing power of final positive (circles) and negative (triangle) pions as function of $z_h$ (left panel) and of $p\T$ (right panel). The open markers represent primary pions when the decays of vector mesons are switched off whereas the closed markers represent all pions when switching on the $\rho$ mesons decay.}\label{fig:effetto rho pioni}
\end{figure}

Switching on the decays of all $\rho$ mesons one gets the results shown in Fig. \ref{fig:effetto rho pioni}. Open points show the analysing power of primary charged pions whereas closed points show the analysing power of primary and secondary pions. The effect is large for both charges and as function of both $z_h$ and $p\T$. The reduction is stronger in the case of positive pions where the secondary ones are characterized by an analysing power with opposite sign with respect to that of the primary ones. From Tab. \ref{tab:collins ap averages} one can see that the analysing power of primary $\pi^+$ is reduced by $55\%$ and that of primary $\pi^-$ is reduced by $35\%$.

\begin{figure}[tb]
\begin{minipage}[t]{.48\textwidth}
  \includegraphics[width=1.0\linewidth]{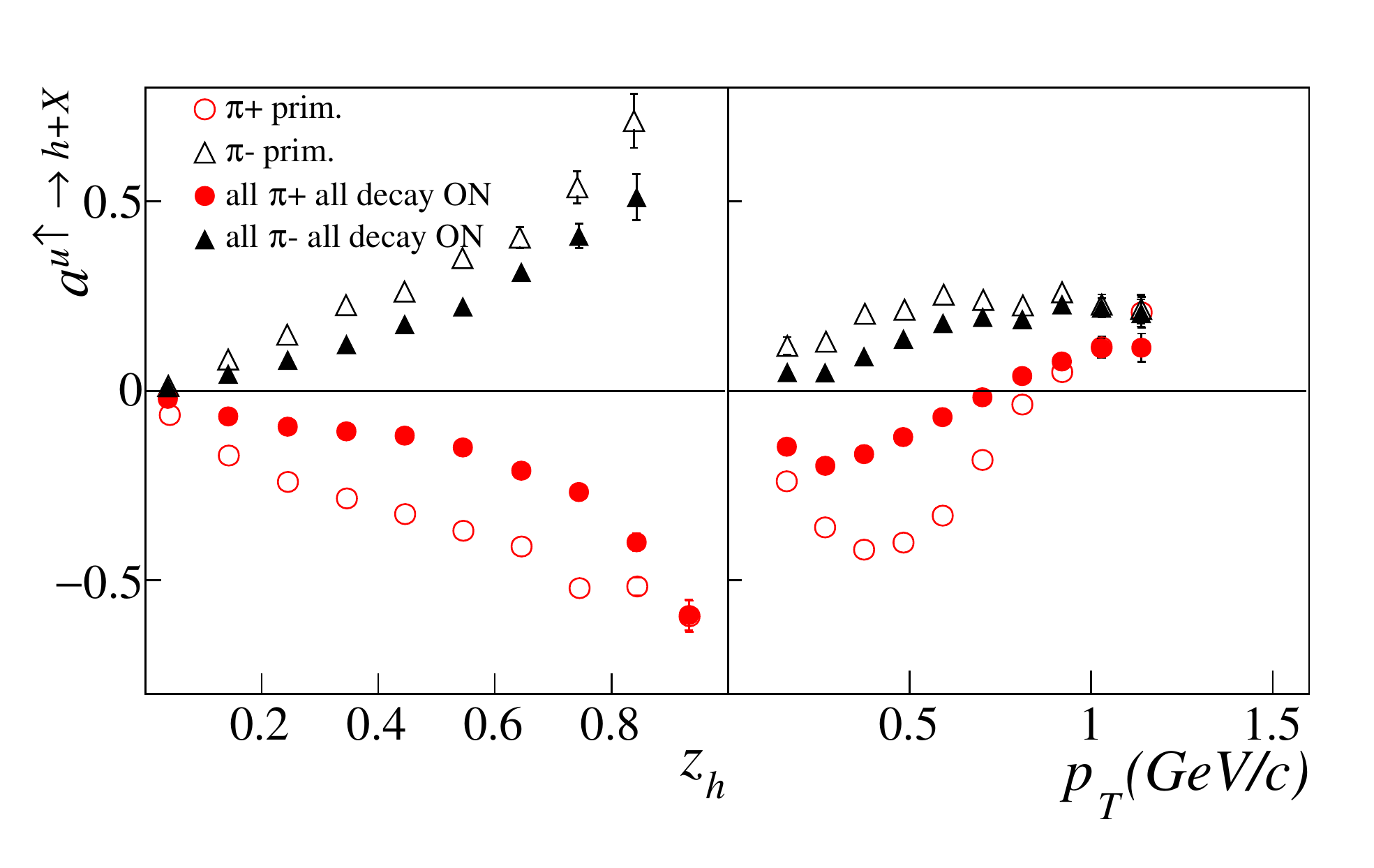}
  \end{minipage}
\begin{minipage}[t]{.48\textwidth}
  \includegraphics[width=1.0\linewidth]{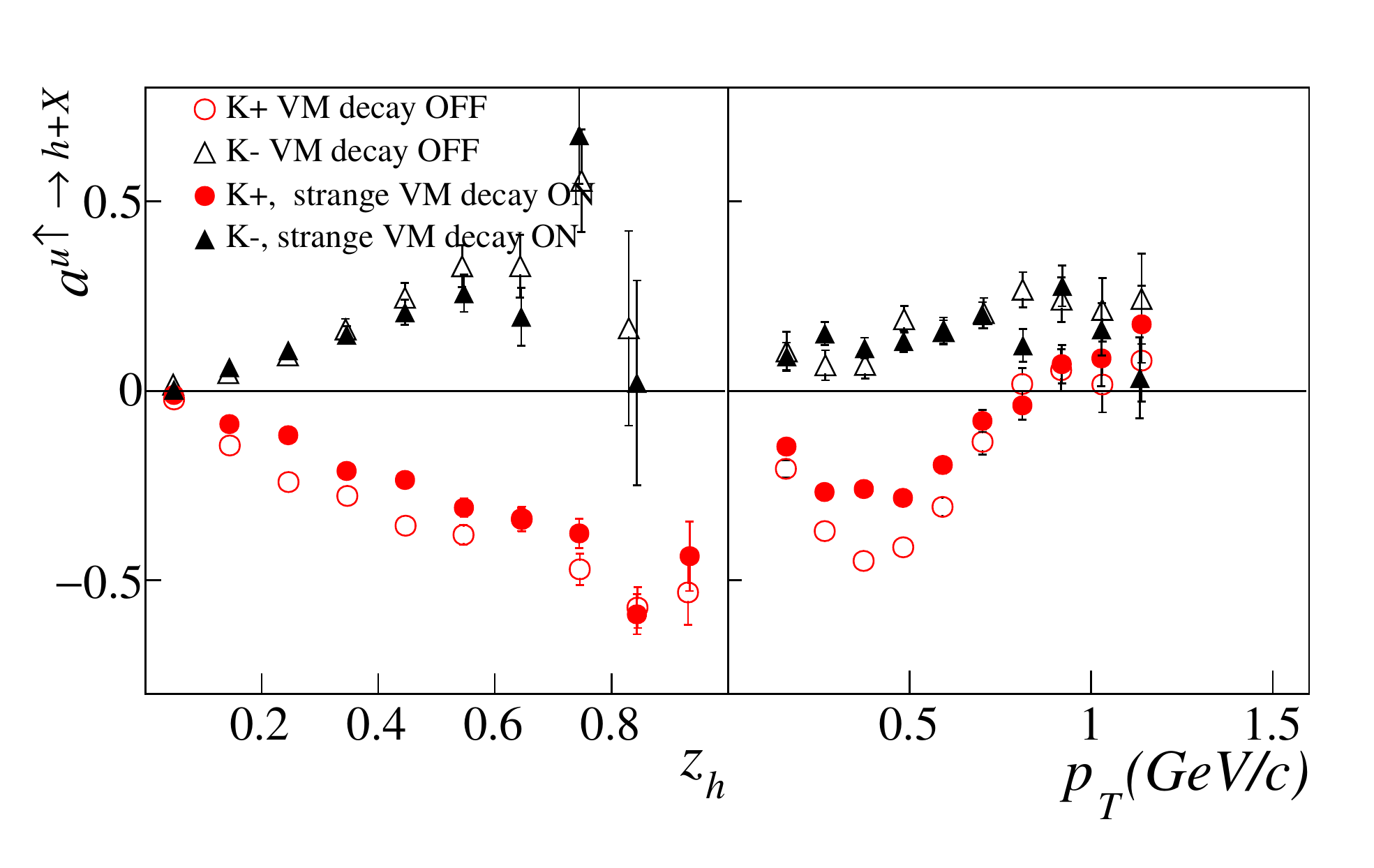}
  \end{minipage}
  \caption{Left: effect of vector mesons decay on the Collins analysing power of $\pi^+$ (circles) and $\pi^-$ (triangle). The open markers refer to primary pions when the decays of vector mesons are switched off whereas the closed markers when switching on all vector mesons decay. Right: same for positive (circles) and negative (triangles) kaons.}\label{fig:effetto all vm pioni}
\end{figure}

Switching on $\omega$ and $\phi$ decays does not change sensibly the Collins analysing power of charged pions. Neither do the decays of strange vector mesons. The final result obtained by switching on all vector meson decays is shown in Fig. \ref{fig:effetto all vm pioni}. For comparison, the analysing power of primary pions is also shown.

It is important to note that M20 produces a different average analysing power for primary $\pi^+$ and for primary $\pi^-$, but after switching on vector meson decays the absolute values of the $\pi^+$ analysing power and the $\pi^-$ analysing power are the same. This similarity, obtained also with M19 and seen in the experimental data \cite{interplay}, is not related to the values of the free parameters used in simulations but it is a prediction of the model itself.

The same analysis has been done for charged kaons and the same considerations as for pions hold. The effect of all vector meson decays on the Collins analysing power is shown in the right panel of Fig. \ref{fig:effetto all vm pioni}. In this case, the effect of vector meson decay is larger for $K^+$ and the mesons that provide the larger contribution are $K^{*+}$ and $K^{*0}$.

\subsection*{Sensitivity to the free parameters}
\begin{figure}[tb]
\centering
\begin{minipage}{0.48\textwidth}
\centering
  \includegraphics[width=1.0\linewidth]{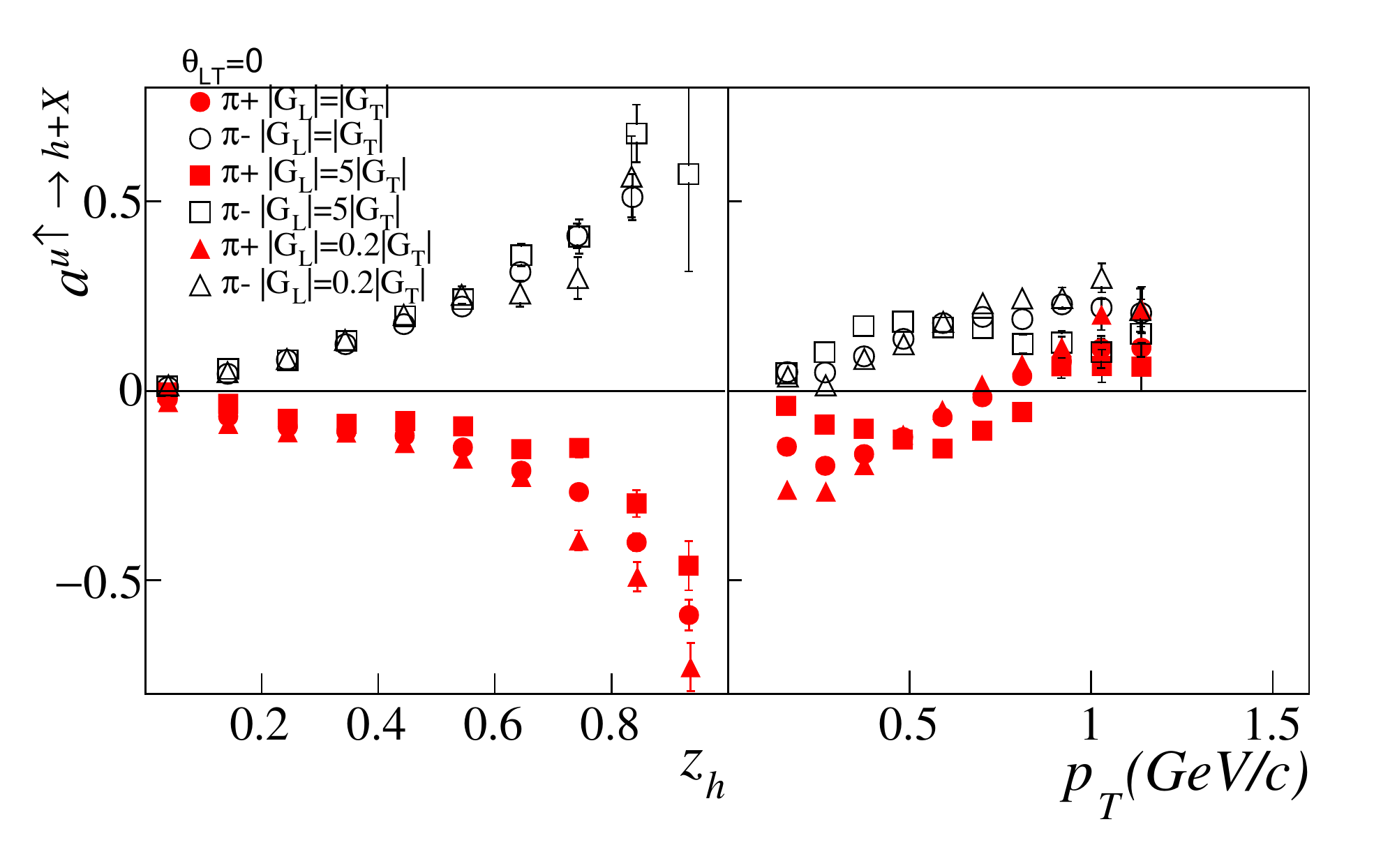}
  \end{minipage}
 \begin{minipage}{0.48\textwidth}
 \centering
    \includegraphics[width=1.0\linewidth]{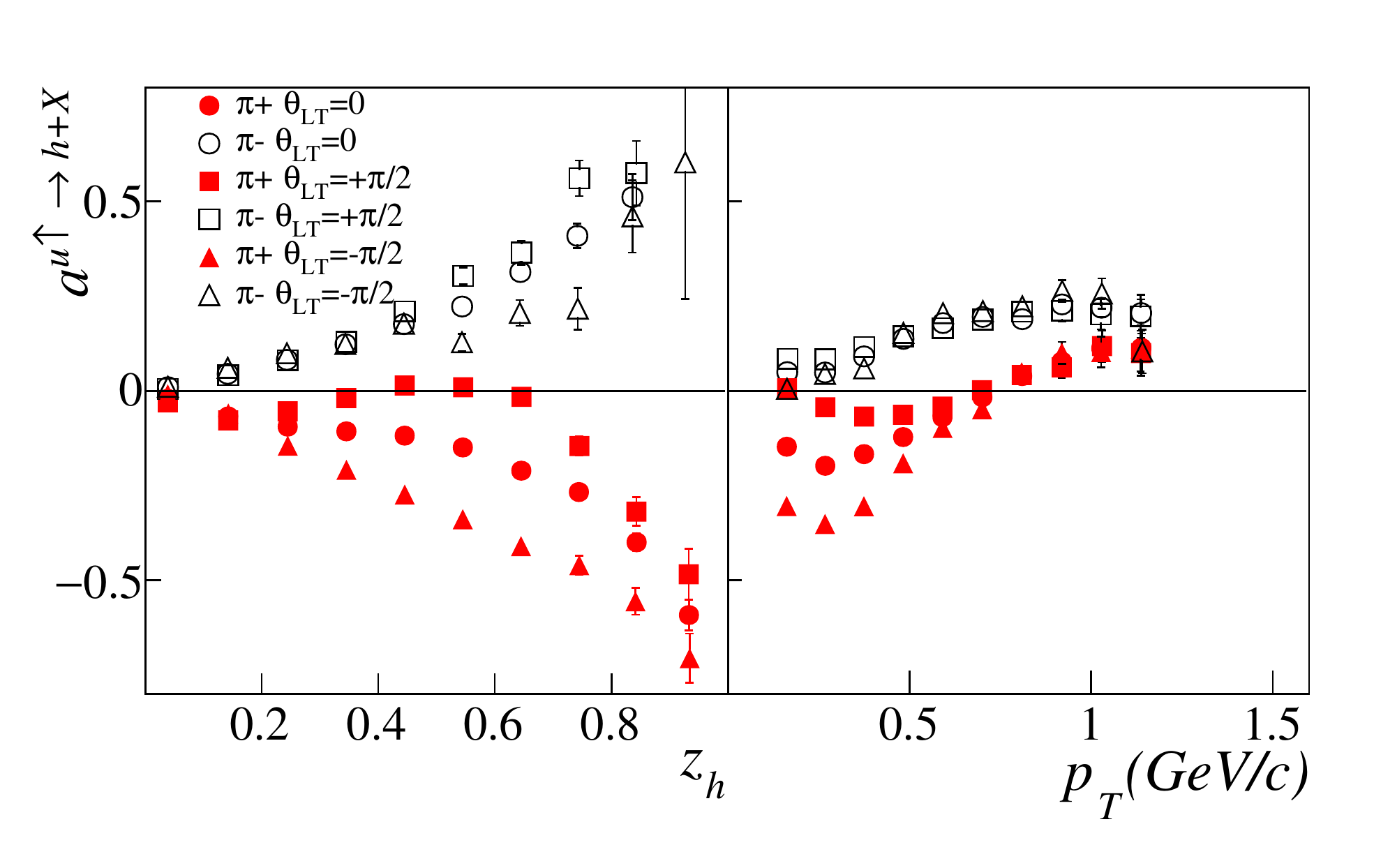}
 \end{minipage}
 \begin{minipage}{0.48\textwidth}
 \centering
    \includegraphics[width=1.0\linewidth]{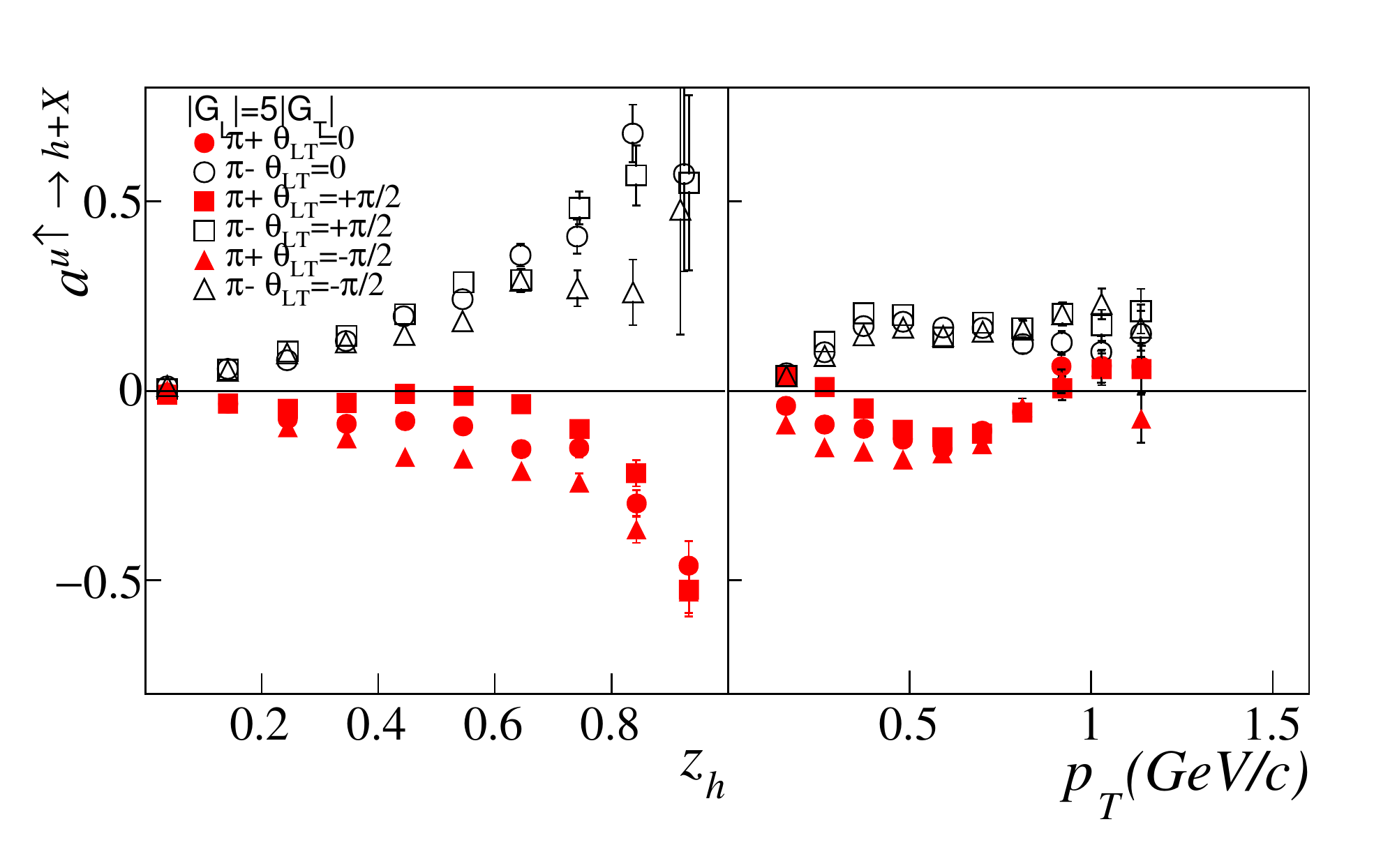}
 \end{minipage}
 \begin{minipage}{.48\textwidth}
  \includegraphics[width=1.0\linewidth]{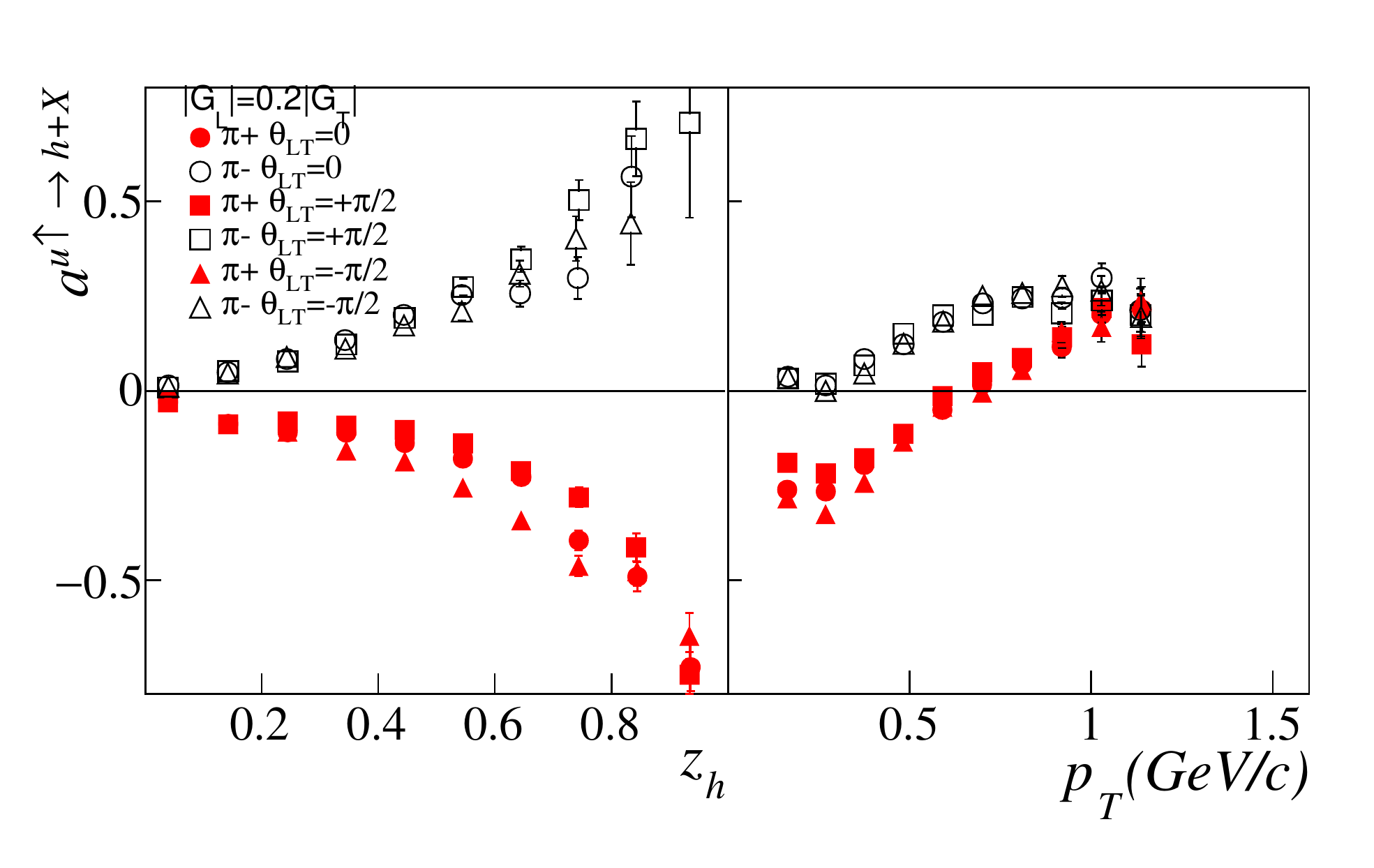}
  \end{minipage}
  \caption{Upper row, left: Collins analysing power for $\pi^+$ (full markers) and $\pi^-$ (open markers) for different values of $|G_L|/|G_T|$. Upper row, right: effect of $\theta_{LT}$ on the Collins analysing power for $|G_L|/|G_T|=1$.  Bottom row, left: effect of $\theta_{LT}$ on the Collins analysing power for $|G_L|/|G_T|=5$. Bottom row, right: effect of $\theta_{LT}$ on the Collins analysing power for $|G_L|/|G_T|=0.2$.}\label{fig:effetto G collins ap}
\end{figure}

All the previous results have been obtained using $|G_L|/|G_T|=1$ and $\theta_{LT}=0$. Here the sensitivity of the analysing power to the variations of these parameters is investigated. For these simulations the decays of all vector meson species are switched on.
The plot in the upper row, left, of Fig. \ref{fig:effetto G collins ap} shows the effect of variations of $|G_L|/|G_T|$ for positive (full markers) and negative (open markers) pions. Triangles are obtained choosing $|G_L|/|G_T|=0.2$, circles with $|G_L|/|G_T|=1$ and squares with $|G_L|/|G_T|=5$. The first choice favors the coupling of quarks to vector mesons with transverse linear polarization. The second one gives an equal weight to longitudinal and transverse linear polarizations. The third favors the coupling to vector mesons with longitudinal linear polarization. The effect of varying $|G_L|/|G_T|$ is seen mostly for positive pions with $z_h>0.6$ or with $p\T<0.5\,\rm{GeV}/c$. For $|G_L|/|G_T|=5$ the analysing power for $\pi^+$ is slightly reduced whereas that for $\pi^-$ is slightly increased. Taking $|G_L|/|G_T|=0.2$ increases slightly the analysing power of $\pi^+$ whereas that of $\pi^-$ does not change.
The plot in the upper row, right, in Fig. \ref{fig:effetto G collins ap} shows the effect of changing $\theta_{\rm{LT}}$ from $0$ to $\pm \pi/2$ for $|G_L|/|G_T|=1$. The effect is very strong for positive pions and it is opposite for positive and negative pions, namely increasing $\theta_{\rm{LT}}$ decreases the analysing power for $\pi^+$ and increases that of $\pi^-$ and vice-versa. The bottom plots in the same figure show the effect of $\theta_{LT}$ on the Collins analysing power for charged pions when $|G_L|/|G_T|=5$ (left) and $|G_L|/|G_T|=0.2$ (right). In the former case the effect is similar to $|G_L|/|G_T|=1$ whereas in the latter case it the effect of the oblique polarization, being proportional to $|G_L|/|G_T|$ as can be seen from the matrix element $\hat{\rho}_{\rm{ml}}(h)$ in Eq. (\ref{eq:rho vm matrix elements}), is smaller.
The average values of $\pi^+$ and $\pi^-$ analysing powers are given in Tab. \ref{tab:effect theta G collins ap averages} for the considered values of $|G_L|/|G_T|$ and $\theta_{LT}$.

\begin{table}
\centering
\begin{tabular}{ |p{2.7cm}|p{0.5cm}|p{2.5cm}|p{2.5cm}|p{2.5cm}| }
 \hline
                       &                       &    $\theta_{LT}=-\pi/2$                                  & $\theta_{LT}=0$                   &  $\theta_{LT}=\pi/2$                   \\
 \hline
 $|G_L|/|G_T|=0.2$     & $\pi^+$               &   $-0.170 \pm 0.004$                                  &  $-0.141 \pm 0.004$               &   $-0.112 \pm 0.004$                      \\
                       & $\pi^-$               &   $\phantom{+}0.121 \pm 0.005$                                   &  $\phantom{+}0.128 \pm 0.005$     &   $\phantom{+}0.127 \pm 0.005$                               \\
 \hline
  $|G_L|/|G_T|=1$      & $\pi^+$               &    $-0.225 \pm 0.004$                                  &  $-0.124 \pm 0.003$               &   $-0.037 \pm 0.004$                    \\
                       & $\pi^-$                &    $\phantom{+}0.121 \pm 0.005$                                    &  $\phantom{+}0.124 \pm 0.003$     &   $\phantom{+}0.140 \pm 0.005$            \\
 \hline
 $|G_L|/|G_T|=5$       & $\pi^+$               &   $-0.140 \pm 0.004$                                 &  $-0.093 \pm 0.004$               &   $-0.043 \pm 0.004$                       \\
                       & $\pi^-$              &   $\phantom{+}0.127 \pm 0.005$                                  &   $\phantom{+}0.136 \pm 0.005$    &   $\phantom{+}0.155 \pm 0.005$             \\
 \hline
\end{tabular}
\caption{Average values of the Collins analysing power for charged pions for different values of $|G_L|/|G_T|$ and $\theta_{LT}$. For each pion the cuts $z_h>0.2$ and $p\T>0.1\,\rm{GeV}/c$ have been applied. All vector meson decays have been switched on.}\label{tab:effect theta G collins ap averages}
\end{table}

\subsection*{Dihadron asymmetry}
For the calculation of the dihadron asymmetry pairs of oppositely charged hadrons in the same data sample are considered, and as angle characterizing the pair the azimuthal angle of the vector $\textbf{R}_T$ has been used. The kinematic cuts $z_{h}>0.1$ and $x_F>0.1$ for each hadron of the pair, $|\textbf{p}_i|>3\,\rm{GeV/c}$ for pions, $|\textbf{p}_i|>10\,\rm{GeV/c}$ for kaons and $R\T>0.07\,\rm{GeV/c}$, are applied in analogy with the COMPASS analysis \cite{compass-dihadron}.
\begin{figure}[tb]
\centering
\begin{minipage}[t]{.5\textwidth}
  \includegraphics[width=1.0\linewidth]{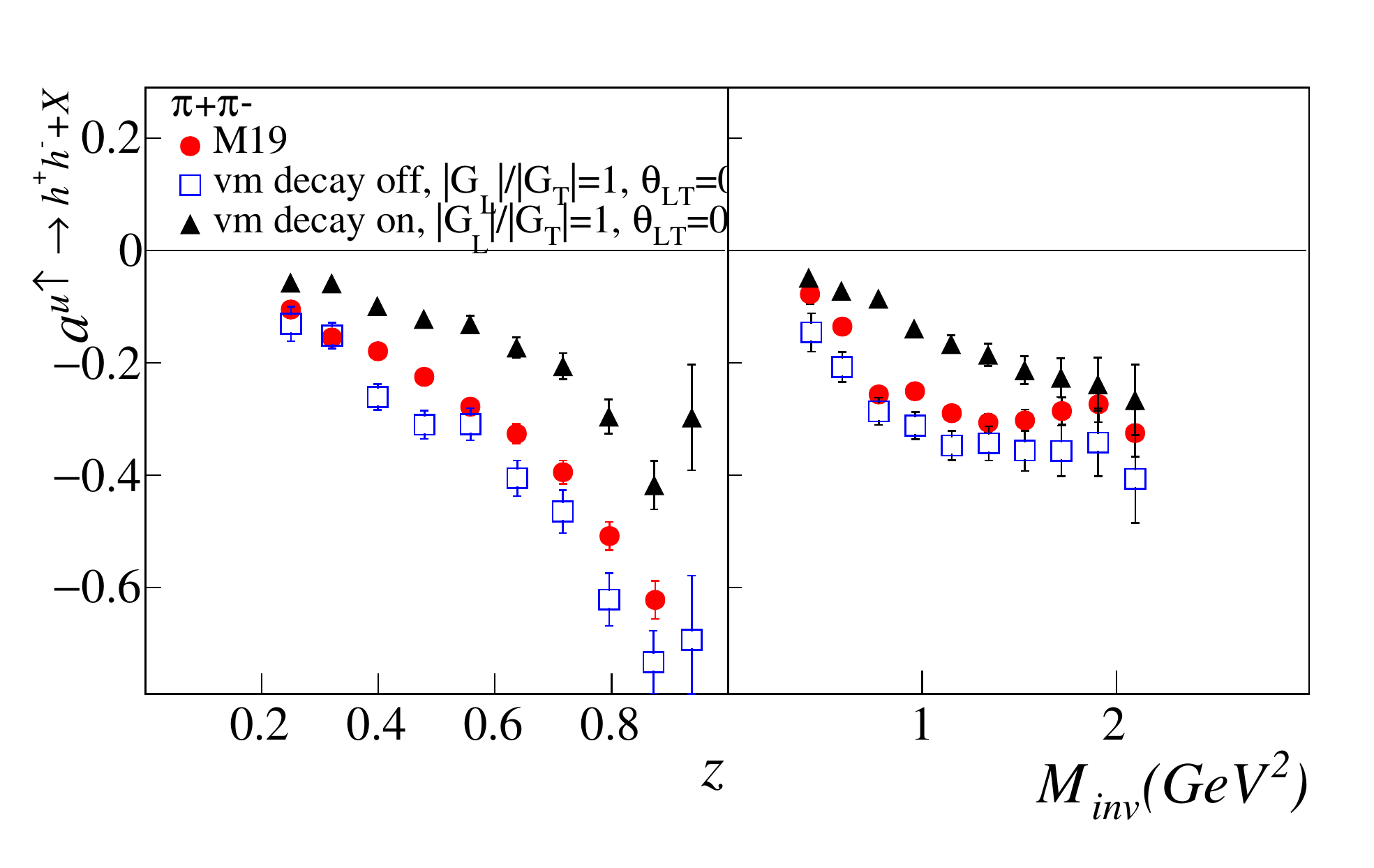}
\end{minipage}
  \caption{Dihadron analysing power for charged pion pairs as obtained with M19 (circles), for primary pions as obtained with M20 (squares) for $|G_L|/|G_T|=1$ and after switching on vector meson decays (triangles).}\label{fig:2h M19 vs M20}
\end{figure}

The dihadron analysing power as function of $z=z_{h_1}+z_{h_2}$ and of the invariant mass $M_{inv}$ is shown in Fig. \ref{fig:2h M19 vs M20} for the pions pairs obtained with M19 (circles), with M20 when vector meson decays are switched off (squares) and after switching on all decays (triangles). The values of the parameters used here are $|G_L|/|G_T|=1$ and $\theta_{LT}=0$.
The vector meson introduction slightly increases the dihadron analysing power as clear also from Tab. \ref{tab:2h ap averages} where the corresponding average values are shown.
\begin{table}
\centering
\begin{tabular}{ |p{3.5cm}|p{3.cm}|p{3.cm}| }
 \hline
        &   $\pi^+\pi^-$     \\
 \hline
 \hline
 M19                                            &   $-0.246 \pm 0.005$       \\
 M20 no decays                                     &   $-0.306 \pm 0.010$       \\
 all VM                                &   $-0.111 \pm 0.005$       \\
 \hline
\end{tabular}
\caption{Average values of dihadron analysing power for charged pions obtained using the parameters $|G_L|/|G_T|=1$ and $\theta_{LT}=0$.}\label{tab:2h ap averages}
\end{table}
When vector meson decays are switched on, the analysing power is diluted due to the fact that the decay process is invariant under $\textbf{R}\T\rightarrow -\textbf{R}\T$ and thus the decay mesons do not contribute to the dihadron analysing power. From Tab. \ref{tab:2h ap averages} one can see the dihadron analysing power for primary pions is diluted of $55\%$.

\begin{figure}[tb]
\centering
\begin{minipage}[t]{.48\textwidth}
  \includegraphics[width=1.0\linewidth]{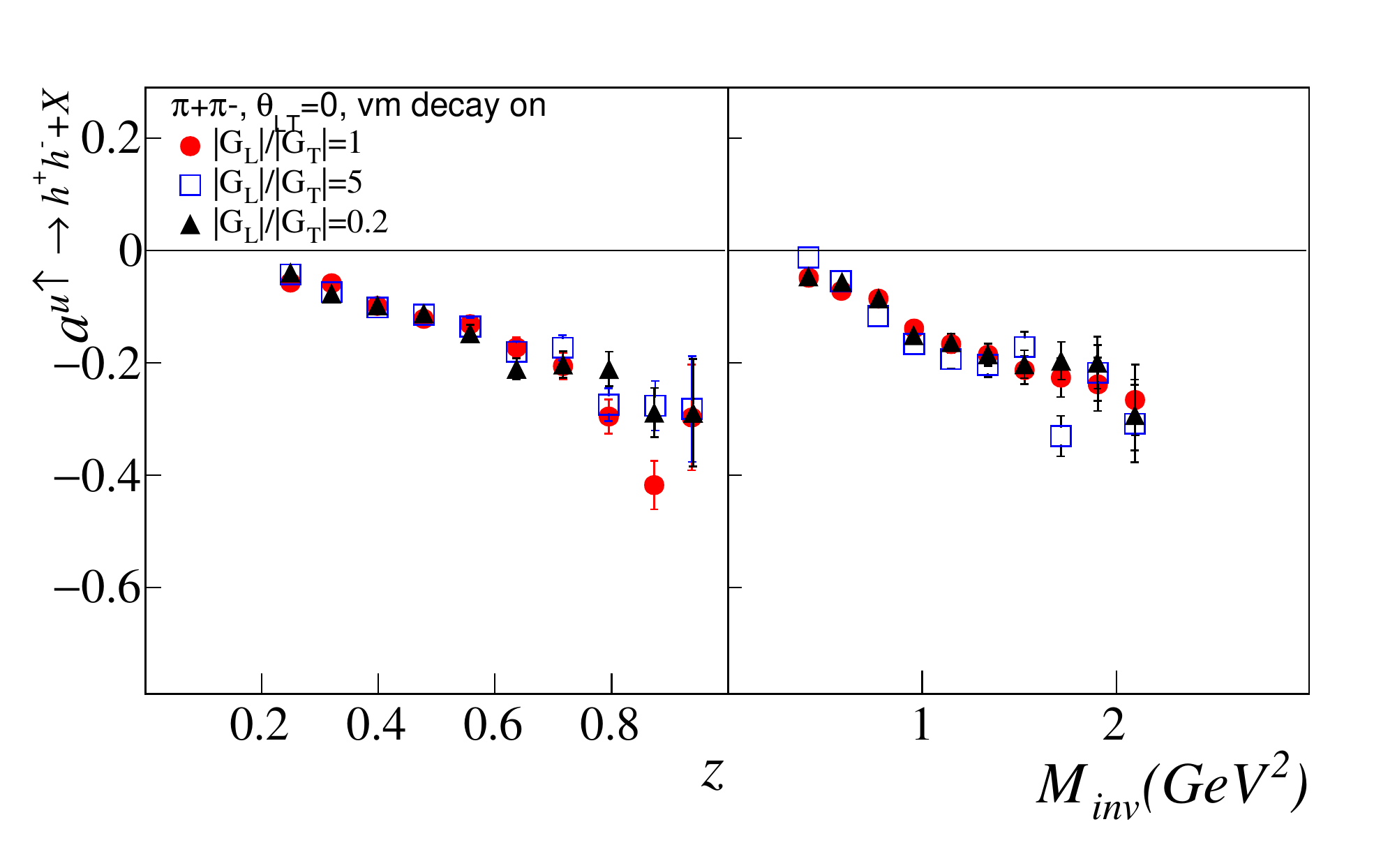}
\end{minipage}
\begin{minipage}[t]{.48\textwidth}
  \includegraphics[width=1.0\linewidth]{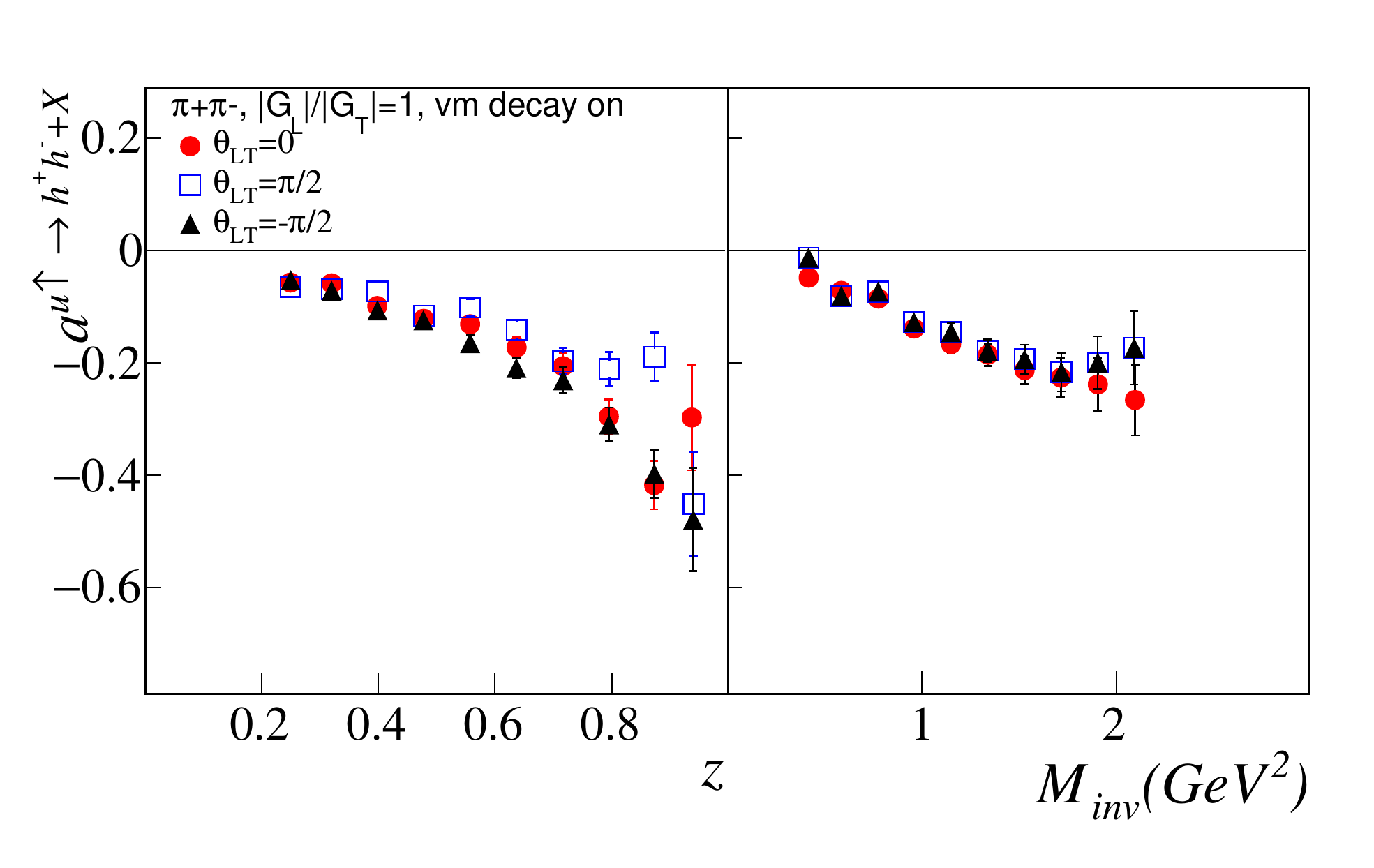}
\end{minipage}  
  \caption{Left: the dihadron analysing power for oppositely charged pions for different values of $|G_L|/|G_T|$. Right: effect of $\theta_{LT}$ on the dihadron analysing power for $|G_L|/|G_T|=1$.}\label{fig:2h ap G theta effect}
\end{figure}

It is also interesting to look at the effect of changing the value of $|G_L|/|G_T|$ and the value of $\theta_{LT}$ on the dihadron analysing power, which is shown in the left and right panels of Fig. \ref{fig:2h ap G theta effect}. The analysing power does not depend much on the value of $|G_L|/|G_T|$. Some effect is seen for $z>0.8$ when $\theta_{LT}=\pi/2$. However the overall dilution effect is not sensitive neither on the size of global Collins effect of the vector meson nor on its polarization.

The average value of the dihadron analysing power turns out to be smaller than that of the Collins analysing power for positive pions with the same parameter setting. This is at variance with the experimental result. Still it must be reminded that these results are obtained with $\kperp=0$. As known, for non vanishing $\kperp$, only the Collins analysing power is reduced. Also, a different setting of the free parameters can change the relative magnitudes of the analysing powers.

\subsection{Comparison with experimental results}
\begin{figure}[tb]
\centering
\begin{minipage}[t]{.48\textwidth}
  \includegraphics[width=1.0\linewidth]{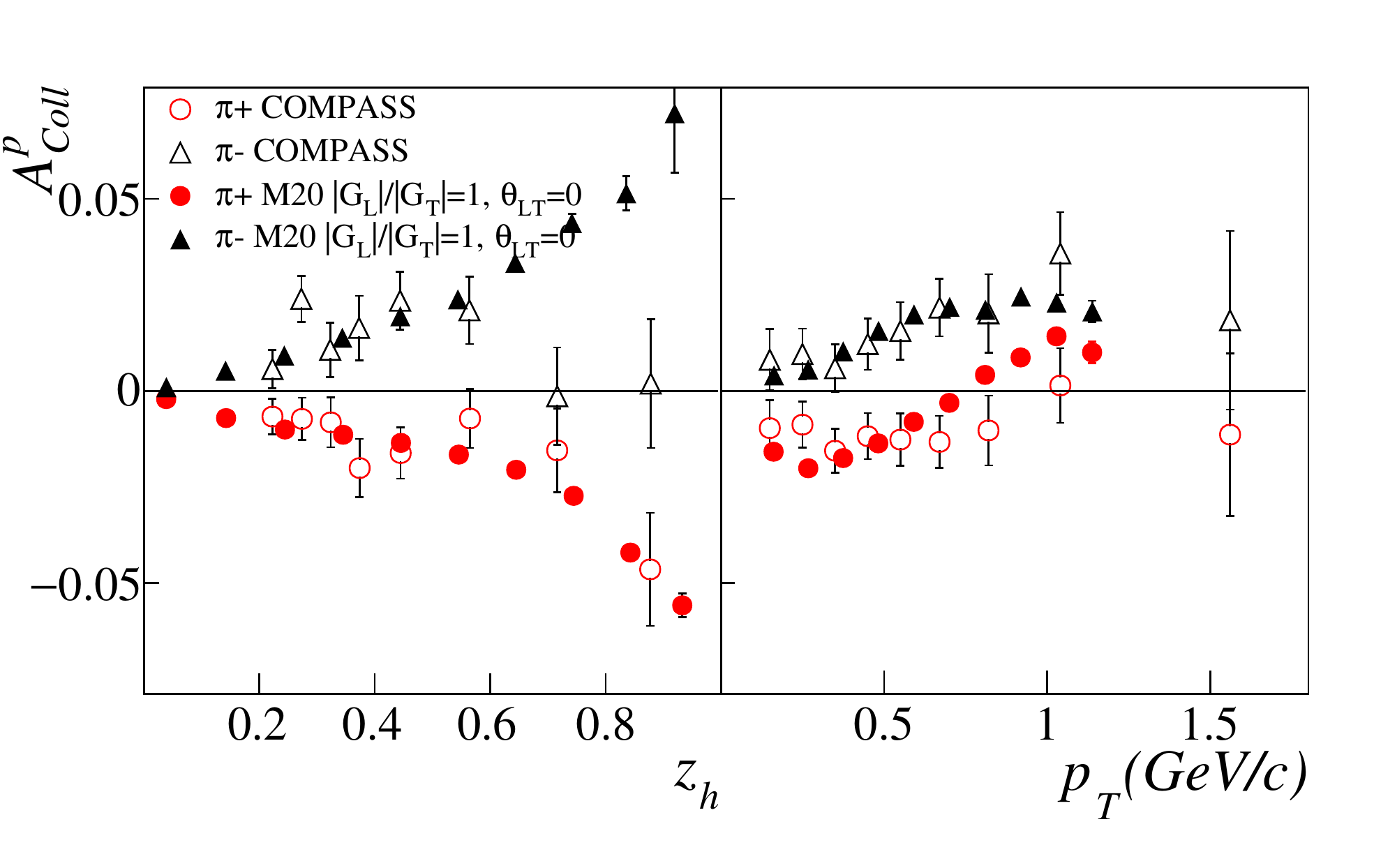}
  \end{minipage}
\begin{minipage}[t]{.48\textwidth}
  \includegraphics[width=1.0\linewidth]{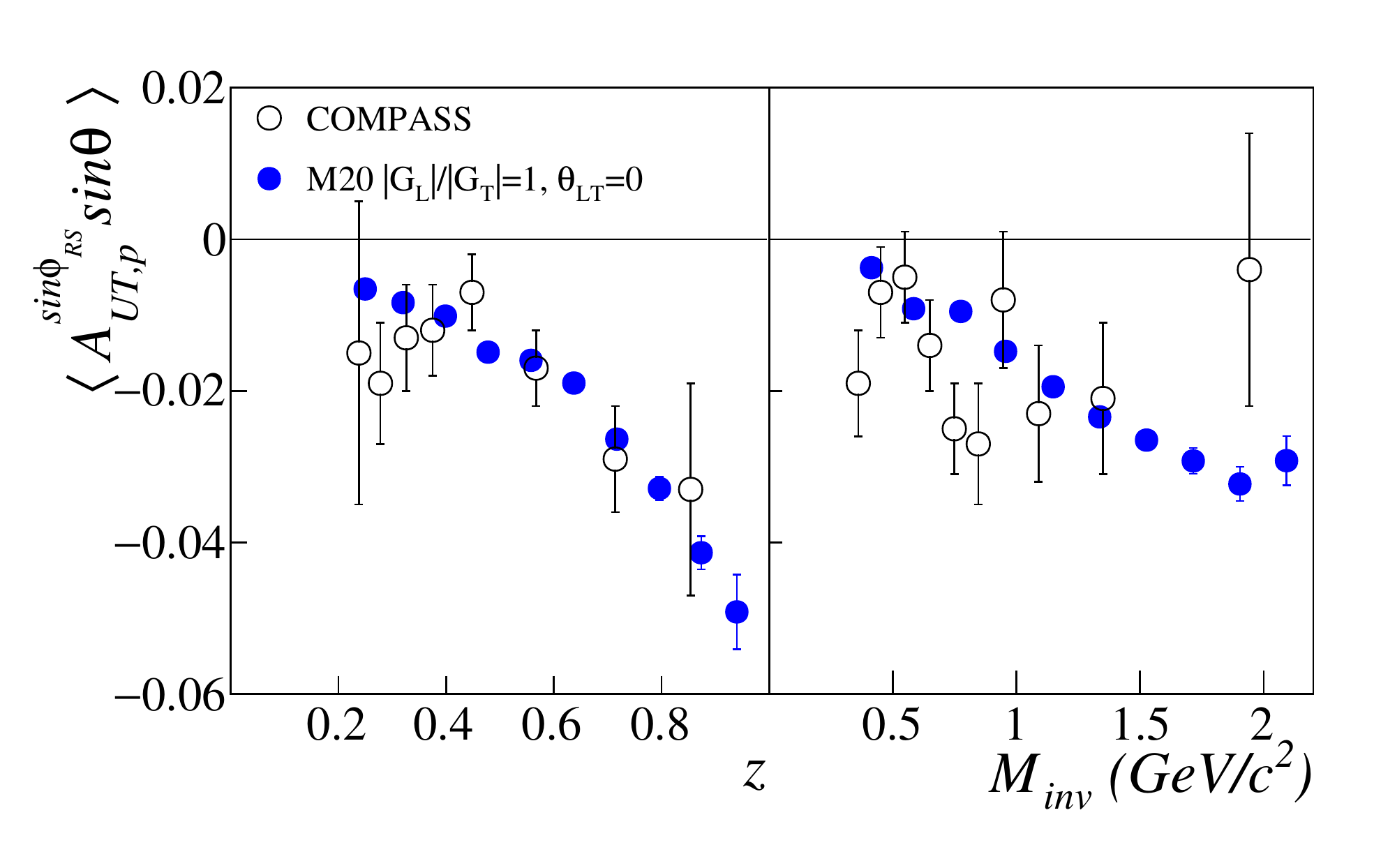}
  \end{minipage}
  \caption{Left: comparison of the Collins analysing power as obtained with M20 (full points) with the Collins asymmetry as measured by COMPASS (open points). Right: comparison of the dihadron analysing power as obtained with M20 (full points) with the dihadron asymmetry measured by COMPASS (open points).}\label{fig:M20 final comparison}
\end{figure}

The left panel of Fig. \ref{fig:M20 final comparison} shows the comparison as function of $z_h$ (left plot) and of $p\T$ (right plot) of the Collins analysing power for charged pions as obtained from the M20 stand alone MC (full points) with the Collins asymmetry measured by COMPASS for protons (open points). The circles indicate the positive pions whereas the triangles indicate the negative pions. The full points are the MC results for the $u$ quark fragmentation simulated with M20 using $|G_L|/|G_T|=1$ and $\theta_{LT}=0$, and without intrisinc quark transverse momentum. The values are rescaled by a factor $\lambda'$ justified by the fact that the transversity distribution is not used in the stand alone MC. The same $\chi^2$ minimization procedure described in sec. \ref{sec:single hadron transverse spin asymmetry M18} has been used, getting $\lambda'=0.11\pm 0.02$. This value is twice as that obtained for M18 as expected from the reduction the Collins analysing power due to the introduction of vector mesons. The overall agreement is satisfactory, in particular for the asymmetry of positive pions as function of $z_h$ and the asymmetry for positive and negative pions as function of $p\T$. Surprisingly, the analysing power for negative pions in M20 increases with $z_h$ whereas it decreases in the data. This is an interesting point to be understood. A different setting of the free parameters could change the values of the asymmetries but other effects have to be considered. For instance, a reduction of the analysing power at large $z_h$ could be obtained by switching on the intrinsic quark transverse momentum which would also shift the point where the analysing power for positive pions as function of $p\T$ changes sign. A further reduction could occur when considering the realistic mixture of the struck quarks in the SIDIS process, as seen for the Collins asymmetry obtained with \verb|PYTHIA|+${}^3P_0$.

In the right panel of Fig. \ref{fig:M20 final comparison} the comparison of the dihadron analysing power for $h^+h^-$ pairs obtained with M20 (full points) with the dihadron asymmetry measured by COMPASS (open points), as function of $z$ (left plot) and as function of the invariant mass (right plot) is shown. The MC points have been obtained from the same sample of simulated events used for the calculation of the Collins analysing power and have been rescaled by the same factor $\lambda'$. The agreement is quite satisfactory.

The success of the M20 model in reproducing the transverse spin asymmetries is an important point. The goal of having a quantum mechanical consistent model of the polarized quark fragmentation process implemented in a MC program has been achieved. M20 is the most complete model one can build with the string+${}^3P_0$ fragmentation model, and no major theoretical improvements are presently foreseen. By now, a systematic work to tune the three additional free parameters can start.

Input on $|G_L|/|G_T|$ and $\theta_{LT}$ can come from the measurements of the spin density matrix elements of vector mesons produced in inclusive processes. For instance, in the case of the $e^+e^-$ annihilation to hadrons process at the $Z^0$ mass, the spin density matrix is investigated by the DELPHI Collaboration \cite{DELPHI-alignment} and by the OPAL Collaboration \cite{OPAL-alignment-rho,OPAL-alignment-K,OPAL-alignment-phi}. These data have already been extensively studied in the literature, e.g. in Refs. \cite{Anselmino-spin-effects-LEP,Anselmino-off-diagonal-SDME}. Using the SIDIS results, in addition to the Collins and dihadron asymmetries, a dedicated comparison with the $z_h$ and $\ptpt$ distributions has also to be performed, and new measurements could also be proposed to investigate specific aspects of the model. Last but not least, the use of the $e^+e^-$ annihilation to hadrons data below the $Z^0$ mass will also give constraints and new input.

\chapter{Summary and outlook}

During the research years that brought to this thesis, a recursive model for the fragmentation of polarized quarks which combines the Lund Model of string fragmentation and the ${}^3P_0$ model of quark pair creation at string breaking has been studied in detail and extended from pseudoscalar meson production to both pseudoscalar and vector meson production. The model is fully quantum-mechanical concerning the spin degree of freedom and preserves the left-right symmetry of the Lund string fragmentation model. The polarized splitting function and the formula governing the transfer of polarization from a recurrent quark to the next one are derived explicitly and written in a form suitable for a Monte Carlo implementation. In the model generating only pseudoscalar mesons, the source of spin asymmetries and spin correlations along the fragmentation chain lies at the level of the quark-antiquark pair production, which is assumed to occur in the ${}^3P_0$ state and involves a complex mass parameter $\mu$. In particular, the imaginary part of $\mu$ is responsible for the transverse spin effects, whereas the longitudinal spin effects like jet handedness are proportional to $\IM(\mu^2)=2\RE(\mu)\IM(\mu)$.

Two variants of the model (M18 and M19), which differ by the choice of an input function, have been formulated. Both variants are restricted to the production of pseudoscalar mesons. They have been implemented in stand alone Monte Carlo programs which have been used to investigate the model predictions. M18 is more general and takes into account possible spin-independent correlations between the transverse momenta of the recurrent quark and the next one at each elementary splitting. From the practical point of view it requires, as preliminary step, the tabulation of some functions and the Monte Carlo drawing of the hadrons species is somewhat complicated if one wants to preserve the left-right symmetry exactly. M19 does not foresee the above mentioned correlations and does not necessitate the preliminary tabulation. It provides analytic expressions for the coefficients which govern the decay of spin information along the fragmentation chain.
The two variants give essentially the same results, in spite of the simplicity of M19. The main advantage of M19 is that it is more suitable for the implementation in event generators in which the fragmentation process is based on the Lund Model. For both variants, the Collins effect has been deeply studied as function of the relevant variables. After fixing the free parameters by comparison with experimental data, the resulting Collins and the dihadron transverse spin asymmetries from simulations have been compared with COMPASS and BELLE data, finding a satisfactory qualitative agreement. A remarkable property of the model is that with the same mechanism it reproduced both the Collins and the dihadron asymmetries.

 Thanks to its formal simplicity, the model M19 could be interfaced with the \verb|PYTHIA 8.2| event generator for the simulation of the polarized SIDIS process. This allowed for the first time to introduce spin effects in the fragmentation process of a complete event generator. The strategy that has been followed consists in looking at each hadron emitted by \verb|PYTHIA| during string fragmentation and to bias its azimuthal distribution according to the rules of M19, thus taking into account the ${}^3P_0$ mechanism. Also, the fragmentation and the quark spin propagation along the chain are forced to evolve from the quark side toward the side of the nucleon remnant. The introduction of spin effects in \verb|PYTHIA| has been validated by comparing the simulation results with those of the stand alone M19 Monte Carlo. Moreover, a parameterization for the quark transversity distribution has been implemented in \verb|PYTHIA|, allowing for the complete simulation of both the Collins and the dihadron asymmetries. These asymmetries have been calculated for transversely polarized proton and deuteron targets and the agreement with the COMPASS data is quite satisfactory, giving confidence in the model and motivating further developments.
 
As a last development, the vector meson production has been grafted to M19, leading to the M20 model. 
The new splitting function depends essentially on two more free parameters, in addition to the complex mass. One of them, $|G_L|/|G_T|$, governs the ratio of longitudinal to transverse vector mesons. It also affects the strength of the Collins effect for a vector meson taken as one particle. The other one, $\theta_{LT}= \arg(G_L/G_T )$, is responsible for the oblique polarization of vector mesons, a new source of Collins effect among the decay hadrons not yet tested experimentally. These two parameters enter also the spin density matrix of the vector meson, which is used to define the angular distribution of the decay products in the vector meson rest frame for the cases of two and three body hadronic decay processes. At each elementary splitting, the vector meson is produced in a correlated spin state with the left-over quark. This correlation is taken into account by using the formalism of the decay matrices.
The M20 model has been implemented in a stand alone Monte Carlo which has been used for a detailed study of the effect of vector meson production and decay on the Collins and the dihadron analysing powers. With the present choice of the two new free parameters, the dependence of the Collins analysing power on the fractional energy and on the transverse momentum of the hadron is different from that of M19. Also, the average value is decreased, as expected. The same reduction is seen in the dihadron analysing power. From the study of the Collins analysing power as function of rank it is clear that the polarization of the initial quark decays faster in chains where both pseudoscalar and vector mesons are produced.
The sensitivity of the Collins and dihadron analysing powers and of the kinematic distributions on the free parameters has also been investigated. All the tests and the studies performed have allowed to understand the features of the model and no critical point showed up.

The model presented in this thesis is obviously far from being complete. Baryon production is not included and interferences between resonance and background, or between particles of permuted ranks, are not treated. Nevertheless, it gives a new insight in the quark fragmentation process, polarized or not. It satisfies the basic rules (positivity, entanglement,$\dots$) of quantum information theory applied to spin. It explicits the link between Collins and di-hadron asymmetries and is able to generate a nonzero jet handedness. The physics ingredients that it contains are sufficient to perform realistic simulations when interfaced with a complete event generator. The results obtained here indicate that the model has a promising predictive power, stable against changes on the free parameters.

In conclusion, the goal of the research project has been reached. The work done is a first important step towards a more complete inclusion of the spin effects in the hadronization part of Monte Carlo event generators and in particular in \verb|PYTHIA|. After a more refined tuning of the parameters, the existing event generator can already be used for the systematic study of the quark spin in the fragmentation process in SIDIS and $e^+e^-$ annihilation to hadrons. 
In the future the code for polarized quark fragmentation with vector meson production will be incorporated in \verb|PYTHIA| for SIDIS and $e^+e^-$ annihilation.
The work done is important not only for a better understanding of hadronization, which is related to the still unexplained phenomenon of confinement, but also in view of the realization of new proposed experiments at EIC and LHC dedicated to the understanding of the nucleon structure.



\begin{acknowledgements}
\addchaptertocentry{\acknowledgementname} 

This work would not have been possible without the kind, careful and close supervision of Prof. Anna Martin and Prof. Xavier Artru, to whom I am very grateful for time they dedicated to me, for teaching me about physics, and for showing me how nicely experiment and theory can be combined together to reach the common goal of understanding some the shades of nature.

I sincerely thank Prof. Franco Bradamante, who first suggested this topic and introduced me to Prof. Artru, for the numerous discussions and for the kind encouragements.

I would like thank Prof. John Collins and Prof. Torbj\"orn Sj\"ostrand for having accepted to be the referees of this thesis. I'm deeply honoured that such important physicists read this work.

I thank Dr. Markus Diefenthaler and to the whole LDRD group for the enlightening discussions and for having introduced me to a community of briliant physicists. In particular I thank Prof. Leif L\"onnblad for teaching me how to deal with hadronization in \verb|PYTHIA| and for the continous discussions during my visit in Lund. 
Thanks to Prof. Vincenzo Barone and to Prof. Zouina Belghobsi for the many discussions and suggestions they gave me.

A big thank to the Trieste COMPASS group for their warm company and the daily laughs. Also for the infinite caffeine taken each day at the group coffee breaks, and through the delightful coffee produced by the coffee machine of Andrea Moretti with capsules chosen by Jan Matousek. Thanks to the Trieste section of INFN who made possible my missions abroad and to the COMPASS colleagues for the many conversations.

Thanks to Giuseppe D'Auria who many times helped me to find the solution of several puzzles without even speaking. Thanks to Eli Yak for the hours spent at gym. Thanks to Gina Raus for being such a close fried since a long time. Thanks to all my friends, to my cousins and to all flat-mates (mine and not) who eased my days, made me laugh and cooked delicious cakes. A special thank to Irene, the young lady with very curly brown hairs who teached me the light heartedness and that there exists an entire beautiful world outside physics.

Dulcis in fundo, there are no words to say thank you to my father Trifon, to my mother Th\"ell\"eza and to my brother Xhoi who have always been a warm hug.

\end{acknowledgements}


\appendix 


\chapter{Lightcone components of a four-vector}\label{Appendix:lightcone}
The lightcone decomposition of four-vectors is used throughout the whole thesis for the parameterization of particle momenta in a splitting process. There are two processes considered. The splitting $N(P_N)\rightarrow q(k) +X(P_X)$ of the nucleon $N$ with momentum $P_N$ in a quark $q$ with momentum $k$ and the remnant $X$ with momentum $P_X$. In this case, the quark takes a fraction $x=k^+/P_N^+$ of the nucleon \textit{"plus"} (or \textit{forward}) \textit{lightcone momentum}.
The second splitting process considered is $q(k)\rightarrow h(p) +q'(k')$. Namely, the quark $q$ with momentum $k$ is split in a hadron $h$ with momentum $p$ and a left-over quark $q'$ with momentum $k'$. Here the hadron takes a fraction $Z=p^+/k^+$ of the quark "plus" lightcone momentum.
The other component along the lightcone is called \textit{"minus"} or \textit{backward lightcone momentum}.

In general, given a reference frame where the longitudinal direction is defined as the $\zu$ axis, a four-vector $v$ is written as $v=(v^0,\textbf{v}\T,v^z)$. The transverse vector $\textbf{v}\T$ indicates the component of $v$ in the plane orthogonal to $\zu$.
The light cone components of $v$ are defined as
\begin{equation}
    v^{\pm} = v^0\pm v^z.
\end{equation}

In terms of the light cone and transverse components, the Lorentz invariant scalar product between two four-vectors $v$ and $w$ is given by
\begin{equation}
    v_{\mu}\cdot w^{\mu} \equiv v\cdot w = \frac{v^+w^- + v^-w^+}{2}-\textbf{v}\T\cdot\textbf{w}\T.
\end{equation}
When considering the scalar product of a four-vector with itself, this relation reduces to
\begin{equation}
    v\cdot v\equiv v^2 = v^+v^- - \textbf{v}^2\T.
\end{equation}
For the four-momentum $p$ of a hadron $h$ with mass $m_h$, it is $p^0=E=\sqrt{|\textbf{p}|^2+m_h^2}$ and the mass-shell constraint $p^2=m_h^2$ can also be written as
\begin{eqnarray}
    p^+p^- = \varepsilon_h^2, & \varepsilon_h^2=m_h^2 +\ptpt
\end{eqnarray}
The quantity $\varepsilon_h^2$ is the squared transverse energy of the hadron.
\chapter{The explicit expression of the correlation matrix}\label{appendix:R}
The correlation matrix $R$ is used in Chapter 5 to describe the spin correlations between $h$ and $q'$ in the elementary splitting $q\rightarrow h + q'$. It is introduced in sec. \ref{sec:h-q' spin correlations} and defined according to Eq. (\ref{eq:R definition}). It is a $6\times 6$ matrix and can be expanded in the quark $\otimes$ vector meson spin basis as
\begin{equation}
    R(q',h) = \sum_{J,A}\,C_{JA}\,\boldsymbol{\sigma}_J^{q'}\otimes \boldsymbol{\Sigma}_A^{h}.
\end{equation}
The \textit{correlation coefficients} $C_{JA}$ describe all the spin correlations between $q'$ and $h$ as allowed by parity conservation. They can be calculated as
\begin{eqnarray}\label{eq:C_IA coefficients}
\nonumber    C_{JA} &=& g_A^{-1}\textrm{Tr} \left( R(q',h) \,\sigma_I^{q'}\otimes \lambda_A^h\right)\\
&=& R_{j\alpha;j'\alpha'} (\sigma_J^{q'})_{j'j}\,(\lambda_A^h)_{\alpha'\alpha},
\end{eqnarray}
where a summation over the repeated spin indices is understood. The trace operation is performed both on the quark and hadron spin indices. The quark polarization states are labelled by $J=0,\rm{m},\rm{n},\rm{l}$ and the meson polarization states by the index $A=0,1,\dots,8$. The $(\rm{\textbf{m}},\rm{\textbf{n}},\rm{\textbf{l}})$ states are defined as $\rm{\textbf{m}}=\hat{\textbf{k}}'\T$, $\rm{\textbf{n}}=\zu\times\hat{\textbf{k}}'\T$ and $\rm{\textbf{l}}=\zu$. $\kpt$ is the transverse momentum of $q'$ with respect to the string axis. The $\boldsymbol{\sigma}$ matrices refer to the Pauli matrices for $q'$.

The $\boldsymbol{\Sigma}$ matrices are defined to be \cite{X.A_et_al_spin_observables}
\begin{eqnarray}
\nonumber \boldsymbol{\Sigma}_1&=&\frac{|\Nx\rangle\langle\Lx|-|\Lx\rangle\langle\Nx|}{i}\equiv \boldsymbol{\lambda}_7 \text{,\,\,\,\,\,\,\,\,\,\,\,\,\,\,\,\,\,\,\,\,\,} \boldsymbol{\Sigma}_2=\frac{|\Lx\rangle\langle\Mx|-|\Mx\rangle\langle\Lx|}{i}\equiv -\boldsymbol{\lambda}_5 \\
\nonumber \boldsymbol{\Sigma}_3&=&\frac{|\Mx\rangle\langle\Nx|-|\Nx\rangle\langle\Mx|}{i}\equiv -\boldsymbol{\lambda}_2 \\
\nonumber \boldsymbol{\Sigma}_4&=&|\Nx\rangle\langle\Lx|+|\Lx\rangle\langle\Nx|\equiv \boldsymbol{\lambda}_6\text{,\,\,\,\,\,\,\,\,\,\,\,\,\,\,\,\,\,\,\,\,\,\,}
\boldsymbol{\Sigma}_5=|\Lx\rangle\langle\Mx|+|\Mx\rangle\langle\Lx|\equiv \boldsymbol{\lambda}_4 \\
\nonumber \boldsymbol{\Sigma}_6&=&|\Mx\rangle\langle\Nx|+|\Mx\rangle\langle\Nx|\equiv \boldsymbol{\lambda}_1 \text{,\,\,\,\,\,\,\,\,\,\,\,\,\,\,\,\,\,}
\boldsymbol{\Sigma}_7=|\Mx\rangle\langle\Mx|-|\Nx\rangle\langle\Nx|\equiv \boldsymbol{\lambda}_3 \\
 \boldsymbol{\Sigma}_8&=&|\Lx\rangle\langle\Lx|-\frac{|\Mx\rangle\langle\Mx|+|\Nx\rangle\langle\Nx|}{2}\equiv \frac{\sqrt{3}}{2}\boldsymbol{\lambda}_8,
\end{eqnarray}
where we have shown also the relation with the $3\times 3$ Gell-Mann $\boldsymbol{\lambda}$ matrices. The identity $3\times 3$ matrix is defined as $\Sigma_0^h$.
The $\boldsymbol{\Sigma}_A$ matrices obey the normalization condition 
\begin{equation}
    \delta_{AA'}=g_A^{-1}\Tr \boldsymbol{\Sigma}_A\boldsymbol{\Sigma}_{A'},
\end{equation}
$g_A$ being the normalization constant.

Using the Eq. (\ref{eq:R definition}) and Eq. (\ref{eq:C_IA coefficients}) the expression for $R$ is
\begin{eqnarray}\label{eq:R explicit}
\nonumber R(q',h) &=& \frac{|C_{q',h,q}|^2}{\sum_H |C_{q',H,q}|^2}\,|D_h(M^2)|^2\, \left(\frac{1-Z}{\varepsilon_h^2}\right)^a \frac{\exp{(-\bl \varepsilon_h^2/Z)}}{N_a(\varepsilon_h^2)} \\
\nonumber &\times& \frac{|\mu|^2+\kptkpt}{|\mu|^2+\langle \ktkt\rangle_{f\T}} f_{\T}^2(\kptkpt) \times \frac{1}{2}\times \frac{1}{2|G_T|^2+|G_L|^2}\\
\nonumber &\times& \bigg\{ \bigg[\left(|\mu|^2+\kptkpt\right)\frac{2|G_T|^2+|G_L|^2}{3}+2\IM(\mu)k'\T\frac{|G_L|^2}{3}S_{q\rm{n}}\bigg]\textbf{1}^{q'}\otimes \textbf{1}^h \\
\nonumber &+& \bigg[\left(|\mu|^2+\kptkpt\right)\frac{|G_T|^2-|G_L|^2}{\sqrt{3}}-2\IM(\mu)k'\T\frac{|G_L|^2}{\sqrt{3}}S_{q\rm{n}}\bigg]\textbf{1}^{q'}\otimes \frac{2}{\sqrt{3}}\boldsymbol{\Sigma}_8^h \\
\nonumber &-& |G_L||G_T|\sin\theta_{LT} \bigg[\left(|\mu|^2+\kptkpt\right) S_{q\rm{n}}+2\IM(\mu)k'\T \bigg]\textbf{1}^{q'}\otimes \boldsymbol{\Sigma}_5^h\\
\nonumber &+& |G_L||G_T| \bigg[\left(|\mu|^2+\kptkpt\right)\sin\theta_{LT}\, S_{q\rm{m}}+2\IM(\mu)k'\T \cos\theta_{LT} S_{q\rm{l}}\bigg]\textbf{1}^{q'}\otimes \boldsymbol{\Sigma}_4^h\\
\nonumber &-& \left(2\IM(\mu)k'\T |G_T|^2\,S_{q\rm{m}}\right)\,\textbf{1}^{q'}\otimes \boldsymbol{\Sigma}_6^h + \left(2\IM(\mu)k'\T |G_T|^2\,S_{q\rm{n}}\right)\,\textbf{1}^{q'}\otimes \boldsymbol{\Sigma}_7^h\\
\nonumber &-& \left(|\mu|^2-\kptkpt\right)|G_T|^2\,S_{q\rm{n}}\,\boldsymbol{\sigma}_{\rm{m}}^{q'}\otimes \boldsymbol{\Sigma}_6^h - \left(|\mu|^2-\kptkpt\right)|G_T|^2\,S_{q\rm{m}}\,\boldsymbol{\sigma}_{\rm{m}}^{q'}\otimes \boldsymbol{\Sigma}_7^h \\
\nonumber &+& \bigg[\left(|\mu|^2-\kptkpt\right)\frac{|G_L|^2}{3}\,S_{q\rm{m}}-2\RE(\mu)k'\T\frac{|G_L|^2-2|G_T|^2}{3}\,S_{q\rm{l}}\bigg]\,\boldsymbol{\sigma}_{\rm{m}}^{q'}\otimes \boldsymbol{1}^h \\
\nonumber &+& |G_T||G_L|\cos\theta_{LT}\,\bigg[\left(|\mu|^2-\kptkpt\right)\,S_{q\rm{l}}+2\RE(\mu)k'\T\,S_{q\rm{m}}\bigg]\,\boldsymbol{\sigma}_{\rm{m}}^{q'}\otimes \boldsymbol{\Sigma}_5^h \\
\nonumber &+& |G_T||G_L|\,\bigg[\left(|\mu|^2-\kptkpt\right)\sin\theta_{LT}+2\RE(\mu)k'\T\cos\theta_{LT}\,S_{q\rm{n}}\bigg]\,\boldsymbol{\sigma}_{\rm{m}}^{q'}\otimes \boldsymbol{\Sigma}_4^h \\
\nonumber &-& \bigg[\left(|\mu|^2-\kptkpt\right)\frac{|G_L|^2}{\sqrt{3}}S_{q\rm{m}}+2\RE(\mu)k'\T(|G_T|^2+|G_L|^2)\frac{S_{q\rm{l}}}{\sqrt{3}}\bigg]\,\boldsymbol{\sigma}_{\rm{m}}^{q'}\otimes \frac{2}{\sqrt{3}}\boldsymbol{\Sigma}_8^h \\
\nonumber &-& \left(|\mu|^2+\kptkpt\right) |G_T|^2\,S_{q\rm{m}} \,\boldsymbol{\sigma}_{\rm{n}}^{q'}\otimes \boldsymbol{\Sigma}_6^h + \left(|\mu|^2+\kptkpt\right) |G_T|^2\,S_{q\rm{n}} \,\boldsymbol{\sigma}_{\rm{n}}^{q'}\otimes \boldsymbol{\Sigma}_7^h\\
\nonumber &-& \left(|\mu|^2+\kptkpt\right) \sin\theta_{LT}|G_T||G_L| \,\boldsymbol{\sigma}_{\rm{n}}^{q'}\otimes \boldsymbol{\Sigma}_5^h \\
\nonumber &+& \left(|\mu|^2+\kptkpt\right) |G_T||G_L|\cos\theta_{LT}\,S_{q\rm{l}} \,\boldsymbol{\sigma}_{\rm{n}}^{q'}\otimes \boldsymbol{\Sigma}_4^h\\
\nonumber &+& \left(|\mu|^2+\kptkpt\right) \frac{|G_L|^2}{3}S_{q\rm{n}} \,\boldsymbol{\sigma}_{\rm{n}}^{q'}\otimes \boldsymbol{1}^h -\left(|\mu|^2+\kptkpt\right) \frac{|G_L|^2}{\sqrt{3}}\,S_{q\rm{n}} \,\boldsymbol{\sigma}_{\rm{n}}^{q'}\otimes \frac{2}{\sqrt{3}}\boldsymbol{\Sigma}_8^h\\
\nonumber &+&\bigg[2\RE(\mu)k'\T\,S_{q\rm{m}}\frac{|G_L|^2}{3}+\left(|\mu|^2-\kptkpt\right)\frac{|G_L|^2-2|G_T|^2}{3}S_{q\rm{l}}\bigg]\,\boldsymbol{\sigma}_{\rm{l}}^{q'}\otimes \boldsymbol{1}^h\\
\nonumber &-& 2\RE(\mu)k'\T\,|G_T|^2\,S_{q\rm{n}}\,\boldsymbol{\sigma}_{\rm{l}}^{q'}\otimes \boldsymbol{\Sigma}_6^h-2\RE(\mu)k'\T\,S_{q\rm{m}}\,|G_T|^2\,\boldsymbol{\sigma}_{\rm{l}}^{q'}\otimes \boldsymbol{\Sigma}_7^h\\
\nonumber &+& |G_L||G_T|\cos\theta_{LT}\bigg[2\RE(\mu)k'\T\,S_{q\rm{l}}-\left(|\mu|^2-\kptkpt\right)S_{q\rm{m}}\bigg]\boldsymbol{\sigma}_{\rm{l}}^{q'}\otimes \boldsymbol{\Sigma}_5^h\\
\nonumber &+& |G_L||G_T|\bigg[2\RE(\mu)k'\T\,\sin\theta_{LT}-\left(|\mu|^2-\kptkpt\right)\cos\theta_{LT}\,S_{q\rm{n}}\bigg]\boldsymbol{\sigma}_{\rm{l}}^{q'}\otimes \boldsymbol{\Sigma}_4^h\\
\nonumber &-&\bigg[2\RE(\mu)k'\T\,S_{q\rm{m}}\frac{|G_L|^2}{\sqrt{3}}+\left(|\mu|^2-\kptkpt\right)\frac{|G_T|^2+|G_L|^2}{\sqrt{3}}S_{q\rm{l}}\bigg]\,\boldsymbol{\sigma}_{\rm{l}}^{q'}\otimes \frac{2}{\sqrt{3}}\boldsymbol{\Sigma}_8^h \bigg\}.\\
\end{eqnarray}
In this expression only the correlation coefficients between the quark polarization and the tensor polarization of the vector meson have been considered. The terms concerning the vector meson axial polarization which involve the $\boldsymbol{\Sigma}_1$, $\boldsymbol{\Sigma}_2$ and $\boldsymbol{\Sigma}_3$ matrices have been neglected.

This equation shows explicitly that the vector meson and the quark $q'$ are produced in a correlated state. The correlation coefficients depend on the quark variables such as $k'\T$ and $\mu$ and also on the quark couplings to the vector meson, namely on the ratio between $|G_L|/|G_T|$ and on the phase $\theta_{LT}=\arg{G_L/G_T}$.

\chapter{Decay of the fragmenting quark polarization in M20}\label{appendix:polarization transfer}
M20 is the model of polarized quark fragmentation with both pseudoscalar and vector meson production introduced in Chapter 5. Since this model includes the production of more than one particle species, the initial quark, during the fragmentation chain, is depolarized faster than in M19. The depolarization coefficients $D^{vm}_{\rm{TT}}$ and $D^{vm}_{\rm{LL}}$ are introduced in Eq. (\ref{eq:DTT and DLL for vm emission}).

For a more quantitative description of the depolarization, consider f.i. a jet initiated by the quark $q_A$ made of $N_h^{tot}$ hadrons. $N_h^{tot}$ is also the number of elementary splittings $q\rightarrow h +q'$. Let $f_{ps}$ be the probability that in each splitting the meson $h$ is a pseudoscalar and $f_{vm}$ the probability that it is a vector. Then the probability of producing $n_h$ pseudoscalar mesons and $N_h^{tot}-n_h$ vector mesons is given by the binomial distribution
\begin{eqnarray}\label{eq: binomial distribution}
    P_{n_h} = \begin{pmatrix} N_h^{tot} \\ n_h \end{pmatrix} f_{ps}^{n_h} f_{vm}^{N_h^{tot}-n_h}.
\end{eqnarray}
Defining $\textbf{S}_A$ the polarization vector of the fragmenting quark, the transverse polarization of the left-over quark $q_{N_h^{tot}+1}$ after $N_h^{tot}$ splittings is
\begin{eqnarray}
\nonumber     \textbf{S}_{\rm{T}\,q_{N_h^{tot}+1}} &=& \left(D_{\rm{TT}}^{\rm{vm}}\right)^{N_h^{tot}-n_h}\left(D_{\rm{TT}}^{ps}\right)^{n_h} \textbf{S}_{\rm{T}\,A}\\
    &\equiv& D_{\rm{TT}}^{M20}\,\textbf{S}_{\rm{T}\,A}.
\end{eqnarray}
Hence the average depolarization after $N_h^{tot}$ splittings is
\begin{eqnarray}\label{eq: polarization decay in M20}
\nonumber    \langle D_{\rm{TT}}^{M20} \rangle &=& \sum_{n_h}\,P_{n_h}\,\left(D_{\rm{TT}}^{\rm{vm}}\right)^{N_h^{tot}-n_h}\left(D_{\rm{TT}}^{ps}\right)^{n_h} \\
    &=&(f_{ps}-f_{vm}\ \frac{|G_L|^2}{2|G_T|^2+|G_L|^2})^{N_h^{tot}}\,\left(D_{\rm{TT}}^{ps}\right)^{N_h^{tot}}.
\end{eqnarray}
This equation shows that the depolarization coefficient of M20 is smaller than that of M19, which would be $\left(D_{\rm{TT}}^{ps}\right)^{N_h^{tot}}$. Having in mind that it is $N_h^{tot}\propto \log W$, $W$ being the invariant mass of the final hadronic system in DIS or the center of mass energy in $e^+e^-$ annihilation to hadrons, Eq. (\ref{eq: polarization decay in M20}) gives also the depolarization of the initial quark as function of the energy stored in the string.


\printbibliography[title = {Bibliography}]

\end{document}